\documentclass[a4paper,12pt]{report}
\usepackage[T1]{fontenc}
\usepackage[brazil]{babel}
\usepackage[latin1,utf8]{inputenc}
\usepackage{float}
\usepackage{morefloats}
\usepackage{amsmath}
\usepackage{amsfonts}
\usepackage{amssymb}
\usepackage{amsthm}

\usepackage{setspace}                   
\usepackage{indentfirst}                
\usepackage{makeidx}                    
\usepackage{courier}                    
\usepackage{type1cm}                    
\usepackage{titletoc}
\usepackage[font=small,format=plain,labelfont=bf,up,textfont=it]{caption}
\usepackage[usenames,svgnames,dvipsnames]{xcolor}
\usepackage[a4paper,top=2cm,bottom=2cm,left=2cm,right=1.25cm]{geometry} 
\addto\captionsbrazil{%
 \renewcommand{\contentsname}%
   {Índice}%
}
\addto\captionsbrazil{%
 \renewcommand{\bibname}%
   {Referências}%
}

\usepackage{graphicx}
\usepackage{multirow,colortbl,array} 
\usepackage{subfig} 
\usepackage{longtable}
\usepackage{rotating}
\usepackage{rotfloat}
\graphicspath{{./figuras/}}
\usepackage{appendix}

\begin{document}

\begin{titlepage}
\begin{center}
\includegraphics[angle=0,width=0.2\textwidth]{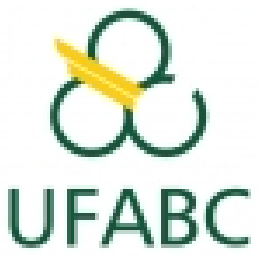}
\end{center}
\vspace*{0.1cm}
\centerline{\large UNIVERSIDADE FEDERAL DO ABC}
\vspace*{0.2cm}
\centerline{\large CENTRO DE CIÊNCIAS NATURAIS E HUMANAS}
\vspace*{0.4cm}
\centerline{\large CURSO DE PÓS-GRADUAÇÃO EM FÍSICA}

\vspace*{1.50cm}

\centerline{\large TESE DE DOUTORADO}

\vspace*{1.50cm}

{\Large \bf \centerline{Física além do Modelo Padrão no Setor Eletrofraco}}
\medskip
{\large \centerline{Acoplamentos Quárticos Anômalos com Fusão de Bósons Vetoriais}}
\medskip
\centerline{e}
\medskip
{\large \centerline{FCNC no Modelo $SU(3)_{C}\otimes SU(3)_{L}\otimes U(1)_{X}$}}

\vspace*{1.40cm}

{\large {\bf \centerline{Patricia Rebello Teles}}}

\vspace*{1.20cm}

\begin{flushright}
\begin{minipage}[l]{9cm}
Trabalho apresentado como requisito parcial para obtenção do título de Doutora em Física, sob orientação do Professor Doutor José Kenichi Mizukoshi.
\end{minipage}
\end{flushright}




\vspace*{3cm}

\centerline{Santo André}
\centerline{2012}

\newpage
\thispagestyle{empty}
\vspace*{7cm}
{\bf Este exemplar foi revisado e alterado em relação à versão original, de acordo com as observações levantadas pela banca no dia da defesa, sob responsabilidade única do autor e com anuência de seu orientador.

\vspace*{2cm}

Santo André,\hspace*{1.5cm}de\hspace*{5cm}de 2012.      

\vspace*{1.5cm}

Assinatura do autor:

\vspace*{1.5cm}

Assinatura do orientador:}

\newpage
\thispagestyle{empty}
\mbox{}
\vfill
O presente trabalho contou com o apoio financeiro da Fundação Universidade Federal do ABC (UFABC), da Coordenação e Aperfeiçoamento de Pessoal de Nível Superior (CAPES) e da Universidade Federal Rural do Rio de Janeiro (UFRRJ).

\end{titlepage}

\doublespacing
\pagenumbering{roman}
\chapter*{Resumo}

Nesta tese trabalhamos com dois enfoques distintos de nova física no setor eletrofraco do Modelo Padrão da Física de Partículas Elementares.

Em primeiro momento investigamos, em nível partônico e em
\emph{leading order}, o potencial do LHC operando em energia de
$\sqrt{s}=14$ TeV e luminosidade $L=100\;\mbox{fb}^{-1}$ para testar
vértices anômalos puramente quárticos entre os bósons de
\textit{gauge} massivos no canal semileptônico $pp \rightarrow
l^{\pm}\nu_{l}jjjj$ onde $l^{\pm}$ são os léptons
$e^{\pm}\;\mbox{e}\;\mu^{\pm}$, e $j$ são jatos compostos por (anti)quarks leves, dentro da topologia de fusão de bósons vetoriais.

Parametrizamos desvios nos acoplamentos quárticos do MP, em uma
análise independente de modelo, escolhendo operadores efetivos com
contribuições puramente quárticas, sem modificar os vértices tríplices
previstos pelo modelo padrão (MP). Essa parametrização efetiva pode realizar-se
``não-linearmente'', caso em que teremos operadores de ordem 4 e que
um bóson escalar leve não aparece no espectro ou não é um dubleto de
isospin (na verdade podemos inserir esse bóson escalar \emph{ad hoc}
como uma flutuação do campo composto), ou então ``linearmente'', caso
em que operadores de ordem 8 supõem a presença de um bóson escalar
leve com as características do bóson de Higgs padrão no espectro.

Tratamos a representação não-linear com maiores detalhes e
apresentamos de forma aproximada a abordagem linear devido aos últimos
resultados do LHC em relação à possível descoberta de um bóson não
vetorial com as características do bóson de Higgs padrão. 

Sendo essa
primeira análise uma questão que envolve basicamente contagem de
eventos então, para mensurar adequadamente nosso sinal, cuidamos dos
possíveis \textit{backgrounds} considerando-os em três diferentes
ordens em teoria de perturbação: (i) irredutível, incluindo processos
de produção de três bósons, em ${\cal O}(\alpha_{em}^{6})$; (ii) QCD
em ${\cal O}(\alpha_{em}^{4}\alpha_{s}^{2})$ e ${\cal
  O}(\alpha_{em}^{2}\alpha_{s}^{4})$ e (iii) processos envolvendo
quarks \textit{top}: $pp\to t\bar{t} +
0\;\mbox{até}\;2\;\mbox{jatos}$. 

A partir das análises dos canais $pp
\rightarrow l^{\pm}\nu_{l}jjjj$, devido aos vértices anômalos $WWWW$ e
$WWZZ$ e levando em conta os \textit{backgrounds} supracitados,
obtivemos vínculos às constantes de acoplamento $\alpha_{4}$ e
$\alpha_{5}$.

Posteriormente, apresentamos uma outra abordagem de fenomenologia em uma extensão do MP sem conexão direta com os acoplamentos anômalos. Usando
um modelo com simetria $SU(3)_{C}\otimes SU(3)_{L} \otimes
U(1)_{X}$ com neutrinos de mão-direita (denominado aqui $331_{RHN}$),
estudamos os vínculos dessa nova física em processos que permitem
correntes neutras com trocas de sabor (processos FCNC).  

Embora, em nível de árvore, tais
processos sejam altamente suprimidos no MP, a estrutura do
modelo $331_{RHN}$ permite operadores que levam a esse tipo de processo neste nível. Sendo assim, usando medidas experimentais de oscilações em sistemas de mésons
neutros $K^{0}-\bar{K}^{0}$, $D^{0}-\bar{D}^{0}$ e
$B^{0}_d-\bar{B^0_d}$, mostramos que o novo bóson de \textit{gauge}
neutro $Z^{\prime}$, que emerge nesse modelo, não é a única fonte
que possibilita FCNC em nível de árvore. De fato, duas novas personagens
entram em cena: as partículas escalares $CP$-par e $CP$-ímpar,
denotadas por $S_2$ e $I_3$ respectivamente. 

Nossas análises mostram
que os processos FCNC produzem vínculos fortes para as massas
dessas novas partículas. Em particular, esses vínculos tornam em princípio a
observação do $Z^{\prime}$ do modelo $331_{RHN}$ bastante improvável
na escala de energia do LHC.

\chapter*{Agradecimentos}

A HISTÓRIA DE UMA TESE.

Apesar do processo solitário a que qualquer cientista se destina, uma tese é antes de tudo o fruto de amizades sinceras, da confiança nos mestres e amigos, enfim do apoio de várias pessoas e instituições sem as quais este trabalho não seria possível. Nada na vida conquistamos sozinhos. Sempre precisamos de outras pessoas para alcançar os nossos objetivos. Muitas vezes um simples gesto pode mudar a
nossa vida e contribuir para o nosso sucesso.

Ao Professor Doutor José Kenichi Mizukoshi, orientador da tese, agradeço o compartilhamento do saber e as valiosas contribuições estimulando meu interesse e continuação na vida acadêmica.

Ao Professor Doutor Oscar Éboli, mentor do projeto, agradeço a objetividade e a voz da experiência com valiosas contribuições em momentos decisivos.

Ao Professor Doutor José Abdallah Helayel Netto e à sua esposa Magali, agradeço a acolhida calorosa no CBPF por ocasião do meu retorno ao Rio de Janeiro na etapa de finalização desta tese. Nossas conversas acompanhadas com cafés e bolos naturais adoçaram minha vida nesses últimos meses. \emph{Shookran kathyiran, yah khalil! Allaihi wa'hayla alaikum!}

Ao Setor de Informática do Departamento de Fisica Matemática do IFUSP e ao CAT--Coordenação de Atividades Técnicas do CBPF, especialmente à Sybelle Guedes e ao Marcelo Giovanni, por tornarem possíveis os pesados cálculos numéricos desse trabalho.

Aos meu amigos e irmãos de coração, Doutor Rone Peterson Galvão Andrade e sua esposa Thaís, pela cumplicidade e companheirismo, e por torcerem e me apoiarem sempre visando meu sucesso profissional e pessoal.

Aos meus queridos amigos e amigas, companheiros e companheiras de toda hora, Alex Gomes Dias e sua mãe Dna. Vani, Luis Cabral, Bertha Quadros Melgar, Karin Fornazier Guimarães, Paulo Reimberg, Martha, Ana, Guilherme, Dna. Marina, Jaqueline Morgan e Farinaldo da Silva Queiroz, por iluminarem meu caminho.

Aos meus pais Manoel João Teles e Rosalina Rebello Teles, pelo amor incondicional, por terem cuidado de mim todos esse anos, pela paciência e pelo apoio integral aos meus estudos. 

À minha família paulista de coração José Luiz, Rose Angela, Luiz Eduardo e Ricardo, pelo carinho incomensurável, pelo apoio nos momentos alegres e, principalmente, em alguns dos mais difíceis da minha vida.

Aos meus irmãos Rafael e Michele, ao meus cunhados Eloyse e Fernando, ao meu afilhado Lucas e à minha sobrinha Sophia, que entenderam carinhosamente quando a irmã, a cunhada, a madrinha e a tia não podia estar presente nos encontros e aniversários da família.

Ao meu filho Michel agradeço o amor, a dedicação, a amizade e a companhia nesses 18 anos. Você é a razão de tudo que eu faço, de tudo que eu penso, de tudo que eu vivo, de tudo que eu luto e de tudo que busco. Essa tese eu dedico à você!

A mim restaram três certezas: a certeza de que estou somente começando, a certeza de que é preciso continuar e a certeza de que podemos ser interrompidos antes de terminar.

Mas aprendi também a fazer da interrupção um novo caminho. Das pedras, um trampolim para alcançar sempre mais longe. Da queda, um passo de dança (ao som de Hossam Ramzy). Do medo, uma escada. Do sonho, uma ponte. E da procura, um encontro!

\emph{An'noor al-qâmár} 

(Luz da Lua)

\newpage
\thispagestyle{empty}
\mbox{}
\vfill
\begin{center}
\emph{We start by observing reality... we try to select solid (unchanging) observations that are not affected by how we perceive (measure) them. We then proceed by increasing our research and measurement, subjecting premises to criticism, and being cautious in drawing conclusions... In all we do, our purpose should be balanced not arbitrary, the search for truth, not support of opinions. Hopefully, by following this method, this road to the truth that we can be confident in, we shall arrive at our objective, where we feel certain that we have, by criticism and caution, removed discord and suspicion. Yet we are but human, subject to human frailties, against which we must fight with all our human might. God help us in all our endeavors}. Ibn al-Haytham, Kitab Al-Manadhir or The Book of Optics, pp. 5-6.
\end{center}

\tableofcontents

\listoffigures
\listoftables

\chapter{Introdução}
\label{chap:intro}

\pagenumbering{arabic}
\setcounter{page}{1}

O conhecimento das ``leis'' da Física em escala subatômica é proveniente em
grande parte da
análise de dados fornecidos pelos colisores de partículas elementares
de altas energias.

A interpretação teórica dos
dados experimentais é feita, grosso modo, escolhendo-se um ``candidato'' à teoria,
calculam-se seções de choque de processos nessa teoria e as comparamos com dados experimentais. Caso essa comparação não seja satisfatória passamos para um próximo ``candidato''.

Nos últimos 30 anos, o ``candidato'' conhecido como Modelo
  Padrão das Partículas Elementares (MP) se mantém no topo do
\textit{ranking} de disputa, já que nenhum desvio estatisticamente
significante em suas previsões foi observado até hoje. No entanto, não há dúvida que o MP se mostra como uma teoria
incompleta na descrição dos processos subatômicos, já que ele não
esclarece o mecanismo da violação de $CP$, a replicação de famílias,
além de não apresentar um candidato à matéria escura, entre outras
questões em aberto.

Com o início das operações do Grande Colisor de Hádrons (LHC) no CERN, o MP passou recentemente por mais um escrutínio: a possível observação do bóson de Higgs, a única partícula fundamental prevista
pelo MP até então não detectada. A observação dessa
partícula daria credibilidade ao chamado mecanismo de Higgs,
que elucida a quebra espontânea de simetria no setor eletrofraco responsável pela
geração de massa para suas partículas.

De fato, dados recentes dos experimentos CMS e ATLAS confirmaram a existência de um bóson com massa em torno de 125 GeV. Diante desse fato os seguintes cenários são
projetados para este colisor hadrônico nos próximos anos: (a) nenhum
indício de ressonâncias novas ou (b) observação de várias novas
partículas.

Na opção (a) precisaremos medir os acoplamentos desse bóson leve com
bastante precisão para podermos discriminar, dentre tantas outras
possibilidades, se de fato se trata do bóson de Higgs com as
propriedades previstas pelo MP. 

Sob o ponto de vista teórico e experimental, o cenário (b) é bem mais
excitante do que a aparente confirmação do MP. Caso sejam
observadas novas partículas então uma espectroscopia detalhada e
medidas de acoplamentos serão necessários para a compreensão da
estrutura fundamental da teoria. Neste cenário emergem, como fortes
candidatos, modelos que se baseiam em supersimetria (SUSY). Embora SUSY
seja uma forte candidata à uma física além do MP, outras idéias também
devem ser consideradas. Dentre elas, uma classe de modelos que
modifica a simetria $SU(2)_L$ do MP para uma estrutura de tripleto,
exibindo assim um simetria $SU(3)_C \otimes SU(3)_L \otimes U(1)_X$,
conhecida genericamente como modelos $331$.

Dentro do cenário (a), além da descoberta do Higgs, o LHC deve testar a estrutura não-abeliana da simetria
$SU(2)_L\otimes U(1)_Y$, em particular medindo os acoplamentos
quárticos $WWWW$ e $WWZZ$. Do
ponto de vista fenomenológico, um estudo independente de modelo é uma
estratégia viável para abordar uma nova física no setor
eletrofraco. 

No primeiro assunto dessa tese tratamos a quebra
espontânea de simetria através de teorias efetivas considerando o MP
na ausência do Higgs padrão. De forma mais específica, buscamos
inferir desvios na medição dos acoplamentos quárticos entre os bósons
de \textit{gauge} massivos do MP escrevendo uma parametrização geral do modelo
onde o Higgs não existe (ou não é parte de um dubleto de
$SU(2)_{L}$). Essa abordagem resulta em uma representação não-linear
da teoria efetiva. Por outro lado pode-se também manter o Higgs como
parte de um dubleto de $SU(2)_{L}$ construindo a teoria linearmente
através da expansão da lagrangiana do MP. Nesse caso os acoplamentos quárticos puros surgem somente em
operadores de ordem 8. Embora haja diferenças nos dois formalismos,
existe uma correspondência direta entre os operadores responsáveis pela
modificação da estrutura quártica entre os bósons de \textit{gauge}.

Ainda na procura de uma nova física no setor eletrofraco, abordamos também a física a ser estudada no cenário (b) nesta tese. Conforme já mencionado, a classe de
modelos $331$ modifica o conteúdo das representações em diversos setores exibindo uma estrutura de simetria
$SU(3)_{C}\otimes SU(3)_{L}\otimes U(1)_{X}$, que é quebrada
primeiramente para $SU(3)_{C}\otimes SU(2)_{L}\otimes U(1)_{Y}$ e
posteriormente para $U(1)_{EM}$. Introduzidos no início da década de 90,
esse modelos abordam questões fundamentais não explicadas pelo MP,
como por exemplo a restrição do número de famílias limitado a três (ou
múltiplos de três), além de na versão com neutrinos de mão-direita possuir candidato à matéria escura.  

Uma característica interessante é que esses
modelos prevêem a existência de novas partículas que permitem
processos, em nível de árvore, com troca de sabor via correntes neutras que são fortemente suprimidos na Natureza. Obtemos vínculos aos
parâmetros desse novo modelo mediante dados experimentais de violação
de sabor em diversos sistemas de mésons neutros, a saber $K^{0}-\bar{K}^{0}$, $D^{0}-\bar{D}^{0}$ e $B^{0}_d-\bar{B^0_d}$.

Com a intenção de fornecer ao leitor uma visão ampla sobre o contexto
experimental atual e sobre a teoria que regem a Física de Partículas
Elementares, abordamos os assuntos nessa tese visando nosso objetivo
final, sem no entanto esgotarmos a abrangência e profundidade formal
dos vários tópicos aqui mencionados\footnote{Nesse aspecto mencionaremos os diferentes experimentos do LHC, sem nos aprofundarmos em especificações técnicas. Além disso, em relação à
  construção do MP, não consideramos explicitamente a
  simetria de cor $SU(3)_{C}$ da QCD que, apesar de importante, foge do foco deste trabalho.}.

O Capítulo~\ref{chap:lhc_mp_higgs} engloba o ``estado d'arte'' da
Física de Partículas Elementares. Apresenta o LHC com uma visão bastante geral de sua história e seus
experimentos, além do seu papel essencial na corroboração do MP e seu
potencial para novas descobertas. Em paralelo ao experimento, esboçamos
a estrutura teórica do MP. Suas características, o mecanismo da quebra
espontânea de simetria, o bóson de Higgs e sua importância no modelo, seus
canais de produção e decaimento e os resultados mais recentes do
experimento CMS e ATLAS sobre esse bóson são abordados.

O Capítulo~\ref{cap:efetivas} introduz o leitor no formalismo teórico
das lagrangianas quirais e na construção de modelos efetivos. Como
motivação histórica para validade desses modelos discutimos primeiro 
o modelo de Fermi, que surgiu para explicar o decaimento $\beta$. 

Dentro dessa abordagem podemos considerar o MP como sendo um modelo
efetivo, válido na escala de energia $\Lambda$, mesmo
com a confirmação experimental do bóson de Higgs. Acima dessa escala existem diversas especulações teóricas sobre a física que se
manifestará. Em particular, se $\Lambda \approx 1$ TeV, espera-se que o LHC consiga 
levantar indícios de física além do MP.

Através da descrição de teoria de campo efetiva podemos supor
que o setor eletrofraco apresenta interações fortes entre os
bósons de \textit{gauge}, já que não há princípio fundamental que
obrigue a física responsável pela quebra de simetria eletrofraca
interagir fracamente. No MP, a intensidade da interação entre os bósons de \textit{gauge}
longitudinalmente polarizados depende do valor da massa do bóson
de Higgs. Para quantificar possíveis desvios na intensidade dos vértices
puramente quárticos previstos pelo MP, introduzimos vértices anômalos
através do formalismo de lagrangianas quirais. 

Investigamos dentro desse contexto, em nível partônico e \textit{leading
  order} (LO), o potencial do LHC, operando em energia no centro de massa de $\sqrt{s}=14$ TeV
e luminosidade $L=100\;\mbox{fb}^{-1}$, para testar os vértices anômalos
puramente quárticos $WWWW$ e $WWZZ$ em processos de colisão com
topologia de fusão de bósons vetoriais (VBF), no canal
semileptônico $pp \rightarrow l^{\pm}\nu_{l}jjjj$, onde $l^{\pm}$ são
os léptons $e^{\pm}$ e $\mu^{\pm}$, e $j$ são jatos compostos por
(anti)quarks leves. Os possíveis \emph{backgrounds} foram
considerados em três diferentes ordens em teoria de perturbação: (i)
irredutível, incluindo processos de produção de três bósons, em ${\cal
  O}(\alpha_{em}^{6})$; (ii) QCD em ${\cal
  O}(\alpha_{em}^{4}\alpha_{s}^{2})$ e ${\cal
  O}(\alpha_{em}^{2}\alpha_{s}^{4})$ e (iii) processos envolvendo
quarks \emph{top}: $pp\to t\bar{t} + 0\;\mbox{até}\;2\;\mbox{jatos}$.

O Capítulo~\ref{chap:331} aborda características de processos que permitem correntes neutras com troca de sabor (FCNC), um fenômeno fortemente suprimido no MP mas
que pode ocorrer, em nível de árvore, em algumas propostas de física além do MP. Por exemplo,
nos modelos com simetria $SU(3)_{C}\otimes
SU(3)_{L}\otimes U(1)_{X}$ os processos FCNC surgem naturalmente em nível de árvore no setor de quarks, posto que um novo bóson de \textit{gauge}
neutro $Z'$ acopla-se de forma distinta com a terceira família gerando interações não-universais nesse setor. Nesta tese mostramos
que, além do $Z'$, no modelo $331$ que apresenta neutrinos de
mão-direita ($331_{RHN}$) existem duas novas fontes de violação CP: os
escalares $S_2$ e $I_{3}^{0}$. Essas partículas escalares contribuem para as oscilações de mésons neutros $K^{0}-\bar{K}^{0}$,
$D^{0}-\bar{D}^{0}$ e $B^{0}_d-\bar{B^0_d}$, que são processos do tipo
FCNC. Usando dados experimentais relacionados a esses sistemas,
obtemos vínculos importantes aos parâmetros do modelo $331_{RHN}$. 

Dentro do contexto espécífico de cálculos numéricos, o leitor encontrará nos
Apêndices alguns detalhes úteis sobre configurações do espaço de fase
e obtenção da seção de choque através da integração por Monte Carlo no
Apêndice~\ref{apendice1}. As regras de Feynman implementadas em nossas
rotinas Fortran se encontram no Apêndice~\ref{ap:regrasfeynman} e
tabelas listando todos os subprocessos utilizados em nossas análises apresentamos no Apêndice~\ref{subs}.

\chapter{O ``estado d'arte'' na Física de Partículas Elementares}
\label{chap:lhc_mp_higgs}
\section{O Grande Colisor de Hádrons}

\begin{flushright}
{\small \it The reasonable man adapts himself to the world;\\
the unreasonable one persists in trying to adapt the world to himself.\\
Therefore all progress depends on the unreasonable man.}\\ \medskip
                {\small George Bernard Shaw}
\end{flushright}
\medskip

A comunidade científica, baseada em razões teóricas, prevê que a
hegemonia do MP será abalada na escala TeV de energia, explorada com o
início das operações do Grande Colisor de Hádrons
(LHC)~\cite{pagina_lhc}, localizado no CERN~\cite{pagina_cern}.

O LHC é um colisor e acelerador hadrônico, com dois anéis
supercondutores, instalado em túnel de aproximadamente 26.7 km,
construído na década de 1980 para abrigar o Grande Colisor Elétron-Pósitron (LEP).

Nos colisores
hadrônicos os processos são iniciados pelos
pártons $q$, que compõem os nucleons e carregam somente uma fração $x$ do
momento inicial destes.
\begin{figure}
\centering
\includegraphics[scale=0.65]{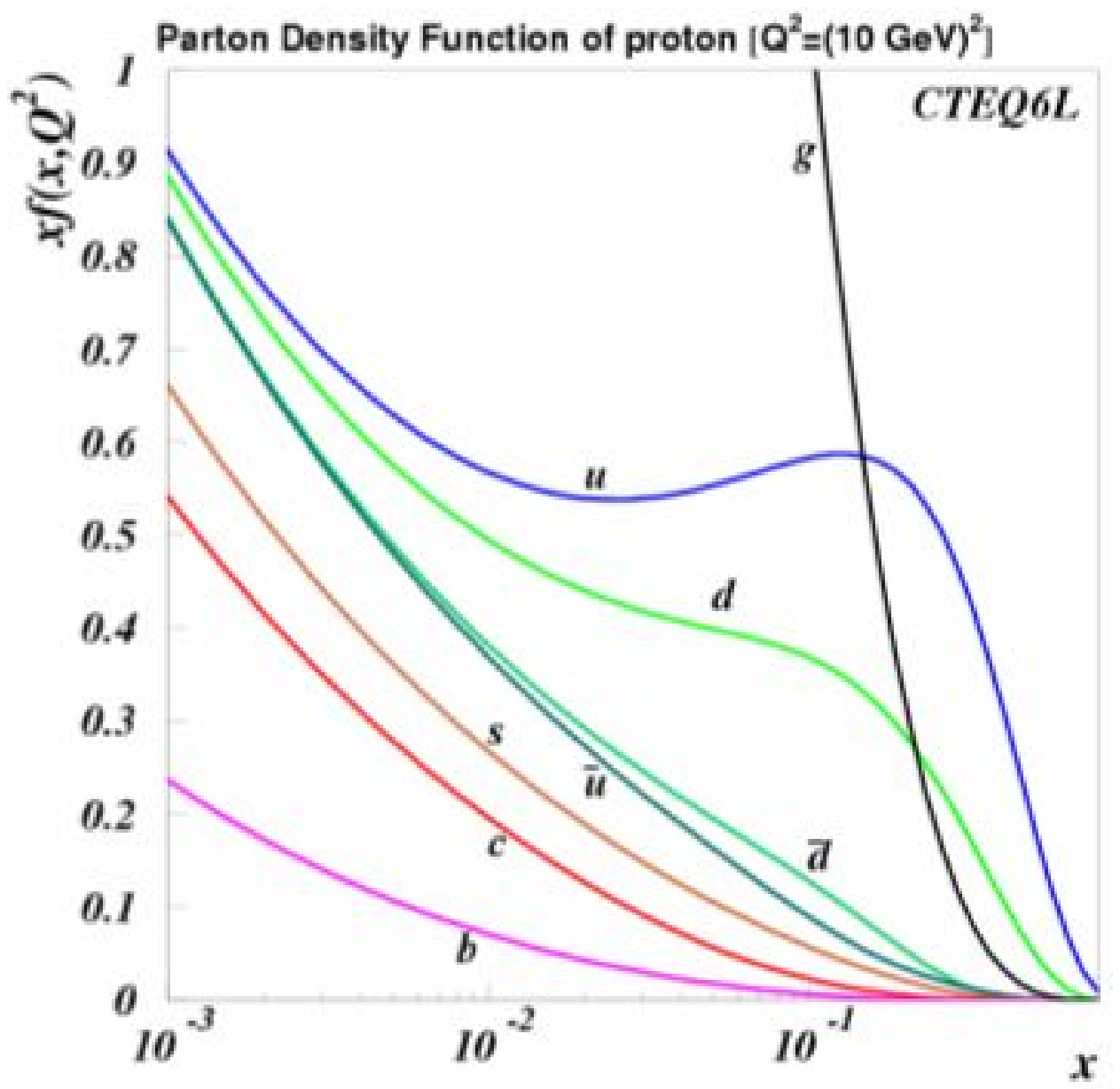}
\caption{Perfil das funções de distribuição dos pártons (PDF) de
  acordo com a opção CTEQ6L~\cite{cteq} utilizada nesse trabalho.}
\label{pdf}
\end{figure}

Cada parton $q$ tem uma função de distribuição $f_{q}(x)$ diferente,
como ilustra a Fig.~\ref{pdf}, e essas funções dependem da energia
de teste $Q^{2}$ do próton.  Sendo assim, as escalas de energia que
podem ser testadas em colisores hadrônicos são substancialmente
menores do que a sua energia do centro de massa (CM) nominal $\sqrt{s}$.

As colisões acontecem entre dois feixes não-contínuos, que
denominaremos $a$ e $b$, contendo um grande número de partículas
agrupadas em ``pacotes'' denominados \emph{bunches}.

O número de colisões (ou eventos) é proporcional ao número de
partículas em cada feixe, $N_{a}$ e $N_{b}$ respectivamente, e
inversamente proporcional à área $S$ da seção transversal de cada
feixe. O coeficiente de proporcionalidade denomina-se \emph{seção de
  choque de espalhamento} ($\sigma$) para um determinado estado final.

Quantitativamente, se os feixes colidem com uma frequência $f$ então a taxa de
ocorrência ${\cal R}$ dos eventos, ou seja o número de eventos de determinado tipo
gravados por segundo pelos detetores, é dada por
\begin{equation}
 {\cal R}=L\times \sigma_{exp},
\end{equation}
onde
\begin{equation}
L=\dfrac{N_{A}N_{B}}{S}\;f
\end{equation}
é a luminosidade integrada.

A taxa ${\cal R}$ é medida diretamente pelos físicos experimentais. Na
verdade mede-se ${\cal R}\times {\cal \varepsilon}$, onde
${\cal \varepsilon}$ é a \emph{eficiência do detetor}~\footnote{A probabilidade de
que um evento real com determinadas propriedades seja identificado pelo detetor.}.

O LHC foi projetado para colidir \emph{bunches} de até $10^{11}$
prótons com energia de $\sqrt{s}=14\;\mbox{TeV}$, em intervalos
discretos de 25 ns, atingindo uma luminosidade nominal de
$L=10^{34}\;\mbox{cm}^{-2}\mbox{s}^{-1}$ ou $100\;\mbox{fb}^{-1}$ por
ano\footnote{À título de curiosidade, esse valor de $100\;\mbox{fb}^{-1}$ por
ano se deve ao fato do LHC não operar durante 12 meses ininterruptos. Fosse esse o caso teríamos uma luminosidade nominal de 315.36 $\mbox{fb}^{-1}$ por ano.}.

O número acentuado de partículas por \emph{bunch} aumenta a
luminosidade e, consequentemente, a quantidade de eventos que podem
ser aproveitados. A seção de choque $pp$ total em colisores hadrônicos
é alta e vale aproximadamente $98\;\mbox{mb}$, sendo que
aproximadamente $73\;\mbox{mb}$ correspondem somente à seção de choque
inelástica~\cite{totem}. Sendo assim, a probabilidade de ocorrerem múltiplas
interações (\emph{pile up}) também aumenta e estima-se que haverá
aproximadamente 19 interações inelásticas por cruzamento de
\emph{bunch}. O \emph{pile up} inerente aos experimentos complica
bastante a análise dos dados, pois necessita-se distinguir a reação
estudada desses eventos adicionais que contaminam o sinal.

Além dos prótons, o projeto do LHC também permite colisões de íons pesados, em particular
núcleos de chumbo, com energia de $\sqrt{s}=5.5\;\mbox{TeV}$ por par
de núcleons e luminosidade de
$L=10^{27}\;\mbox{cm}^{-2}\mbox{s}^{-1}$.

Alguns parâmetros do LHC podem ser visualizados na Tabela~\ref{parametros}.

\begin{figure}
\centering
\includegraphics[scale=0.30]{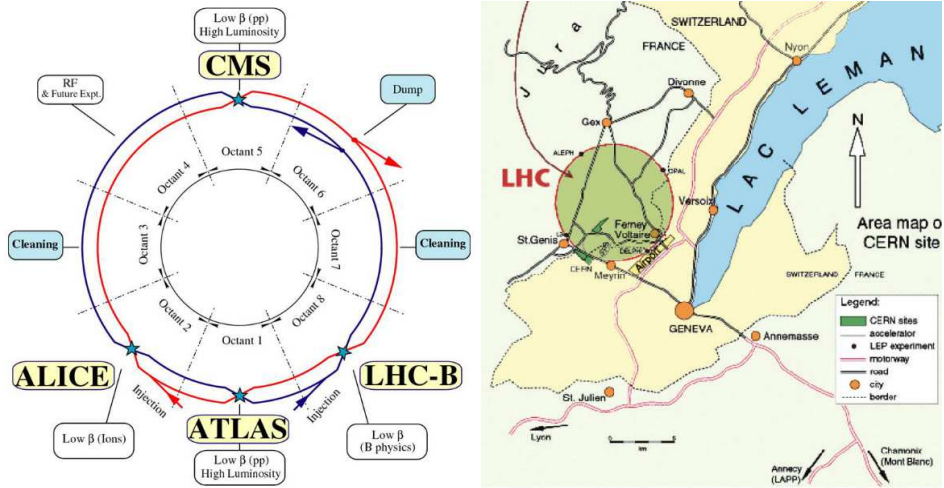}
\caption{Grande Colisor de Hádrons (LHC): layout (à esquerda) e
  localização (à direita). Figura extraída da Ref.~\cite{lhc}.}
\label{lhc_experimento_local}
\end{figure}

{\footnotesize \begin{table}[ht]
\centering
\begin{tabular}{c|c|c|c}
\hline \hline
&&&\\
Parâmetros do feixe & & pp & PbPb \\
&&&\\
\hline
&&&\\
Energia por nucleon & [TeV]& 7 & 2.76 \\
&&& \\
\emph{Bunches} & & 2808 & 592 \\
&&& \\
Partículas por \emph{bunch} & & $1.15\times 10^{11}$ & $7.0\times 10^{7}$\\
&&& \\
\emph{Bunch Crossing} &[ns]& 25 & 100 \\
&&& \\
Frequência & [MHz] & 40 & 0.008 \\
& &&\\
Luminosidade &[$\mbox{cm}^{-2}\mbox{s}^{-1}$]&$1.0\times 10^{34}$&$1.0\times 10^{27}$ \\
&&&\\
\hline \hline
\end{tabular}
\vspace{0.3cm}
\caption{Parâmetros nominais do LHC para colisões próton-próton e chumbo-chumbo. Maiores detalhes na Ref.~\cite{lhc}.}
\label{parametros}
\end{table}}

\subsection{Detetores}
\label{sub:detetores}
O LHC dispõe de seis detetores - ATLAS, CMS, ALICE, LHCb, LHCf e TOTEM
- direcionados à analise de diferentes fenômenos
físicos~\cite{experimentos}.

O ATLAS (``A large Toroidal LHC ApparatuS'') e o CMS (``Compact Muon
Solenoid'') são os dois maiores detetores do LHC, direcionados a
propósitos gerais, que cobrirão um amplo leque de possibilidades de
fenômenos físicos acessíveis à sua escala de energia, além de serem
construídos para analisar tanto colisões $pp$ quanto colisões de íons
pesados.
\begin{figure}[!ht]
\centering
\includegraphics[scale=0.4]{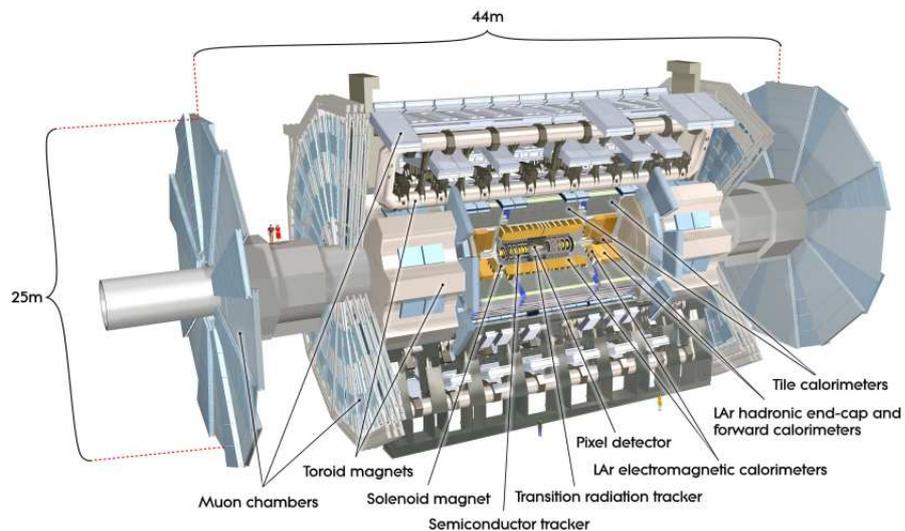}
\caption{Visão interna do detetor ATLAS. Dimensões: 25m de altura e 44m de comprimento; Peso: $\approx$ 7000 toneladas. Figura obtida da Ref.~\cite{experimentos}.}
\label{atlas}
\end{figure}

\begin{figure}[!ht]
\centering
\includegraphics[scale=0.4]{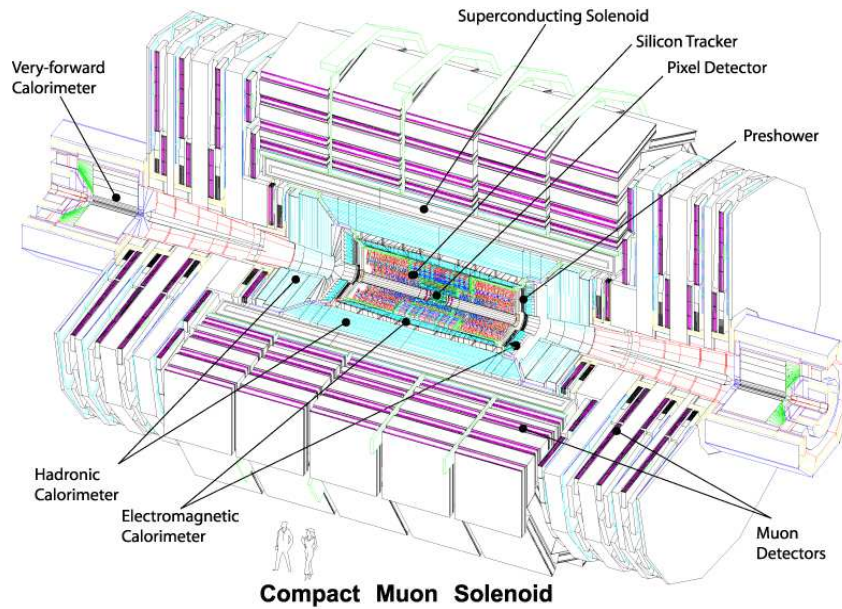}
\caption{Visão interna do detetor CMS. Dimensões: 15m de altura, 15m de
	       largura e 21m de comprimento. Figura obtida de
``http://cms.web.cern.ch/cms''.}
\label{cms}
\end{figure}

De forma geral, o ATLAS e o CMS, cujos detectores estão esquematizados
nas Figs.~\ref{atlas} e~\ref{cms} respectivamente, foram preparados para busca do bóson
de Higgs e de sinais de nova física, entre os quais SUSY, dimensões
extras e outras extensões do MP.

O experimento ALICE (``A Large Ion Collider Experiment'') foi
construído especificamente para analisar a colisão de íons
pesados. Seu objetivo é estudar a física da matéria que interage
fortemente em densidades de energia extremas, situação na qual espera-se
a formação de uma nova fase da matéria, o plasma de quarks e glúons. A
existência dessa fase de matéria e suas propriedades é essencial para
entender o confinamento e a simetria quiral na QCD. Esse experimento
 também analisará colisões $pp$, para comparação tanto com os
resultados das suas colisões $Pb$--$Pb$ quanto com os outros
experimentos.

Outros experimentos bem menores que os anteriores e mais específicos em
suas análises são:
\begin{itemize}
\item o LHCb (``Large Hadron Collider beauty'') para medidas de
  precisão de violação $CP$ e decaimentos raros de mésons
  $B$~\footnote{A título de curiosidade, o LHCb foi o primeiro experimento a obter resultados
    de nova física no LHC através da medição direta de violação $CP$ nos
    decaimentos $D \to K^{+}K^{-}$ e $D \to \pi^{+}\pi^{-}$. Maiores detalhes na Ref.~\cite{lhcb}};
\item o TOTEM (``Total Cross Section, Elastic Scattering and
  Diffraction Dissociation''), locado próximo ao CMS, foi projetado para estudar partículas emitidas na direção frontal (pequenos ângulos e altos valores de pseudorapidez $\eta$),
  bem próximas ao eixo dos feixes, focando a física em regiões de $\eta $ não acessível ao experimento CMS;
\item o LHCf (``Large Hadron Collider forward'') consiste de dois
      pequenos calorímetros, posicionados cada um a 140 metros do ponto
      de interação do detetor ATLAS, também com o propósito de estudar a
      produção \emph{forward} de partículas nas colisões $pp$ em
      ângulos extremamente baixos.
\end{itemize}

\subsubsection{Sistema de coordenadas}

Tendo em vista os assuntos abordados nesse trabalho, nos parece
relevante mencionar alguns detalhes sobre a padronização do sistema de
coordenadas adotado, por exemplo pelo experimento CMS.

\begin{figure}[!ht]
\centering 
\includegraphics[scale=0.5]{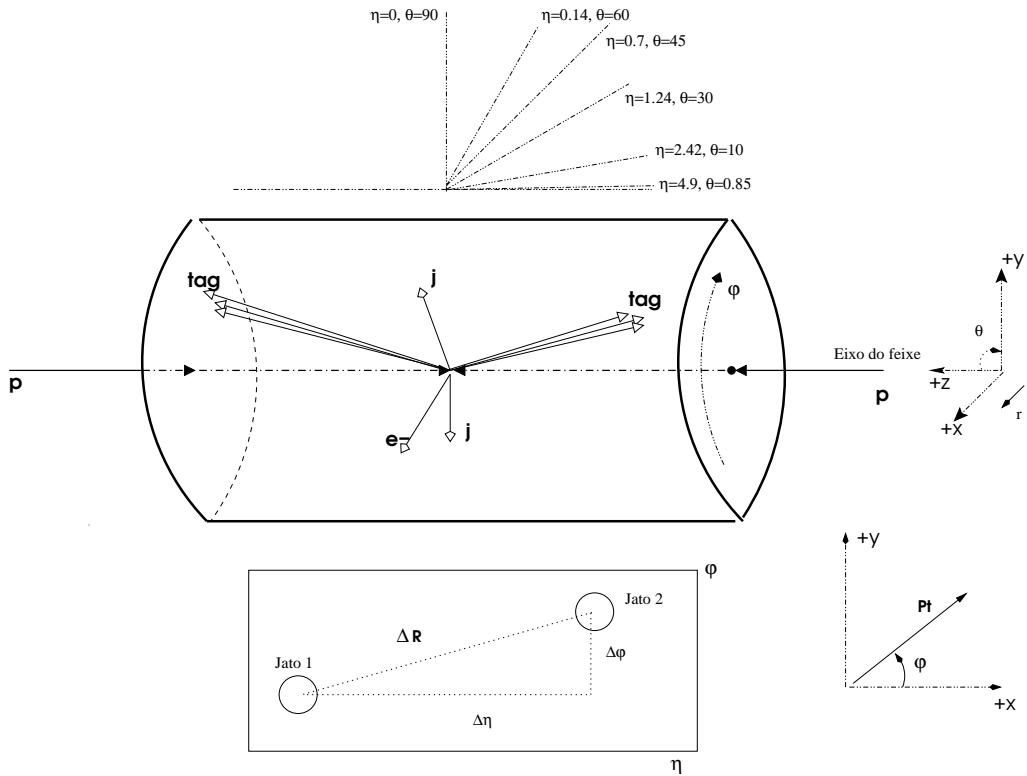} 
\caption{Esboço de um detetor genérico destacando regiões de pseudorapidez $\eta$ e caracterização de uma reação com topologia de fusão de bósons vetoriais.} 
\label{vbf_plano} 
\end{figure}

Neste experimento os feixes de prótons circulam em sentido horário e
anti-horário, sendo o sentido anti-horário definido como sentido
positivo na direção do eixo-$z$. A origem do sistema de coordenadas
$xyz$ é o ponto de colisão, com os eixos $x$ e $y$ apontando 
radialmente para o centro do anel de colisão, conforme mostrado na Fig.~\ref{vbf_plano}.

O ângulo azimutal $\varphi $ é medido a partir do eixo $x$ no plano
$xy$ e o ângulo polar $\theta$ é medido a partir do eixo $z$ em
direção ao plano $xy$.

Em colisões costuma-se utilizar a variável cinemática
pseudorapidez $\eta $, definida como
\begin{equation}
\eta \equiv -\ln{\left(\tan{\dfrac{\theta}{2}}\right)},
\end{equation}
no lugar do ângulo polar $\theta$ para medir a inclinação do eixo de
emissão das partículas em relação ao eixo do feixe, pois diferenças $\Delta \eta$ são invariantes sob \emph{boosts} de Lorentz ao longo deste eixo.

Na Fig.~\ref{vbf_plano} visualiza-se também a variável cinemática
$\Delta R$, que mede a distância (ou separação) entre dois objetos, jatos por
exemplo, no plano ($\eta, \varphi$) e é definida como
\begin{equation}
\Delta R = \sqrt{(\Delta \eta)^{2}+(\Delta \varphi)^{2}},
\label{deltaR}
\end{equation} 
onde $\Delta \eta \equiv \eta_{2}-\eta_{1}$ e $\Delta \varphi \equiv
\varphi_{2}-\varphi_{1}$ são medidos em relação ao eixo $z$ do
feixe. Essa variável é fundamental para estabelecer critérios de
isolamento entre objetos e na definição de jatos.

\newpage

\section{O Modelo Padrão da Física de Partículas Elementares}
\label{sec:mp}

\begin{flushright}
{\small \it By convention there is color, by convention sweetness,\\
       By convention bitterness, but in reality there are atoms and
 space}\\
                           {\small Democritus (aprox. 400 BCE)}
\end{flushright}
\medskip

Na verdade sabemos que \emph{there are leptons, quarks, gluons, bosons...}

O MP é uma teoria elegante, que descreve todas as partículas
conhecidas na Natureza e as suas interações até o momento.

Ele nos diz que toda a
matéria conhecida no Universo é composta por partículas elementares, os
{\it férmions}, subdivididos nos tipos {\it léptons} e {\it quarks}, sujeitas
aos quatro tipos de interações fundamentais - eletromagnética,
fraca (que descreve por exemplo o decaimento $\beta$), forte (responsável por manter os quarks dentro do próton) - mediadas pelos bósons de \textit{gauge}
(respectivamente, o fóton $\gamma$ para a força eletromagnética, $W^{\pm}$ e $Z$ para
força fraca e o glúon $g$ para força forte) e gravitacional (esta não é abordada teoricamente pelo MP), seguindo algumas
regras básicas determinadas por conceitos
matemáticos formais baseados nas noções de grupos de simetria e
invariância de \textit{gauge}. Direcionamos o leitor à algumas
referências~\cite{modelo} que tratam características, sucessos e
problemas do MP com maiores detalhes não considerados neste trabalho devido
à extensão do assunto.

O MP evoluiu de uma descrição do eletromagnetismo pela teoria de
campos quântica, a eletrodinâmica quântica (QED), para uma teoria mais
ampla que engloba a força fraca e a força forte. O arcabouço teórico
do MP se baseia no grupo de simetria de \textit{gauge}
$SU(3)_{C}\otimes SU(2)_{L}\otimes U(1)_{Y}$, onde $C$ refere-se ao
grupo de cor da cromodinâmica quântica (QCD), $L$ refere-se à
estrutura de mão-esquerda (\emph{left-handed}) do grupo de isospin $SU(2)$ e $Y$ refere-se
ao grupo de hipercarga $U(1)$.

À essa estrutura de \textit{gauge} adicionou-se minimalmente um
dubleto escalar de $SU(2)_{L}$ com hipercarga $Y=+1$, o dubleto de
Higgs, que quebra a simetria $SU(2)_{L}\otimes U(1)_{Y} \to U(1)_{EM}$
dando origem às massas dos bósons de \textit{gauge} $W^{\pm}$ e $Z$,
mantendo o fóton $\gamma $ sem massa. Além disso, a interação desse dubleto com os
férmions, através da lagrangiana de Yukawa, é responsável pela
geração de massa das partículas de matéria.

Usando a mesma notação da Ref.~\cite{rainwater} observa-se
a existência de três famílias no setor fermiônico
{\footnotesize
\begin{equation}
\left[\begin{array}{ccc}
\left(\begin{array}{c}
\nu_{e} \\
e
\end{array}
\right)_{L} & e_{R} &
\\
\left(\begin{array}{c}
u \\
d
\end{array}
\right)_{L} & d_{R} & u_{R}
\end{array}
\right]\;\;\left[
\begin{array}{ccc}
\left(\begin{array}{c}
\nu_{\mu} \\
\mu
\end{array}
\right)_{L} & \mu_{R} &
\\
\left(\begin{array}{c}
c \\
s
\end{array}
\right)_{L} & s_{R} & c_{R}
\end{array}
\right]\;\;\left[
\begin{array}{ccc}
\left(\begin{array}{c}
\nu_{\tau} \\
\tau
\end{array}
\right)_{L} & \tau_{R} &
\\
\left(\begin{array}{c}
t \\
b
\end{array}
\right)_{L} & b_{R} & t_{R}
\end{array}
\right]
\label{mp}
\end{equation}}
onde nota-se a estrutura de dubletos \emph{left-handed}, acompanhada dos
singletos \emph{right-handed}.

Das quatro interações fundamentais existentes na Natureza, o MP
consegue unificar as interações eletromagnética e fraca, mas é incapaz
de unificar a interação forte com as demais, além de não descrever a interação gravitacional.

As interações forte, fraca e eletromagnética surgem devido à troca de
bosóns vetoriais entre os férmions (spin-1/2) que constituem a
matéria. Os bósons de \textit{gauge} vetoriais (spin-1) são associados
aos geradores da álgebra dos grupos que compõem a estrutura do MP.

Observa-se o seguinte em relação aos bósons vetoriais:
\begin{itemize}
\item associados ao grupo de simetria $SU(3)_{C}$ existem oito campos geradores $G_{\mu}^{\alpha}$ ($\alpha =1,\cdots,8$) cujos estados físicos são denominados glúons. Os glúons não possuem massa, carregam carga de cor e são responsáveis pela interação forte. Deste modo, qualquer partícula que se acople com os glúons é caracterizada como sendo fortemente interagente.
\item associados ao grupo de simetria de isospin $SU(2)_{L}$ aparecem
  três campos geradores $W^{a}_{\mu}$ ($a=1,2,3$) e ao grupo de
  simetria de hipercarga $U(1)_{Y}$ um campo gerador $B_{\mu}$ (o
  índice $Y$ é utilizado para diferenciar o grupo relacionado à
  hipercarga fraca do grupo eletromagnético, $U(1)_{EM}$, associado à
  carga elétrica $Q$). Esses quatro campos vetoriais associados ao
  grupo de simetria $SU(2)_{L}\otimes U(1)_{Y}$ geram, após o
  mecanismo de quebra dessa simetria, os bósons de \textit{gauge} massivos que mediam as interações eletrofracas e o
  fóton, não massivo, descrito pela QED.
\end{itemize}

Apesar de observarmos experimentalmente todo o conteúdo das famílias
descritas em~\eqref{mp}, somente a primeira família, ou \emph{primeira
  geração} (composta pelo elétron $e^{-}$, neutrino do elétron
$\nu_{e}$, quarks \emph{up} $u$ e \emph{down} $d$), é encontrada na
matéria usual. As duas gerações restantes (compostas pelo múon $\mu$,
neutrino do múon $\nu_{\mu}$, quarks $c$ e $s$, tau $\tau$, neutrino
do tau $\nu_{\tau}$, quarks $t$ e $b$) contém partículas instáveis, que podem
ser criadas diretamente e/ou através do decaimento de
partículas pesadas nos aceleradores de partículas.

As três gerações de férmions mostram uma diferença muito significativa entre
suas massas. Deste modo, as partículas da segunda e terceira gerações
são mais pesadas e instáveis e decaem em partículas da primeira
geração.

De forma geral, podemos escrever a lagrangiana do MP como
\begin{equation}
{\cal L}_{MP}={\cal L}_{G} + {\cal L}_{f} + {\cal L}_{Y} +{\cal L}_{GF} + {\cal L}_{FP},
\label{lagrangiana_sm}
\end{equation}
onde ${\cal L}_{G}$ contém os termos cinéticos dos bósons de
\textit{gauge}, ${\cal L}_{f}$ refere-se aos férmions e aos seus
acoplamentos com os bósons de \textit{gauge}, ${\cal L}_{Y}$ é a lagrangiana de Yukawa, ${\cal L}_{GF}$ e ${\cal
  L}_{FP}$ referem-se aos termos de fixação de \textit{gauge} e à
lagrangiana de \emph{Faddeev-Popov}.

Detalhes da forma e conteúdo da lagrangiana~\eqref{lagrangiana_sm}, além de sua
construção, podem ser obtidos em várias referências~\cite{modelo}. Aqui apresentaremos somente os termos significantes para nossas análises.

O setor fermiônico do MP é regido pela seguinte lagrangiana,
\begin{equation}
{\cal L}_{f}=i\sum_{a=1}^{3}\;[\bar{\psi}^{a}_{L}\gamma^{\mu}D_{\mu}^{L}\psi^{a}_{L}+\bar{\psi}^{a}_{R}\gamma^{\mu}D_{\mu}^{R}\psi^{a}_{R}],
\label{lagrangiana_fermions}
\end{equation}
onde os spinores $\psi_{L}$ representam os férmions de mão-esquerda e
$\psi_{R}$ os férmions de mão-direita. $D_{\mu}$ são as derivadas
covariantes
\begin{align}
D_{\mu}^{L} &\equiv \partial_{\mu} + ig\dfrac{\tau^{a}}{2}W^{a}_{\mu}+ig'\dfrac{Y}{2}B_{\mu}, \nonumber \\
D_{\mu}^{R} &\equiv  \partial_{\mu} +ig'\dfrac{Y}{2}B_{\mu},
\label{covariante}
\end{align}
que se acoplam com os campos de mão-esquerda e mão-direita,
respectivamente. Em~\eqref{covariante}, $W^{a}_{\mu}$ e $B_{\mu}$ são
os bósons de \textit{gauge} do grupo de simetria de isospin
$SU(2)_{L}$ e $U(1)_Y$, respectivamente. A mistura entre eles dá origem
ao bósons de \textit{gauge} físicos, conforme veremos mais adiante.
$\tau_{a}$ são as matrizes de Pauli usuais, geradoras do grupo de
isospin, e $Y$ é a hipercarga.

O setor eletrofraco se apresenta com a lagrangiana
\begin{equation}
{\cal L}_{G} =
 -\dfrac{1}{4}W_{\mu\nu}^{a}W^{a\;\mu\nu}\;-\;\dfrac{1}{4}B_{\mu\nu}B^{\mu\nu},
\label{lagrangiana_gauge}
\end{equation}
onde $a=1,2,3$ e os tensores $W_{\mu\nu}^{a}$ e $B_{\mu\nu}$ têm a forma
\begin{align}
W^{a}_{\mu\nu}&\equiv 
 \partial_{\mu}W^{a}_{\nu}-\partial_{\nu}W^{a}_{\mu}+g\epsilon^{abc}W_{\mu}^{b}W_{\nu}^{c}
\label{tensorw}
\\
B_{\mu\nu}&\equiv  \partial_{\mu}B_{\nu}-\partial_{\nu}B_{\mu}
\label{tensor}
\end{align}

Observa-se que a estrutura não-abeliana de~\eqref{tensorw} na lagrangiana~\eqref{lagrangiana_gauge} origina vértices tríplices e quárticos entre os bósons de
\textit{gauge}.

Na lagrangiana~\eqref{lagrangiana_sm} não existem
termos de massa para os campos\footnote{Massa nula caracteriza forças
  com alcance infinito. No entanto, as interações fracas têm alcance de
  $10^{-15}$cm sugerindo massa para os bósons de \textit{gauge} da
  ordem de $100\;\mbox{GeV}$.} pois termos da forma
$m_{\psi}\bar{\psi}\psi
=m_{\psi}(\bar{\Psi}_{L}\Psi_{R}+\bar{\Psi}_{R}\Psi_{L})$ não são
invariantes sob transformações do grupo de simetria
$SU(2)_{L}$. Desta forma, um mecanismo que gere massa
para esses campos, mantendo a renormalizabilidade da teoria\footnote{Condição
que não seria satisfeita inserindo termos de massa ``à mão''}, se faz
necessário. Conforme abordaremos a seguir, as massas das partículas no
MP são geradas através do chamado mecanismo de Higgs.

\subsection{Quebra de simetria eletrofraca e mecanismo de Higgs}
\label{sec:higgs}

Conforme exposto, a invariância por transformações do grupo de simetria
$SU(2)_{L}\otimes U(1)_{Y}$ proíbe que as partículas do MP adquiram massa. Por outro lado, como os bósons e os férmions observados na Natureza possuem
massa (com exceção do fóton), então essa simetria no setor
eletrofraco precisa ser quebrada. Note que a inexistência de massa
para o fóton pode ser traduzida como uma invariância por transformações do grupo $U(1)_{EM}$, o grupo de simetria da QED.

Esse comportamento de quebra de simetria pode ser explicado no MP pelo mecanismo de Higgs\footnote{Também chamado de mecanismo de
  Brout–Englert–Higgs ou mecanismo de
  Englert-Brout-Higgs-Guralnik-Hagen-Kibble ou ainda mecanismo de
  Anderson–Higgs. Foi inicialmente proposto em 1962 por Philip Warren
  Anderson, que adotou um modelo não-relativístico para aplicação em
  física de partículas. Em 1964, Peter Higgs e independentemente o
  grupo de Robert Brout e Francois Englert, assim como o grupo de
  Gerald Guralnik, C.\ R.\ Hagen e Tom Kibble trabalharam em sua versão
  relativística atual.}, cujo precedimento será esquematizado
abaixo\footnote{Escolha original feita por Weinberg em 1967, que
  complementa a especificação ``padrão'', ou minimal, do modelo para
  interações eletrofracas.}.
\begin{itemize}
\item[1.] No modelo de Weinberg-Salam-Glashow introduz-se quatro
  campos escalares reais, formando um
  dubleto escalar complexo $\Phi$ de $SU(2)_{L}$, com hipercarga
  $Y=+1$, denominado \emph{dubleto de Higgs}, tal que
\begin{equation}
\Phi = \left(\begin{array}{c}
	\phi^{+} \\ \phi^{0}
       \end{array}\right) = \sqrt{\dfrac{1}{2}}\left(\begin{array}{c}
\phi_{1}+i\phi_{2} \\ \phi_{3} + i\phi_{4}
\end{array}
\right).
\label{dubleto}
\end{equation}
\item[2.] Constrói-se e insere-se, na teoria representada pela
	  lagrangiana~\eqref{lagrangiana_sm}, a lagrangiana escalar
\begin{equation}
{\cal L}_{H} = (D_{\mu}^{L}\Phi)^{\dagger}\;(D^{\mu L}\Phi) - V(\Phi^{\dagger}\Phi),
\label{higgslagrangiana}
\end{equation}
sendo que o acoplamento do dubleto escalar com os campos de \textit{gauge} se
dá através da derivada covariante $D^{L}_{\mu}$. Em forma matricial, a Eq.~\eqref{covariante}
se escreve como
\begin{align}
D_{\mu}^{L} &= \partial_{\mu} + i\dfrac{g}{2}\left\{ \left( \begin{tabular}{cc} 0 & $W^{1}_{\mu}$ \\  
$W^{1}_{\mu}$ & 0   
\end{tabular} \right) +\left( \begin{tabular}{cc} 
0 & $-iW^{2}_{\mu}$ \\  
$iW^{2}_{\mu}$ & 0   
\end{tabular} \right) +\left( \begin{tabular}{cc} 
$W^{3}_{\mu}$ & 0 \\  
0 & $-W^{3}_{\mu}$  \end{tabular} \right) \right\} + i\dfrac{g'}{2}Y\mathbf{I}_{2\times 2}B_{\mu} 
&  \nonumber \\
& \nonumber \\ 
       &= \left( \begin{tabular}{cc} 
$\partial_{\mu}+i\dfrac{g}{2}W^{3}_{\mu}+ i\dfrac{g'}{2}YB_{\mu}$ & $i\dfrac{g}{\sqrt{2}}W^{+}_{\mu}$ \\  
$i\dfrac{g}{\sqrt{2}}W^{-}_{\mu}$ & $\partial_{\mu}-i\dfrac{g}{2}W^{3}_{\mu}+ i\dfrac{g'}{2}YB_{\mu}$\end{tabular} \right)
\label{covariante_matricial}
\end{align}
onde definiu-se
\begin{equation}
W^{\pm}_{\mu}\equiv \dfrac{1}{\sqrt{2}}(W^{1}_{\mu}\mp W^{2}_{\mu}).
\label{campos_fisicos}
\end{equation}

O  potencial escalar $V(\Phi^{\dagger}\Phi)$ tem a forma
\begin{align}
V(\Phi^{\dagger}\Phi)&=\mu^{2}\;\Phi^{\dagger}\Phi + \lambda
 (\Phi^{\dagger}\Phi)^{2} \nonumber \\
&= \dfrac{\mu^{2}}{2}\;\left[ \phi_{1}^{2}+\phi_{2}^{2}+\phi_{3}^{2}+\phi_{3} ^{2}\right]+\dfrac{\lambda}{4}\left[ \phi_{1}^{2}+\phi_{2}^{2}+\phi_{3}^{2}+\phi_{4}^{2}  \right]^{2}.
\label{potencial}
\end{align}

 Este potencial escalar é o mais simples possível de forma que, sendo
 $\mu^{2}<0$ e $\lambda>0$, possui valor mínimo global diferente de zero
para (ver Fig.~\ref{higgs})
\begin{equation}
|\Phi|^{2}=\dfrac{1}{2}(\phi_{1}^{2}+\phi_{2}^{2}+\phi_{3}^{2}+\phi_{4}^{2})\;=\;-\dfrac{\mu^{2}}{2\lambda}=\dfrac{v^{2}}{2}.
\label{global}
\end{equation}

\begin{figure}[!ht]
\centering
\includegraphics[scale=0.40]{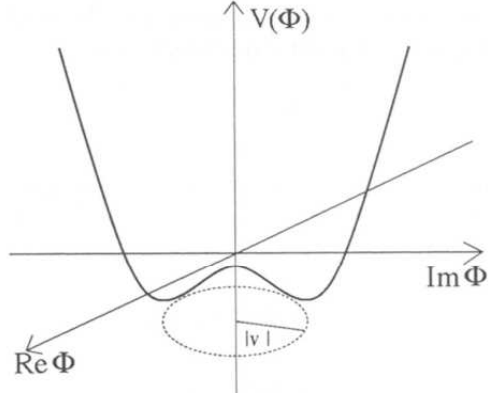}
\caption{Potencial para um campo escalar complexo $\Phi$ com
 $\mu^{2}<0$ e $\lambda>0$.}
\label{higgs}
\end{figure}

\item[3.] Escolhe-se de acordo com a expressão~\eqref{global} qual
  componente do dubleto de Higgs $\Phi$ não se anulará na configuração
  mínima do potencial. Por conveniência, escolhe-se a componente
  $\phi_{3}$ e sendo assim impõe-se que
\begin{equation}
\phi_{1}=\phi_{2}=\phi_{4}=0\;,\;\phi_{3}^{2}=-\dfrac{\mu^{2}}{\lambda}\equiv v^{2}.
\end{equation}

Esse será o valor esperado do vácuo (\emph{vev}) do campo de Higgs, ou seja
\begin{equation}
\langle \Phi \rangle_{0} = \left(\begin{array}{c}
	0 \\ \phi^{0}
       \end{array}\right)=\left(\begin{array}{c}
0 \\
v/\sqrt{2}  \end{array}\right),
\label{vev}
\end{equation}
onde
\begin{equation}
v=\sqrt{-\dfrac{\mu^{2}}{\lambda}}.
\end{equation}
\item[4.] Qual razão da escolha efetuada nos itens anteriores? Ou
  seja, porque um dubleto de $SU(2)_{L}$ de campos escalares complexos
  com hipercarga $Y=+1$ e \emph{vev} $\langle \Phi\rangle_{0}$ dado
  por~\eqref{vev} é aplicável?

Na verdade, qualquer escolha para $\langle \Phi^{0}\rangle$, que
quebre a simetria, gerará massa para o bóson de \textit{gauge}
correspondente. Entretanto se o vácuo $\langle \Phi\rangle_{0}$ se
mantiver invariante em um subgrupo das transformações de
\textit{gauge} então os bósons associados a esse subgrupo se mantém
sem massa. Em outras palavras, para preservar a simetria
eletromagnética exata de modo a manter a carga elétrica conservada,
deve-se quebrar o grupo original de simetria como $SU(2)_{L}\otimes
U(1)_{Y} \to U(1)_{EM}$. Deste modo, após a quebra espontânea da
simetria, o subgrupo $U(1)_{EM}$ se mantém como uma simetria do
vácuo. Nesse caso, o bóson de \textit{gauge} correspondente, o fóton,
se mantém sem massa e a escolha feita para $\langle \Phi\rangle_{0}$
deixa o vácuo invariante sob $U(1)_{EM}$. De fato, tal invariância
requer que o estado de vácuo $\langle \Phi \rangle_{0}$ se transforme
como
\begin{equation}
\exp{(i\theta Q)}\;\langle \Phi \rangle_{0}\approx (1+i\theta Q)\langle \Phi \rangle_{0}=\langle \Phi \rangle_{0},
\label{transf}
\end{equation}
onde $Q$ é o operador carga elétrica\footnote{Importante lembrar que a
  representação de um modelo baseado em determinado grupo de
  \emph{gauge} pode ser conhecido através do operador carga elétrica
  $Q$.}  que satisfaz a relação de Gell-Mann-Nishima,
$Q=T_{3}+\dfrac{1}{2}Y$ com os operadores $T_{3}$ e hipercarga $Y$ da
álgebra do grupo $SU(2)_{L}\otimes U(1)_{Y}$, ou seja
$[T^{i},T^{j}]=i\epsilon^{ijk}T^{k}\;\;\mbox{e}\;\;[T^{i},Y]=0$.

Vemos então em~\eqref{transf} que o operador $Q$ aniquila o vácuo, ou seja $Q\langle \Phi \rangle_{0}=0$. De fato,
\begin{align}
Q\langle \Phi \rangle_{0}&=\left(T_{3}+\dfrac{1}{2}Y\right)\langle \Phi
 \rangle_{0} \nonumber \\
&=\dfrac{1}{2}\left(\tau_{3}+Y\right)\langle \Phi \rangle_{0} \nonumber \\
&=\dfrac{1}{2}\left[\left(\begin{array}{cc}
1 & 0 \\ 0 & -1
\end{array}\right)+\left(\begin{array}{cc}
1 & 0 \\ 0 & 1
\end{array}\right)
\right]\left(\begin{array}{c}
0 \\ v/\sqrt{2}
\end{array}\right)=\left(\begin{array}{c}
0 \\ 0
\end{array}\right).
\end{align}

Note que para os geradores de $SU(2)_{L}\otimes U(1)_{Y}$ temos explicitamente
\begin{align}
\tau_{1}\langle \Phi \rangle_{0} &= \left(\begin{array}{cc}
1 & 1 \\ 1 & 0 \end{array}\right)\;\left(\begin{array}{c}
0 \\ v/\sqrt{2} \end{array}\right) = \left(\begin{array}{c}
v/\sqrt{2}\\0 \end{array}\right) \neq 0, \\
& & \nonumber \\
\tau_{2}\langle \Phi \rangle_{0} &= \left(\begin{array}{cc}
0 & -i \\ i & 0 \end{array}\right)\;\left(\begin{array}{c}
0 \\ v/\sqrt{2} \end{array}\right) = \left(\begin{array}{c}
-iv/\sqrt{2}\\0 \end{array}\right) \neq 0,  \\
& & \nonumber \\
\tau_{3}\langle \Phi \rangle_{0} &= \left(\begin{array}{cc}
1 & 0 \\ 0 & -1 \end{array}\right)\;\left(\begin{array}{c}
0 \\ v/\sqrt{2} \end{array}\right) = \left(\begin{array}{c}
0 \\ -v/\sqrt{2} \end{array}\right) \neq 0, \\
&  \nonumber \\
Y\langle \Phi \rangle_{0} &= 1\langle \Phi \rangle_{0} \neq 0.
\end{align}
\item[5.] Para obtermos os campos de \textit{gauge} massivos,
  desloca-se o campo de Higgs em relação ao seu vácuo e, como o campo
  $\Phi$ e as suas componentes respeitam a simetria $SU(2)_{L}$, para
  obter os campos físicos explicitamente, adota-se o \textit{gauge}
  unitário. Fazendo uma transformação de \textit{gauge} desse grupo,
  tem-se que
\begin{equation}
 \Phi \to \Phi' = \exp{\left(-i\dfrac{\tau^{a}}{2}\dfrac{\theta_{a}}{v}\right)}\;\Phi = \sqrt{\dfrac{1}{2}}\left(\begin{array}{c}
0 \\ v+H \end{array}\right),
\label{phi_transf}
\end{equation}
onde passamos a usar a notação $\phi_{3}\equiv H$.
\end{itemize}

Usa-se então a expressão do campo transformado~\eqref{phi_transf} na lagrangiana~\eqref{higgslagrangiana} tal que 
\begin{equation}
D^{L}_{\mu}\Phi^{\prime}=\left( \begin{tabular}{cc} 
$\partial_{\mu}+i\dfrac{g}{2}W^{3}_{\mu}+ i\dfrac{g'}{2}YB_{\mu}$ & $i\dfrac{g}{\sqrt{2}}W^{+}_{\mu}$ \\  
$i\dfrac{g}{\sqrt{2}}W^{-}_{\mu}$ & $\partial_{\mu}-i\dfrac{g}{2}W^{3}_{\mu}+ i\dfrac{g'}{2}YB_{\mu}$\end{tabular} \right) \sqrt{\dfrac{1}{2}}\left(\begin{array}{c}
0 \\ v+H \end{array}\right),
\end{equation}
e escrevendo
\begin{align}
Z_{\mu}&=(gW^{3}_{\mu}-g'B_{\mu})/(g^{2}+g^{\prime 2})^{1/2} \nonumber \\
A_{\mu}&=(gW^{3}_{\mu}+g'B_{\mu})/(g^{2}+g^{\prime 2})^{1/2},
\label{rotacao}
\end{align}
obtém-se após algum trabalho algébrico 
\begin{align}
{\cal L}_{H}&=\dfrac{1}{2}(\partial_{\mu}H\partial^{\mu}H+2\mu^{2}H^{2})+\dfrac{g^{2}}{4}(v^{2}+2vH+H^{2})W^{\mu}_{+}W_{-\mu}
\nonumber \\ &  +\dfrac{(g^{2}+g^{\prime
    2})}{8}(v^{2}+2vH+H^{2})Z^{\mu}Z_{\mu}-\dfrac{\lambda
}{4}(4vH^{3}+H^{4})
\label{completa}
\end{align}

Note que os quatro campos adicionais inseridos através de $\Phi$ foram
re-arranjados da seguinte forma: os três campos escalares $\phi_{1}$, $\phi_{2}$ e
$\phi_{4}$ desapareceram dando lugar aos três graus de liberdade que
surgem como as componentes longitudinais dos bósons de \textit{gauge}
massivos e o quarto campo escalar $\phi_{3}$ se tornou uma partícula
real e interagente, o bóson de Higgs $H$.

Os bósons de \textit{gauge} físicos $W^{\pm}$ e $Z$ adquirem massa 
\begin{equation}
M_{W}=\dfrac{gv}{2}\;\;\;\mbox{e}\;\;\;M_{Z}=(g^{2}+g^{\prime 2})^{1/2}\dfrac{v}{2},
\end{equation}
enquanto o fóton $A^{\mu}$ permanece sem massa como esperado.

Sabendo que $e=g\sin \theta_{W}=g^{\prime}\cos \theta_{W}$ obtemos a
relação $M_{W}=M_{Z}\cos \theta_{W}$ onde $\theta_{W}$ é o ângulo de
mistura eletrofraco.

Considerando resultado obtido em baixa energia e relacionando a
constante de acoplamento eletrofraco $g$ com a constante de Fermi
$G_{F}$
\begin{equation}
\dfrac{g}{2\sqrt{2}}\;=\;\left(\dfrac{M_{W}^{2}G_{F}}{\sqrt{2}}\right)^{1/2},
\end{equation}
obtém-se para o \emph{vev} $v$
\begin{equation}
 v\;=\;(\sqrt{2}G_{F})^{1/2} \approx 246\;\mbox{GeV}.
\label{vev2}
\end{equation}

Identificamos a massa do Higgs $M_{H}$ através da
expressão~\eqref{completa} reescrita como
\begin{equation}
-\dfrac{1}{2}(-2\mu^{2})H^{2},
\end{equation}
de onde obtemos
\begin{equation}
 M_{H}=\sqrt{-2\mu^{2}}=v\sqrt{2\lambda}.
\label{massa}
\end{equation}

Apesar de prever a existência do bóson de Higgs, o MP não consegue
explicar sua massa já que os parâmetros $\mu^{2}$ e $\lambda $ são
desconhecidos. Esses parâmetros só aparecem na teoria nos termos de
acoplamentos entre os bósons de Higgs e no seu termo de massa.

Já no setor fermiônico, a quebra de simetria se dá através da lagrangiana de Yukawa,
\[
{\cal L}_{Y}^{MP}=-g_{e}[(\bar{\psi}_{L}\Phi)\psi_{R}+ \bar{\psi}_{R}(\Phi^{\dagger})\psi_{L}]
= -\dfrac{g_{e}}{\sqrt{2}}(v+H)(\bar{\psi}_{L}\psi_{R}+
\bar{\psi}_{R}\psi_{L}),
\]
onde $g_{e}$ é o acoplamento de Yukawa para um determinado
férmion.

Um fato importante proveniente do mecanismo de Higgs é uma relação entre as massas do $W$ e do $Z$,
\begin{equation}
\rho \equiv \dfrac{M_{W}^{2}}{\cos^{2}{\theta_{W}}M_{Z}^{2}}.
\label{rho}
\end{equation}

No MP, em nível de árvore, esse parâmetro adquire o valor $\rho
=1$. Esse valor reflete uma simetria global aproximada e acidental do
MP, denominada \emph{simetria custodial} sobre a qual falaremos a
seguir. De modo geral, o parâmetro $\rho$ representa um ótimo teste
para a estrutura de isospin do setor de Higgs sendo sensível às
correções radiativas oriundas de contribuições de uma nova física.

\subsubsection{Simetria custodial}
\label{sec:simetriacustodial}

Seguindo o exposto na Ref.~\cite{willenbrock}, tomando como base o
dubleto de Higgs~\eqref{dubleto}, podemos construir um outro dubleto
de $SU(2)_{L}$ como
\begin{equation}
\phi^{c} = \left(\begin{array}{c}
	\phi^{0*} \\ -\phi^{-}
       \end{array}\right),
\label{dubleto2}
\end{equation}
onde $\phi^{-}=\phi^{+*}$.

Define-se assim um campo bi-dubleto (ou uma matriz de Higgs) tal que
\begin{equation}
\Phi =\dfrac{1}{\sqrt{2}}\left(\phi^{c},\phi \right) = \sqrt{\dfrac{1}{2}}\;\left( \begin{array}{cc}
\phi^{0*} & \phi^{+} \\
-\phi^{-} & \phi^{0}
\end{array} \right),
\label{bidubleto}
\end{equation}
que se transforma sob a simetria de isospin e hipercarga $SU(2)_{L}\otimes U(1)_{Y}$ como
\begin{align}
SU(2)_{L}&:\;\;\;\Phi \to  U_{L}\Phi \nonumber \\
U(1)_{Y}&:\;\;\;\Phi \to  \Phi \exp{(-\dfrac{i}{2}\tau_{3}\theta)}
\end{align}
sendo que o aparecimento de $\tau_{3}$ se deve aos valores opostos de hipercarga entre $\phi$ e $\phi^{c}$.

Com essa modificação podemos escrever a lagrangiana do setor de Higgs~\eqref{higgslagrangiana} em função do campo~\eqref{bidubleto} como
\begin{equation}
{\cal L}_{H} = \mbox{Tr}\left[ (D_{\mu}\Phi)^{\dagger}\;(D^{\mu}\Phi)\right] + \mu^{2}\mbox{Tr}[\Phi^{\dagger}\Phi] - \lambda(\mbox{Tr}[\Phi^{\dagger}\Phi])^{2},
\label{higgslagrangianamatricial}
\end{equation}
onde a derivada covariante atua sobre o bi-dubleto $\Phi$ seguindo
\begin{equation}
D_{\mu}\Phi = \partial_{\mu} \Phi + ig\dfrac{\tau^{a}}{2}W^{a}_{\mu}\Phi -i\dfrac{g'}{2}B_{\mu}\Phi \tau_{3}\;\;\;(a=1,2,3),
\label{covariantecustodial}
\end{equation}
sendo que mais uma vez a matriz $\tau_{3}$ é necessária, já que os
campos~\eqref{dubleto} e~\eqref{dubleto2} possuem valores opostos de
hipercarga.

Para manifestar a simetria global aproximada considere o limite de
hipercarga nula, $g^{'}=0$. Desta forma, a derivada covariante~\eqref{covariantecustodial} resulta em
\begin{equation}
D_{\mu}\Phi = \partial_{\mu} \Phi + ig\dfrac{\tau^{a}}{2}W^{a}_{\mu}\Phi \;\;\;(a=1,2,3).
\label{covariantecustodial2}
\end{equation}

Nota-se que nesse caso a derivada covariante~\eqref{covariantecustodial2} é simétrica sob transformações de $SU(2)_{R}$, pois
\begin{equation}
SU(2)_{R}:\;\;\;\Phi \to  \Phi U_{R}^{\dagger}
\end{equation}
e daí concluímos que no limite $g^{'}=0$ o setor de Higgs do MP
apresenta uma simetria global $SU(2)_{L}\otimes SU(2)_{R}$, onde
$SU(2)_{L}$ é a versão global da simetria de \textit{gauge} e
$SU(2)_{R}$ é uma simetria global aproximada. Na verdade $U(1)_{Y}$ é um subgrupo de $SU(2)_{R}$~\cite{willenbrock}.

O bi-dubleto de Higgs~\eqref{bidubleto} se transforma sob a simetria global $SU(2)_{L}\otimes SU(2)_{R}$ da seguinte forma
\begin{equation}
SU(2)_{L}\otimes SU(2)_{R}:\;\;\;\Phi \to  U_{L}\Phi U_{R}^{\dagger}.
\end{equation}

Com a quebra de simetria, o bi-dubleto de Higgs adquire um \emph{vev}, em analogia com a Eq.~\eqref{vev}, dado por
\begin{equation}
\langle \Phi_{0} \rangle =\dfrac{1}{2}\;\left( \begin{array}{cc}
v & 0 \\
0 & v
\end{array} \right).
\label{vevbidubleto}
\end{equation}

O \emph{vev}~\eqref{vevbidubleto} quebra as simetrias $SU(2)_{L}$ e $SU(2)_{R}$ separadamente, já que
\begin{equation}
U_{L}\langle \Phi_{0} \rangle \neq \langle \Phi_{0} \rangle \;\;\;\mbox{e}\;\;\;\langle \Phi_{0} \rangle U_{R}^{\dagger} \neq \langle \Phi_{0} \rangle,
\end{equation}
mas mantém intacto o subgrupo $SU(2)_{L+R}$, que corresponde às
transformaçoes sob $SU(2)_{L}$ e $SU(2)_{R}$ simultaneamente,
supondo-se que $U_{L}=U_{R}$. Ou seja,
\begin{equation}
U_{L}\langle \Phi_{0} \rangle U_{L}^{\dagger} = \langle \Phi_{0} \rangle.
\end{equation}

Assim temos a quebra de simetria
\begin{equation}
SU(2)_{L}\otimes SU(2)_{R} \rightarrow SU(2)_{L+R}
\end{equation}
onde $SU(2)_{L+R}$ é conhecida como simetria custodial.

Os geradores $W_{\mu}^{a}$ ($a=1,2,3$) se transformam como
um tripleto sob transformações globais do grupo $SU(2)_{L}$ e como um singleto sob transformações do grupo $SU(2)_{R}$,
e sendo assim se transformam como um tripleto sob
$SU(2)_{L+R}$~\cite{willenbrock}. Disso resulta que, no limite
$g^{'}=0$, os bósons de \textit{gauge} $W^{\pm}$ e $Z$ formam um
tripleto de uma simetria global não-quebrada e portanto terão a mesma
massa. Quando igualamos os termos de massa obtemos de acordo
com~\eqref{rotacao}
\begin{equation}
\dfrac{1}{2}M_{Z}^{2}Z_{\mu}Z^{\mu} = \dfrac{1}{2}M_{Z}^{2}(\cos{\theta_{W}}W_{\mu}^{3}-\sin{\theta_{W}}B_{\mu})^{2}=\dfrac{1}{2}M_{Z}^{2}\cos^{2}{\theta_{W}}W_{\mu}^{3}W^{3\mu}+\cdots
\label{massaw3}
\end{equation}

Sendo assim, observando a Eq.~\eqref{massaw3} obtida pela imposição da simetria custodial, e lembrando que nesse caso todos os bósons têm a mesma massa, obtemos
\begin{equation}
M_{W}^{2}=M_{Z}^{2}\cos^{2}{\theta_{W}}\;\;\Longrightarrow \;\;\rho =1.
\end{equation}

No entanto o MP quebra explicitamente a simetria custodial, já que
necessariamente $g^{'}\neq 0$, e por isso aparecerão correções no
parâmetro $\rho $, tais que $\hat{\rho}=1+\delta \rho$, provenientes de
\emph{loops} do bóson de Higgs e de quarks pesados, dadas de acordo com a Ref.~\cite{willenbrock} por
\begin{align}
\delta \rho_{Higgs}& =  -\dfrac{11G_{F}M_{Z}^{2}\sin^{2}{\theta_{W}}}{24\pi^{2}\sqrt{2}}\ln{\dfrac{M_{h}^{2}}{M_{Z}^{2}}}
\nonumber \\
\delta \rho_{quarks}& =  +\dfrac{3G_{F}}{8\pi^{2}\sqrt{2}}\left( m_{t}^{2}+ m_{b}^{2}-2\dfrac{m_{t}^{2}m_{b}^{2}}{m_{t}^{2}-m_{b}^{2}}\ln{\dfrac{m_{t}^{2}}{m_{b}^{2}}}  \right).
\label{correcoes}
\end{align}

Note que a simetria ``custodial'' protege (ou guarda) a relação em
nível de árvore $\rho =1$ de correções radiativas, já que $\delta
\rho_{Higgs} \to 0$ no limite $g^{'}\to 0\;(\sin^{2}{\theta_{W}}\to
0)$. Isso é essencial, pois qualquer proposta de nova física, sendo
uma extensão do MP, deve possuir a simetria custodial de um modo
aproximado, satisfazendo os vínculos experimentais obtidos por medidas
de precisão eletrofracas. O valor experimental atual para o parâmetro
$\rho $ apresentado pelo Particle Data Group (PDG)~\cite{pdg} é
\[
\rho_{exp}=1.0008_{-0.0007}^{+0.0017}
\]
que está bem de acordo com o MP.

Apesar da simetria custodial ser necessária, esse não é exatamente o
caso do Higgs padrão, já que pode-se desenvolver uma teoria de campo
efetiva que quebra a simetria eletrofraca mantendo a simetria
custodial. Um modo de fazer isso será tratado no
Cap.~\ref{cap:efetivas}.

\section{O Bóson de Higgs do Modelo Padrão}

Vimos que o MP unificou com bastante sucesso as interações
eletromagnética e fraca. No entanto, seja como for, a existência de um
setor de Higgs nesse modelo é essencial para a quebra de simetria
e geração de massa de suas partículas. 

Diferentes cenários foram propostos para explicar a dinâmica do
mecanismo de quebra de simetria eletrofraca. Para uma revisão sobre
essas abordagens indicamos a Ref.~\cite{bhatta}. O modo mais simples de
implementar a quebra espontânea de simetria é através da introdução de
campos bosônicos escalares na teoria, como foi apresentado na
Sec.~\ref{sec:higgs}.

Quando o mecanismo de Higgs é acionado, três bósons escalares são
absorvidos e se tornam as componentes longitudinais dos bósons
vetoriais massivos, enquanto que o bóson de Higgs permanece no espectro do modelo como um grau de liberdade físico. Esta partícula escalar se acopla com todas
as partículas do MP que adquirem massa através do mecanismo de quebra
espontânea de simetria. Ao leitor interessado recomendamos consultar as Refs.~\cite{modelo} para observar a forma completa da
lagrangiana do MP após o mecanismo de Higgs. Além disso, a fenomenologia relacionada ao bóson
de Higgs foi objeto de análise em diversos trabalhos~\cite{higgsfeno}.

 Na Tabela~\ref{acop} esquematizamos a intensidade dos acoplamentos
 entre as partículas do MP e o Higgs, conforme apresentado na
 Ref.~\cite{novaes}. É importante observar que esses acoplamentos são
 diretamente proporcionais à massa das partículas.  {\footnotesize
\begin{table}[ht]
\centering
\begin{tabular}{cc}
\hline \hline
& \\
Acoplamento & Intensidade \\
& \\
\hline
\\
$Hf\bar{f}$ & $M_{f}/v$ \\
& \\
$HW^{+}W^{-}$ & $2M_{W}^{2}/v$ \\
& \\
$HZ^{0}Z^{0}$ & $M_{Z}^{2}/v$ \\
& \\
$HHW^{+}W^{-}$ & $M_{W}^{2}/v^{2}$ \\
& \\
$HHZ^{0}Z^{0}$ & $M_{Z}^{2}/v^{2}$ \\
& \\
$HHH$ & $M_{H}^{2}/2v$ \\
& \\
$HHHH$ & $M_{H}^{2}/8v^{2}$\\
& \\
\hline \hline
\end{tabular}
\vspace{0.3cm}
\caption{Acoplamentos do Higgs com as partículas do MP. Maiores informações na Ref.~\cite{modelo}.}
\label{acop}
\end{table}}

O MP não especifica a massa do Higgs. Contudo a consistência deste modelo
através de medidas de precisão eletrofraca impõe limites
teóricos à sua massa como veremos adiante. Além disto, a procura direta pelo Higgs no colisor LEP estabeleceu
por muito tempo que $M_H$ fosse aproximadamente maior que 114 GeV.

\subsection{Limites teóricos à $M_{H}$}
\label{vinculos_teo}

Existem diversas análises teóricas que impõem vínculos à massa do
Higgs. Um desses vínculos está relacionado à unitaridade das ondas
parciais nos processos de espalhamento de bósons de \textit{gauge}
longitudinalmente polarizados. Em outras palavras, um processo de
espalhamento não pode dar probabilidade maior do que 1.
 
Os autores da Ref.~\cite{lee} mostraram que a unitaridade é respeitada
caso a massa do bóson de Higgs não exceda um valor crítico
\begin{equation}
 M_{H}\,\leq M_{c}=(\dfrac{8\pi\sqrt{2}}{3G_{F}})^{1/2} \approx 1\,\mbox{TeV},
\label{lee}
\end{equation}

É interessante observar que dentro do MP sem o Higgs (ou com o Higgs
mas supondo $M_H \to \infty$), a amplitude de espalhamento cresce
indefinidamente com a energia, ou seja as interações fracas se tornam
fortes na escala de altas energias eventualmente violando a
unitaridade da matriz $S$, o que não faz sentido fisicamente. 

O bóson
de Higgs relativamente leve resolve este problema. De fato, a estrutura de \textit{gauge} do MP eletrofraco possibilita 
cancelamentos importantes entre as amplitudes de espalhamento dos
bósons de \textit{gauge} longitudinalmente polarizados em altas
energias, onde o bóson de Higgs desempenha papel
essencial. 

\begin{figure}[!ht]
\centering
\includegraphics[scale=0.6]{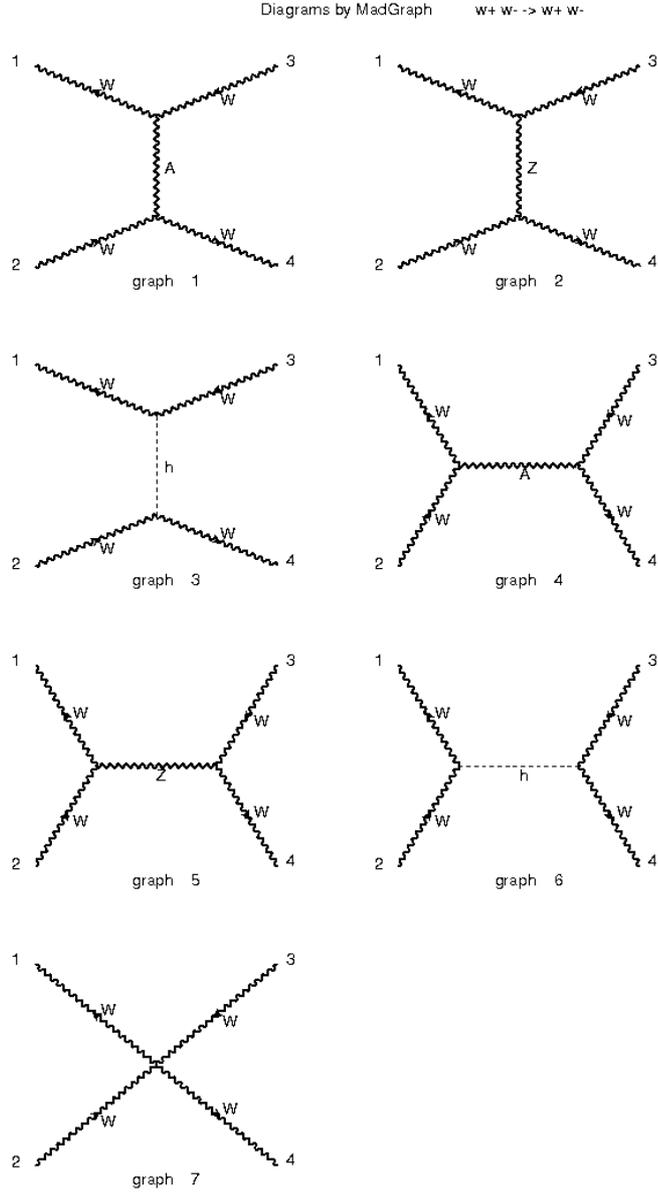}
\caption{Diagramas de Feynman, em nível de árvore, para o espalhamento
 $W^{+}+ W^{-}\to W^{+}+W^{-}$.}
\label{espalhamento_ww}
\end{figure}

Consideremos o processo de espalhamento entre bósons de
\textit{gauge}~\cite{proceedings}
\begin{equation}
W^{+}(k_{+})+W^{-}(k_{-})\to W^{+}(q_{+})+W^{-}(q_{-})
\end{equation}
no referencial do centro de massa (CM), com os bósons $W^{\pm}$ na direção do eixo $z$ (polarizados longitudinalmente) cujos diagramas de Feynman esboçamos na Fig.~\ref{espalhamento_ww}.

Neste referencial podemos escrever os momentos inicial e final tais que
\begin{align}
k_{\pm}&=(E,0,0,\pm\,k_{z})\;\;\mbox{e}\nonumber \\
q_{\pm}&=(E,0,\pm\,k_{z}\,\sin{\theta},\pm\,k_{z}\,\cos{\theta}),
\end{align}
onde $E^{2}=M^{2}_{W}+\mathbf{k}^{2}$ e $\theta$ é o ângulo de
espalhamento no CM.

Considerando o comportamento do espalhamento dos bósons longitudinais em altas energias, os vetores de polarização longitudinal dos bósons $W^{\pm}$ são
\begin{align}
\epsilon_{L}(k_{\pm})&=\left(\dfrac{k}{M_{W}},0,0,\pm\,\dfrac{E}{M_{W}}\right),\nonumber \\
\epsilon_{L}(q_{\pm})&=\left(\dfrac{k}{M_{W}},0,\pm\,\dfrac{E}{M_{W}}\sin{\theta},\pm\,\dfrac{E}{M_{W}}\cos{\theta}\right),
\end{align}
satisfazendo a condição de Lorentz $\epsilon(q)\cdot q=0$ e normalizados tais que $\epsilon^{2}=-1$.

No limite de altas energias, a amplitude de
espalhamento envolvendo os vértices tríplices com o fóton $A$ e bóson
neutro $Z$ (ver diagramas 1, 2, 4 e 5 na
Fig.~\ref{espalhamento_ww}), considerando termos que crescem com
fatores de $\dfrac{q^{2}}{M_{W}^{2}}$, é tal que
\begin{equation}
{\cal M}_{1,2,4,5}=g^{2}\left[\dfrac{q^{2}}{M_{W}^{2}}\left(\dfrac{9}{2}-\dfrac{11}{2}\cos{\theta}-2\cos^{2}{\theta}\right)+\dfrac{q^{4}}{M_{W}^{4}}\left(3-6\cos{\theta}-\cos^{2}{\theta}\right)\right].
\end{equation}

Do mesmo modo, o diagrama com vértice quártico contribui com o termo (ver diagrama 7 na Fig.~\ref{espalhamento_ww})
\begin{equation}
{\cal M}_{7}=g^{2}\left[\dfrac{q^{2}}{M_{W}^{2}}\left(-4+6\cos{\theta}+2\cos^{2}{\theta}\right)+\dfrac{q^{4}}{M_{W}^{4}}\left(-3+6\cos{\theta}+\cos^{2}{\theta}\right)\right].
\end{equation}

E finalmente, os diagramas que envolvem troca dos bósons de Higgs fornecem (ver diagramas 3 e 6 na Fig.~\ref{espalhamento_ww})
\begin{equation}
{\cal M}_{3,6}=g^{2}\left[\dfrac{q^{2}}{M_{W}^{2}}\left(-\dfrac{1}{2}-\dfrac{1}{2}\cos{\theta}\right)-\dfrac{M_{H}^{2}}{4M_{W}^{4}}\left(\dfrac{s}{s-M_{H}^{2}}+\dfrac{t}{t-M_{H}^{2}}\right)\right],
\end{equation}
onde $s$ e $t$ são as variáveis de Mandelstam, definidas como
\begin{align}
s&=(k_{+}+k_{-})^{2}=(q_{+}+q_{-})^{2}=(2E)^{2}=E_{CM}^{2}\nonumber \\
t&=(k_{+}-q_{+})^{2}=(k_{-}-q_{-})^{2}=2k^{2}(1-\cos{\theta}).
\end{align}

Note então que podemos escrever de forma geral que
\begin{equation}
{\cal M}(s,t)=A\left(\dfrac{q^{2}}{M_{W}^{2}}\right)^{2}+B\left(\dfrac{q^{2}}{M_{W}^{2}}\right)+C
\end{equation}
onde, somando-se todas as contribuições, o termo $A$ se cancela sem a
necessidade dos diagramas do Higgs.

No entanto, o cancelamento completo do termo $B$ envolve de forma
direta o bóson de Higgs sobrando a amplitude
\begin{align}
{\cal M}(s,t)&=-g^{2}\dfrac{M_{H}^{2}}{4M_{W}^{4}}\left(\dfrac{s}{s-M_{H}^{2}}+\dfrac{t}{t-M_{H}^{2}}\right)\nonumber \\
&\equiv  -4\left(\lambda +\dfrac{(\lambda v)^{2}}{s-2\lambda
	     v^{2}}+\dfrac{(\lambda v)^{2}}{t-2\lambda v^{2}}\right),
\label{amp1}
\end{align}
usando que $M_{H}=\sqrt{2}\mu =\sqrt{2\lambda}v$ e $M_{W}=\dfrac{g}{2}v$.

Então, apesar da amplitude de espalhamento $WW$ não divergir em altas
energias, ainda é necessário verificar se ela satisfaz o limite de
unitaridade para a matriz de espalhamento. Nesse caso, é conveniente
analisar o comportamento fazendo a expansão da amplitude ${\cal M}$ em
ondas parciais decompondo as amplitudes de acordo com as
contribuições dos valores do momento angular total $J$ tal que
\begin{equation}
{\cal M}(s,t)=16\pi\,\sum_{J}(2J+1)a_{J}(s)P_{J}(\cos{\theta}),
\end{equation}
onde $P_{J}$ são os polinômios de Legendre ($P_{0}(x)=1$, $P_{1}(x)=x$, $P_{2}(x)=\dfrac{1}{2}(3x^{2}-1)$, $\cdots$ ) com $\cos{\theta}=1+\dfrac{2t}{s}$ em altas energias.

Nesse contexto, a seção de choque diferencial no limite de massa nula é
dada por
\begin{equation}
\dfrac{d\sigma}{d\Omega}=\dfrac{1}{64\pi^{2}s}|{\cal M}(s,t)|^{2},
\end{equation}
de onde resulta, usando a propriedade de ortogonalidade dos polinômios de
Legendre
\begin{equation}
\int^{1}_{-1}\,dx\,P_{J}(x)P_{K}(x)=\delta_{JK}\,\dfrac{2}{2J+1},
\end{equation}
a seção de choque total na forma
\begin{equation}
\sigma = 16\pi\,\sum_{J}(2J+1)|a_{J}(s)|^{2}.
\end{equation}

Usando o Teorema Ótico,
\begin{equation}
\sigma = \dfrac{1}{s}\mbox{Im}\,T(s,0),
\end{equation}
temos que
\begin{equation}
|a_{J}|^{2}=\mbox{Im}\,a_{J} \leq 1.
\end{equation}

Como $J$ é conservado, o módulo de cada amplitude de onda parcial não pode exceder seu valor inicial, que para um estado de onda plana incidente é igual à 1, e então o resultado para a amplitude de onda parcial para $J=0$ será
\[
a_{0}(s)=-\dfrac{G_{F}M_{H}^{2}}{8\pi\sqrt{2}}\left[2+\dfrac{M_{H}^{2}}{s-M_{H}^{2}}-\dfrac{M_{H}^{2}}{s}\ln{\left(1+\dfrac{s}{M_{H}^{2}}\right)}\right]
\]
o que leva a $a_0  \approx  -\dfrac{G_{F}M_{H}^{2}}{4\pi\sqrt{2}}$,
no limite de energia $\sqrt{s}\gg M_{H}$.

Como a unitaridade é violada para $|a_{0}|>1$, temos então que
\begin{equation}
M_{H}<\left(\dfrac{4\pi\sqrt{2}}{G_{F}}\right)^{1/2} .
\end{equation}

Esse resultado difere um pouco da expressão~\eqref{lee} pois nela
foram considerados vários canais de espalhamento elástico entre os
diferentes bósons de \textit{gauge}.

\begin{figure}[!ht]
\centering
\subfloat[ ]{\includegraphics[scale=0.55]{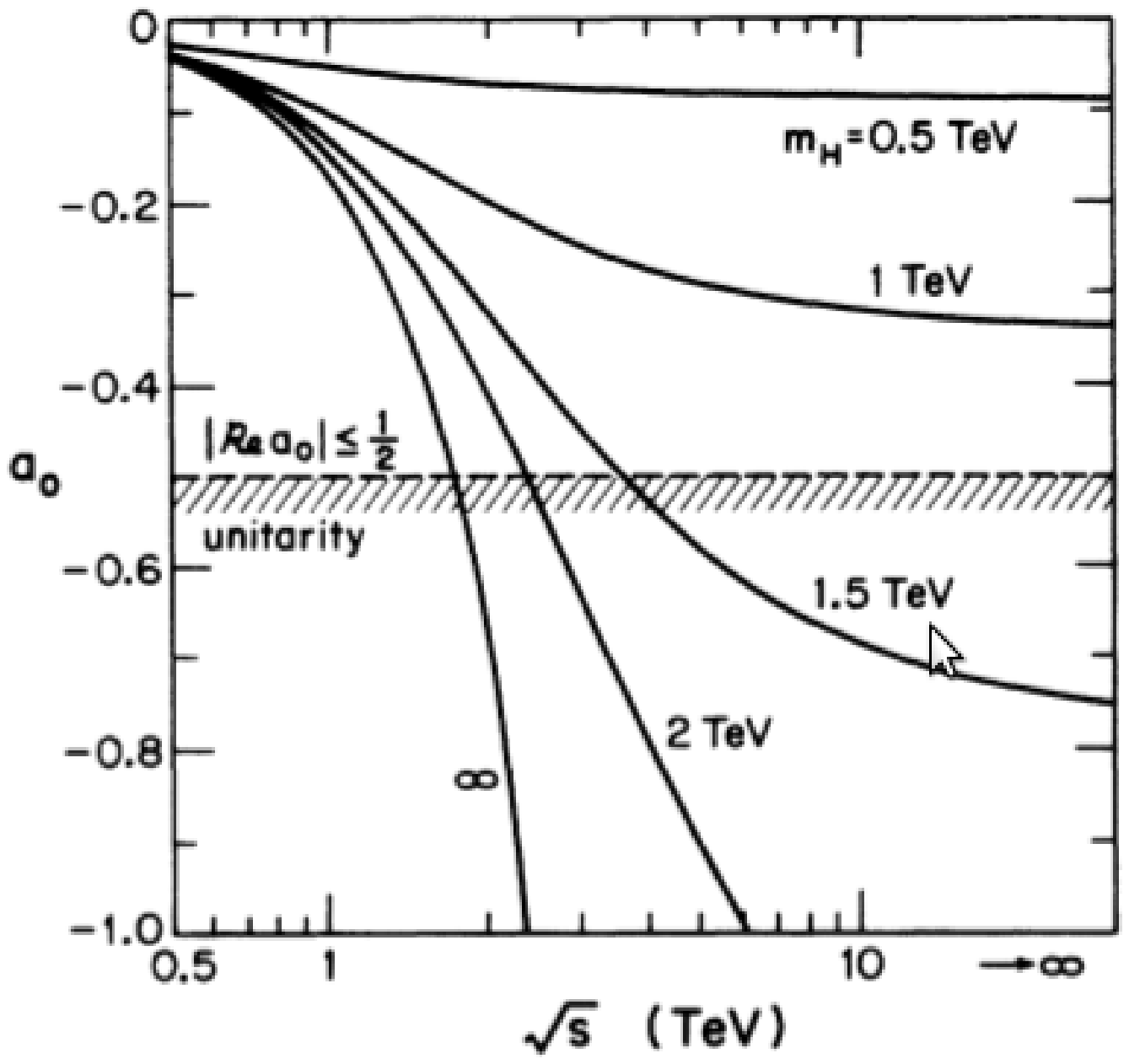}\label{unitarizacao}}
\hspace{0.4cm}
\subfloat[ ]{\includegraphics[scale=0.55]{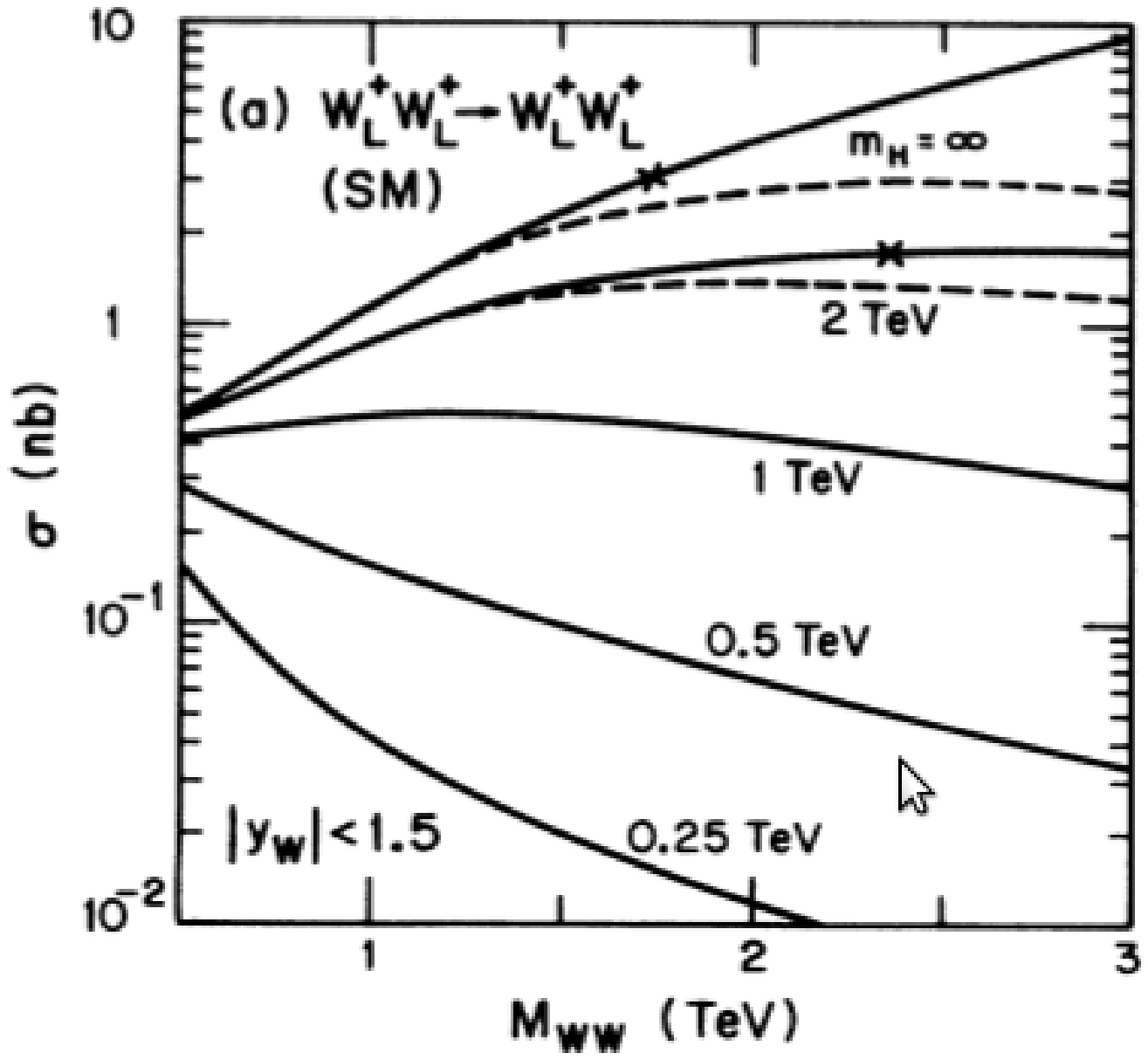}\label{unitarizacao2}}
\caption{Em (a) vemos que a restrição $|\mbox{Re}a_{0}|\leq 1/2$, que
  garante a unitaridade, é mantida para $M_{H}=1$ TeV. Em (b)
  temos a seção de choque versus $\sqrt{s}=M_{WW}$ para vários valores
  de $M_{H}$. A curva contínua para $M_{H}=\infty$ mostra o caso onde
  a condição de unitaridade $|\mbox{Re}a_{0}|\leq 1/2$ é violada e a
  curva pontilhada que a acompanha se refere à unitariedade
  restabelecida. Note o valor $M_{WW}\approx 1.25$ TeV no qual as duas
  curvas bifurcam. Figuras obtidas da Ref.~\cite{barger2}.}
\label{unit}
\end{figure}

Uma outra análise realizada na Ref.~\cite{barger2} mostra que, se $M_H$
for menor que aproximadamente 1 TeV, a unitariedade é assegurada para
todos os valores de energia no CM, conforme ilustra a
Fig.~\ref{unitarizacao}. Já a Fig.~\ref{unitarizacao2} exibe o comportamente da seção
de choque do espalhamento $W^+_L W^+_L \to W^+_L W^+_L$ em função
da massa invariante $M_{WW}$. Observa-se que na ausência do Higgs (o
que equivale a $M_H \to \infty$), a unitariedade não é violada para $M_{WW}\;<\;1.25$ TeV.

Outra forma teórica de impor um limite à massa do Higgs é através da
análise da trivialidade do potencial $V(\Phi)$. A equação do grupo de
renormalização, em nível de um \emph{loop}, para a constante de acoplamento
$\lambda$ do potencial puro (desconsiderando os outros acoplamentos na
lagrangiana) $\lambda \Phi^{4}$ da Eq.~\eqref{potencial} tem, de acordo com a Ref.~\cite{campos}, a forma
\begin{equation}
\dfrac{d\bar{\lambda}}{d\log{(Q^{2}/\mu^{2})}}=\dfrac{1}{16\pi^{2}}\,(12\bar{\lambda}^{2}),\;\;\;\;\mbox{com}\;\;\;\;\bar{\lambda}(\mu^{2};\lambda)=\lambda.
\label{eqrenor}
\end{equation}

Integrando a Eq.~\eqref{eqrenor} obtemos
\begin{equation}
 \left(\dfrac{3}{4\pi^{2}}\right)^{-1}\,\left[\dfrac{1}{\lambda}-\dfrac{1}{\bar{\lambda}}\right]=\log{\left(\dfrac{Q^{2}}{\mu^{2}}\right)},
\end{equation}
de onde resulta
\begin{equation}
\bar{\lambda}(Q^{2})=\dfrac{\lambda}{1-\left(\dfrac{3\lambda}{4\pi^{2}}\right)\log{\left(\dfrac{Q^{2}}{\mu^{2}}\right)}}.
\label{lambda_bar}
\end{equation}

Como a estabilidade do potencial de Higgs requer
$\bar{\lambda}(Q^{2})\geq 0$ então pode-se escrever que
\begin{equation}
\lambda(\mu^{2})\leq \dfrac{4\pi^{2}}{3\log{\left(\dfrac{Q^{2}}{\mu^{2}}\right)}}.
\label{lambda}
\end{equation}

De acordo com~\eqref{lambda}, para grandes valores de $Q^{2}$, temos $\lambda(\mu^{2})\to 0$ e a teoria se torna trivial (ou seja não interagente).

Escrevendo~\eqref{lambda} como $Q^{2}\leq \mu^{2}\exp{\left(\dfrac{4\pi^{2}}{3\lambda(\mu^{2})}\right)}$ e considerando a escala $\mu^{2}=M_{H}^{2}$, sabendo que a relação entre $M_{H}$ e $\lambda$ é dada por~\eqref{massa}, então
\begin{equation}
Q^{2}\leq M_{H}^{2}\exp{\left(\dfrac{8\pi^{2}v^{2}}{3M_{H}^{2}}\right)}.
\end{equation}

Assim, temos um limite superior $Q^{2}=\Lambda^{2}$, dado um valor para a
massa do bóson de Higgs, até o qual o MP é válido.
\begin{figure}[!ht]
\centering
\includegraphics[scale=0.5]{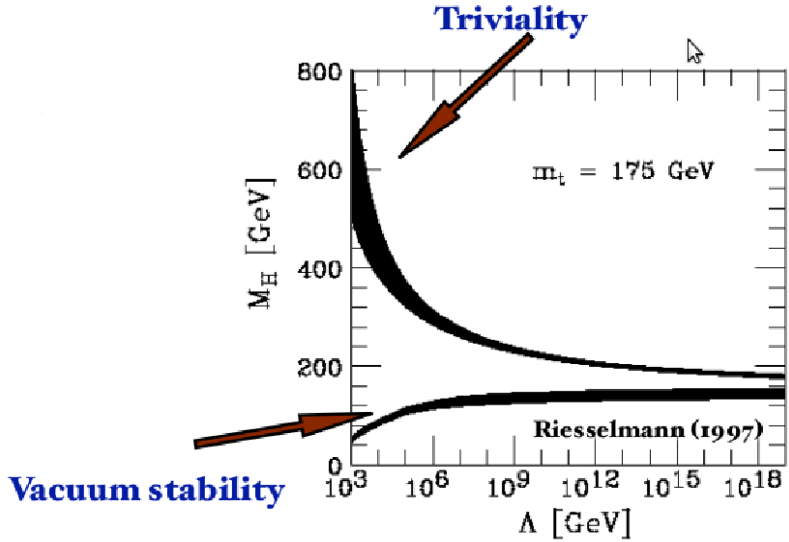}
\caption{Limites teóricos para massa do Higgs de acordo com a
  estabilidade do vácuo (curva inferior) e com a trivialidade do
  potencial escalar (mede a consistência da teoria eletrofraca até a
  escala de energia $\Lambda$; curva superior). As áreas sólidas
  expressam as incertezas nos limites. Ver detalhes de cálculo na Ref.~\cite{riesselmann}.}
\label{trivialidade}
\end{figure}

\begin{figure}[!ht]
\centering
\includegraphics[scale=1]{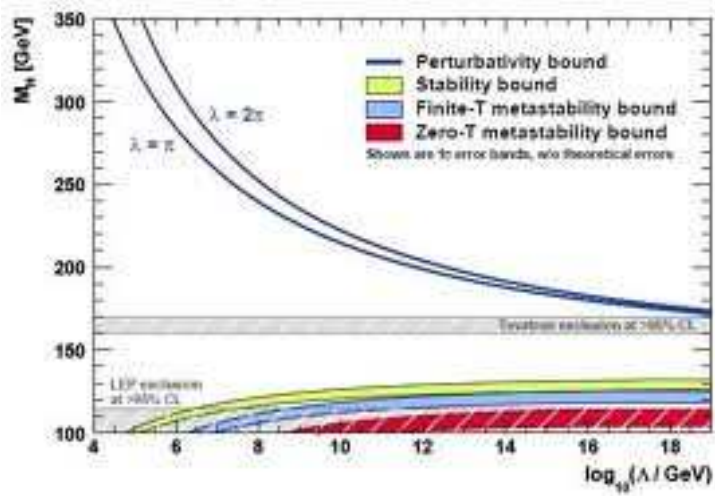}
\caption{O limite de trivialidade é dado para $\lambda = \pi$ e
  $\lambda = 2\pi$. A diferença entre ambas indica a magnitude das
  incertezas teóricas nesse limite. O limite de estabilidade do vácuo
  está em verde-limão. Ver detalhes de cálculo na Ref.~\cite{ellis}.}
\label{trivialidade2}
\end{figure}

Se esperarmos que o MP seja válido até uma escala de energia de
$\Lambda_{GUT}\approx 10^{16}\,\mbox{GeV}$ (escala de Planck) então a
massa do Higgs deve se encontrar no intervalo $130\,\mbox{GeV}\lesssim
\,M_{H}\,\lesssim 200\,\mbox{GeV}$, como pode ser verificado na
Fig.~\ref{trivialidade}. Caso sua massa seja de aproximadamente 500 GeV, verifica-se que a teoria será válida até 1 TeV. Uma análise mais
recente encontra-se na Fig.~\ref{trivialidade2}. 

\subsection{Modos de decaimento do Higgs}

Embora o MP não preveja a massa do bóson de Higgs, ele fornece todos
os acoplamentos dessa partícula escalar com o restante do conteúdo do
modelo. Sendo assim, o MP é capaz de prever todos os modos de
decaimento e quantificá-los através do cálculo da largura de
decaimento $\Gamma_H$, desde que $M_H$ seja fornecida. 

\begin{figure}[!ht]
\centering
\includegraphics[scale=1]{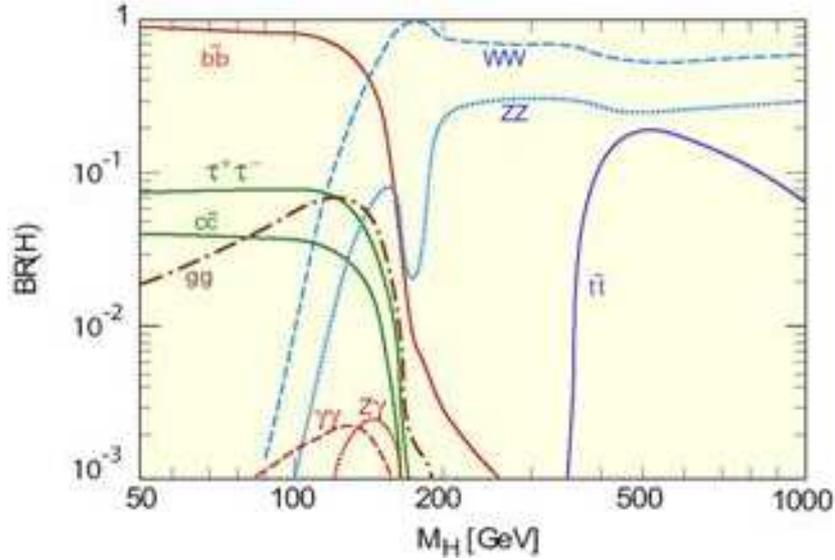}
\caption{Canais de decaimento do Higgs e taxas de ramificação (ou
  \emph{branching ratios}) em função de sua massa. Figura extraída da Ref.~\cite{proceedings}.}
\label{higgs_br}
\end{figure}

A Fig.~\ref{higgs_br} mostra as taxas de ramificação (BR, do inglês
\textit{branching ratio}) para o decaimento do Higgs em função de sua
massa. Para massas intermediárias, de 114 GeV (limite de exclusão
imposto pelo LEP2) até aproximadamente 130 GeV, o decaimento é dominado
completamente pelo canal $b\bar{b}$. Para o segundo maior valor de BR
existe uma certa competição entre vários modos, a saber $\tau
\bar{\tau}$, $c \bar{c}$, $WW^\ast$ e $gg$. É interessante observar que, embora ocorra em nível de \textit{loop}, o canal $gg$ é bastante significativo. Finalmente, embora seja relativamente pequeno, devemos destacar
os modos de decaimento $ZZ^\ast$ e em dois fótons $\gamma \gamma$. Nesta janela de massa a largura de decaimento do Higgs é $\Gamma (H)\leq 10$ MeV, ou seja por ser bem pequena o sinal do Higgs exibe nesse canal um pico bem pronunciado.

Para $M_H \approx 2 M_Z$ a $2 m_t$, os modos de decaimento têm como
estados finais pares de $W$'s ou $Z$'s reais. Para massas um pouco
acima de $2 m_t$, o modo $t\bar{t}$ tem certa importância mas os
modos dominantes continuam sendo os decaimentos em pares de bósons de
\textit{gauge}.

\subsection{Mecanismos de produção do Higgs no LHC}

\begin{figure}[!ht]
\centering
\includegraphics[scale=0.3]{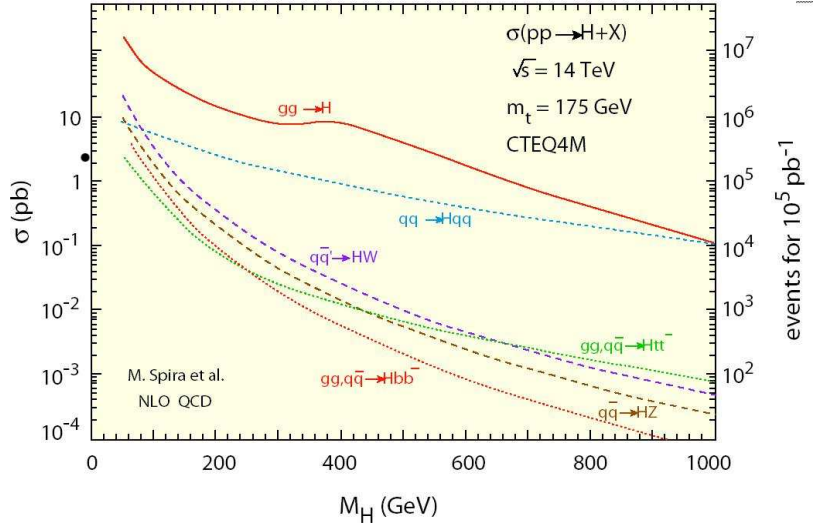}
\caption{Seções de choque dos mecanismos de produção do Higgs padrão
  no LHC em função de sua massa. Nota-se a predominância do processo
  de fusão de glúons $gg\to H$. Figura obtida da Ref.~\cite{spira}.}
\label{higgs_prod}
\end{figure}

Assim como nos modos de decaimento, os processos de produção do Higgs no LHC 
dependem crucialmente do valor de sua massa como mostra a Fig.~\ref{higgs_prod}.

\begin{figure}[!ht]
\centering
\includegraphics[scale=0.8]{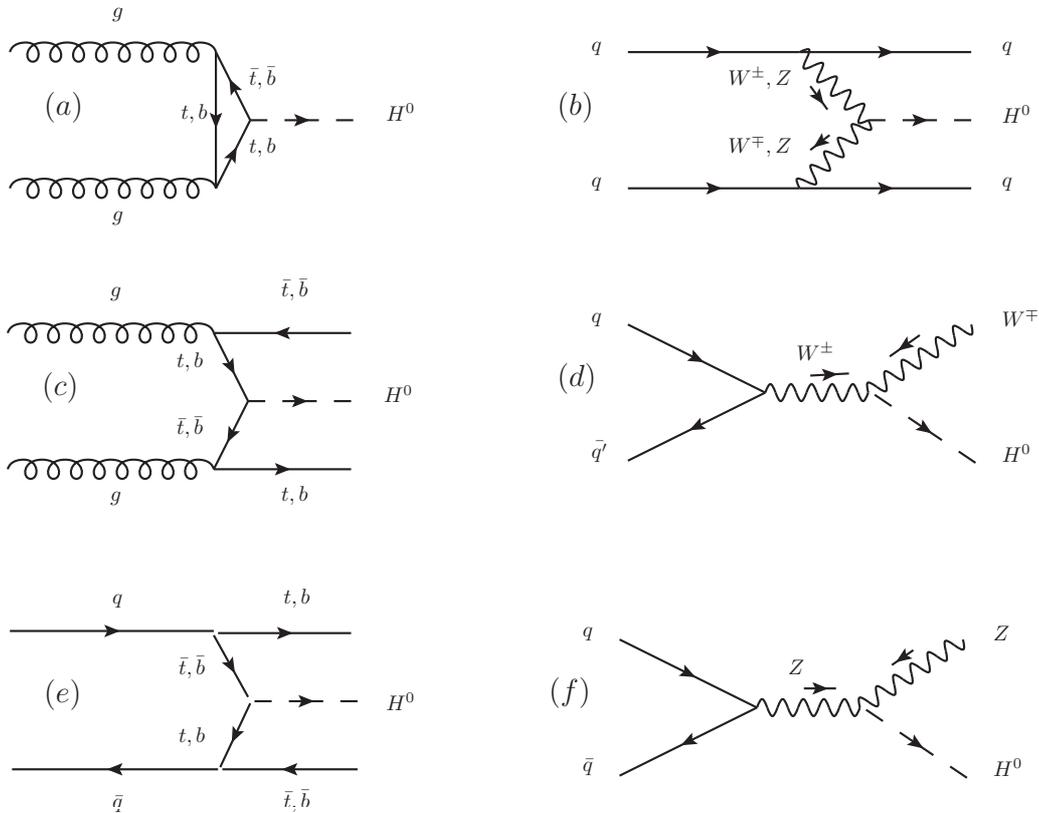}
\caption{Os processos mais importantes de produção do Higgs em
  colisores hadrônicos. Em (a) temos a fusão de glúons; em (b) temos
  fusão de bósons vetoriais; em (c) e (e) temos fusão de quarks
  \emph{top} ou \emph{bottom}; em (d) e (f) temos a produção associada
  com bósons $W^{\pm}$ e $Z$ respectivamente.}
\label{higgsproducao}
\end{figure}

A Fig.~\ref{higgs_prod} mostra claramente que o canal de produção
mais relevante é a fusão de glúons~\cite{gluons} $gg\to H$,
possibilidade essa que se estende por ampla faixa de valores da massa
do Higgs. Este processo, que se dá ao nível de \textit{loop} (veja
Fig.~\ref{higgsproducao}a), é totalmente dominado pelo quark $t$ tendo em vista o forte
acoplamento do Higgs com este quark pesado\footnote{Lembrar que o
  Higgs se acopla com os férmions proporcionalmente às suas
  respectivas massas.}.

O canal de produção através de VBF $qq\to qqH$ se torna competitivo para valores maiores de massa do
Higgs~\cite{vbfhiggs}. Em particular, esse canal fornece uma perspectiva
interessante quando o Higgs decai em pares de bósons e estes decaem
leptonicamente (conhecido como modo \emph{gold-plated}). Esse modo possui baixa atividade de jatos na região central do detetor,
devido à ausência de troca de cor nessa região, possibilitando eventos menos contaminados pelo \textit{background} de QCD. 

\subsection{Limites experimentais}
\label{sec:limiteexperimental}

No período de 2010 a 2011 o LHC operou com a energia no centro de
massa de 7 TeV (configuração denominada nessa tese como LHC7), metade do seu valor projetado,
disponibilizando uma luminosidade de aproximadamente 5 fb$^{-1}$ para cada experimento ATLAS e CMS. Esses dados permitiram a procura pelos
sinais do bóson de Higgs através de diferentes canais na janela de massa do Higgs de 114 GeV (limite inferior imposto pelo
colisor LEP2) até 600 GeV~\cite{higgs_results}.

Conforme mostrado nas Figs.~\ref{limites}, resultados iniciais produziram
um limite de exclusão de 95\% C.L. para massas na
região 127 GeV até 600 GeV e de 99\% CL para massas na região 128 GeV
até 525 GeV. No entanto, de acordo com a Fig.~\ref{limite2}, um
excesso de eventos em relação ao previsto pelo MP já surgia na região
entre 115 GeV e 127 GeV, permitindo aos experimentais em princípio não
excluir essa região de massa do Higgs em 95\% CL.
\begin{figure}[!ht]
\centering
\includegraphics[scale=0.35]{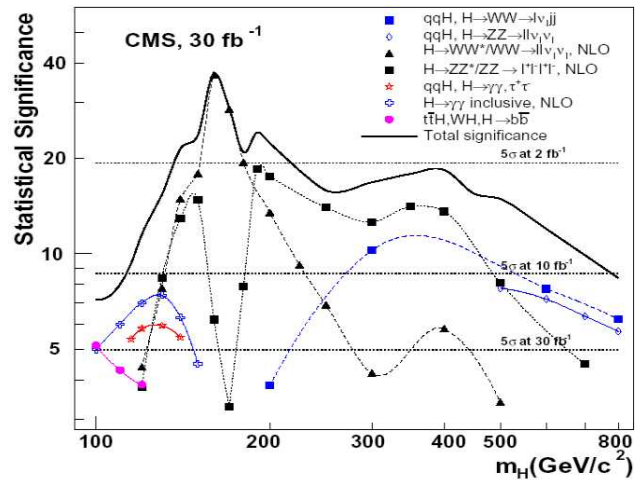}
\caption{Significância Estatística esperada para luminosidade de $30\;\mbox{fb}^{-1}$ no CMS. Detalhes na Ref.~\cite{cms}.}
\label{significance}
\end{figure}

\begin{figure}[!ht]
\centering
\subfloat[ ]{\includegraphics[scale=0.25]{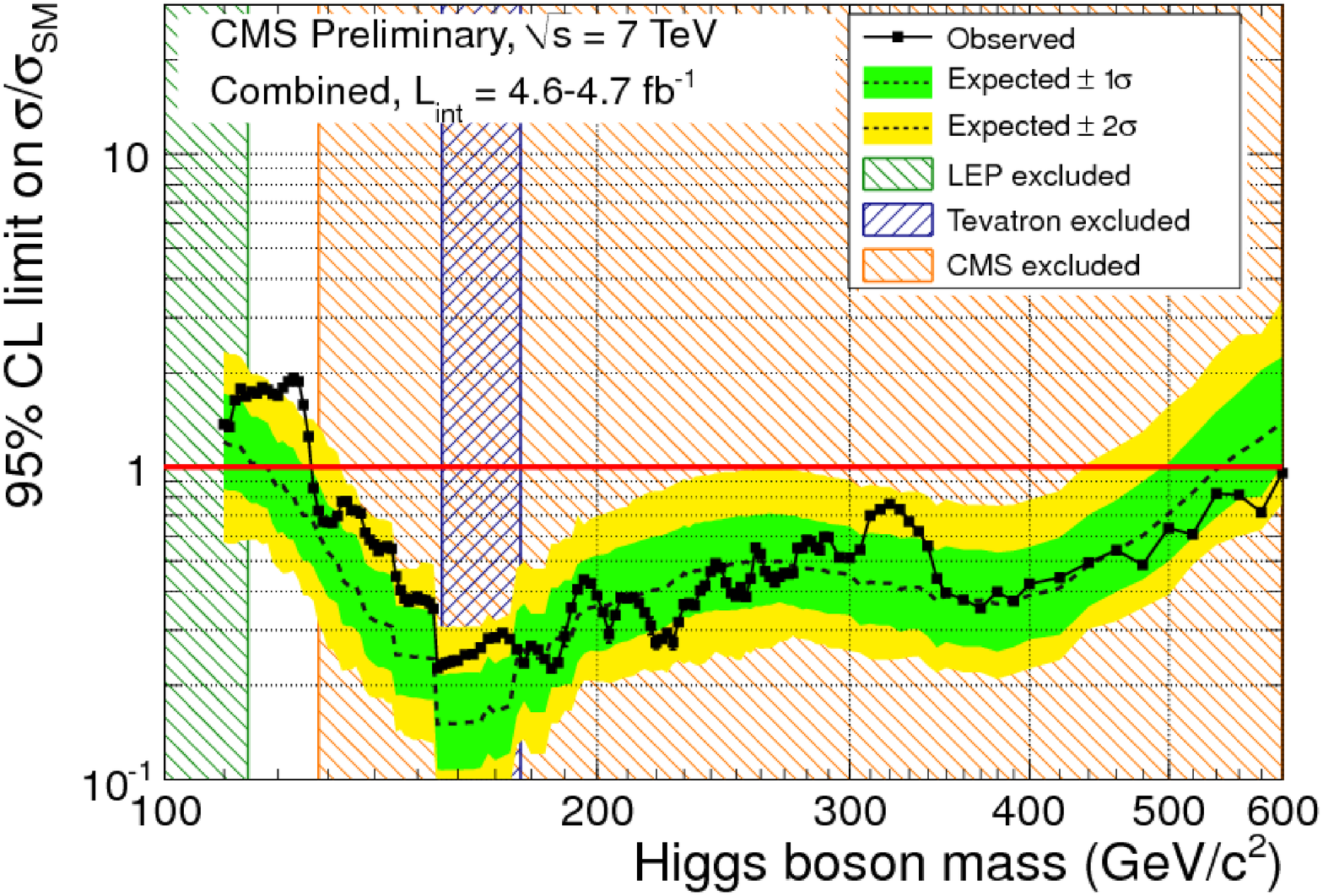}\label{limite1}}
\vspace{0.3cm}
\subfloat[ ]{\includegraphics[scale=0.25]{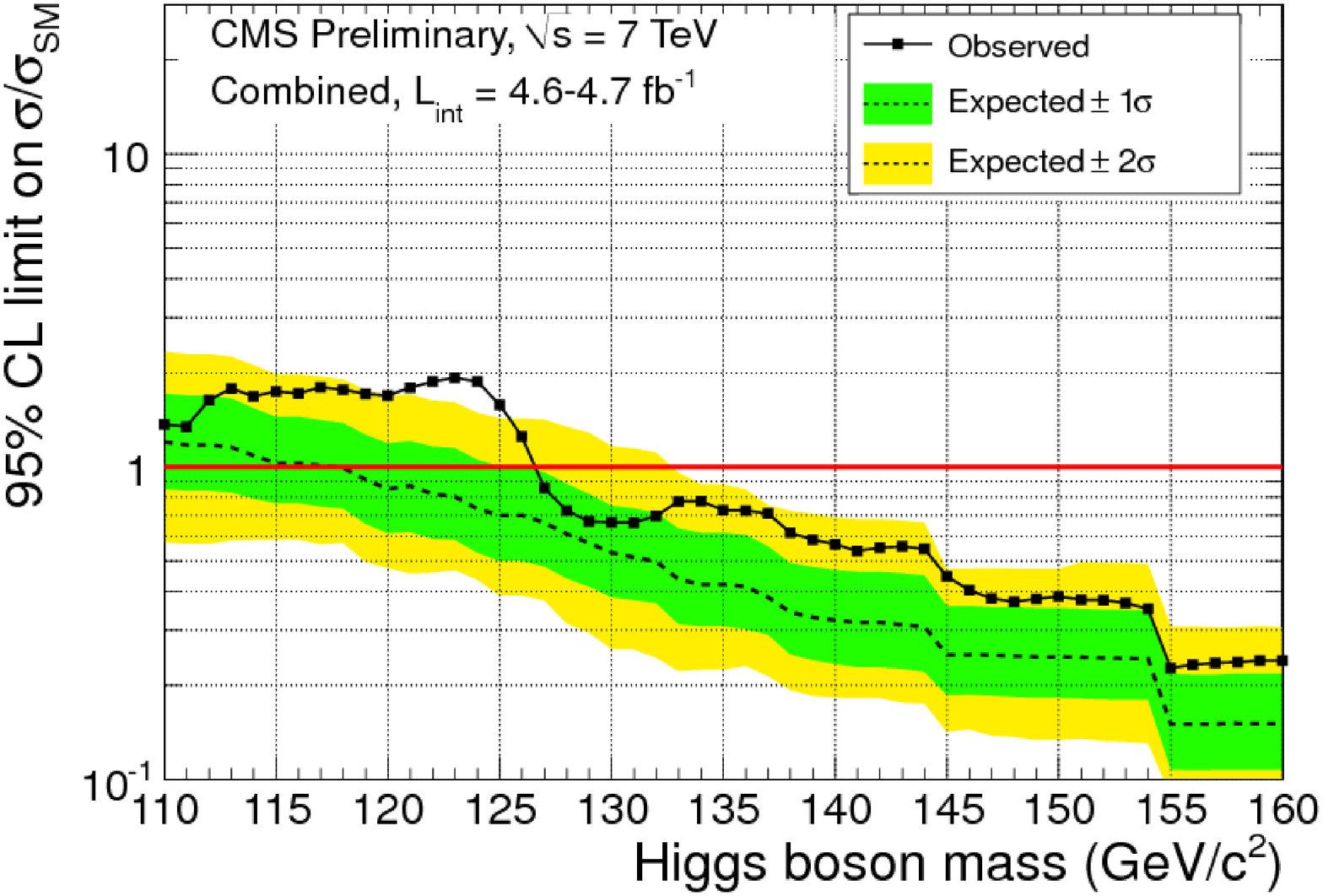}\label{limite2}}
\caption{Limites de exclusão em 95\% C.L. (abaixo da linha vermelha)
  para massa do bóson de Higgs. Análise baseada em dados coletados com
  luminosidade de $4.7\;\mbox{fb}^{-1}$ pelo experimento CMS no
  período 2010-2011. Detalhes na Ref.~\cite{cms}.}
\label{limites}
\end{figure}

Apesar desse excesso de eventos ser compatível com um Higgs padrão de
massa aproximadamente 125 GeV, essa hipótese se apresentava à época
com significância estatística menor que 2$\sigma$ em relação ao
\emph{background}~\footnote{Um efeito qualquer em uma análise de dados
  deve alcançar no mínimo significância estatística de $5\sigma$ para
  ser considerado como uma descoberta.} do MP. Nessa situação as
análises dos dados devem levar em conta outros
efeitos~\footnote{Look-Elsewhere Effect~\cite{Gross:1265134}} nessa
região de massa, já que tal excesso de eventos pode ter origem tão
somente em flutuações estatísticas do \textit{background}.

\subsubsection{Novo bóson descoberto no LHC}

Por ocasião da finalização desta tese, mais precisamente no dia 04 de
julho de 2012, cientistas dos experimentos CMS e ATLAS anunciaram no
CERN a descoberta de uma nova partícula escalar. Embora haja
necessidade de uma análise mais profunda, os primeiros indícios apontam que
tal partícula é consistente com o bóson de Higgs do MP.

Tais resultados combinaram dados do LHC7 com o
\textit{run} de 2012, quando o LHC iniciou suas operações com energia de 8 TeV
(LHC8). Os seguintes canais foram explorados~\cite{novo_higgs}:
\[
H \to  \gamma \gamma,\; b\bar{b},\; \tau^{+}\tau^{-},\; W^{+}W^{-},\; ZZ.
\]

\begin{figure}[!ht]
\centering
\subfloat[ ]{\includegraphics[scale=0.43]{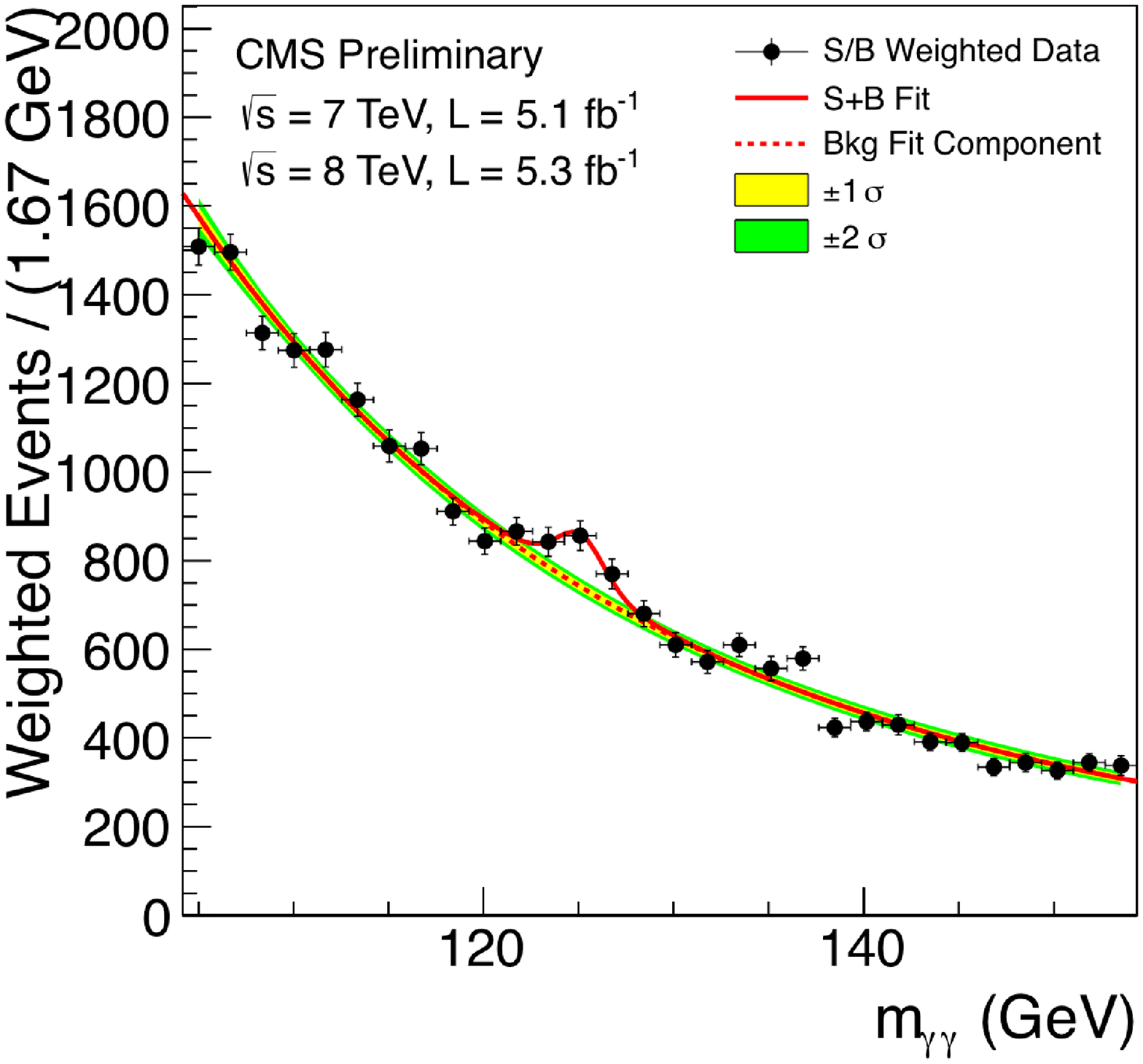}}
\hspace{0.2cm}
\subfloat[ ]{\includegraphics[scale=0.4]{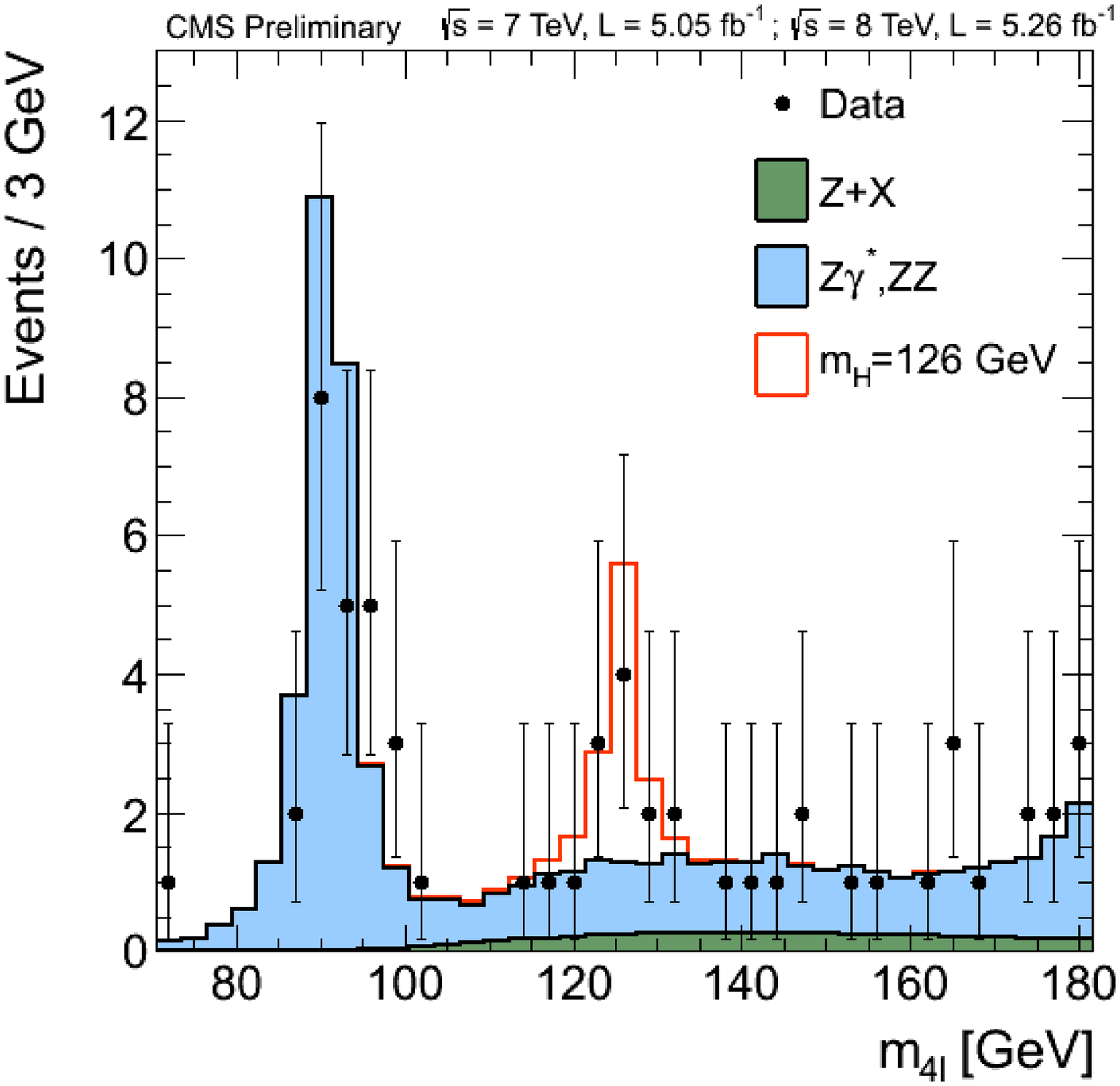}}
\caption{Excesso de eventos observado pelo CMS nos
  canais $\gamma \gamma$ e $ZZ\to \mbox{4 leptons}$,
  respectivamente. Maiores detalhes em
  ``http://cms.web.cern.ch/news/observation-new-particle-mass-125-gev''
  e Ref.~\cite{novo_higgs}.}
\label{higgs_descoberta}
\end{figure}

Ambas as colaborações conseguiram observar, com $L=5\;\mbox{fb}^{-1}$ por experimento, uma nova partícula nos moldes
do Higgs padrão em dois canais de decaimento, $H\to \gamma
\gamma$ e $H\to ZZ^{*}\to 4\; \mbox{léptons}$, cuja análise combinada
alcançou 5$\sigma$ de significância estatística no pico de
ressonância (ver Fig.~\ref{higgs_descoberta}) de massa na ordem de
$M_{H}=125.3\;\pm\;0.4\;(\mbox{stat})\;\pm\;0.5\;(\mbox{sist})$
GeV.

Entretanto, apesar das fortes evidências em associar essa descoberta ao bóson de Higgs do MP, mais análises serão necessárias para testar as propriedades desse novo estado e corroborar, ou não, a hipótese de se tratar do escalar fundamental previsto pelo modelo.


\chapter{Acoplamentos Anômalos entre os Bósons de \textit{Gauge}}
\label{cap:efetivas}

Embora o LHC tenha anunciado a descoberta de uma nova partícula
escalar, estamos ainda em estágio preliminar para associá-la ao bóson
de Higgs padrão, apesar dos sinais serem compatíveis, pois não conhecemos com precisão o
acoplamento dessa nova partícula.

Sendo assim as discussões sobre mecanismos alternativos para
a quebra de simetria eletrofraca ainda são válidas. Além disso, do
ponto de vista teórico, podemos considerar o MP como uma teoria
efetiva e acreditar que exista uma teoria mais fundamental, do ponto
de vista conceitual, regendo os fenômenos na escala subatômica. Logo,
muitos aspectos discutidos no cenário $M_H \to \infty$ podem ser úteis
na construção de novas ideias além do MP.

\section{Espalhamento de bósons vetoriais e cenários de Higgs no LHC}

Conforme visto na Seção~\ref{vinculos_teo}, os processos de
espalhamento entre bósons vetoriais com polarização longitudinal
desempenham um papel importante no setor de quebra de simetria
eletrofraca já que uma possível ausência do campo de Higgs implica em
uma amplitude de espalhamento quadraticamente crescente com a energia,
violando a unitaridade em teoria de perturbação para energias da
ordem de 1.0 até 1.5 TeV~\cite{Chanowitz:1998wi}. Essa violação torna
o MP uma teoria fortemente interagente para energias maiores que 1
TeV.

De fato, caso $M_{H}>1\;\mbox{TeV}$, as amplitudes em ondas parciais
divergem transformando as interações fracas em fortes. Nesse contexto
as interações entre os bósons de \textit{gauge} para energias da ordem
de TeV se assemelham às interações de QCD no regime GeV e portanto não
seria mais possível aplicar teoria de perturbação para representar a
física, que passaria a exibir atributos de uma teoria fortemente
acoplada~\cite{lee}, quais sejam estados ligados (ressonâncias).

Após algumas manipulações algébricas de relações anteriores obtém-se que a relação entre
o parâmetro de expansão ($\approx \lambda/\pi^{2}$) do campo escalar e a
massa do bósons de Higgs~\cite{killian} adquire a forma
\begin{equation}
\dfrac{\lambda}{\pi^{2}}=\dfrac{g^{2}M_{H}^{2}}{8\pi^{2}M_{W}^{2}}=\dfrac{1}{\sqrt{2}\,\pi^{2}}\,G_{F}M_{H}^{2}.
\end{equation}

No regime de acoplamento forte $\lambda/\pi^{2}\approx 1$, e portanto para
$M_{H}\approx 1\,\mbox{TeV}$, temos uma outra indicação de escala de
massa para ressonâncias no setor de
Higgs. Nesse regime, a expansão em ordens de $\lambda$ não seria mais
possível e não se aplicaria mais a teoria de perturbação.

Vários mecanismos alternativos ao MP para explicar a quebra espontânea
de simetria eletrofraca foram propostos nas últimas décadas. Dentre
eles, lembramos das teorias que envolvem estados compostos para explicar o mecanismo
de quebra de simetria que são há algum tempo discutidas em vários
trabalhos~\cite{gu}. 

No entanto, medidas de precisão de experimentos
como o LEP favorecem modelos com um escalar de massa na ${\cal
  O}(100)$ GeV, números quânticos e acoplamentos compatíveis com o
Higgs padrão~\cite{lep}. Modelos como \emph{little Higgs}~\cite{lh} e
\emph{holographic Higgs} pressupõem tal partícula, eventualmente com
desvios nos acoplamentos em relação ao MP. Tais modelos prevém a
produção intrínseca de novas partículas, o que faz com que 
tenhamos grandes expectativas para os próximos anos de operação do LHC.

Algumas análises~\cite{ballestrero} mostram que o LHC operando com
energia $\sqrt{s}=14$ TeV será capaz de determinar se o setor
eletrofraco interage fortemente, com ou sem o aparecimento de novas
ressonâncias. Ressonâncias pesadas com massa da ordem de 1 TeV poderão
ser observadas no LHC a partir de processos envolvendo fusão de bósons
vetoriais com luminosidade em torno de $50\;\mbox{fb}^{-1}$. Já os
cenários sem ressonâncias, ou com ressonâncias bem mais pesadas do que
1 TeV, requerem luminosidade aproximada maiores do que
$100\;\mbox{fb}^{-1}$. Além disso, seria necessário uma luminosidade
de $400\;\mbox{fb}^{-1}$ para distinguir um Higgs composto de um
elementar fracamente acoplado. Enfatizamos que todas essas estimativas supõem canais
envolvendo fusão de bósons vetoriais~\cite{ballestrero02}.

Considerando que ``baixa energia'' se refere à região onde $E \ll
1\,\mbox{TeV}$, pode-se analisar qual impacto que um Higgs não
elementar (composto), ou com interação forte, teria no setor de
\textit{gauge} do MP. O estudo dos efeitos de um bóson de Higgs
``pesado'' no MP, em associação com vértices anômalos quárticos entre
os bósons de \textit{gauge}, tem sido aprimorado há algumas décadas.

Como nesta tese trabalhamos com o formalismo das lagrangianas
efetivas, vamos apresentar na próxima seção uma discussão sobre
teorias efetivas. Em particular, mostramos como elas são úteis a
partir do exemplo histórico da teoria de Fermi para a
explicação do decaimento $\beta$.

\section{Teorias efetivas}

Sabemos que uma das características marcantes do modelo de Glashow,
Weinberg e Salam é a sua
renormalizabilidade~\cite{'tHooft:1971rn}. Assim, nesse modelo e em
muitas de suas extensões, os observáveis podem ser expressos em termos
de um certo número de parâmetros e calculados em várias ordens em
teoria de perturbação, possibilitando previsões em escalas de energias
muito além do alcance tecnológico dos experimentos em uso na época de sua
proposta.

Apesar da condição de renormalizabilidade ser essencial para a
construção de teorias de campos fundamentais, ela pode ser dispensável
quando precisamos descrever uma física ainda não explorada. 

Ao trabalharmos com uma teoria de campo renormalizável, consideramos
implicitamente que os graus de liberdade (partículas) incluídos em sua
lagrangiana geram integralmente o espectro do modelo e que essas
partículas são objetos fundamentais. Ao aceitarmos isso, o número de
parâmetros livres do modelo (ou seja, parâmetros que não são fixados
pelo modelo e só podem ser determinados experimentalmente) é finito,
mesmo que da ordem de algumas dezenas, como é o caso do MP que possui
19 parâmetros livres.

Entretanto, quem dá a resposta final é a Natureza e algumas condições,
impostas por modelos que de alguma forma restringem ou direcionam seu
espectro, podem se mostrar inconsistentes quando temos os dados experimentais ao nosso dispor.

As ideias e princípios que norteam o entendimento de uma física
em nível de energia ou precisão ainda não acessado podem se basear
em teorias de campo não-renormalizáveis, que abarquem todos os fatos
já estabelecidos, mas que contenham um conjunto infinito \emph{a
  priori} de novos parâmetros observáveis que, de alguma forma, possam
mapear um conjunto também infinito de possíveis teorias subjacentes.

A técnica de construção de lagrangianas efetivas (LE) é muito útil
na ausência de modelos específicos para uma nova física que poderia
aparecer em escalas de energia ainda não exploradas. Essa abordagem,
cuja semente encontramos no modelo de Fermi para as interações fracas, foi
desenvolvida com detalhes formais no contexto das interações fortes por
Weinberg \emph{et al.}~\cite{quiral}, e mais
tarde foi adaptada à descrição em baixa energia das interações
eletrofracas por outros autores~\cite{appelquist}.

De modo geral, independente de qualquer modelo, uma LE parametriza os
efeitos de uma nova física que se manifestaria em altas energias para uma
escala de energia mais baixa, bastando para isso especificar o conteúdo de
partículas e as simetrias da teoria em baixa energia.

Ao construirmos uma LE genérica para um dado espectro a única
restrição para a forma dos operadores (não-renormalizáveis)
relaciona-se à conservação de algumas grandezas fundamentais, como
a carga elétrica~\cite{killian}. 

Os operadores na LE são ordenados em ordem crescente de dimensão
canônica de energia. Uma característica desse tipo de abordagem é que, por construção, uma
LE apresenta um número infinito de termos organizados em potências de
$\dfrac{1}{\Lambda}$, onde $\Lambda$ é a escala da nova física. Sendo
assim, para energias muito menores do que essa escala de energia,
$E\ll \Lambda$, somente os primeiros termos da expansão são
relevantes.

\subsection{Modelo de Fermi}

O exemplo mais conhecido de uma abordagem conceitual sobre um fenômeno em
física de partículas através da técnica de construção de uma LE foi a
proposta de Fermi para explicar as interações fracas~\cite{killian}.

Abaixo do limiar de energia para qual os bósons $W^{\pm}$ e $Z$ são
produzidos diretamente, os graus de liberdade básicos do modelo são os
férmions leves e os bósons vetoriais sem massa (glúons e fótons). 

Os efeitos da troca de bósons vetoriais com massa são considerados quando
insere-se operadores não-renormalizáveis (inseridos como campos
auxiliares, ou seja sem termos cinéticos, já que em baixas energias os
bósons de \textit{gauge} da interação fraca não são observáveis) na LE adequadamente construída,
a qual generaliza o modelo original de Fermi das interações fracas. Esse
modelo efetivo funciona em escala de energia $\sqrt{s}\ll M_{W}$. 

No MP a interação de corrente carregada entre dois férmions, representados pelos spinores $\psi$, é descrita
pela troca de um bóson de \textit{gauge} carregado $W$ na forma
\begin{equation}
\dfrac{g^{2}}{8}\;\bar{\psi}\gamma_{\mu}(1-\gamma_{5})\psi\;\dfrac{1}{q^{2}-M_{W}^{2}}\;\bar{\psi}\;\gamma^{\mu}\;(1-\gamma_{5})\psi,
\label{interacao_W}
\end{equation}
onde $q^{2}$ é o momento transferido (escala de energia) da interação.

Considerando a escala de energia $q^{2}\ll M_{W}^{2}$, ao expandirmos o
propagador do W em potências de $\dfrac{q^{2}}{M_{W}^{2}}$ obtemos
\begin{equation}
\dfrac{1}{q^{2}-M_{W}^{2}}\approx
 -\dfrac{1}{M_{W}^{2}}\left[1+\dfrac{q^{2}}{M_{W}^{2}}+ \cdots \right].
\label{expansao_W}
\end{equation}

Usando a expansão~\eqref{expansao_W} podemos re-escrever~\eqref{interacao_W} como uma soma de um número infinito de termos.

Para energias $q^{2}\ll M_{W}^{2}$, o segundo termo
$\dfrac{q^{2}}{M_{W}^{2}}\rightarrow 0$ e somente o primeiro termo é
importante.  Este termo é exatamente o que representa a interação fraca entre
quatro férmions na teoria originalmente proposta por Fermi (modelo
$V-A$), dado por~\cite{quigg}
\begin{equation}
-\dfrac{G_{F}}{\sqrt{2}}\;\bar{\psi}\gamma_{\mu}(1-\gamma_{5})\psi\;\bar{\psi}\;\gamma^{\mu}\;(1-\gamma_{5})\psi,
\label{interacao_fermi}
\end{equation}
onde
\begin{equation}
 \dfrac{G_{F}}{\sqrt{2}}=\dfrac{g^{2}}{8M_{W}^{2}}.
\end{equation}

Deste modo, o modelo fenomenológico de Fermi se apresenta como um
modelo efetivo, válido para energias bem menores que $M_{W}$, onde os
 graus de liberdade ``pesados'' (neste caso, o bóson $W$) da teoria original
podem ser eliminados.

A medida que a escala de energia $q^{2}$ se aproxima de $M_{W}^{2}$, não
podemos mais ``truncar'' os termos da expansão em $\dfrac{q^{2}}{M_{W}^{2}}$
e cada um dos
infinitos termos vai se tornando importante. Nesse regime,
os graus de liberdade ``pesados'' não podem ser
desconsiderados do espectro do modelo e a abordagem efetiva começa a falhar.

As limitações da abordagem da LE (ou seja, da fenomenologia em baixa
energia do modelo de Fermi) para as interações eletrofracas descrita acima podem
ser expostas analisando a reação de decaimento inverso do múon $\mu$,
ou seja $\nu_{\mu}e\to\mu\nu_{e}$ de acordo com~\cite{quigg}. Neste
caso, o modelo só funciona se a energia do centro de massa for da ordem
de $\sqrt{s}\leq 617\,\mbox{GeV}$ e o elemento de matriz é
quadraticamente divergente em segunda ordem em teoria de perturbação.

Além disso, a escala na qual a unitaridade é aparentemente violada fornece um
limite superior para as massas dos graus de liberdade pesados (no caso
analisado, para a massa do bóson $W$).

No entanto, em uma teoria fracamente acoplada como é o caso do MP, esse
limite superior superestima as massas dos graus de liberdade pesados e então é
necessário supor novas ressonâncias (novas partículas) e/ou pelo menos adotar uma
abordagem supondo interações fortes para obter as
seções de choque que se aproximam do limite de unitaridade da
teoria e refinar esses resultados.

Por exemplo, no caso da reação de decaimento inverso do múon
teríamos um limite superior de $M\leq 309\,\mbox{GeV}$. Ao considerar a
opção mais simples para resolver o problema (em outras palavras para
``salvar o modelo'', já que
o mesmo se comporta adequadamente em baixa energia), considera-se que a
interação fraca seria mediada pela troca de um bóson vetorial, em analogia
com o eletromagnetismo, com algumas propriedades
particulares de acordo com~\cite{quigg}, resultando em uma
seção de choque bem comportada, constante em altas energias
\begin{equation}
\lim_{E\to \infty}\,\sigma(\nu_{\mu}e\,\to \mu \nu_{e})=\dfrac{G_{F}^{2}M_{W}^{2}}{\pi},
\end{equation}
com a unitaridade sendo respeitada até energias da ordem de
\begin{equation}
s\,\leq \,M_{W}^{2}\left[\exp{\left(\dfrac{\pi\,\sqrt{2}}{G_{F}M_{W}^{2}}\right)}-1\right]\approx
 7.3\times 10^{29}\,\mbox{GeV}^2,
\end{equation}
considerando $M_{W}\approx 80\,\mbox{GeV}$.

A medida que a escala de energia aumenta, novos canais se abrem, por
exemplo os canais de produção de $W$ e $Z$ na teoria de Fermi. Mas as seções de choque desses novos canais podem ser muito pequenas para serem observadas, especialmente se a teoria subjacente tiver natureza perturbativa. Esses aspectos, facilmente entendidos no contexto do MP e na teoria de Fermi, são propriedades gerais do formalismo das LE's.

Para o desenvolvimento formal da construção de uma LE e sua utilização em
várias situações indicamos o leitor às referências~\cite{quiral, killian, appelquist}.

\section{Setor eletrofraco com interações fortes}
\label{sec:int_forte}

De forma geral, a construção canônica do MP se baseia em três hipóteses fundamentais:
\begin{itemize}
\item invariância sob a simetria de \textit{gauge} não-abeliana $SU(2)_{L}\otimes
      U(1)_{Y}$;
\item renormalização;
\item quebra espontânea da simetria através de campos escalares
      elementares inseridos no modelo como um dubleto de isospin
      complexo e do mecanismo de Higgs.
\end{itemize}

Nesta seção, seguindo
basicamente o exposto na Ref.~\cite{killian,appelquist}, abordaremos a
construção das lagrangianas quirais que originam acoplamentos anômalos\footnote{Acoplamentos com estrutura de \textit{gauge} diferente à encontrada nas
interações do MP.} tríplices e quárticos, cujos efeitos são relevantes no
sinal de processos sujeitos à fusão de bósons vetoriais.

\subsection{Formalização da abordagem efetiva}
\label{sub:quiral}

Como vimos, se $m_{H}\approx 1$TeV, o que implica que
$\dfrac{\lambda}{\pi^{2}}\approx 1$, podemos considerar o MP, com seu
campo escalar elementar, um modelo efetivo válido para baixas
energias.

A massa do Higgs seria um \emph{cut-off} de energia natural nesse
contexto e, como $\dfrac{\lambda }{\pi^{2}}\to 1$, nos aproximamos da
região de acoplamento forte, testando assim a sensibilidade dos
observáveis que surgem no regime de baixas energias ($\lambda \ll 1$)
nessa região.

Seguindo a revisão apresentada na Ref.\cite{killian}, pretendemos nessa
seção apresentar qual o impacto que um bóson de Higgs relativamente
pesado exerce na estrutura do MP.

Para tanto, a estrutura usual do MP pode ser reformulada de maneira
útil para descrição em baixas energias. Essa reformulação manifesta
uma simetria quiral global aproximada $SU(2)_{L}\otimes SU(2)_{R}$ do
setor de Higgs.

Para analisarmos o
caso particular que não contempla um campo de Higgs padrão, a
sensibilidade do modelo efetivo é tratada formalmente considerando o
limite $M_{H}\rightarrow \infty$, e daí resulta uma lagrangiana análoga
à do modelo $\sigma $ não-linear acoplada à lagrangiana de
Yang-Mills. Essa nova lagrangiana é não-renormalizável.

Os novos efeitos
físicos relacionados às hipóteses acima aparecem nos ``contratermos''
necessários para tornar a teoria não-linear finita, que gerarão termos anômalos permitindo interações ausentes no MP, como é o caso do vértice $ZZZZ$.

Sendo um pouco mais formal na construção da lagrangiana quiral, a partir
do modelo de Fermi para as
interações fracas em ordem mais baixa, pode-se escrever os operadores
de interação entre quatro férmions, que intermediam as interações em
baixas energias entre léptons e quarks, como a soma de produtos de
correntes carregadas e neutras na forma geral~\cite{killian}
\begin{equation}
-4\sqrt{2}G_{F}(2J_{\mu}^{+}J^{\mu\,-}+\rho_{*}J_{\mu}^{0}J^{\mu\,0}),
\label{lag}
\end{equation}
sendo que essa estrutura corrente-corrente indica a presença de uma
simetria subjacente (idéia original que levou à elaboração do Modelo Padrão das
interações eletrofracas).

A simetria local da expressão~\eqref{lag} se manifesta com a introdução
dos campos vetoriais $W^{\pm}$ e $Z$ tais que~\eqref{lag} pode ser escrita como
\begin{equation}
-g_{W}(W^{+\,\mu}J^{+}_{\mu}+W^{-\,\mu}J^{-}_{\mu})-g_{Z}(Z^{\mu}J^{0}_{\mu})+M_{W}^{2}\,(W^{+\,\mu}W^{-}_{\mu})+\dfrac{1}{2}\,M_{Z}^{2}\,Z^{\mu}Z_{\mu}.
\label{lag2}
\end{equation}

No modelo de Fermi, em baixas energias, os bósons de \textit{gauge}
não são graus de liberdade observáveis e portanto são introduzidos
como campos auxiliares (campos sem termo cinético) e podem ser
calculados algebricamente como
\begin{equation}
W^{\pm}_{\mu}=\dfrac{g_{W}}{M_{W}}\,J^{\mp}_{\mu}\;\;\mbox{e}\;\;Z^{\pm}_{\mu}=\dfrac{g_{Z}}{M_{Z}}\,J^{0}_{\mu}
\end{equation}
onde
\begin{equation}
g^{2}_{W}=\dfrac{4M_{W}^{2}}{v^{2}}\;\;\;\;\mbox{e}\;\;\;\;g^{2}_{Z}=\rho^{*}\dfrac{4M_{Z}^{2}}{v^{2}},
\label{acoplamentos}
\end{equation}
sendo definido aqui $v\equiv (\sqrt{2}G_{F})^{-1/2}$ (note que não se trata do mesmo procedimento de obtenção do \emph{vev} do bóson de Higgs).

A medida que os campos $W$ e $Z$ não correspondem aos graus de liberdade
físicos, os valores de suas massas e constantes de acoplamento são
arbitrários.

Para torná-los graus de liberdade físicos, precisamos adicionar termos
cinéticos à expressão~\eqref{lag2}, ou seja termos invariantes na forma
\begin{equation}
-\dfrac{1}{2}\,\mbox{Tr}\left[\mathbf{W}_{\mu\nu}\mathbf{W}^{\mu\nu}+\mathbf{B}_{\mu\nu}\mathbf{B}^{\mu\nu}\right].
\label{cineticos}
\end{equation}

A lagrangiana resultante de~\eqref{lag2} e de~\eqref{cineticos} será
válida para escalas de energia $E\approx M_{W}, M_{Z}$, onde os
tensores de intensidade de campo têm a forma
\begin{align}
\mathbf{W}_{\mu\nu}&=\partial_{\mu}\mathbf{W}_{\nu}-\partial_{\nu}\mathbf{W}_{\mu}+ig\left[\mathbf{W}_{\mu},\mathbf{W}_{\nu}\right], \\
\mathbf{B}_{\mu\nu}&=\partial_{\mu}\mathbf{B}_{\nu}-\partial_{\nu}\mathbf{B}_{\mu},
\end{align}
com
\begin{equation}
\mathbf{W}_{\mu}=\dfrac{\tau^{a}}{2}W^{a}_{\mu}\;\;\;\mbox{e}\;\;\;\mathbf{B}_{\mu}=\dfrac{\tau^{3}}{2}B_{\mu}.
\end{equation}

De forma geral, a lagrangiana do setor eletrofraco do MP para os
bósons de \textit{gauge} pode ser escrita antes da quebra espontânea
de simetria como
\begin{align}
{\cal L}_{WS}=&-\dfrac{1}{2}\,\mbox{Tr}\,\left[\mathbf{W}_{\mu\nu}\mathbf{W}^{\mu\nu}\right]-\dfrac{1}{2}\,\mbox{Tr}\left[\mathbf{B}_{\mu\nu}\mathbf{B}^{\mu\nu}\right]\nonumber \\
& +\left[D_{\mu}\Phi(x)\right)^{\dagger}D_{\mu}\Phi(x)-\mu^{2}\,\Phi^{\dagger}(x)\Phi(x)-\lambda\,\left(\Phi^{\dagger}(x)\Phi(x)\right]^{2}.
\label{lws}
\end{align}

\subsection{Simetria quiral}

Nesse ponto, desenvolveremos o que mencionamos na
Seção~\ref{sec:simetriacustodial} em relação à construção de uma
teoria de campo efetiva que quebra a simetria eletrofraca de forma
simples mantendo a simetria custodial.

Uma forma de fazer isso é substituir a matriz de Higgs~\eqref{bidubleto} por um campo matricial $\Sigma$ que contenha os bósons de Goldstone $\pi_{i}$ ($i=1,2,3$) mas não contenha o bóson de Higgs físico. No entanto, aqui iremos manter o campo escalar de Higgs $\phi $ para sermos o mais geral possível.

Sendo assim, para manifestar a simetria de \textit{gauge} $SU(2)_{L}\otimes SU(2)_{R}$ em~\eqref{lws},
introduz-se um campo matricial $\Sigma (x)$ tal que
\begin{equation}
 \Sigma(x)=\phi(x)+i\vec{\tau} \cdot \vec{\pi}(x)
\label{campo}
\end{equation}
onde $\vec{\tau}$ são as matrizes de Pauli, $\phi(x)$ é o campo de Higgs e $\vec{\pi}(x)\equiv (\pi_{1}(x),\pi_{2}(x),\pi_{3}(x))$
é um tripleto de bósons de Goldstone (escalares sem massa). Todos os campos
são hermitianos.

De forma explícita,
\begin{equation}
\Sigma(x)= \left(\begin{array}{cc}
\phi(x)+i\pi_{3}(x) & \pi_{2}(x)+i\pi_{1}(x) \\
-(\pi_{2}(x)-i\pi_{1}(x)) & \phi(x)-i\pi_{3}(x)
\end{array}\right)
\label{campo2}
\end{equation}
onde se vê claramente que
\begin{equation}
\mbox{det}\,(\Sigma)=\mbox{det}\,(\Sigma^{\dagger})=\Sigma\,\Sigma^{\dagger}=\Sigma^{\dagger}\,\Sigma
 = \phi^{2}+\pi^{2}.
\label{detsigma}
\end{equation}

Em analogia com o que foi exposto anteriormente na Seção~\ref{sec:simetriacustodial} podemos escrever~\eqref{campo} como
\begin{equation}
\Sigma(x)= \left(\begin{array}{cc}
\phi^{0\,*}(x) & \phi^{+}(x) \\
-\phi^{-}(x) & \phi^{0}(x)
\end{array}\right) \equiv \sqrt{2}\,(\tilde{\Phi}(x)\,\Phi(x)),
\label{campo3}
\end{equation}
onde $\tilde{\Phi}(x)\equiv i\tau_{3}\Phi^{*}(x)$.

Sabendo que os campos $\Phi(x)$ e $\tilde{\Phi}(x)$ se transformam
localmente sob a simetria de \textit{gauge} $SU(2)_{L} \otimes U(1)_{Y}$ tal que
\begin{align}
\Phi(x)&\to  \exp{\left[(\epsilon_{0}(x)+\vec{\epsilon}(x)\cdot
 \vec{\tau})/2\right]}\,\Phi(x)
\\
\tilde{\Phi}(x)&\to  \exp{\left[(-\epsilon_{0}(x)+\vec{\epsilon}(x)\cdot
 \vec{\tau})/2\right]}\,\tilde{\Phi}(x),
\label{gauge}
\end{align}
infere-se que o campo $\Sigma(x)$ se transforma localmente sob a mesma
simetria de
acordo com~\cite{quiral}\footnote{A presença explícita da
matriz de Pauli $\tau_{3}$ nas leis de transformação de $\Sigma(x)$
refletem o fato que, apesar dos campos $\Phi(x)$ e $\tilde{\Phi}(x)$
compartilharem transformações idênticas sob o grupo de simetria local
$SU(2)_{L}$, eles se transformam de modo oposto sob o grupo de simetria
local $U(1)_{Y}$.}
\begin{equation}
\Sigma(x)\to \exp{\left[\vec{\epsilon}(x)\cdot
 \vec{\tau}/2\right]}\;\Sigma(x)\;\exp{\left[\epsilon_{0}(x)\tau_{3}/2\right]}.
\end{equation}

A derivada covariante aplicada ao campo auxiliar terá a
seguinte forma
\begin{align}
D_{\mu}\Sigma(x)&=\partial_{\mu}\Sigma(x)+ig\dfrac{\vec{\tau}}{2}\cdot
 \vec{W}_{\mu}\,\Sigma(x)-i\dfrac{g^{'}}{2}\,B_{\mu}(x)\Sigma(x)\tau_{3}\nonumber \\
&=\partial_{\mu}\Sigma(x)+ig\,\mathbf{W}_{\mu}\,\Sigma(x)-ig^{'}\,\mathbf{B}_{\mu}(x)\Sigma(x).
\label{covariantesigma}
\end{align}

A lagrangiana do setor escalar, antes de torná-la localmente invariante
pela substituição $\partial_{\mu}\to D_{\mu}$, re-escrita em função do campo
$\Sigma(x)$ é exatamente a lagrangiana do modelo $\sigma$-linear. De
fato,
\begin{align}
{\cal
 L}_{S}(x)&=\left(\partial_{\mu}\Phi(x)\right)^{\dagger}\,\partial_{\mu}\Phi(x)-\mu^{2}\,\Phi^{\dagger}(x)\Phi(x)-\lambda\,\left(\Phi^{\dagger}(x)\Phi(x)\right)^{2}\nonumber \\
&\equiv 
 \dfrac{1}{4}\,\mbox{Tr}\left[\partial_{\mu}\Sigma^{\dagger}(x)\partial^{\mu}\Sigma(x)\right]-\dfrac{\mu^{2}}{4}\,\mbox{Tr}\left[\Sigma^{\dagger}(x)\Sigma(x)\right]-\dfrac{\lambda}{16}\;\left(\mbox{Tr}\left[\Sigma^{\dagger}(x)\Sigma(x)\right]\right)^{2} 
\label{ls}
\end{align}
é explicitamente invariante sob transformações da simetria quiral,
$SU(2)_{L}\otimes SU(2)_{R}$
\begin{equation}
\Sigma(x)\to \exp{\left[i\vec{\epsilon}_{L}\cdot
				 \dfrac{\vec{\tau}}{2}\right]}\;\Sigma(x)\;\exp{\left[-i\vec{\epsilon}_{R}\cdot
				 \dfrac{\vec{\tau}}{2}\right]}.
\end{equation}

Portanto, o potencial escalar
mais geral consistente com a invariância de \textit{gauge} $SU(2)_{L}\otimes
U(1)_{Y}$ do modelo minimal de Weinberg-Salam é uma função de
$\Sigma^{\dagger}(x)\Sigma(x)$ invariante por transformações em
$SU(2)_{L}\otimes SU(2)_{R}$.

\subsection{Quebra espontânea da simetria eletrofraca no cenário de interação forte}

Ao aplicarmos o mecanismo de Higgs, escolhendo $\mu^{2}<0$, na lagrangiana do setor eletrofraco~\eqref{lws} re-escrita em função do campo de \textit{gauge} matricial $\Sigma (x)$~\eqref{ls}
\begin{align}
{\cal L}_{WS}=&-\dfrac{1}{2}\,\mbox{Tr}\,\left(\mathbf{W}_{\mu\nu}\mathbf{W}^{\mu\nu}\right)-\dfrac{1}{2}\,\mbox{Tr}\,\left[\mathbf{B}_{\mu\nu}\mathbf{B}^{\mu\nu}\right]
 \nonumber \\
&+\dfrac{1}{4}\,\mbox{Tr}\left[\partial_{\mu}\Sigma^{\dagger}(x)\partial^{\mu}\Sigma(x)\right]-\dfrac{\mu^{2}}{4}\,\mbox{Tr}\left[\Sigma^{\dagger}(x)\Sigma(x)\right]-\dfrac{\lambda}{16}\;\left(\mbox{Tr}\left[\Sigma^{\dagger}(x)\Sigma(x)\right]\right)^{2} 
\label{lwsigma}
\end{align}
força-se o potencial a adquirir um mínimo
\begin{equation}
\Sigma^{\dagger}(x)\Sigma(x)=\Sigma(x)\Sigma^{\dagger}(x)=-\dfrac{\mu^{2}}{\lambda}\equiv v^{2},
\end{equation}
ou seja, o campo matricial $\Sigma(x)$ possui um \emph{vev} não nulo
\begin{equation}
\langle 0|\Sigma(x)|0 \rangle = v.
\end{equation}

Para aplicarmos teoria de perturbação na vizinhança de um estado de
vácuo assimétrico, porém estável, desloca-se o campo $\Sigma(x)$
definindo-se um novo campo
\begin{equation}
\Sigma^{'}(x)\equiv \Sigma(x)-v,
\end{equation}
tal que
\begin{equation}
\langle 0|\Sigma^{'}(x)|0 \rangle = 0.
\end{equation}

Efetuando a substituição $\Sigma(x)= \Sigma^{'}(x)+v$ em~\eqref{lwsigma}, aparece explicitamente o campo escalar $\phi^{'}$, uma das componentes de
$\Sigma^{'}(x)$, sendo o Higgs físico com massa
$M_{H}=\sqrt{2\lambda}v$.

Fazendo a substituição $\partial_\mu\Sigma(x) \to D_{\mu}\Sigma(x)$
e introduzindo de acordo do a Ref.~\cite{quiral} as abreviações
\begin{equation}
V_{\mu}\equiv
 \Sigma\left(D_{\mu}\Sigma\right)^{\dagger}\;\;\;\;\mbox{e}\;\;\;\;T\equiv \Sigma\tau_{3}\Sigma^{\dagger}
\label{abrev}
\end{equation}
temos que o termo de massa na expressão~\eqref{lag2} será substituído pelo termo cinético para o
campo $\Sigma$
\begin{equation}
{\cal L}^{(2)}=-\dfrac{v^{2}}{4}\mbox{Tr}\left[V_{\mu}V^{\mu}\right]-\beta^{'}\dfrac{v^{2}}{8}\mbox{Tr}\left[TV_{\mu}\right]\mbox{Tr}\left[TV^{\mu}\right],
\label{l2}
\end{equation}
com o parâmetro livre $\beta^{'}$.

Adotando-se o \textit{gauge} unitário $\Sigma(x)=1$, o campo vetorial composto $V_{\mu}$ é
uma combinação linear dos campos $W$ e $Z$ (sem contribuição do fóton),
ou seja
\[
V_{\mu}=-ig\mathbf{W}_{\mu}+ig^{'}B_{\mu}
=-i\dfrac{g_{W}}{\sqrt{2}}\left(W^{+}_{\mu}\tau^{+}+W_{\mu}^{-}\tau^{-}\right)-ig_{Z}Z_{\mu}\dfrac{\tau_{3}}{2}
\]
onde os acoplamentos $g_{W}$ e $g_{Z}$ foram definidos em~\eqref{acoplamentos}.

Expandindo~\eqref{l2} no \textit{gauge} unitário, obtém-se
\begin{equation}
{\cal L}^{(2)}=v^{2}\dfrac{g_{W}^{2}}{2}W^{+}_{\mu}W^{-\,\mu}+v^{2}\dfrac{g_{Z}^{2}}{4}(1+\beta^{'})Z_{\mu}Z^{\mu},
\end{equation}
de onde tiramos que as massas do bósons vetoriais são
\begin{equation}
M_{W}=\dfrac{v}{2}g_{W}\;\;\;\;\mbox{e}\;\;\;\;M_{Z}=\dfrac{v}{2}g_{Z}(1+\beta^{'}).
\end{equation}

O parâmetro $\rho^{*}$ das interações neutras~\eqref{acoplamentos} é dado por
\begin{equation}
\rho^{*}=1-\beta^{\prime}
\end{equation}
e $\rho^{*}=1$ implica $\beta^{\prime}=0$.

\subsection{O que acontece quando $M_{H}\to \infty$ ?}

Quando consideramos $M_{H}\to \infty$, o bóson de Higgs é removido do
espectro físico e a teoria não-renormalizável resultante será o modelo
$\sigma$ não-linear acoplado de modo invariante de \textit{gauge} à teoria de
Yang-Mills $SU(2)_{L}\otimes U(1)_{Y}$. 

De fato, no que concerne ao
setor escalar, com o vínculo
\begin{equation}
\Sigma^{\dagger}\Sigma=\Sigma
 \Sigma^{\dagger}=\dfrac{(2M_{W})^{2}}{g^{2}}=v^{2}
\label{condicao}
\end{equation}
a lagrangiana~\eqref{ls} adquire a
forma em termos dos campos de Goldstone $\vec{\pi}$~\cite{quiral}
\begin{equation}
{\cal L}_{S}^{M_{H}\to
 \infty}=\dfrac{1}{2}\left(\partial_{\mu}\vec{\pi}\right)^{2}+\dfrac{1}{2}\left(\vec{\pi}\cdot
		    \partial_{\mu}\vec{\pi}\right)^{2}\,\dfrac{1}{v^{2}-\vec{\pi}^{2}}.
\end{equation}

Deste modo, a lagrangiana efetiva demonstra uma teoria
não-linear. Mas embora seja não-renormalizável, ela pode se tornar
finita em qualquer ordem na expansão em \emph{loops} pela adição de uma
quantidade suficiente de contratermos de acordo com~\cite{appelquist}. De modo contrário ao modelo
renormalizável de Weinberg-Salam, a não-renormalizabilidade desta teoria
significa que alguns destes contratermos apresentarão novas estruturas, com
diferentes formas em relação aos termos que aparecem em nível de árvore na
lagrangiana não-linear, e com dependências de parâmetros que em princípio
são mensuráveis.

\subsection{Contratermos no modelo não-linear}
\label{quiral}

De forma geral, para obter o espectro físico sem o bóson de Higgs, como
mencionado anteriormente, vinculados que estamos pela condição~\eqref{condicao}, podemos impor que
\begin{equation}
\Sigma^{\dagger}\Sigma \equiv 1,
\end{equation}
implicando que a matriz $\Sigma$ seja necessariamente unitária.

Escreve-se convenientemente a matriz $\Sigma(x)$ em termos de uma matriz unitária
$U(x)$ tal que
\begin{equation}
\Sigma(x) = vU(x).
\end{equation}

Sendo assim a análise se torna mais simples, pois trabalhamos com o campo
escalar adimensional $U(x)\equiv \Sigma(x)/v$. Pode-se então adotar a parametrização~\cite{quiral}
\begin{equation}
U(x)=\sqrt{1-\dfrac{\vec{\pi}(x)^{2}}{v}}+i\vec{\tau}\cdot
 \dfrac{\vec{\pi}(x)}{v},
\label{u}
\end{equation}
onde o campo $\dfrac{\vec{\pi}}{v}$ é adimensional.

Em termos da parametrização de campo $U(x)$, a lagrangiana efetiva
não-linear no \textit{gauge} de Landau, $\xi\to 0$, é escrita como~\cite{quiral}
\begin{equation}
{\cal L}_{eff}=\dfrac{v^{2}}{4}\;\mbox{Tr}\left[(D_{\mu}U)^{\dagger}D^{\mu}U\right]-\dfrac{1}{2}\mbox{Tr}\left[\mathbf{W}_{\mu\nu}\mathbf{W}^{\mu\nu}\right]-\dfrac{1}{2}\mbox{Tr}\left[\mathbf{B}_{\mu\nu}\mathbf{B}^{\mu\nu}\right]+{\cal
 L}_{FG}+{\cal L}_{FP}.
\label{nl}
\end{equation}

Todos os contratermos invariantes por $SU(2)_{L}\otimes U(1)$ da
teoria não-linear, com expansão em nível de \emph{loops} em várias ordens em teoria de perturbação, podem
ser construídos como traços de objetos covariantes em $SU(2)_{L}$ conforme nos orienta a Ref.~\cite{appelquist}.

Detalhar essas deduções está fora do escopo do nosso trabalho por ser
um assunto devidamente tratado em diversas referências já citadas. Nos
restringimos aqui a listar os contratermos invariantes por $CP$, de
dimensão $D=4$, adicionados à lagrangiana efetiva
não-linear~\eqref{nl}, obtidos pelo algoritmo desenvolvido
na Ref.~\cite{appelquist}, considerando a parametrização~\eqref{u}, a saber
\begin{align*}
{\cal L}_{1}&=\dfrac{g^{2}}{2}\;\alpha_{1}\;B_{\mu
 \nu}\;\mbox{Tr}(T\mathbf{W}^{\mu \nu}) \nonumber \\
{\cal L}_{2}&=i\;\dfrac{g}{2}\;\alpha_{2}\;B_{\mu
 \nu}\;\mbox{Tr}(T[V^{\mu},V^{\nu}]) \nonumber \\
{\cal L}_{3}&=i\;g\;\alpha_{3}\;\mbox{Tr}(\mathbf{W}_{\mu
 \nu}[V^{\mu},V^{\nu}]) \nonumber \\
{\cal L}_{4}&=\alpha_{4}\;[\mbox{Tr}(V_{\mu}V_{\nu})]^{2}
 \nonumber \\
{\cal L}_{5}&=\alpha_{5}\;[\mbox{Tr}(V_{\mu}V^{\mu})]^{2}
 \nonumber \\
{\cal L}_{6}&=\alpha_{6}\;\mbox{Tr}(V_{\mu}V_{\nu})\;\mbox{Tr}(TV^{\mu})\;\mbox{Tr}(TV^{\nu}) \nonumber\\
{\cal L}_{7}&=\alpha_{7}\;\mbox{Tr}(V_{\mu}V^{\mu})\;[\mbox{Tr}(TV_{\nu})]^{2}
\end{align*}
\begin{align}
{\cal L}_{8}&=\dfrac{g^{2}}{4}\;\alpha_{8}\;[\mbox{Tr}(T\mathbf{W}_{\mu
 \nu})]^{2} \nonumber \\
{\cal L}_{9}&=i\;\dfrac{g}{2}\;\alpha_{9}\;\mbox{Tr}(T\mathbf{W}_{\mu
 \nu})\;\mbox{Tr}(T[V^{\mu},V^{\nu}]) \nonumber \\
{\cal L}_{10}&=\dfrac{1}{2}\;\alpha_{10}[\mbox{Tr}(TV_{\mu})\;\mbox{Tr}(TV_{\nu})]^{2}
 \nonumber \\
{\cal L}_{11}&=\alpha_{11}\;\mbox{Tr}([D_{\mu}V^{\mu}]^{2})
 \nonumber \\
{\cal L}_{12}&=\dfrac{1}{2}\;\alpha_{12}\;\mbox{Tr}(TD_{\mu}\;D_{\nu}V^{\nu})\;\mbox{Tr}(TV^{\mu})
 \nonumber \\
{\cal L}_{13}&=\dfrac{1}{2}\;\alpha_{13}\;[\mbox{Tr}(TD_{\mu}\;V_{\nu})]^{2}
\label{efetivas}
\end{align}

Nas lagrangianas listadas em~\eqref{efetivas} temos
\begin{equation}
 T(x)\equiv U(x)\tau_{3}U^{\dagger}(x)\;\;\;,\;\;\;V_{\mu}(x)\equiv
 (D_{\mu}U(x))U^{\dagger}(x)
\end{equation}
com derivada covariante definida como
$D_{\mu}O(x)=\partial_{\mu}O(x)+ig[W_{\mu},O(x)]$.

As regras de Feynman para obtenção dos acoplamentos
anômalos no \textit{gauge} unitário $U(x)=1$ aplicado aos termos da lagrangiana~\eqref{efetivas} foram listados no Apêndice~\ref{ap:regrasfeynman}.

No \textit{gauge} unitário, no que se refere aos acoplamentos
puramente quárticos, existem contribuições anômalas provenientes das
lagrangianas ${\cal L}_{4},\;{\cal L}_{5},\;{\cal L}_{6},\; {\cal
  L}_{7},\;{\cal L}_{10}$. O vértice quártico anômalo $ZZZZ$ emerge de
todos esses operadores, mas neste trabalho eles são irrelevantes
tendo em vista o perfil de nossa análise. O vértice quártico
anômalo do tipo $WWZZ$ também emerge de todos os operadores, com exceção do
operador ${\cal L}_{10}$. Vértice quártico anômalo do tipo $WWWW$
só aparecem nos operadores ${\cal L}_{4}$ e ${\cal L}_{5}$.  Os
operadores ${\cal L}_{6}$ e ${\cal L}_{7}$ violam a simetria custodial
$SU(2)_{C}$. 

Nesse trabalho manipulamos as lagrangianas ${\cal L}_{4}$
e ${\cal L}_{5}$ e sendo assim, está implícito a contribuição geral de
vértices quárticos anômalos do tipo $WWWW$ e $WWZZ$.

Colisores como o LEP ou o Tevatron não possuíam energia no centro de
massa suficientemente alta para produzir múltiplos bósons massivos e
portanto não possibilitaram estudos diretos das interações quárticas
entre os bósons de \emph{gauge}. Em particular, processos de VBF no
Tevatron apresentam seções de choque muito pequenas.

Como as colaborações ATLAS e CMS ainda não deram início às medidas dos
acoplamentos quárticos, os únicos vínculos existentes até o
presente se devem às medidas de precisão eletrofracas advindas do LEP, de acordo com \cite{Brunstein:1996fz}.



Muito antes do início da operação do LHC, inúmeros trabalhos
exploraram o potencial desse colisor para os testes dos acoplamentos
anômalos, em particular os vértices $WWWW$ e $WWZZ$
~\cite{tese_jkm,belyaev,eboli2} e também para o futuro colisor
\textit{International Linear Collider}
(ILC)~\cite{tese_jkm,eboli4,han,Boos:1997gw}, outrora conhecido como
\textit{Next Linear Collider} (NLC), cujo projeto de construção ainda
está em discussão~\cite{ilc}.


Para efeito de comparação com os nossos cálculos, vamos apresentar
aqui alguns vínculos que podem ser obtidos aos acoplamentos $\alpha_4$
e $\alpha_5$ através das análises de processos de VBF. 

Na Ref.~\cite{belyaev}, os autores estudaram os processos $pp\to
VV\;+\;2$ jatos, com $V = W^{\pm}\;\mbox{e}\;Z$. Os resultados foram
obtidos levando-se em conta o estado final com léptons + 2 jatos
\textit{taggings}, mas não houve tentativa de se fazer um cálculo
completo dos elementos de matriz levando-se em conta os decaimentos
dos bósons de \textit{gauge}. Tais implementações foram efetuadas
posteriormente na Ref.~\cite{eboli2}, onde os autores analisaram os processos
completos $pp\to jje^{\pm} \mu^{\pm} \nu \nu$ e
$pp \to jje^{\pm}\mu^{\mp}\nu\nu$, em
${\cal O}(\alpha_{em}^{6})$ e $ {\cal
  O}(\alpha_{em}^{4}\alpha_{s}^{2})$ em teoria de perturbação.

\begin{figure}
\centering
\subfloat[ ]{\includegraphics[scale=0.35]{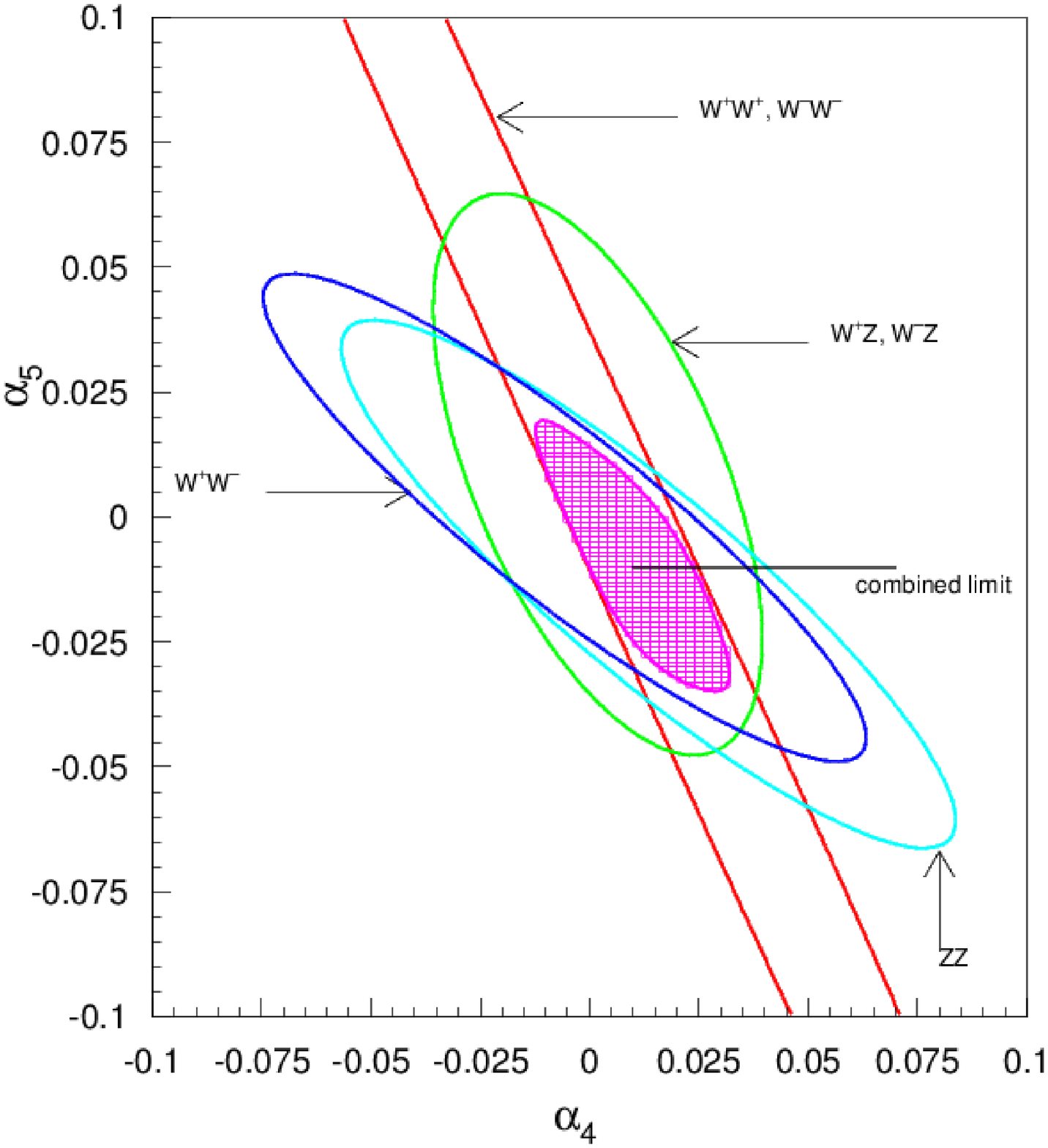}\label{belyaev}}
\hspace{0.4cm}
\subfloat[ ]{\includegraphics[scale=0.35]{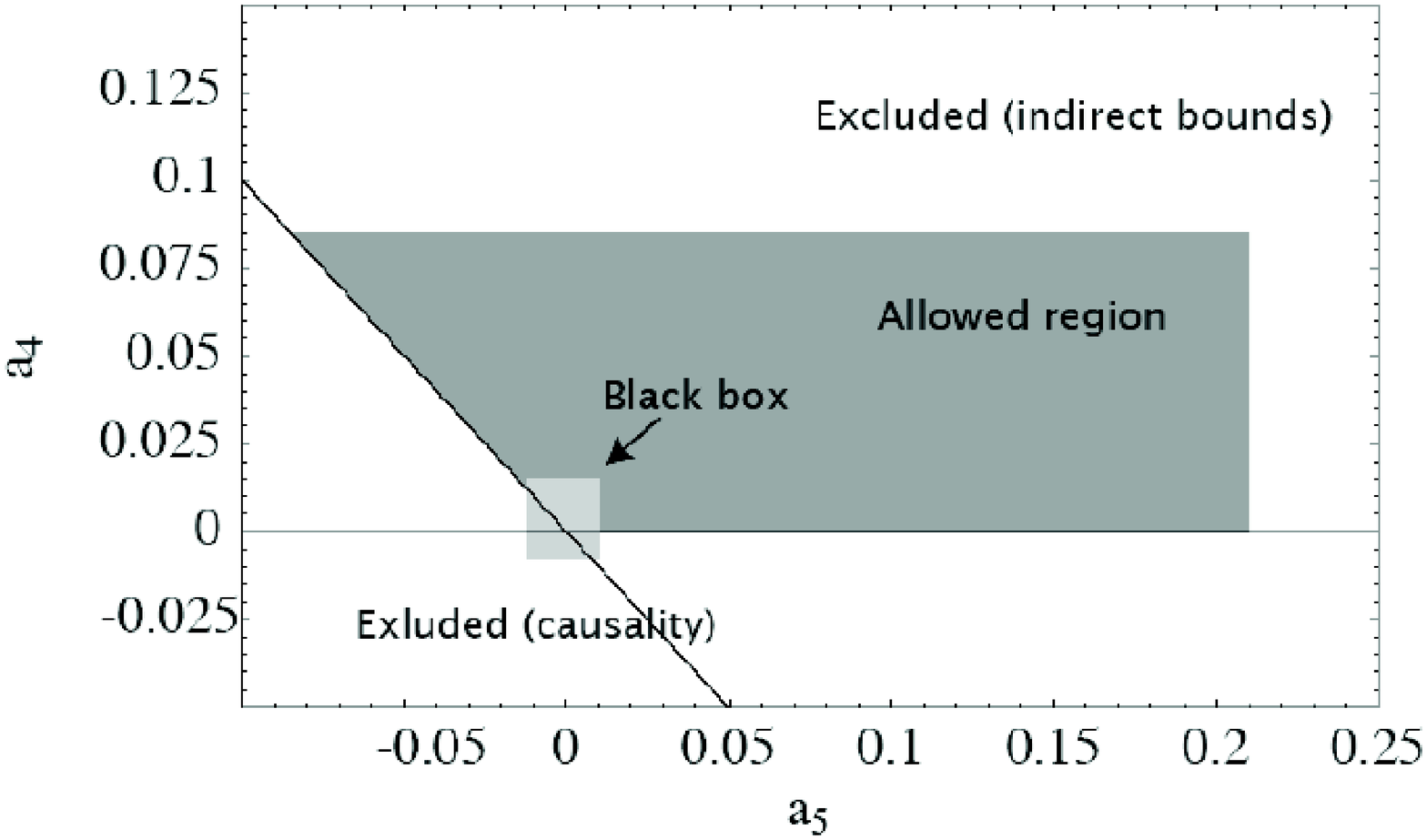}\label{vecchi}}
\caption{(a) indica a região de exclusão no LHC, operando em
  luminosidade integrada de $100\mbox{fb}^{-1}$, no plano
  $(\alpha_{4},\alpha_{5})$ para significância estatística de
  $1\sigma$, de acordo com a Ref. \cite{belyaev}. (b) mostra a região
  de valores permitidos no plano $(\alpha_{4},\alpha_{5})$ (em cinza)
  combinando limites indiretos e causais~\cite{vecchi_tese}. Na
  figura (b) também verifica-se a região de valores dos coeficientes
  $\alpha_{4}$ e $\alpha_{5}$ não acessível ao LHC (\emph{black
    box}).}
\label{vinculos}
\end{figure}

A Fig.~\ref{vinculos}a exibe os vínculos possíveis de serem sondados
pelo LHC, após 100 fb$^{-1}$ de luminosidade integrada, de acordo com
os estudos realizados na Ref.~\cite{belyaev}.
 
Além dos limites atingíveis nos colisores e vínculos de precisão,
pode-se obter também vínculos de natureza teórica para os acoplamentos
anômalos, como os de causalidade~\cite{vecchi_tese,vecchi} conforme
mostra a Fig. \ref{vinculos}b.


Na Tabela~\ref{concha} listamos os vínculos para os acoplamentos
anômalos quárticos $\alpha_{4}$ e $\alpha_{5}$ obtidos em 90\%
C.L. em~\cite{belyaev} e em 99\% C.L. em~\cite{eboli2}, quando somente um deles
contribui por vez. Esses vínculos aprimoram em mais de uma ordem de
magnitude os vínculos indiretos obtidos através de medidas de precisão
eletrofraca dos parâmetros $S$, $T$ e $U$~\cite{Brunstein:1996fz}.

\begin{table}[h!tb]
{\footnotesize
\begin{center}
\begin{tabular}{||c|c|c||}
\hline
&  & \\
 & $\alpha_{5}=0$ & $\alpha_{4}=0$ \\
&  & \\
\hline
&  & \\
Indiretos~\cite{Brunstein:1996fz} & $-0.16 \leq \alpha_{4} \leq 0.05 $ &  $-0.4\leq \alpha_{5} \leq 0.013$ \\
& & \\
Belyaev \emph{et al}.~\cite{belyaev} (90\% CL) & $-0.003 \leq \alpha_{4} \leq 0.015 $ & $-0.007\leq \alpha_{5} \leq 0.013$ \\
& & \\
Éboli \emph{et al}.~\cite{eboli2} (99\% CL) &  $-0.007 \leq \alpha_{4} \leq 0.015 $ & $-0.012\leq \alpha_{5} \leq 0.010$ \\
&& \\
\hline
\end{tabular}
\caption{Vínculos indiretos e diretos para os acoplamentos anômalos
  quárticos $\alpha_{4}$ e $\alpha_{5}$, considerando separadamente um
  dos dois diferente de zero. Os vínculos diretos foram obtidos para
  processos de VBF no LHC, com os pares de bósons de \textit{gauge}
  decaindo leptonicamente.}
\label{concha}
\end{center}}
\end{table}





Recentemente, análises do espalhamento entre bósons de \textit{gauge}
longitudinais confrontando diferentes modelos de unitarização no canal
semi-leptônico, variando os valores de massa do Higgs, através da
abordagem de lagrangianas efetivas foram
disponibilizados~\cite{ballestrero,ballestrero02}. Considerando vários
canais e valores de energia do centro de massa, análises em nível
partônico do processo $pp \to l^{\pm}\nu_{l}\;+\;\mbox{4 jatos}$ comparando
diversos cenários com Higgs leve, sem Higgs e novas ressonâncias foram
consideradas~\cite{ballestrero02,vecchi}. Nesses trabalhos, os autores
adotaram distintos esquemas de unitarização e fixaram os valores das
constantes de acoplamentos $\alpha_{4}$ e $\alpha_{5}$.

\section{Contemplando o Higgs padrão na abordagem efetiva}

Nossos cálculos se basearam na representação
não-linear~\eqref{nl}, na qual o espectro da teoria não possui um
bóson de Higgs leve ou pelo menos essa ressonância escalar não é
considerada como parte de um dubleto de isospin $SU(2)_{L}$.

No caso da representação linear, quando considera-se a existência de um
Higgs padrão (um dubleto de $SU(2)_{L}$) no espectro, os operadores
$\Sigma $ da teoria efetiva podem ser construídos a partir do dubleto
de Higgs e dos campos do setor de \textit{gauge} do MP, como fizemos
na Eq.~\eqref{campo}, por exemplo. 

De forma simples, podemos incluir
operadores de dimensão $D>4$ na lagrangiana do MP. Vértices
quárticos anômalos puros podem ser introduzidos pela derivada
covariante~\eqref{covariante} do campo de Higgs~\eqref{phi_transf}
$D_{\mu}\Phi^{\prime}$, resultando em termos de ordem 8, de acordo
com as Refs.~\cite{eboli2,tese_feigl}, tais que
\begin{align}
{\cal L}_{S,0}&=\dfrac{f_{0}}{\Lambda^{4}}\;\dfrac{}{}\left[ (D_{\mu}\Phi^{\prime})^{\dagger}D_{\nu}\Phi^{\prime}\right]\times \left[ (D^{\mu}\Phi^{\prime})^{\dagger}D^{\nu}\Phi^{\prime}\right] \\
{\cal L}_{S,1}&=\dfrac{f_{1}}{\Lambda^{4}}\;\left[ (D_{\mu}\Phi^{\prime})^{\dagger}D^{\mu}\Phi^{\prime}\right]\times \left[ (D_{\nu}\Phi^{\prime})^{\dagger}D^{\nu}\Phi^{\prime}\right],
\label{operadores_linear}
\end{align}
cujas regras de Feynman dos vértices quárticos estão listados
explicitamente no Apêndice~\ref{ap:regrasfeynman}.

Usaremos esses vértices para relacionar qualitativamente, dentro de
algumas restrições que apresentaremos adiante, os vínculos obtidos
formalmente nos cálculos da representação não-linear, onde temos os
acoplamentos anômalos $\alpha_{4}$ e $\alpha_{5}$, com a representação
linear onde os acoplamentos são $f_{0}/\Lambda^{4}$ e
$f_{1}/\Lambda^{4}$.

Vínculos obtidos na Ref.~\cite{eboli2} em 99\% C.L. e supondo a massa
do Higgs de $M_{H}=120$ GeV são apresentados na
Tabela~\ref{vinculos_linear}, para fins de comparação com nossa
estimativa.

\begin{table}[h!tb]
{\footnotesize
\begin{center}
\begin{tabular}{||c|c|c||}
\hline
&  & \\
 & $f_{S,1}=0$ & $f_{S,0}=0$ \\
&  & \\
\hline
&  & \\
Indiretos~\cite{eboli2} & $ -325 < \dfrac{f_{0}}{\Lambda^{4}}\;(\mbox{TeV}^{-4}) < 562.5 $ &  $ -812.5 < \dfrac{f_{1}}{\Lambda^{4}}\;(\mbox{TeV}^{-4}) < 1375 $ \\
& & \\
Éboli \emph{et al}.~\cite{eboli2}  &  $-22 \leq \dfrac{f_{0}}{\Lambda^{4}}\;(\mbox{TeV}^{-4}) \leq 24 $ & $-25 \leq \dfrac{f_{1}}{\Lambda^{4}}\;(\mbox{TeV}^{-4}) \leq 25 $ \\
&& \\
\hline
\end{tabular}
\caption{Vínculos para os acoplamentos anômalos quárticos na representação linear $f_{1}$ e $f_{0}$, considerando separadamente um dos dois diferente de zero. Obtidos no canal leptônico no LHC para 99\% e indiretamente através de medidas de precisão eletrofraca~\cite{eboli2}.}
\label{vinculos_linear}
\end{center}}
\end{table}

\section{Análise e cálculos dos acoplamentos quárticos em colisões $pp$ no LHC}
\label{analise}

Os valores obtidos anteriormente para os vínculos dos
acoplamentos anômalos quárticos no LHC através dos processos de VBF
analisaram somente o canal leptônico, o chamado modo ``gold-platted''. Embora esse canal possua um sinal razoavelmente limpo, ele
sofre limitação no número de eventos em relação ao canal semileptônico
e hadrônico.

No caso específico em que tenhamos processos $pp\to W^\pm V jj$, com
$V=W^\pm, Z$, é natural questionarmos se canais com decaimentos
hadrônicos dos bósons de \textit{gauge} possam no mínimo ser
complementares aos canais leptônicos.  

Para minimizar a contribuição
dos \textit{backgrounds} de QCD, uma possibilidade é analisar o
canal semileptônico, onde $W^\pm \to l \nu_l$ com $l=e^{\pm},\mu^{\pm}$
e $V$ decai hadronicamente.  

Lembrando que no caso específico de $W^\pm$
a taxa de ramificação do decaimento hadrônico é
\begin{equation}
\mbox{BR}(W^{\pm} \to q q') \approx 3\;\mbox{BR}(W^{\pm} \to l\,\nu_{l}),
\end{equation}
e que podemos reconstruir a massa invariante do bóson $V$ a partir de
dois jatos, não é tão claro que esse tipo de canal seja completamente
dominado pelos \textit{backgrounds} de QCD.

No intuito de averiguar essa questão, efetuamos nesta tese o estudo
dos canais semileptônicos $pp \rightarrow l^{\pm}\nu_{l}jjjj$. Efetuamos as nossas análises em nível partônico e em \emph{leading
  order} (LO), tentando na medida do possível realizar o cálculo
completo, ou seja considerando elementos de matriz completos tanto do sinal
como dos \textit{backgrounds}
\begin{enumerate}
\item  irredutível, incluindo processos de
produção de três bósons, em ${\cal O}(\alpha_{em}^{6})$;
\item QCD em
${\cal O}(\alpha_{em}^{4}\alpha_{s}^{2})$ e ${\cal
  O}(\alpha_{em}^{2}\alpha_{s}^{4})$;
\item  processos envolvendo
quarks \emph{top}: $pp\to t\bar{t} +
0\;\mbox{até}\;2\;\mbox{jatos}$. 
\end{enumerate}

Escolhemos para nossas análises os vértices anômalos puramente
quárticos entre os bósons de \textit{gauge} dados pelos operadores
${\cal L}_{4}$ e ${\cal L}_{5}$ levando em conta a existência de
uma partícula escalar com massa de $120$ GeV\footnote{Na verdade
  incluimos o próprio bóson de Higgs do MP nas nossas análises, já que 
elas são insensíveis a esse detalhe.}.

Conforme discutido na Seção~\ref{vinculos_teo}, no processo de
espalhamento de bósons de \textit{gauge} longitudinalmente polarizados
no MP com a ausência do Higgs (ou $M_H \to \infty$), a unitaridade é
conservada para energias no centro de massa $M_{WV} \approx 1.25$
TeV. Independentemente da presença ou não do Higgs, os operadores
anômalos envolvendo vértices quárticos se tornam fortemente
interagentes nessa escala de energia e eventualmente colocam em risco
a unitaridade da matriz $S$. 

Um procedimento comum existente na literatura é introduzir novas
ressonâncias para resolver esse problema, papel desempenhado pelo
bóson de Higgs leve no MP, ou utilizar o método de
\textit{sharp-cutoff}, o que consiste em tomar o corte na energia do
centro de massa, $M_{WV}$. Acima dessa escala, espera-se o surgimento
de um grau de liberdade não descrito pela teoria efetiva. Seguindo o
critério existente na literatura~\cite{belyaev,eboli2} aplicamos o valor
$1.25$ TeV como o valor máximo.

É importante destacar que, como estamos realizando cálculos completos
dos processos $pp \to l^{\pm}\nu_{l}jjjj$, conseguimos obter
corretamente a interferência entre o sinal e o \emph{background} irredutível,
que se apresentam ambos na mesma ordem em teoria de perturbação.  Não
seria o caso, se estivéssemos adotando a aproximação de $W$
efetivo~\cite{efet} ou se aplicássemos o teorema de
equivalência~\cite{equiv1,equiv2} nas amplitudes de espalhamento.

\subsection{Ambientação do cálculos}
\label{ambientacao}

A análise numérica, tanto do sinal como dos diversos
\emph{backgrounds}, requer o cálculo em nível partônico de
subprocessos com $n=6$ partículas no estado final.


Utilizamos o pacote Madgraph~\cite{mad}, com a implementação dos
vértices quárticos anômalos seguindo o protocolo da biblioteca
HELAS~\cite{helas}, para obtermos os elementos de matriz necessários
no cálculo das seções de choque.

\begin{figure}
\centering
\includegraphics[scale=0.65]{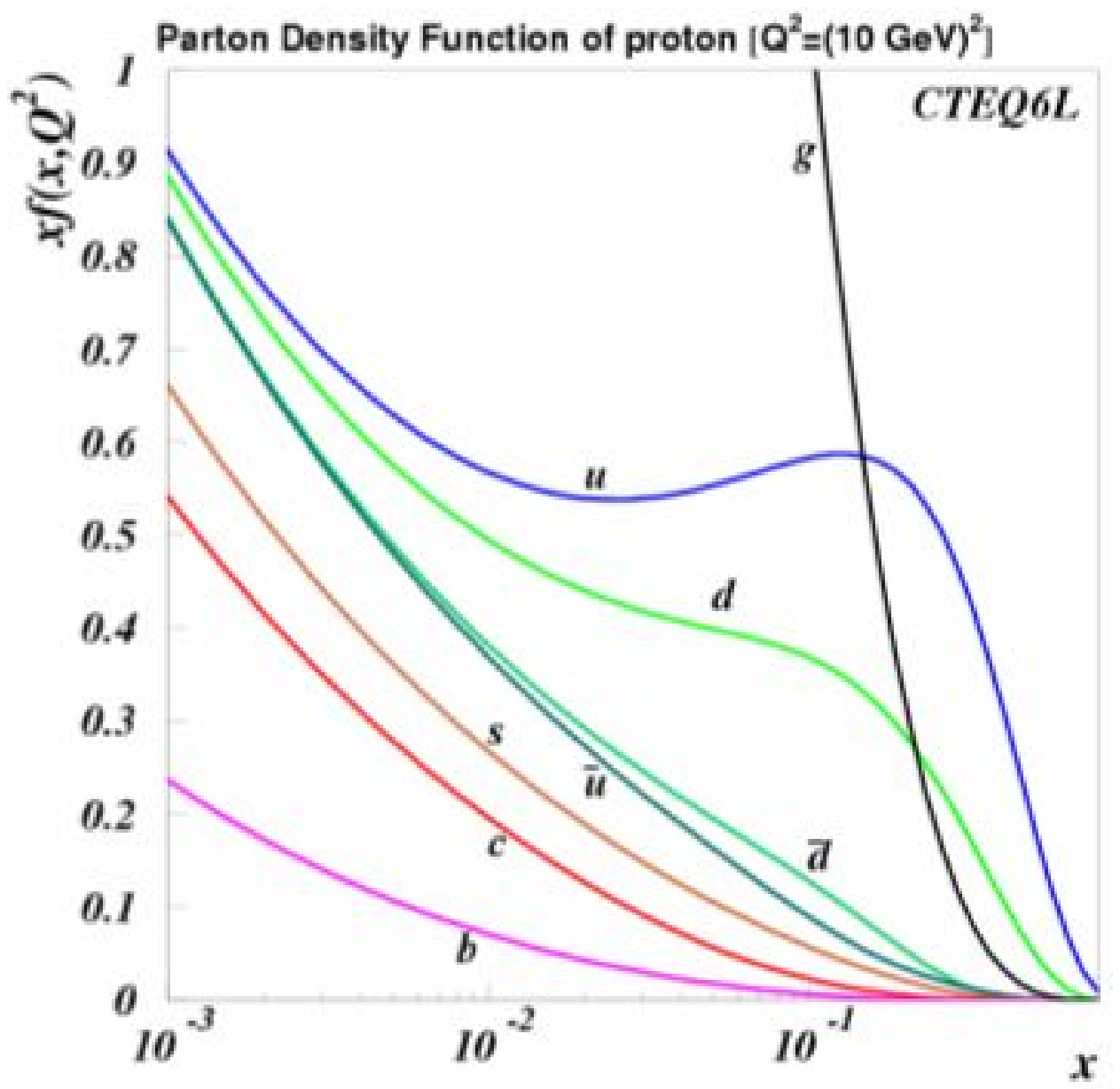}
\caption{Perfil das funções de distribuição dos pártons (PDF) de
  acordo com a opção CTEQ6L~\cite{cteq} utilizada nesse trabalho.}
\label{pdf2}
\end{figure}

A integração multidimensional pelo algoritmo Vegas~\cite{vegas} foi
efetuada nos moldes do exposto no Apêndice~\ref{sec:montecarlo},
usando as amplitudes de helicidade geradas pelo Madgraph e a função de
distribuição dos pártons CTEQ6L~\cite{cteq}, que adota o esquema de
massa zero para os pártons e cujo comportamento pode ser visualizado
na Fig.~\ref{pdf2}. 

Adotamos também os parâmetros eletrofracos
$\sin^{2}{\theta_{W}}=0.23124$, $\alpha_{EM}=1/128.93$,
$M_{Z}=91.1876$ GeV e $M_{W}=80.385$ GeV.

Os subprocessos do sinal e \emph{background} irredutível, listados nas
Tabelas~\ref{wp_qed6} e~\ref{wm_qed6}, tiveram seus elementos de
matriz modificados em sua estrutura. Identificamos os vértices
puramente quárticos da forma $WWWW$ e $WWZZ$ e implementamos nesses
vértices as contribuições anômalas listadas no
Apêndice~\ref{ap:regrasfeynman}. Utilizamos elementos de matriz
completos na medida do possível, conforme explicaremos em detalhes
mais adiante, já que alguns subprocessos
foram tratados de forma a 
contornarmos problemas de convergência do Vegas. Nesses casos
consideramos somente diagramas semi-ressonantes, ou seja o bóson de
\textit{gauge} com decaimento hadrônico está sempre na sua camada
de massa.

Tendo em vista a enorme quantidade de subprocessos, optamos por
desprezar as contribuições dos subprocessos com dois quarks do mar no
estado inicial, já que estes influenciam muito pouco o resultado da
seção de choque final do processo. De fato, subprocessos do tipo
geral $q_{1}q_{2}\to VV q_{3}q_{4}$ ocorrem para $x\gtrsim 0.2$ no
LHC14, sendo a contribuição do mar irrisória conforme ilustra
a Fig.~\ref{pdf}. Por outro lado, todos os subprocessos de \emph{backgrounds} que envolvem glúons no
estado inicial foram considerados, pois os processos de VBF no LHC ocorrem
tipicamente na região de $x$ onde esses pártons são importantes.
 
\begin{figure}
\includegraphics[scale=0.75]{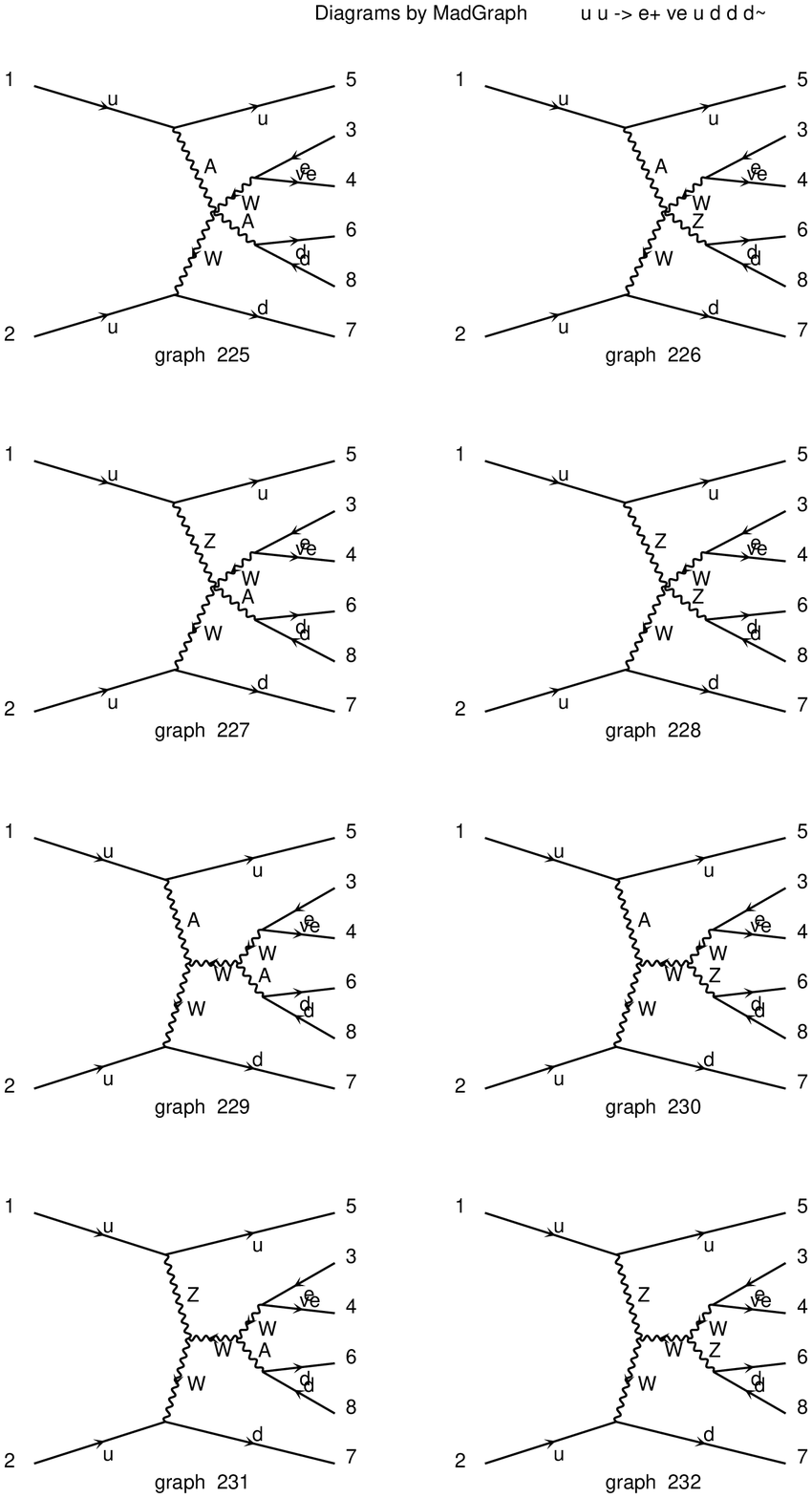}
\caption{Exemplos de diagramas de Feynman mostrando, dentre outros, os
  vértices quárticos do subprocesso $uu\to e^{+}\nu_{e}udd\bar{d}$ em
  ${\cal O}(\alpha_{em}=6)$.}
\label{diagramas1}
\end{figure}

Para fins de ilustração, na Fig.~\ref{diagramas1} temos um ínfimo
exemplo dos possíveis diagramas que surgem quando geramos o elemento
de matriz de um subprocesso dominante, como é o caso do subprocesso $uu\to e^{+}\nu_{e}udd\bar{d}$. Nesta figura,
o diagrama $228$ possui topologia de VBF com vértice quártico
recebendo contribuição do acoplamento quártico anômalo $WZWZ$.

Ao considerarmos a contribuição anômala, a amplitude de espalhamento
total adquire a forma geral
\begin{equation}
\mathcal{M}=\mathcal{M}_{\mbox{\tiny MP}} + \mathcal{M}_{\mbox{\tiny anômalo}}.
\label{elem}
\end{equation}

Como $\mathcal{M}_{\mbox{\tiny anômalo}}$ é proporcional aos
acoplamentos $\alpha_{4}$ e $\alpha_{5}$ então pode-se fatorar as
constantes de acoplamentos anômalas do elemento de matriz anômalo e
escrever a seção de choque explicitamente em função desses parâmetros de tal forma que
\begin{equation}
\sigma = \sigma_{bck} + \alpha_4\sigma_4+ \alpha_5\sigma_{5}
+ \alpha_4^{2}\sigma_{44}+ \alpha_5^{2}\sigma_{55}+
\alpha_4 \alpha_5 \sigma_{45},
\label{cross_section}
\end{equation}
onde $\sigma_{bck}$ contém as contribuições das seções de choque de
todos os \emph{backgrounds} do MP listados anteriormente, adotando
$M_H=120$ GeV. $\sigma_{4}$ e $\sigma_{5}$ descrevem as seções de
choque da interferência entre o MP com os vértices anômalos dados
pelos operadores ${\cal L}_{4}$ e ${\cal L}_{5}$, respectivamente (com
as constantes de acoplamento já devidamente fatoradas). $\sigma_{44}$ e
$\sigma_{55}$ são as seções de choque com contribuições puramente anômalas,
com um único operador contribuindo por vez e finalmente,
$\sigma_{45}$ representa a interferência entre os dois operadores anômalos.

No intuito de estimar as incertezas advindas das escolhas das escalas
de renormalização e fatorização, utilizamos dois conjuntos de valores
para estas escalas
\begin{eqnarray}
\mbox{C1}&:& \mu^{0}_{F}=\mu^{0}_{R}=\varepsilon \sqrt{\sum_{i=1}^{4} p^{2}_{T_{j_{i}}}/2}
\label{c1}\\
& & \nonumber\\
\mbox{C2}&:& \alpha_{s}(\mu^{0}_{R})= \sqrt[n]{ \prod_{i=1}^n \alpha_s (\varepsilon p_{T_{j_i}}) } \quad \textrm{e} \quad
 \mu^{0}_{F}=\varepsilon \times \mbox{min}(p_{T_{j_{i}}},p_{T_{j_{k}}}),
\label{c2}
\end{eqnarray}
onde $p_{T_{j_{i}}}$ ($i=1, \ldots, 4$) são os momentos transversais dos
jatos e consideramos os valores $\varepsilon = 0.25$, $1$ e $4$. No conjunto C2~\eqref{c2}, fizemos
a escolha de $\mu^{0}_{R}$ para cada vértice de QCD. Assim, para o
\textit{background} ${\cal O}(\alpha_{em}^{4}\alpha_{s}^{2})$, como $n=2$, adotamos os momentos transversais dos dois jatos \textit{taggings}. Já
para ${\cal O}(\alpha_{em}^{2}\alpha_{s}^{4})$, como $n=4$, usamos os
$p_T$'s de todos os quatro jatos.

Trabalhamos em nível partônico em \textit{leading order} (LO) e não
consideramos \emph{showering} nem hadronização. Contudo, no intuito de
se produzir resultados mais realísticos, aplicamos um \emph{smear}
gaussiano nas energias $E$ (em GeV), mas não nas direções,
para simular a resolução experimental de todos os quarks (jatos) e léptons. 
Para os jatos, utilizamos de acordo com~\cite{eboli2}
\[
\dfrac{\Delta E}{E} = \left\{ 
\begin{array}{ll}
&\dfrac{0.5}{\sqrt{E}}\oplus 0.02, \; \mbox{se}\;\;|\eta_{j}|\leq 3
\\
&\dfrac{1}{\sqrt{E}}\oplus 0.07, \; \mbox{se}\;\;|\eta_{j}|>3.
\end{array}
\right. 
\] 

Já para os léptons carregados, usamos a resolução também de acordo com~\cite{eboli2} 
\begin{equation}
\dfrac{\Delta E}{E}=\dfrac{0.1}{\sqrt{E}}\oplus 0.01.
\end{equation}

A eficiência na detecção dos jatos foi considerada como
$(0.75)^{4}=0.31$, já que temos 4 jatos no estado final, e do lépton
carregado como $0.9$.

Dos 4 jatos, pela topologia de VBF têm-se 2 na direção
\emph{forward}/\emph{backward}, que caracterizam esse tipo de
processo, denotados como \emph{tagging} jatos, e dois centrais, que no
caso do sinal deveriam ser provenientes de $W$ ou $Z$. Exceto quando
há partículas idênticas no estado final (cujo detalhe de cálculo será
discutido mais adiante), conhecemos de antemão (pelo
elemento de matriz), quais são os jatos \textit{taggings} e quais os
centrais. É a partir desta configuração que se constrói o espaço de
fase para gerar os eventos.

Contudo, após a geração de cada evento pelo método de Monte Carlo, nós
efetuamos o reordenamento desses 4 jatos, no intuito de tornar os
nossos cálculos mais próximo possível da situação real. Fazendo o
reordenamento de acordo com o valor da pseudo-rapidez associamos os jatos com menor e maior $\eta$ aos jatos \emph{tagging} 
(nos sentidos \emph{forward} e \emph{backward}). Jatos com valores de
$\eta$ intermediários foram identificados como os jatos centrais.

Nesse ponto, vale comentar que os \emph{backgrounds} provenientes dos
processos envolvendo quarks \emph{top} ($pp\to t\bar{t}$, $pp\to
t\bar{t}j$ e $pp\to t\bar{t}jj$) foram analisados usando aproximação
de largura fina. Levando em consideração que o estado final desses
processos pode apresentar eventos com mais de quatro jatos, analisamos
cuidadosamente o comportamento desses jatos como descrevemos a seguir.

No processo $pp\to t\bar{t}$, os quarks $b$ produzidos no decaimento
dos quarks $t$ foram identificados como jatos \emph{tagging} nos
nossos cálculos, para efeito de Monte Carlo.

Nos processos $pp\to t\bar{t}j$ e $pp\to t\bar{t}jj$ o(s) jato(s)
adicional(is) são identificados como jato(s) \emph{tagging}, para
efeito de Monte Carlo, enquanto um dos quarks $b$ em $pp\to t\bar{t}j$
e ambos em $pp\to t\bar{t}jj$ são mais \emph{soft} e portanto
implementados como jatos centrais.

Após geração de cada evento os jatos foram reordenados conforme
critério descrito acima. No caso da existência de jatos extras (casos
de $t\bar{t}$ + 1 ou 2 jatos), selecionamos o par que está dentro da
janela de massa do $W^{\pm}$ como sendo os jatos centrais e não
aplicamos veto nesses jatos adicionais, já que não estimamos correções
de \textit{next leading order} (NLO).

No canal semileptônico, estudado nesta tese, não há como separar
experimentalmente o par de jatos oriundos do decaimento hadrônico do
$W$ ou do $Z$. Sendo assim, embora haja um pico na massa do $Z$ no
perfil das distribuições de massa invariante dos jatos centrais, salientamos que se trata apenas
de uma ``contaminação'' e contamos como sendo eventos com $W$'s pois
estes são dominantes.

Como no sinal temos a produção $W^\pm V$ acompanhado de 2 jatos e ainda $V$
decai hadronicamente, então é possível reconstruir o momento do
neutrino $\nu_{l}$, visto que se trata do produto de decaimento do $W$. Impondo que
a massa invariante do par $l\nu_{l}$ seja igual à massa do bóson
$W^{\pm}$, de acordo com~\cite{eboli2}, temos que
\begin{equation}
M_{W}^{2}=(p^{l}+p^{\nu})^{2}.
\label{neutrino}
\end{equation}

A componente longitudinal do momento do neutrino pode assim ser obtida com o
auxílio da Eq.~\eqref{neutrino}, mas com uma ambiguidade, já que
existem duas soluções  
\begin{align}
p_{L}^{\nu} = & \dfrac{1}{2(p_{T}^{l})^{2}} \biggl \{ [M_{W}^{2}+2({\mathbf p}_{T}^{l}\;.\;\not\!\!{\mathbf p}_{T} )] p_{L}^{l} 
\nonumber \\
          &   \pm \sqrt{ [M_{W}^{2}+2({\mathbf p}_{T}^{l}\;.\;\not\!\!{\mathbf p}_{T})]^{2} |{\mathbf p}^{l}|^{2} - 4(p_{T}^{l} E^{l}\not\!\!E_{T})^{2}} \biggr \}.
\label{pzneutrino}
\end{align}

Para quantificar os efeitos dos vértices anômalos que poderão ser
testados pelo LHC, definimos a significância estatística 
para o sinal anômalo ${\cal  S}_{sinal}$ como sendo
\begin{equation} 
{\cal S}\equiv \dfrac{N_{sinal}}{\sqrt{N_{back}}},
\label{significancia}
\end{equation}
onde $N_{sinal}$ é o número de eventos do sinal anômalo e
$N_{\mbox{back}}$ o número de eventos dos \emph{backgrounds}, ambos
obtidos com expressão da seção de choque~\eqref{cross_section} e
considerando as eficiências na detecção dos jatos e lépton carregado.

\subsubsection{Peculiaridades do sinal}
\label{peculiaridades}

Descrevemos nesta subseção algumas observações de natureza técnica. No
início desse projeto, nossa estratégia foi tratar o processo completo
$pp\to l^{\pm}\nu_{l}jjjj$, com
$l^{\pm}=e^{\pm}\;\mbox{e}\;\mu^{\pm}$, sem utilizar quaisquer
aproximações nos elementos de matriz. No entanto, durante essa
abordagem nos deparamos com vários problemas de convergência nos
cálculos numéricos do Vegas, sobretudo nos subprocessos que possuem
duas partículas idênticas no estado final. 

Para contornar esses problemas, simplificamos esses processos considerando apenas os
chamados ``diagramas semi-ressonantes'' $pp\to l^{\pm}\nu_{l}V(\rightarrow
jj)jj$, com $V=W^{\pm},Z$, ou seja avaliamos contribuições nas quais $V$ é produzido e decai em sua camada de
massa. Como esses diagramas dão maiores contribuições então
consideramos que os nossos resultados são próximos do cálculo
completo.

Para entender o motivo das divergências, analisamos cada elemento de
matriz de cada um dos 301 subprocessos da reação completa $pp\to
l^{\pm}\nu_{l}jjjj$ e verificamos diversas sutilezas na elaboração dos
diagramas de Feymann que nos obrigaram a modificar nosso protocolo de
cálculos inicial.

\begin{figure}
\includegraphics[scale=0.75]{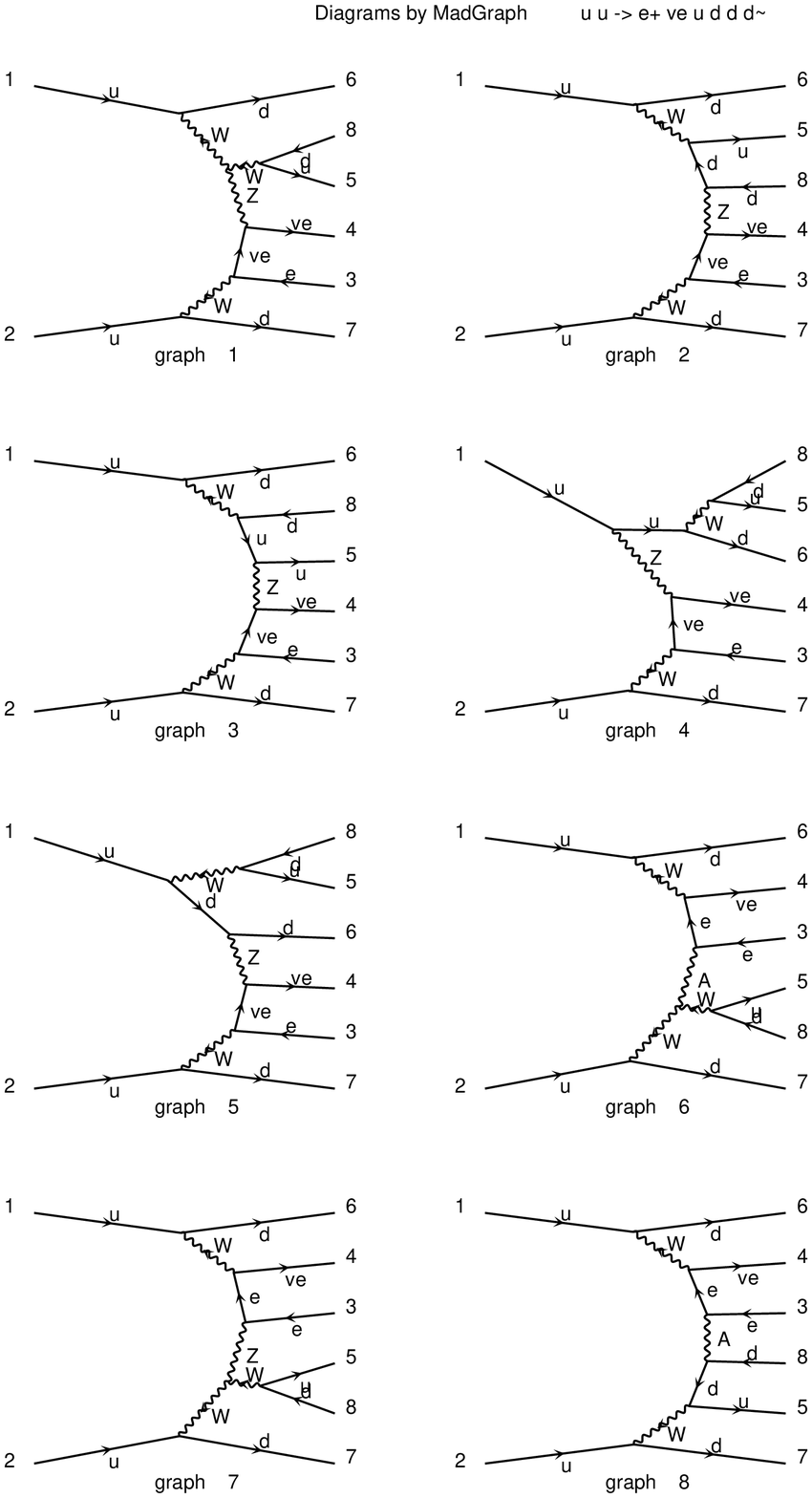} 
\caption{Diagramas de Feynman não-ressonantes do subprocesso $uu\to
  e^{+}\nu_{e}udd\bar{d}$ em ${\cal O}(\alpha_{em}^6)$.}
\label{diagramas2}
\end{figure}

A reação completa $pp\to l^{\pm}\nu_{l}jjjj$ em ${\cal
  O}(\alpha_{em}^{6})$, como esperado, gera subprocessos tanto com
diagramas ``ressonantes'' (produção de pares de bósons $WW$
e $WZ$ e subsequente decaimento destes em léptons e quarks) como mostram os
diagramas $228$ e $232$ da Fig.~\ref{diagramas1}, quanto com
diagramas ``não ressonantes'' (onde um ou nenhum bóson de
\textit{gauge} massivo se encontra na camada de massa), como mostram os
diagramas $1$ e $4$ da Fig.~\ref{diagramas2}.

Em alguns casos, um único subprocesso possui múltiplas configurações de
vértices quárticos com topologia de VBF. Por exemplo, o subprocesso
$uu\to e^{+}\nu_{e}udd\bar{d}$, com seus 1046 diagramas de Feymann,
apresenta duas configurações de vértices quárticos envolvendo bósons
de \textit{gauge}, a saber $WW\to WW$ e $WZ\to WZ$.

Uma característica crucial dos processos semileptônicos se deve à
presença de partículas idênticas no estado final. Essas partículas,
para fins de cálculo numérico, dificultam a identificação dos jatos
\emph{tagging} pelo Vegas, prejudicando o reconhecimento do espaço de
fase, e consequentemente a convergência dos cálculos, como relatado no
Apêndice~\ref{subsec: eftagging}.

\begin{figure}
\includegraphics[scale=0.75]{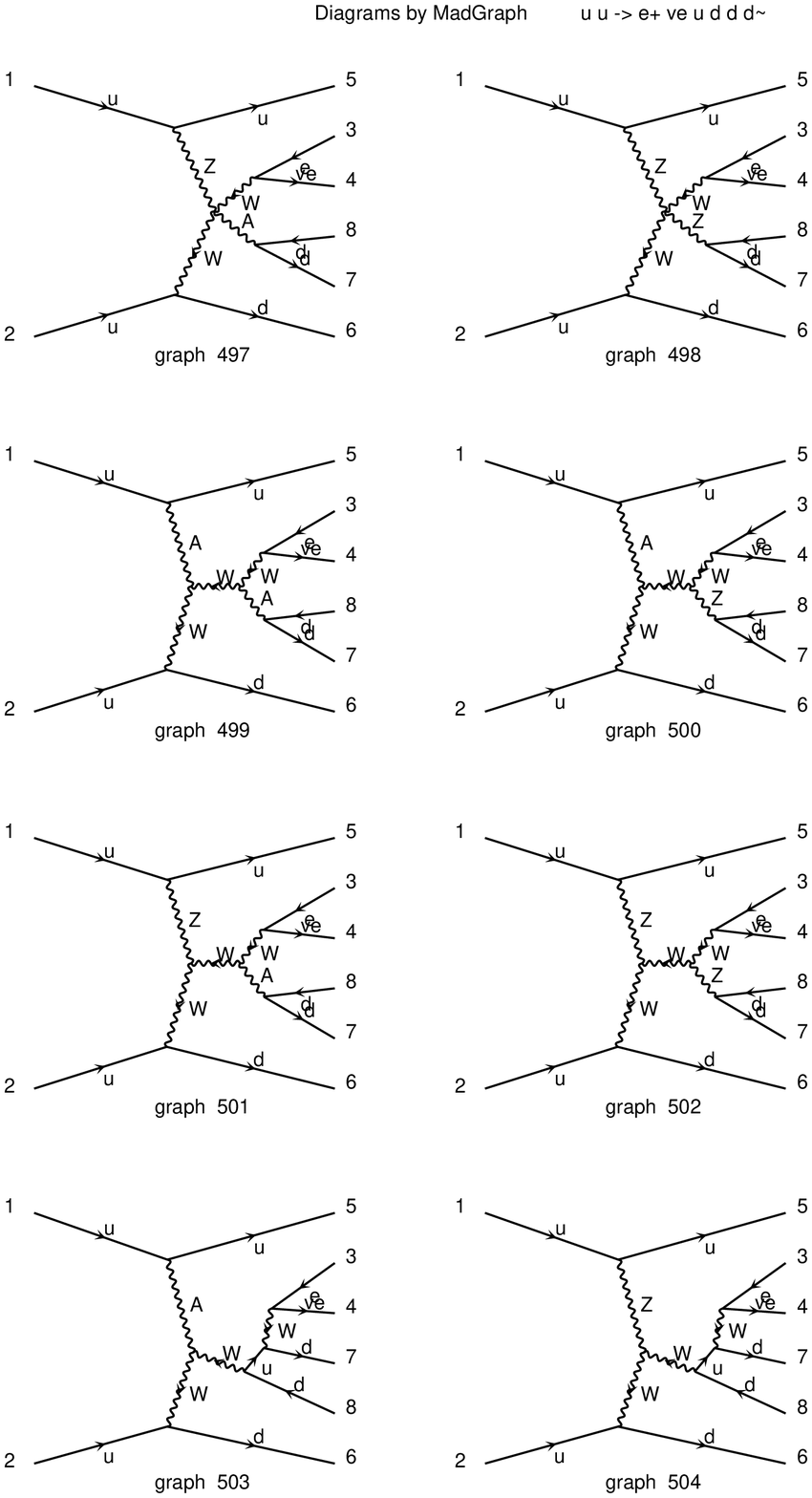} 
\caption{Diagramas do subprocesso $uu\to e^{+}\nu_{e}udd\bar{d}$, onde
  identificam-se partículas idênticas no estado final com diferentes
  funções na reação.}
\label{diagramas3}
\end{figure}

Para ilustrar esse fato, comparemos os diagramas $228$ e $498$ nas
Figs.~\ref{diagramas1} e~\ref{diagramas3}, respectivamente. Note
que se tratam do mesmo tipo de diagramas, ou seja ambos com topologia
de VBF e contribuindo com o vértice quártico anômalo $WZ\to WZ$. No
diagrama $228$ os jatos \emph{tagging} são identificados pelos quarks
$u$ e $d$ no estado final com a numeração $5$ e $7$,
respectivamente. Nesse mesmo diagrama o outro quark $d$, com numeração
$6$, se apresenta como produto de decaimento do bóson $Z$.
Denominemos esses quarks $u_{5}$, $d_{6}$ e $d_{7}$ para facilitar a
identificação. Agora, verificando o diagrama $498$ nota-se que,
enquanto o quark $u_{5}$ se mantém como jato \emph{tagging}, há uma
troca de papéis entre os quarks $d$, pois o quark $d_{7}$ agora aparece
como produto de decaimento do $Z$ enquanto o quark $d_{6}$ aparece
como jato \emph{tagging}.

Além disso, subprocessos com quarks $u$ e $d$ combinados com seus
respectivos antiquarks no estado inicial, possuem da mesma forma
algumas peculiaridades. Citamos aqui o caso do subprocesso
$d\bar{d}\to e^{+}\nu_{e}u\bar{u}\bar{u}d$.

\begin{figure}
\includegraphics[scale=0.75]{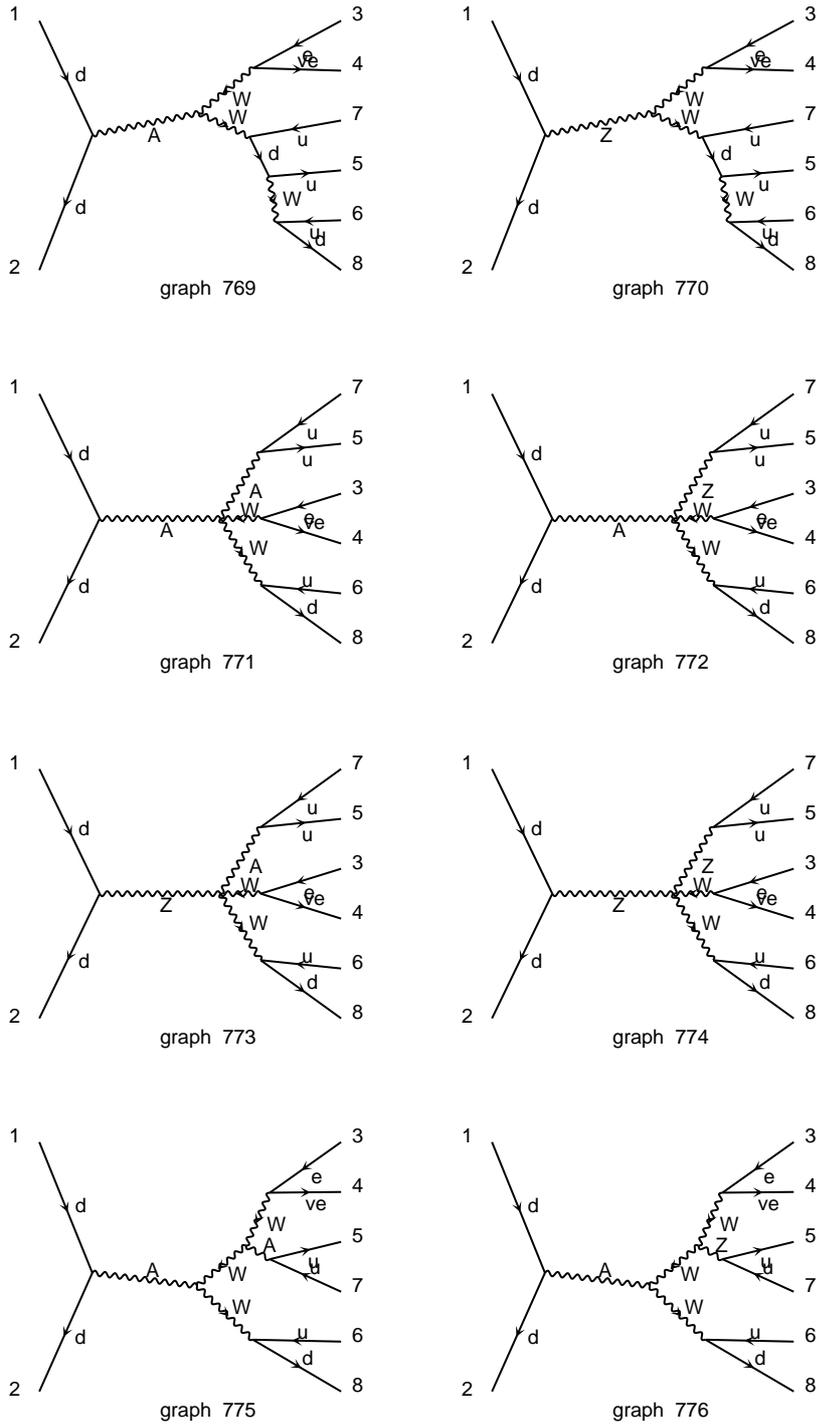}
\caption{Diagramas do subprocesso $d\bar{d}\to e^{+}\nu_{e}udd\bar{d}$ com
  exemplo de acoplamento quártico com produção de três bósons de
  \textit{gauge} com massa.}
\label{diagramas4}
\end{figure}

O subprocesso $d\bar{d}\to e^{+}\nu_{e}u\bar{u}\bar{u}d$ também possui
1046 diagramas de Feymann, mas dentre eles encontramos três distintas
configurações de vértices quárticos, a saber $WW\to WW$, $WZ\to WZ$ e
$Z \to Z WW$. Dentre essas possibilidades observamos a produção de
três bósons de \textit{gauge} massivos, como pode-se ver no diagrama
$774$ da Fig.~\ref{diagramas4}. Esse tipo de diagrama, apesar de
contribuir com vértice anômalo, não apresenta o perfil de VBF e gera
incompatibilidades se tratamos seu espaço de fase considerando essa
topologia. Deste modo, vértices quárticos apresentando processos de
produção de três bósons de \textit{gauge} foram analisados
numericamente usando o espaço de fase ``tradicional'' com uma cadeia
de decaimento do tipo $1 \to 2$, conforme mostrado
no Apêndice~\ref{subsec:eftradicional}. Um estudo detalhado desse tipo de 
contribuição foi efetuado na Ref.~\cite{tese_feigl}.

Tendo como base as características expostas até esse momento,
apresentamos adiante nossa estratégia para manipular os
\textit{backgrounds}, mantendo o máximo possível o nosso sinal anômalo
intacto.

\subsection{Cortes aplicados}

Antes de entrar nos cortes relacionados aos observáveis, reiteramos o corte de natureza teórica
$M_{WV} < 1.25$ TeV para processos envolvendo vértices
anômalos, conforme discutido na Seção~\ref{analise}.

Em primeiro lugar, estabelecemos cortes mínimos para cumprir com critérios básicos de isolamento de jatos e do lépton carregado, além de sua aceitação nos calorímetros hadrônico e
eletromagnético\footnote{Consideramos, por simplicidade, que a detecção
  do múon é similar à do elétron.}, tais que
\begin{align}
&p_{T}^{j}\geq20\;\mbox{GeV}\;\;,\;\;|\eta_{j}|\leq4.9\;\;\mbox{e}\;\;\Delta
  R_{jj}\geq0.4,
\label{cut22}
\\
&p_{T}^{l}\geq 30\;\mbox{GeV}\;,\;|\eta_{l}|\leq 2.5\;\;\mbox{e}\;\;\Delta R_{lj} \geq 0.4
\label{cut23}
\end{align}

Embora não tenhamos aplicado o veto aos jatos extras dos
\textit{backgrounds} com quarks \textit{top}, tomamos o cuidado de
considerá-los nos critérios de isolamento~\eqref{cut22} e~\eqref{cut23}.

Além disto, como o nosso sinal anômalo contém neutrinos que carregam uma parte
da energia transversal da reação, exigimos um mínimo para a energia
transversal perdida (\emph{missing})
\begin{equation}
\not\!\!E_{T} \geq 30\;\mbox{GeV}.
\label{cut24}
\end{equation}



É bem conhecido que processos que ocorrem via VBF são caracterizados
por um par de jatos de grande rapidez, em hemisférios opostos no
detector. Logo, adotamos os cortes básicos específicos para este tipo
de topologia dados por~\cite{barger3}
\begin{equation}
|\eta_{j_{\mbox{\tiny tag1}}}-\eta_{j_{\mbox{\tiny tag2}}}| >
3.8, \quad \mbox{com}\quad \eta_{j_{\mbox{\tiny tag1}}} .\eta_{j_{\mbox{\tiny tag2}}} < 0.
\label{cut25}
\end{equation}




Impomos também que os produtos de decaimento dos
bósons de \textit{gauge} (jatos centrais e lépton carregado) estejam dentro da
região central definida pelos jatos \textit{taggings}. Sendo assim,
os produtos de decaimento devem se manter na região
\begin{equation}
\eta_{\mbox{\tiny min}}^{\mbox{\tiny tag}} < \eta_{l,j} <
\eta_{\mbox{\tiny max}}^{\mbox{\tiny tag}}.
\label{cut27}
\end{equation}

Pelas análises anteriores da topologia de VBF~\cite{eboli2},
os \textit{backgrounds} de QCD preferem regiões de massa
invariante menores quando comparados ao \textit{background} irredutível
eletrofraco. Nas nossas análises, verificamos que tanto
os \textit{backgrounds} de QCD como de $t\bar{t}+n$ jatos possuem esse
comportamento. Deste modo, impomos logo de início uma massa invariante
considerável dos jatos \textit{tagging}, no valor mínimo de
\begin{equation}
M_{\mbox{\tiny tags}}\geq 1000\;\mbox{GeV}.
\label{cut26}
\end{equation}

\begin{figure}[!ht]
\centering
{\includegraphics[scale=0.75]{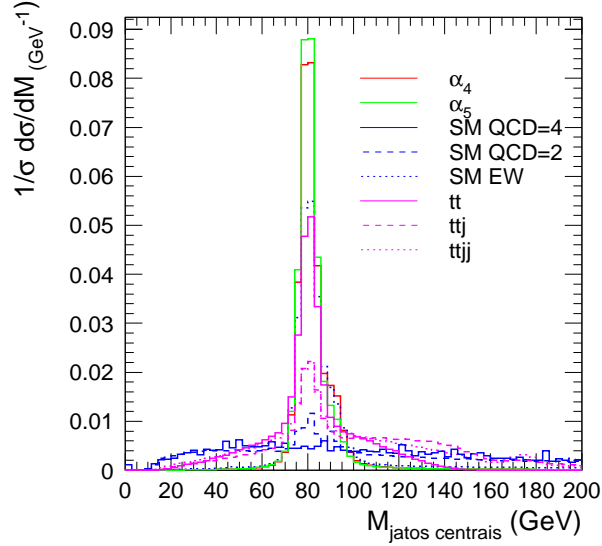}}
\caption{Massa invariante dos jatos
  centrais com cortes~\eqref{cut23} a~\eqref{cut26}.}
\label{Mj2j3}
\end{figure}

Como no sinal anômalo os jatos centrais, oriundos do $W$ ou $Z$, são produzidos na região central do detetor e os \textit{backgrounds} de QCD não costumam
ter essa característica, conforme mostra a Fig.~\ref{Mj2j3},
impomos que a massa invariante entre os jatos centrais esteja dentro
da janela de massa do $W$ de acordo com o critério
\begin{equation}
|M_{\mbox{\tiny jatos centrais}} - M_W| \leq 15\;\mbox{GeV}.
\label{cut:delta_mv}
\end{equation}

Os cortes implementados até esse ponto removem muito pouco o sinal,
deixando espaço para analisar outros perfis de distribuições
cinemáticas até encontrar um conjunto que propicie valor de
significância estatística adequado em relação ao \emph{background} do MP.

\begin{figure}[!ht]
\centering
\subfloat[ ]{\includegraphics[scale=0.75]{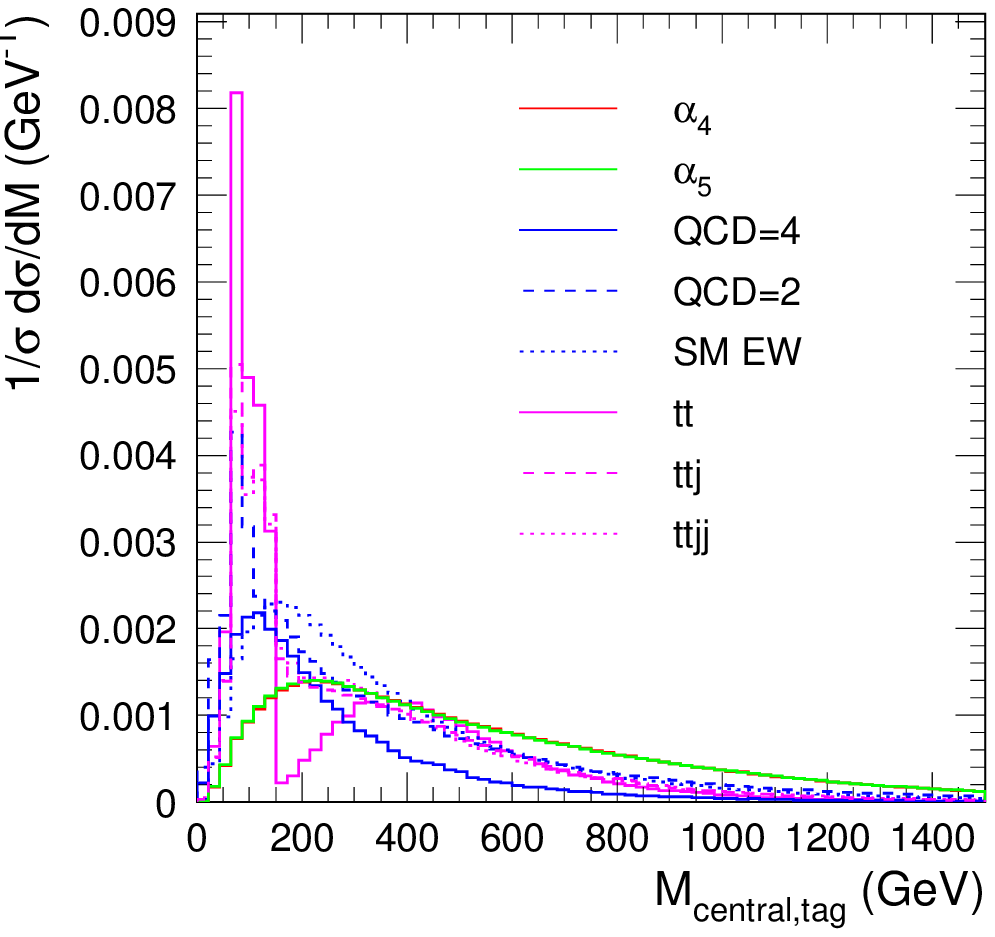}\label{mj1j2}}
\hspace{0.2cm}
\subfloat[ ]{\includegraphics[scale=0.75]{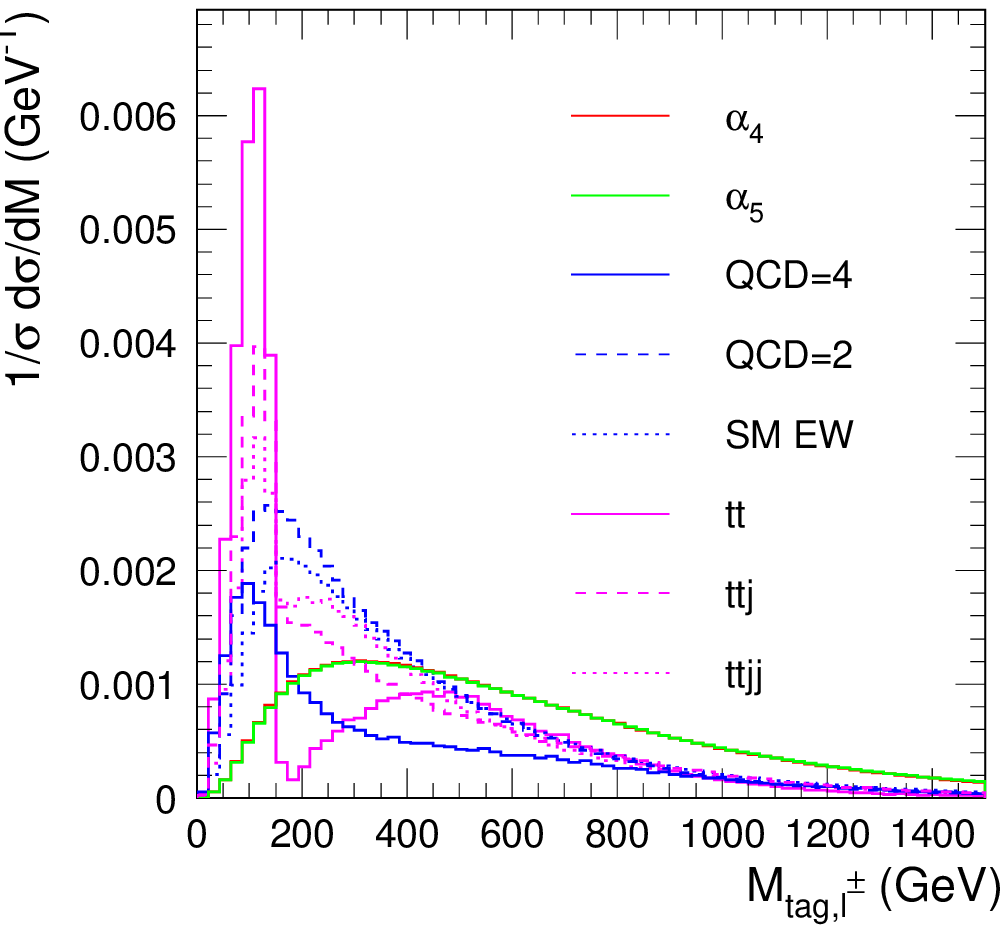}\label{mlj1}}
\caption{Perfis das massas invariantes (a) entre jatos central e 
  \textit{tagging} e (b) entre o lépton carregado e o jato
  \textit{tagging}. Ambas após cortes~\eqref{cut22} a
  \eqref{cut:delta_mv}.}
\end{figure}

\begin{figure}[!ht]
\centering
\subfloat[ ]{\includegraphics[scale=0.75]{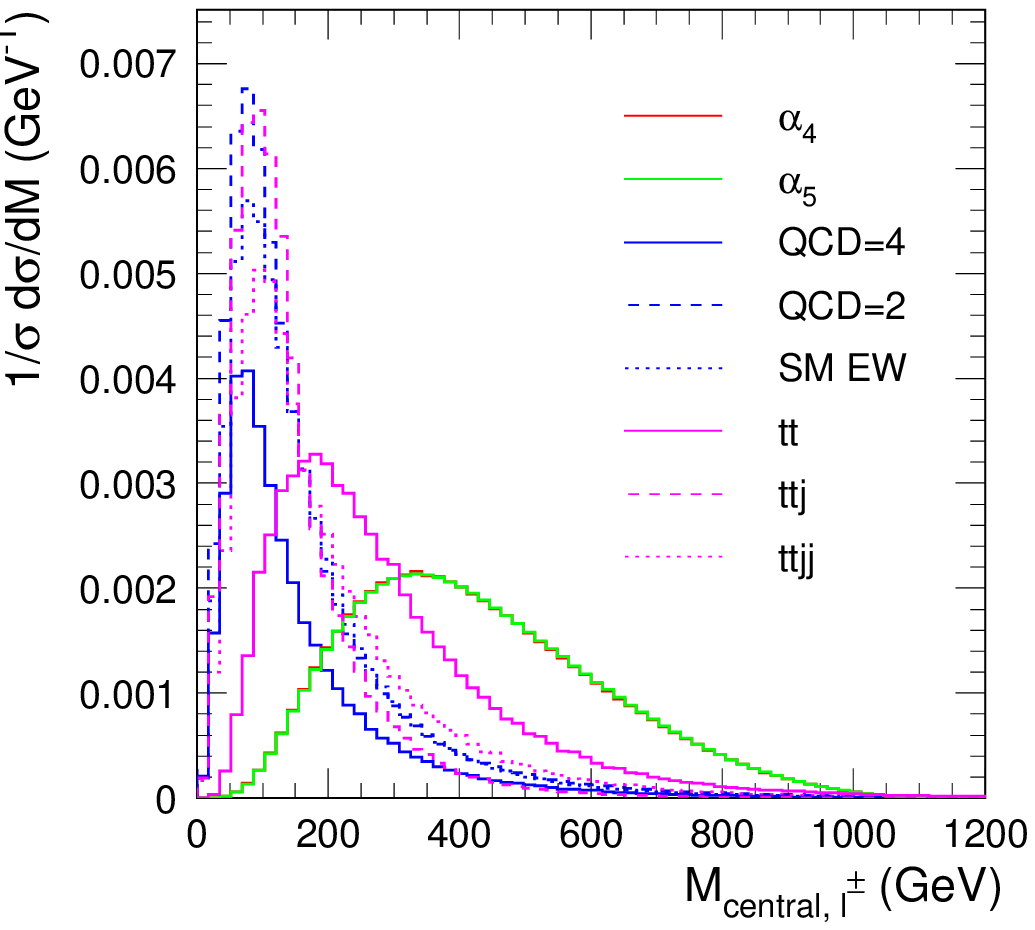}\label{mlj2}}
\hspace{0.2cm}
\subfloat[]{\includegraphics[scale=0.75]{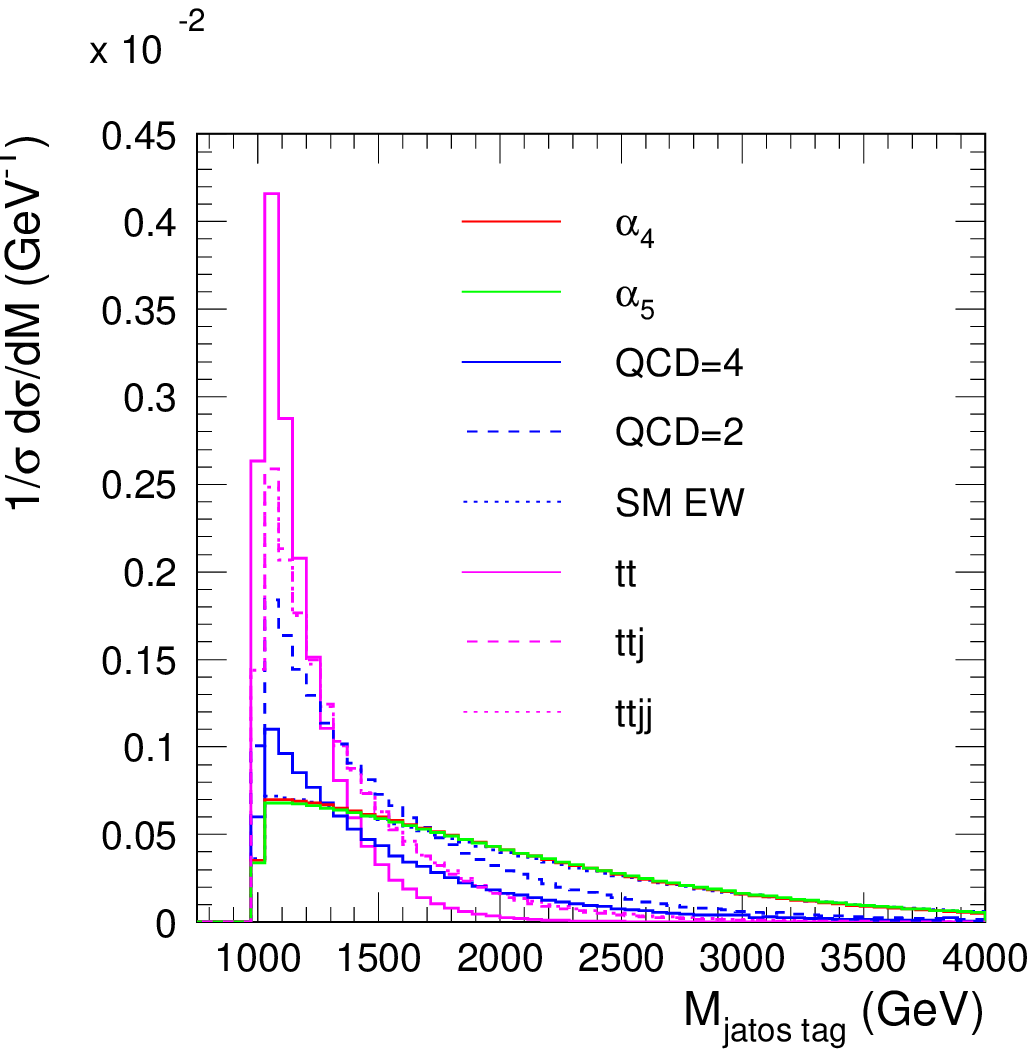}\label{mj1j4}}
\caption{Perfis das massas invariantes (a)
entre o lépton carregado e o jato central e (b) entre os jatos
\textit{taggings}. Ambas após cortes~\eqref{cut23}-\eqref{cut:delta_mv}.}
\end{figure}

Após diversas tentativas concluimos que existem variáveis
cinemáticas promissoras, como as massas invariantes entre um jato
\textit{tagging} e um central, $M_{\mbox{\tiny central, tag}}$
(conforme Fig.~\ref{mj1j2}), entre o lépton carregado e um jato
\textit{tagging}, $M_{\mbox{\tiny tag}, l^{\pm}}$ (conforme Fig.~\ref{mlj1}),
entre o lépton carregado e um jato central, $M_{\mbox{\tiny
    central},l^{\pm}}$ (conforme Fig.\ref{mlj2}) e entre os jatos
\textit{taggings}, $M_{\mbox{\tiny jatos tag}}$ (conforme Fig.~\ref{mj1j4}).

\begin{figure}[!ht]
\centering 
\subfloat[ ]{\includegraphics[scale=0.75]{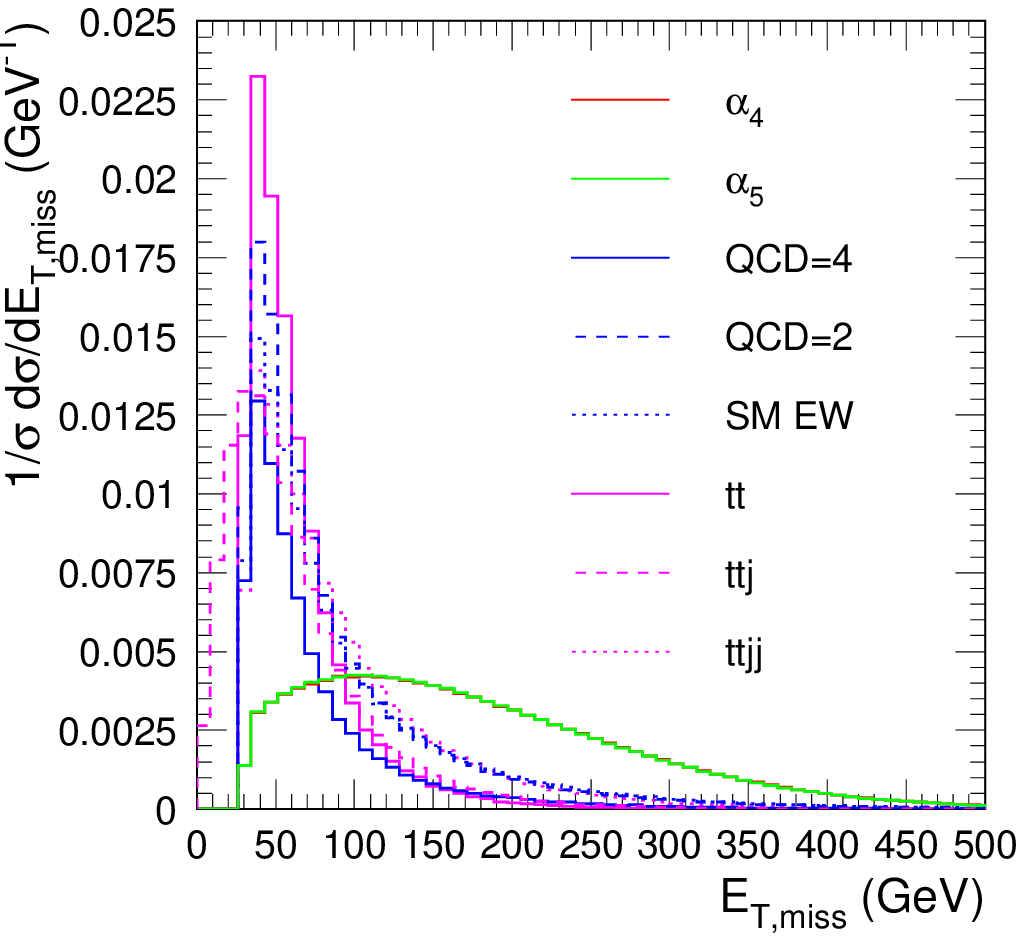}\label{etmiss}} 
\hspace{0.2cm} 
\subfloat[]{\includegraphics[scale=0.75]{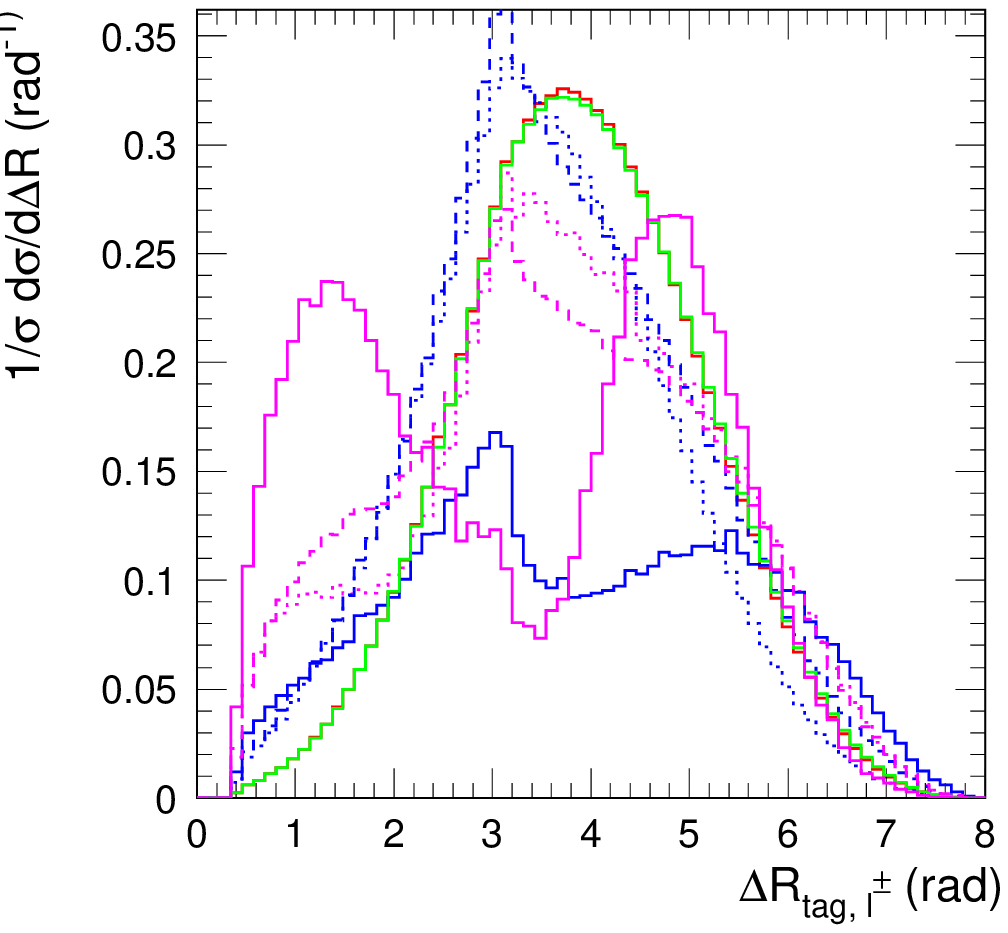}
\label{deltaR_leptag}}
\caption{(a) Energia transversal perdida e (b) $\Delta R$ entre lépton
  carregado e jato \textit{tagging}. Após
  cortes~\eqref{cut22}-\eqref{cut:delta_mv}.}
\end{figure}

Além das massas invariantes listadas acima, as Figs.~\ref{etmiss} e~\ref{deltaR_leptag} mostram que cortes nas variáveis cinemáticas
$\not\!\!E_{T}$ e $\Delta R_{l^{\pm},\mbox{\tiny tag}}$ respectivamente, são bastante eficientes.

\begin{figure}[!ht]
\centering
\includegraphics[scale=0.75]{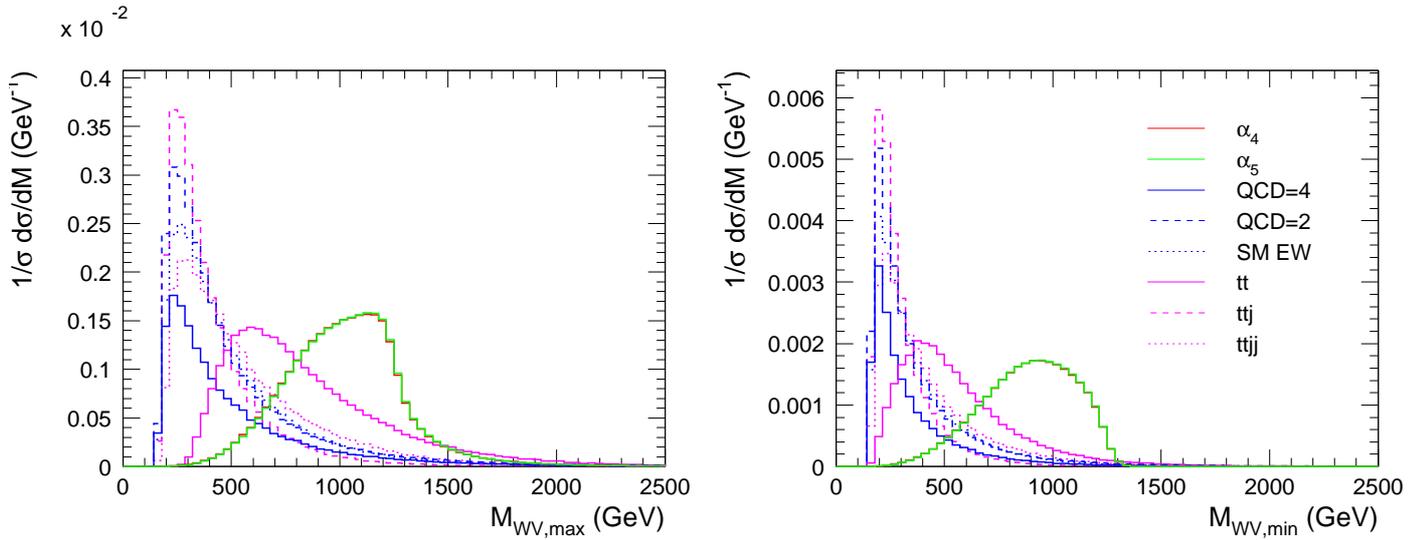}
\caption{Massa invariante do par $M_{WV}$, onde $V=
  W^{\pm}\;\mbox{e}\;Z$, considerando a reconstrução do momento
  longitudinal do neutrino da Eq.~\eqref{pzneutrino}. Após cortes
  \eqref{cut23} a~\eqref{cut:delta_mv}.}
\label{MVV}
\end{figure}

Finalmente, como conseguimos reconstruir o quadrimomento do neutrino
através da Eq.~\eqref{pzneutrino}, podemos reconstruir a massa
invariante do sistema $W^\pm V$ de acordo com duas possibilidades:
$M_{W^{\pm}V}^{\mbox{ \tiny min (máx)}}$ é massa invariante reconstruída a
partir do menor (maior) valor de $p_L$ na Eq.~\eqref{pzneutrino}.  A
Fig.~\ref{MVV} mostra que, qualitativamente, as duas distribuições são
similares e boas discriminantes do sinal anômalo.

Após vários testes, um conjunto de cortes mais eficaz para minimizar
os \textit{backgrounds}, sem prejuízo do sinal, foi definido com base
nas distribuições mostradas nas Figs.~\ref{mj1j2} à~\ref{MVV}.
Sendo assim, além dos cortes dados pelas Eqs.~\eqref{cut23}-\eqref{cut:delta_mv},
adicionamos os seguintes cortes:
\begin{eqnarray}
&M_{\mbox{\tiny tag}, l^{\pm}} \geq 200\;\mbox{GeV}, \quad
  M_{\mbox{\tiny central},l^{\pm}} \geq 150\;\mbox{GeV}, \quad
  M_{\mbox{\tiny central, tag,}} \geq 400\;\mbox{GeV}, & 
\nonumber\\ 
&  M_{\mbox{\tiny jatos tag}} \geq 1600\;\mbox{GeV}, \quad M_{W^{\pm}V}^{\mbox{\tiny min}}
  \geq 600\;\mbox{GeV}, \quad \;\not\!\!E_{T} \geq 50\;\mbox{GeV},
  \quad \Delta R_{\mbox{\tiny tag},l^{\pm}} \geq 2.0.&
\label{cutfinal}
\end{eqnarray}

Durante a etapa de análise da otimização da significância estatística,
testamos diversos valores de cortes para diferentes distribuições
cinemáticas. Embora graficamente muitos deles pareçam promissores,
não foram efetivos para aumentar nossa significância em relação ao \emph{background}. Uma análise mais refinada dessas variáveis, como combinações de cortes, podem
eventualmente ser úteis no estudo do sinal anômalo. A seguir,
discutiremos algumas delas.

\begin{figure}[!ht]
\centering
\subfloat[]{\includegraphics[scale=0.75]{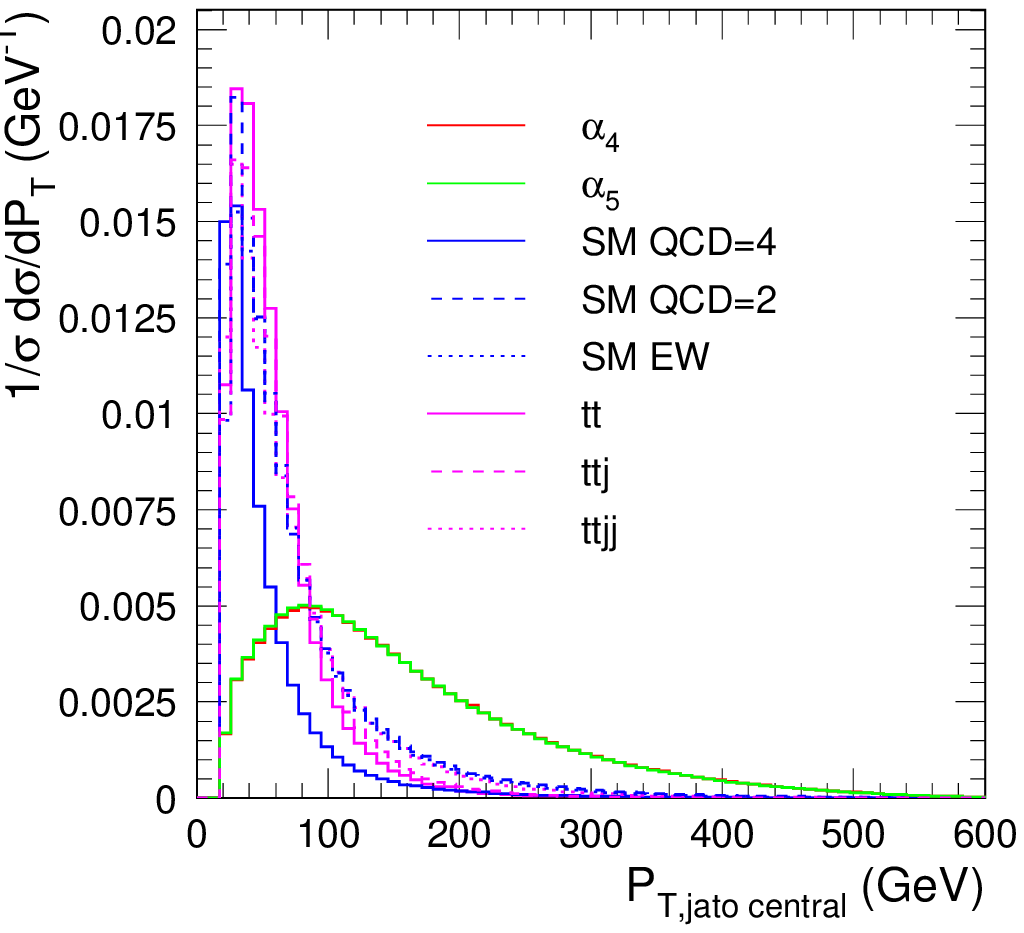}\label{pt_central}}
\hspace{0.2cm}
\subfloat[]{\includegraphics[scale=0.75]{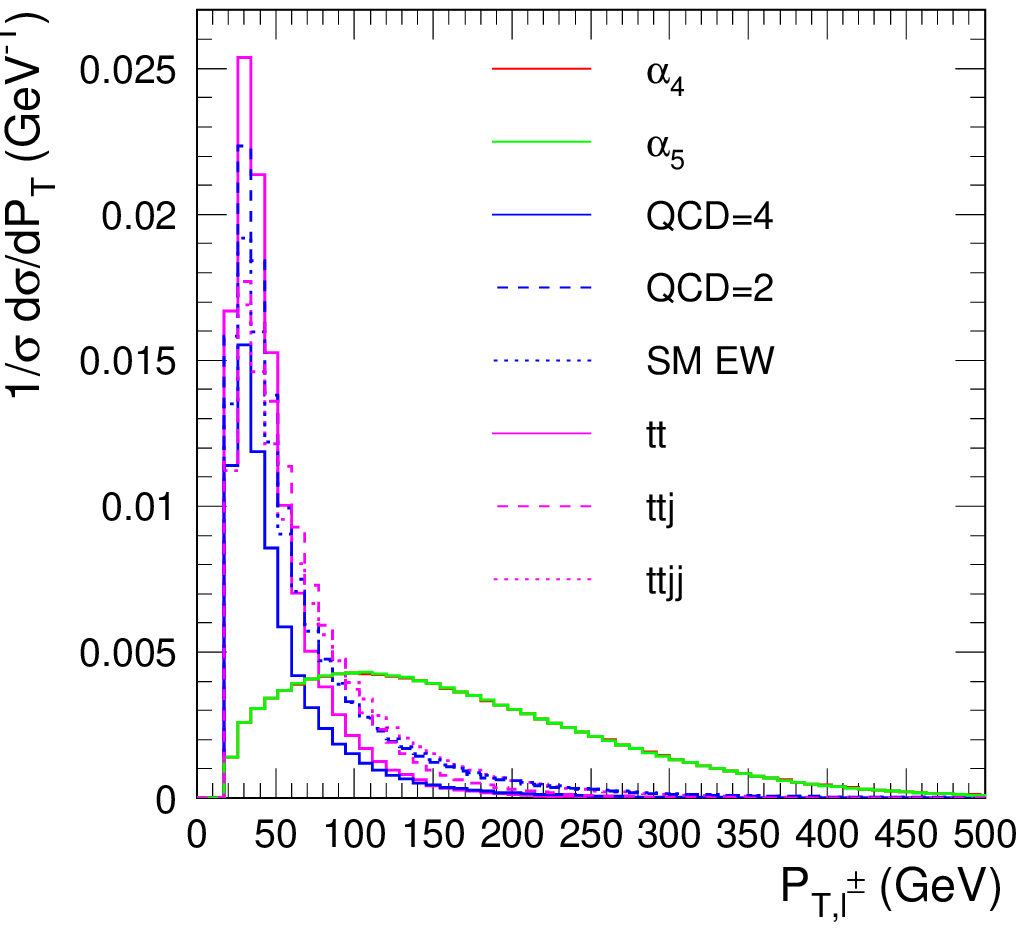}\label{pt_chargedlepton}}
\caption{Distribuições de momento transversal $p_{T}$. Em (a) dos
  jatos centrais e em (b) do lépton carregado. Ambas após
  cortes~\eqref{cut23} a~\eqref{cut:delta_mv}.}
\end{figure}

De acordo com as Figs.~\ref{pt_central} e~\ref{pt_chargedlepton},
cortes mais duros no momento transversal $p_{T}$ dos produtos do
decaimento dos bósons de \textit{gauge} poderiam ser
aplicados. Nota-se nitidamente nessas figuras que há um aumento
significante de eventos na região $p_{T}\gtrsim 30$ GeV para o caso
dos diferentes \textit{backgrounds} considerados. Já o nosso sinal
anômalo se espalha suavemente para regiões com valores de $p_{T}$ mais
altos até aproximadamente 500 GeV.

Alternativamente à massa invariante do sistema $WV$, que possui
ambiguidade devido à reconstrução da componente longitudinal do
neutrino, podemos definir a grandeza 
\begin{equation}
M_{T, WV}^{2}= \left[\sqrt{M^{2}({l^{\pm}jj})+p_{T}^{2}(l^{\pm}jj)} + |p_{T}^{miss}|\right]^{2}-\left[\vec{p}_{T}(l^{\pm}jj)+\vec{p}_{T}^{miss} \right]^{2},
\label{mt}
\end{equation}
cuja raiz quadrada chamamos de massa transversal do par $WV$.  A Fig.~\ref{mt}
mostra um claro domínio do sinal anômalo na região acima de $M_{T,WV}
\approx 500-600$ GeV. 

\begin{figure}[!ht]
\centering
\includegraphics[scale=0.75]{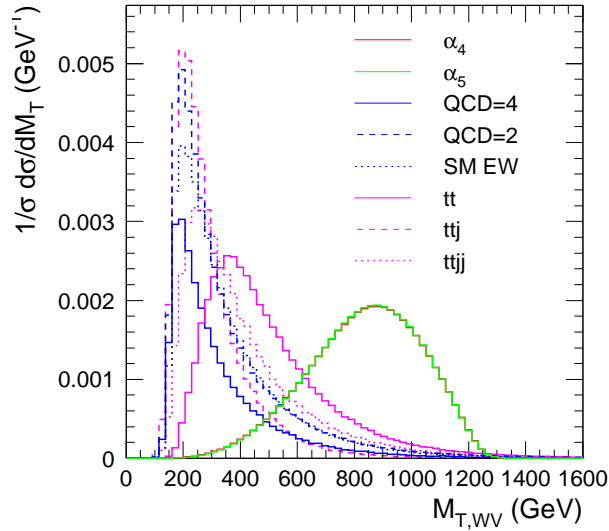}
\caption{Perfil da massa transversal do par $WV$, onde $V=
  W^{\pm}\;\mbox{e}\;Z$. Ambas após
  cortes~\eqref{cut23}-\eqref{cut:delta_mv}.}  
\label{mT} 
\end{figure}

Finalmente, antes de apresentar os nossos resultados das seções de
choque, é preciso observar um detalhe muito importante no que diz
respeito aos jatos centrais. Observa-se da Fig.~\ref{MVV}, que o sinal
está distribuído na região onde a massa invariante $WV$ é alta e na qual
os acoplamentos anômalos começam a interagir fortemente. Assim como em
situações onde ressonâncias pesadas emergem com massa da ordem de
1~TeV ou maior, os bósons $W^\pm$ e $Z$ produzidos centralmente são bem
``boostados'', fazendo com que os seus produtos de decaimento tendam a
sair bem próximos um do outro. No nosso caso, se os jatos centrais forem
emitidos muito próximos, aumenta a possibilidade da formação de jatos
``gordos''\footnote{Jatos decorrentes do decaimento hadrônico de
  \textit{highly boosted massive particles.}}.

\begin{figure}[!ht]
\centering
\subfloat[ ]{\includegraphics[scale=0.75]{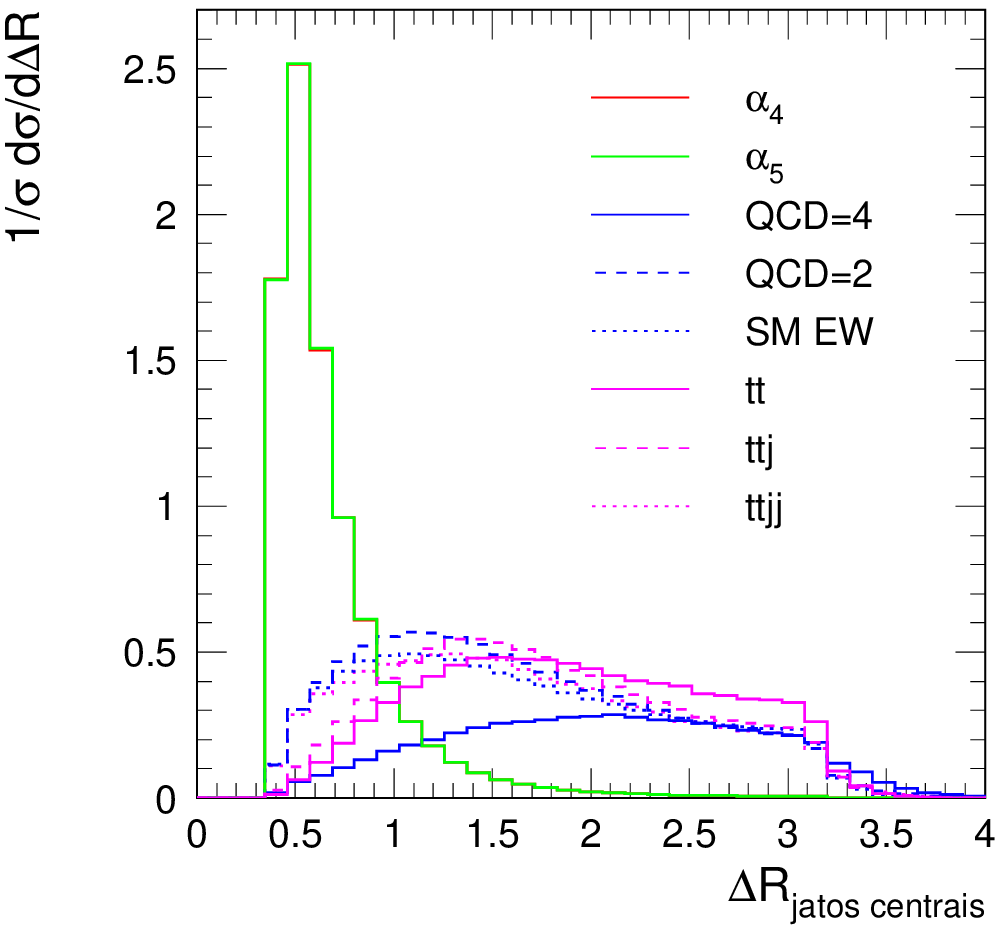}\label{drj2j3}}
\hspace{0.2cm}
\subfloat[ ]{\includegraphics[scale=0.75]{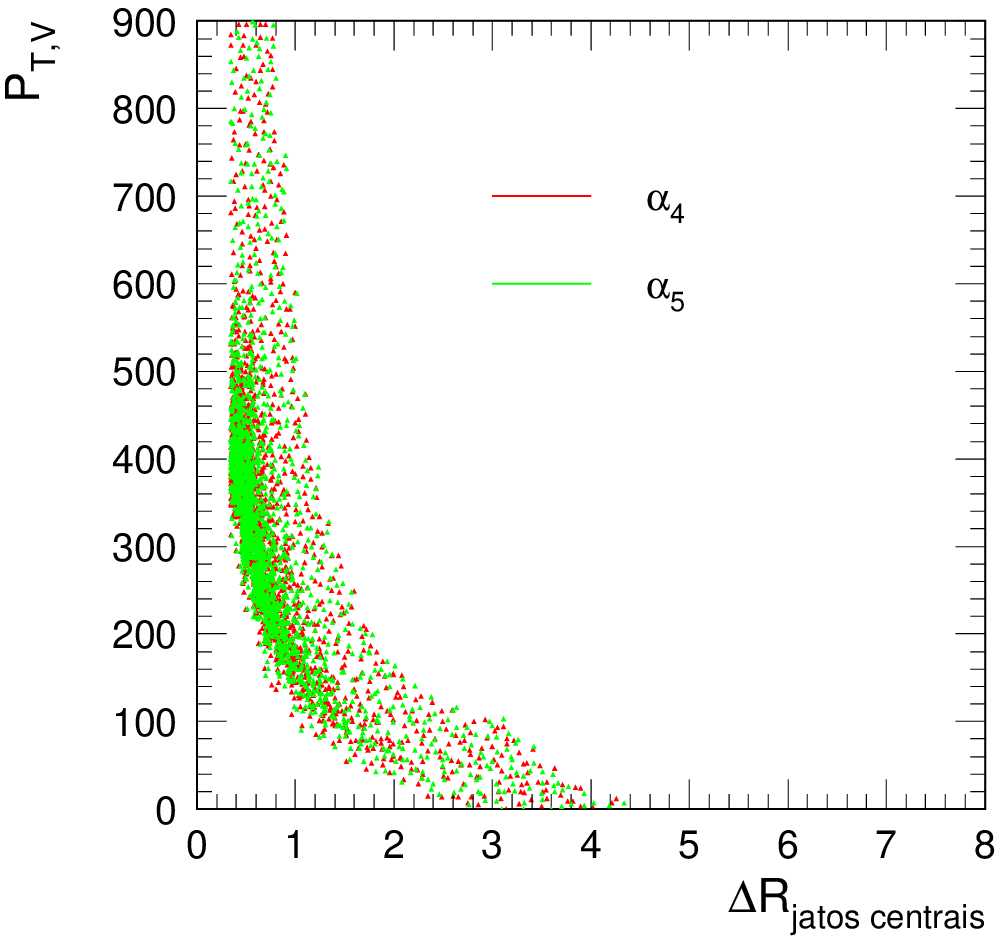}\label{pt_deltaR}}
\caption{ (a) mostra a distribuição do $\Delta R$ entre os jatos
  centrais e (b) a correlação entre essa variável e o momento
  transversal do bóson $V$, com $V=W^{\pm}$ e $Z$,. Após
  cortes~\eqref{cut23} a~\eqref{cut:delta_mv}.}
\end{figure}

Embora o corte em $\Delta R_{jj}$ garanta a separação dos jatos, é
importante investigar a sensibilidade do sinal anômalo frente à
escolha numérica dessa variável. A Fig.~\ref{drj2j3}a mostra que de
fato a separação entre os jatos centrais é bem pequeno, logo é preciso
uma investigação além dos cálculos partônicos para verificar se o
corte $\Delta R_{\mbox{\tiny jatos centrais}} > 0.4$ garante com
eficiência a separação dos jatos. Contudo, uma análise completa com a
hadronização dos pártons está fora do escopo desta tese.

Como informação adicional, a Fig.~\ref{pt_deltaR} mostra a grande
concentração de eventos com jatos centrais emitidos com momento
transversal alto, na faixa de $100\;\mbox{GeV}\lesssim p_{T}\lesssim
500\;\mbox{GeV}$, e bem próximos, com $0.4 \lesssim \Delta R\lesssim
1$.

\section{Resultados}
\label{resultados:anomalo}

Apresentamos nesta seção os resultados principais dos nossos
cálculos. Nas Tabelas~\ref{resultados_wp} e~\ref{resultados_wm}
listamos as seções de choque em unidades de fb dos canais $pp\to
e^{+}\nu_{e}jjjj$ e $pp\to e^{-}\bar{\nu}_{e}jjjj$, respectivamente,
considerando vários conjuntos de cortes e escolhas C1~\eqref{c1} e C2~\eqref{c2} das escalas
de normalização e fatorização, com as variações $\varepsilon=0.25, 1, 4$.
Esses valores de seção de choque foram multiplicados por um fator 2
para levar em conta tanto o canal com $e^{\pm}$ quanto o canal com
$\mu^{\pm}$. 

Observe que, devido à simetria entre o quark $top$ e o anti-$top$ nos
processos $pp \to t\bar{t}+n$ jatos, esses \textit{backgrounds}
contribuem igualmente para os processos $pp\to e^{+}\nu_{e}jjjj$ e
$pp\to e^{-}\bar{\nu}_{e}jjjj$. Por esse motivo eles não foram listados na
Tabela~\ref{resultados_wm}.

\begin{table}[h!tb]
{\footnotesize
\begin{center}
\begin{tabular}{||l|l|l|l|l|l|l|l|l||}
\hline
& & & & & & & & \\
	&\eqref{cut22}-\eqref{cut26}& \eqref{cut22}-\eqref{cut:delta_mv} &C1XIR1&C1XIR025&C1XIR4&C2XIR1&C2XIR025&C2XIR4\\
& & & & & & & & \\
\hline
$\sigma_{44}$&8349 &7630 &2460 &3206 &1876 &3323 &4789 &2491\\
$\sigma_{55}$&6191 &5726 &1840 &2412 &1418 &2505 &3585 &1875\\
$\sigma_{4}$&-	&-	&1.85 &2.8 &1.88 &-4.19 &8.73 &0.85\\
$\sigma_{5}$&-	&-	&.085 &-1.2 &-2.12 &-12.19 &1.73 &0.85\\
$\sigma_{45}$&-  &-	&7755 &10214 &5955 &10588 &15190 &7901\\
SM EW     &11 &8	&0.15 &0.20 &0.12 &0.19	&0.27  &0.15\\
QCD\= 2     &63 &11	&0.07 &0.14 &0.04 &0.13	&0.26  &0.07\\
QCD\= 4     &888 &	136	&0.32	&0.41	&0.23	&4.17	&13	& 2\\
$t\bar{t}$  &3	&2&	0.00&	0.00&	0.00&	0.00&	0.00&	0.00\\
$t\bar{t}$ + j &620&	382&	0.00&	0.00&	0.00&	0.00&	0.00&	0.00\\
$t\bar{t}$ + jj &61&	42&	0.02&	0.06&	0.01&	0.08&	0.36&	0.02\\
\hline
\end{tabular}
\caption{Seções de choque em \emph{fb} do canal $pp\to
  e^{+}\nu_{e}jjjj$ considerando diversos cortes e diferentes escolhas
  para as escalas de normalização e fatorização. A legenda
  ``C1XIR025'' significa escolha do conjunto $C1$~\eqref{c1}, com a escolha $\varepsilon =
  0.25$ e assim por diante, para as diversas opções analisadas.  Nas
  primeiras duas colunas utilizamos o conjunto $C1$, com $\varepsilon =1$, com
  os cortes especificados na tabela. Já para o restante das colunas,
  os resultados foram obtidos com os cortes das
  Eqs.~\eqref{cut23}-\eqref{cutfinal}.}
\label{resultados_wp}
\end{center}}
\end{table}

\begin{table}[h!tb]
{\footnotesize
\begin{center}
\begin{tabular}{||l|l|l|l|l|l|l||}
\hline
& & & & & &\\
	&C1XIR1&C1XIR025&C1XIR4&C2XIR1&C2XIR025&C2XIR4\\
& & & & & &\\
\hline
$\sigma_{44}$&1183&	1615&	941	&1704	&2483	&1281\\
$\sigma_{55}$&774	&1052	&617	&1114	&1607	&833\\
$\sigma_{4}$	&0.95	&-0.07	&2.96	&-2.07	&1.91	&-0.05\\
$\sigma_{5}$	&1.95	&-2.07	&-1.04  &-12.07	&0.91  &0.95\\
$\sigma_{45}$	&3607	&4997	&2917	&5295	&7629	&3938\\
SM EW &0.05	&0.07	&0.04	&0.07	&0.09	&0.05\\
QCD\= 2 &0.02 &	0.04	&0.01	&0.04	&0.08	&0.03\\
QCD\= 4 &0.08 &	0.12	&0.06	&1.23	&5.23	&0.42\\
\hline
\end{tabular}
\caption{Idem à legenda da Tabela~\ref{resultados_wp}, para o 
canal $pp\to e^{-}\bar{\nu}_{e}jjjj$.}
\label{resultados_wm}
\end{center}}
\end{table}

\begin{figure}[!ht]
\centering
\subfloat[ ]{\includegraphics[scale=0.6]{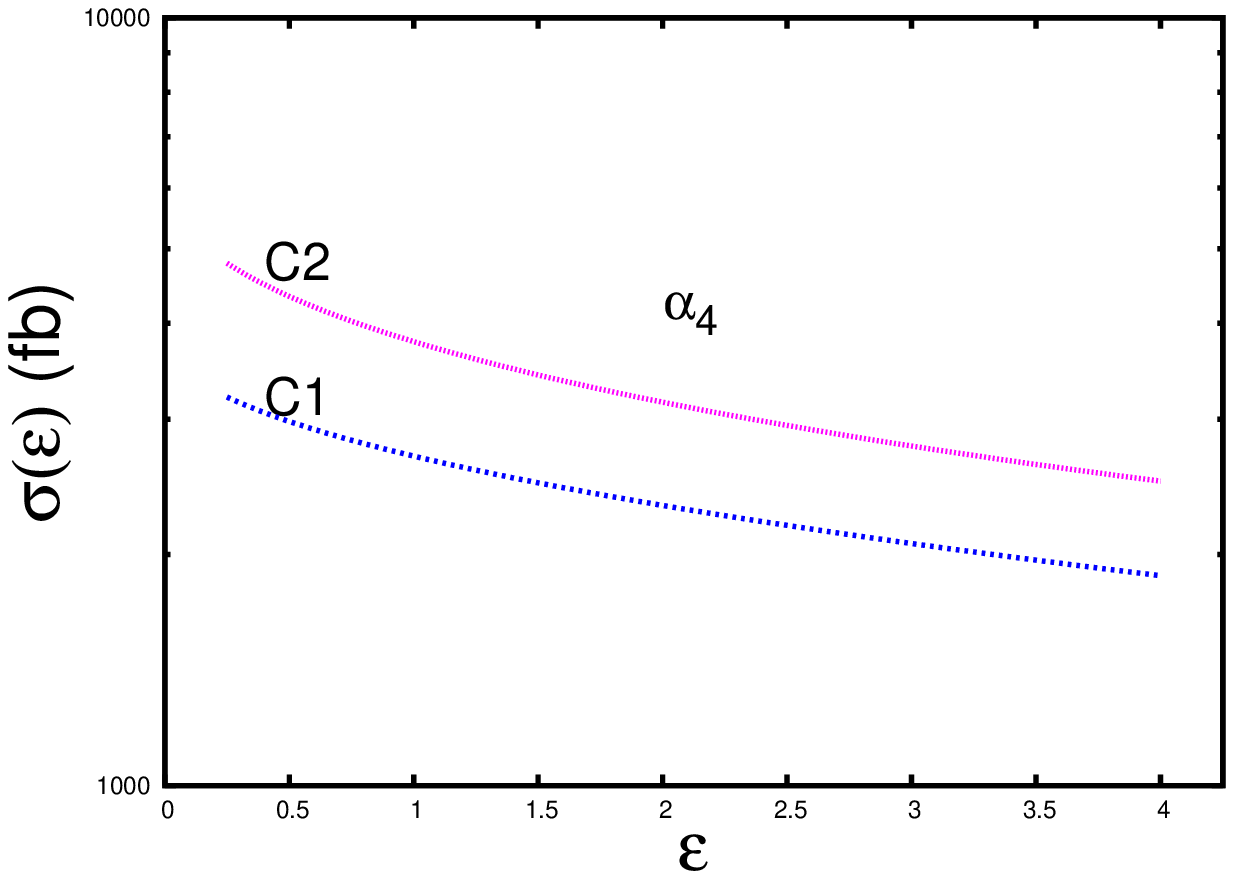}\label{anom}} 
\hspace{0.2cm}
\subfloat[ ]{\includegraphics[scale=0.6]{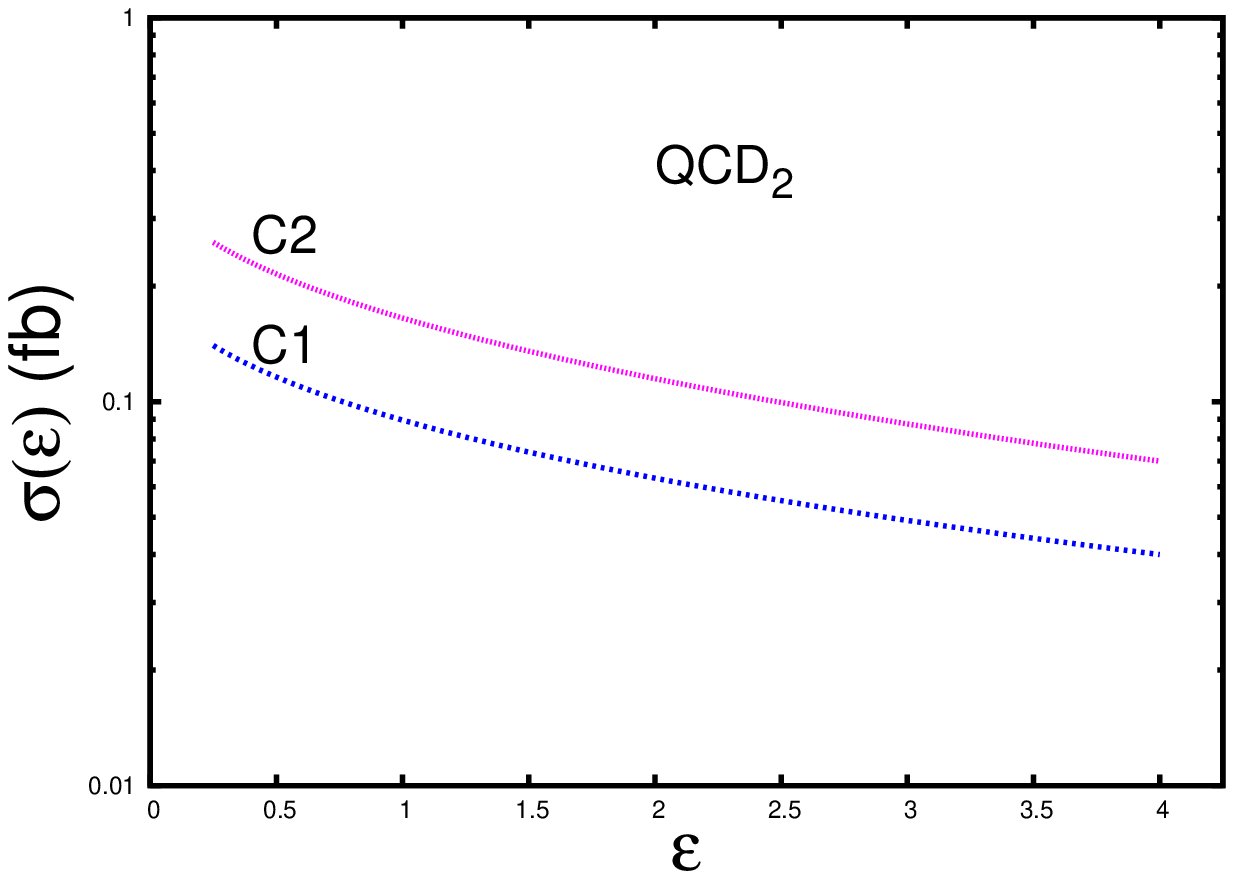}\label{qcd2}}
\hspace{0.2cm}
\subfloat[ ]{\includegraphics[scale=0.6]{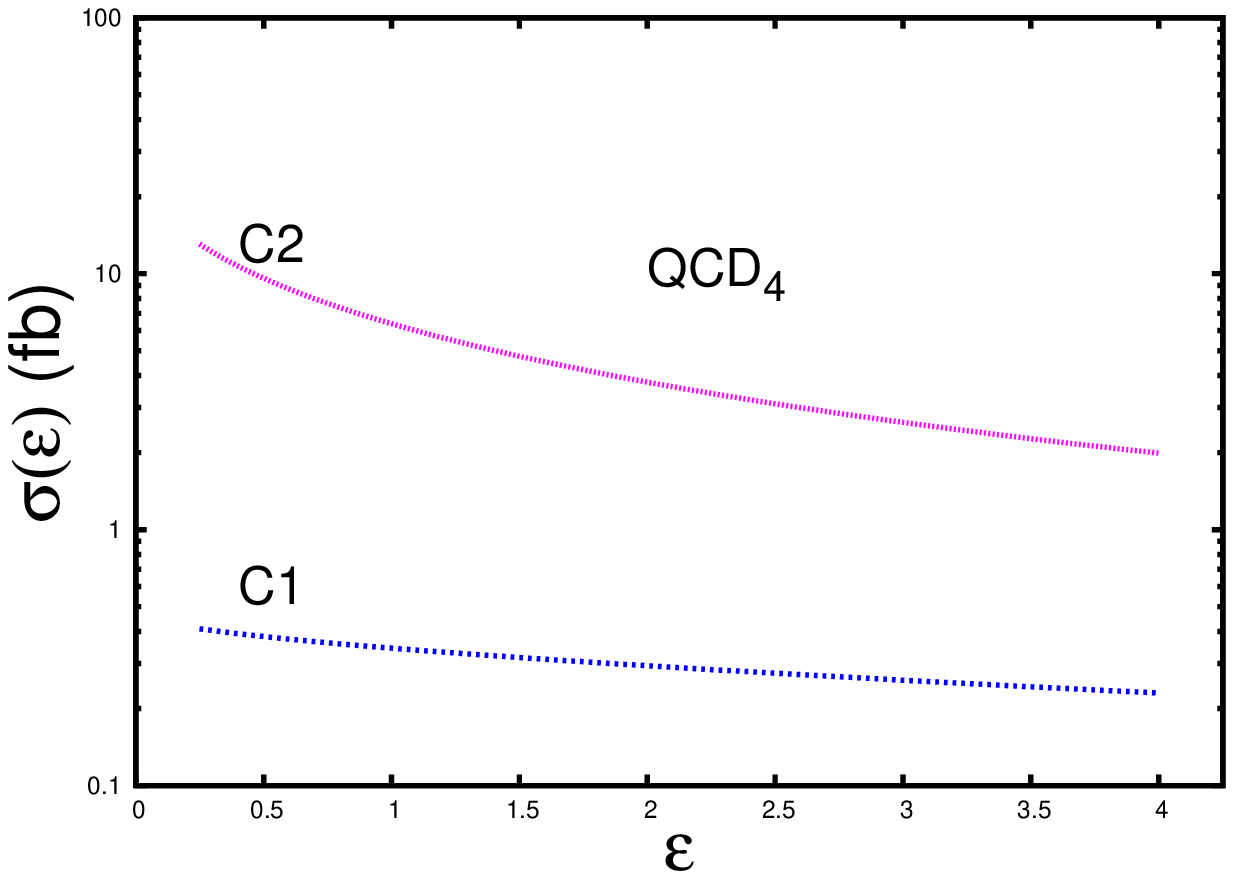}\label{qcd4}}
\caption{Seções de choque de (a) sinal anômalo de ${\cal L}_4$ e (b)
  \emph{backgrounds} de QCD após os cortes~\eqref{cut23}
  à~\eqref{cut:delta_mv} como função da variável $\varepsilon$, para as
  escolhas de escalas de fatorização e normalização C1~\eqref{c1} e C2~\eqref{c2}.}
\label{escalas}
\end{figure}

Apresentamos na Fig.~\ref{escalas} as incertezas na escolha das
escalas de renormalização e fatorização, representadas pelas escolhas
C1~\eqref{c1} e C2~\eqref{c2}, em função do parâmetro $\varepsilon$. Conforme esperado, os
resultados são bem sensíveis às escolhas das escalas, sobretudo no que tange ao
\emph{background} em ordem ${\cal O}(\alpha_{EM}^{2}\alpha_{S}^{4})$.
Notamos também, a partir das Tabelas~\ref{resultados_wp} e
\ref{resultados_wm}, que o padrão de interferência (construtiva ou
destrutiva) do sinal anômalo com os \textit{backgrounds} depende das 
escalas utilizadas.  Somente uma
análise mais refinada ratificaria esse comportamento ou se, na verdade, trata-se tão somente de flutuações oriundas do cálculo
numérico. De qualquer forma, para $\alpha_{4,5} \gtrsim {\cal O}
(10^{-2})$ os termos de interferência não são cruciais para a obtenção
dos vínculos a esses acoplamentos.

Como trabalhamos essencialmente com contagem de eventos, posto que nenhuma ressonância está presente, podemos combinar os canais $l^{+}$ e
$l^{-}$ de forma que a significância
estatística definida na Eq.~\eqref{significancia} será calculada tal que 
\begin{equation} 
{\cal S}_{comb}\equiv \dfrac{N_{l^{+}}+N_{l^{-}}}{\sqrt{N_{back_{l^{+}}}+N_{back_{l^{-}}}}},
\label{significancia2}
\end{equation}
onde $N_{l}$ é o número de eventos do sinal nos canais com lépton
positivo $l^{+}$ e lépton negativo $l^{-}$, e $N_{back}$ é o número de
eventos dos \emph{backgrounds} nesses canais, todos obtidos com
expressão da seção de choque~\eqref{cross_section} e considerando as
eficiências na detecção dos jatos e lépton carregado, para o LHC
operando com energia de 14 TeV e com luminosidade integrada de
$L=100\;\mbox{fb}^{-1}$.

\begin{figure}[!ht]
\centering
\subfloat[C1XIR025]{\includegraphics[scale=0.65]{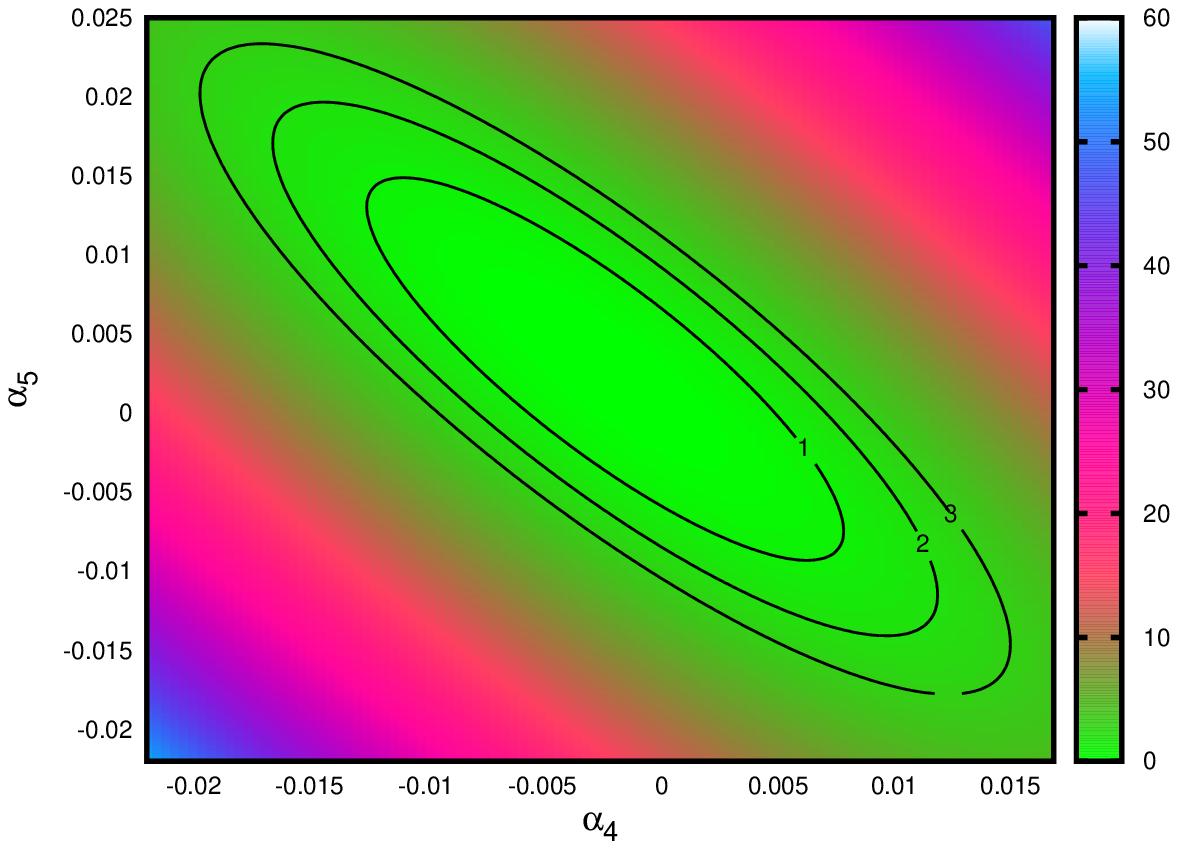}}
\hspace{0.05cm}
\subfloat[C1XIR1]{\includegraphics[scale=0.65]{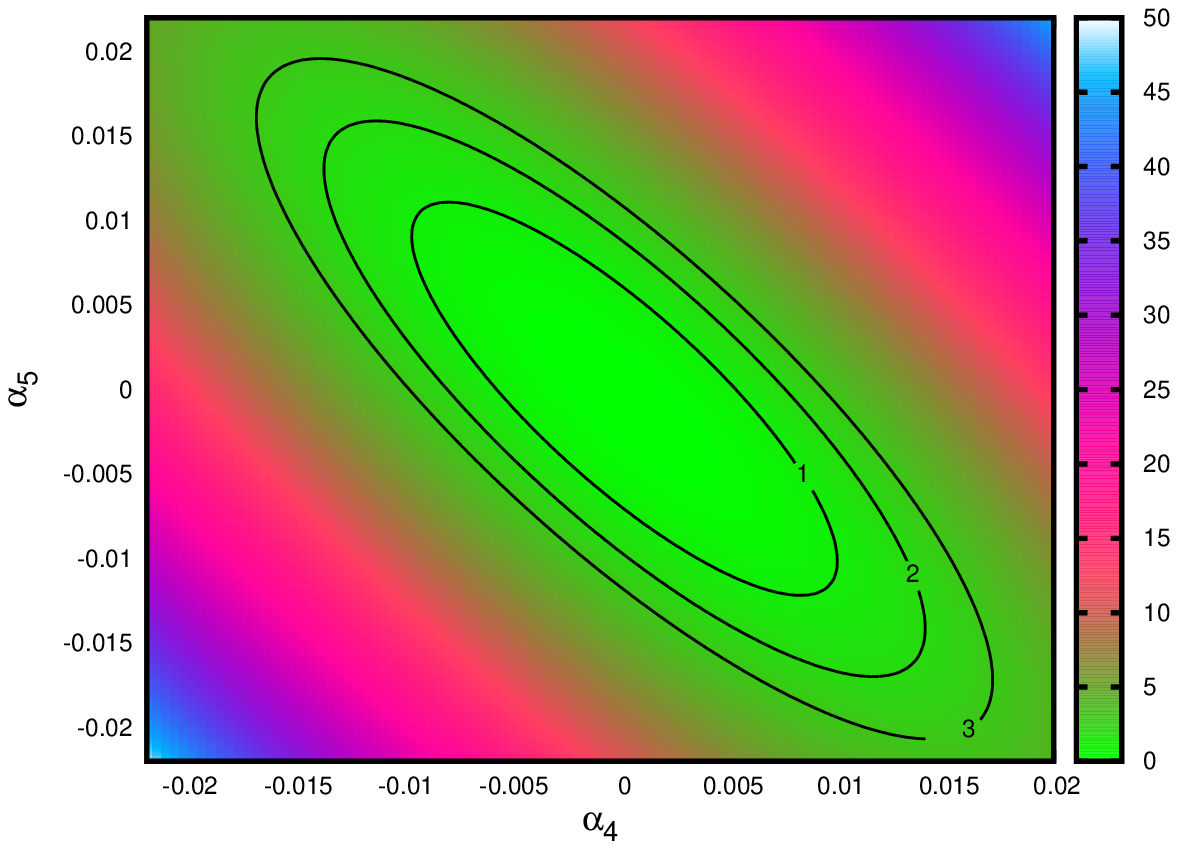}}
\hspace{0.05cm}
\subfloat[C1XIR4]{\includegraphics[scale=0.65]{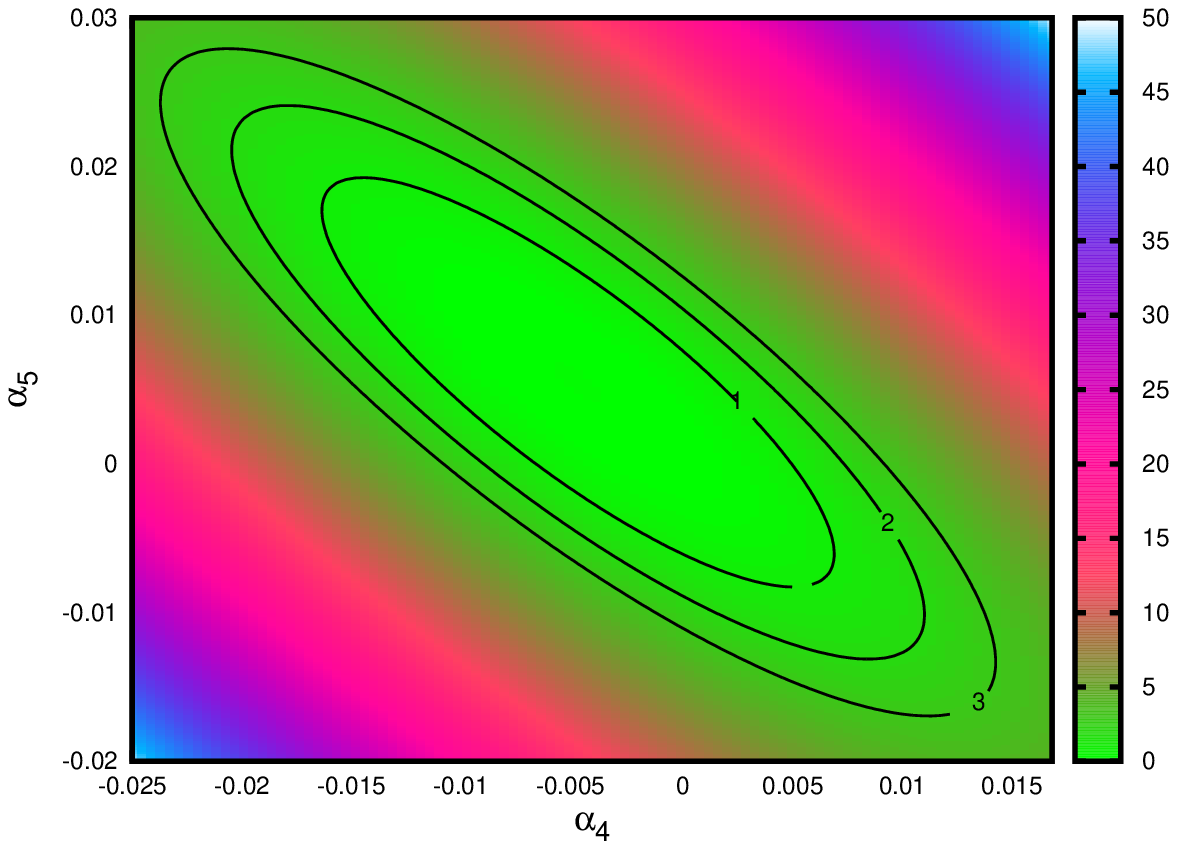}}
\hspace{0.05cm}
\subfloat[C2XIR025]{\includegraphics[scale=0.65]{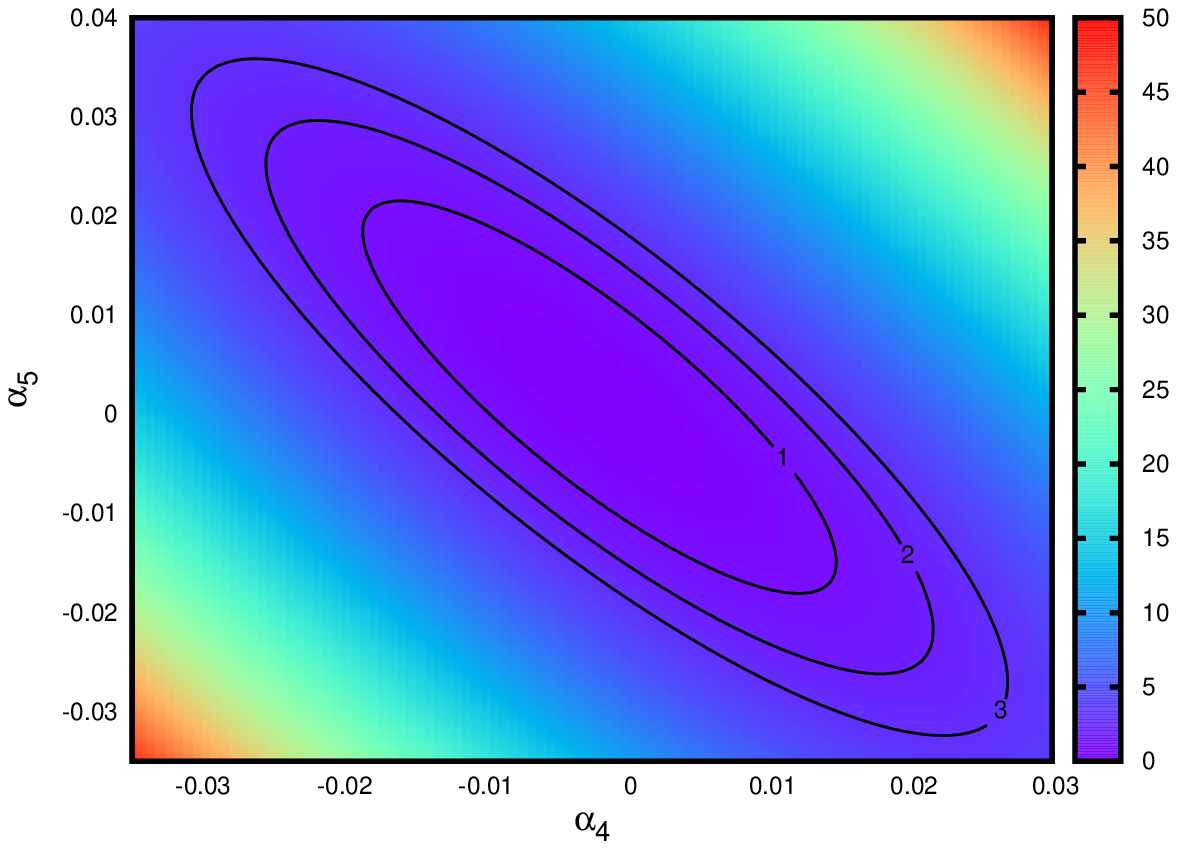}}
\hspace{0.05cm}
\subfloat[C2XIR1]{\includegraphics[scale=0.65]{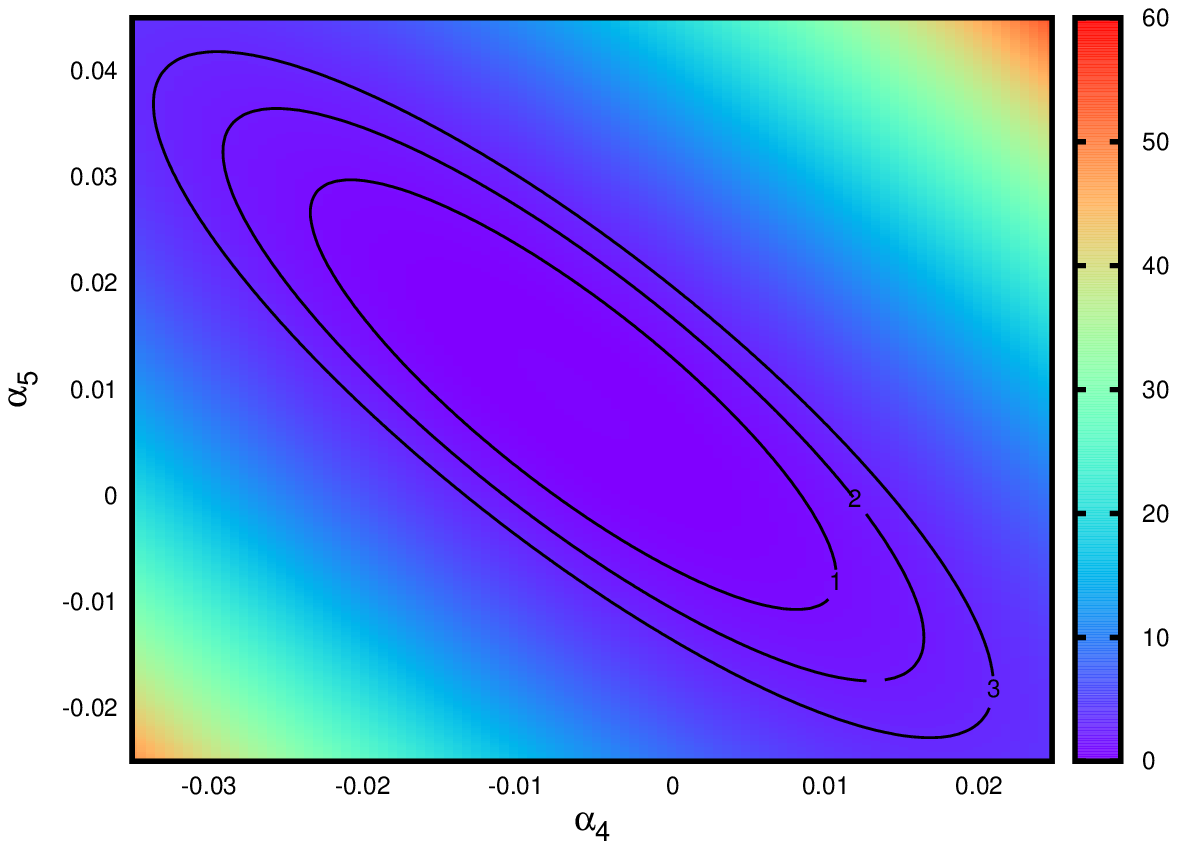}}
\hspace{0.05cm}
\subfloat[C2XIR4]{\includegraphics[scale=0.65]{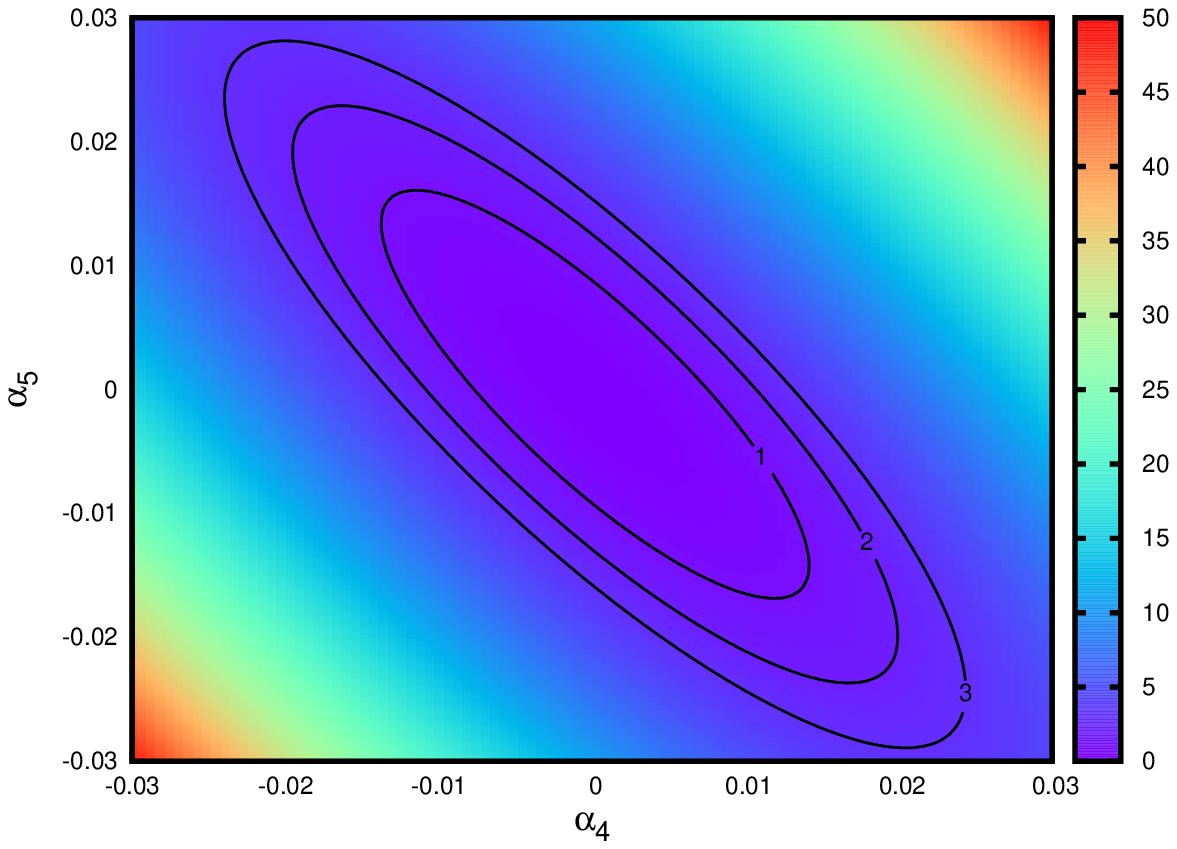}}
\caption{Regiões de exclusão no LHC operando com energia de 14 TeV,
  após uma luminosidade integrada de 100 fb${}^{-1}$. Limites de
  exclusão para diferentes escolhas das escalas de renormalização e
  fatorização. De (a)-(c) para a escolha C1~\eqref{c1} e (d)-(f) para a escolha
  C2~\eqref{c2}.}
\label{contour}
\end{figure}

Utilizando-se os dados das Tabelas~\ref{resultados_wp} e~\ref{resultados_wm}, com o auxílio da Eq.~\eqref{significancia2},
obtemos um dos resultados centrais desta tese. A Fig.~\ref{contour}
mostra as regiões de exclusão para os acoplamentos anômalos puramente
quárticos no plano $(\alpha_{4},\alpha_{5})$, levando-se em conta o
canal $pp\to l^{\pm}\;+\not\!\!E_{T}\;+\;\mbox{4 jatos}$. Para quantificar a dependência dos resultados com as escalas de
fatorização e renormalização, mostramos as regiões de exclusão para
todas as escolhas analisadas. Qualitativamente, vemos que não há diferença
significativa entre os diferentes vínculos obtidos. 

Para ilustrar a sensibilidade da significância estatística em função dos valores
de $\alpha_{4}$ e $\alpha_{5}$, destacamos as elípses de 1$\sigma$ até
3$\sigma$, que correspondem à 68\%, 95.4\% e 99.7\% C.L. respectivamente.

\begin{figure}[!ht]
\centering
\includegraphics[scale=0.55]{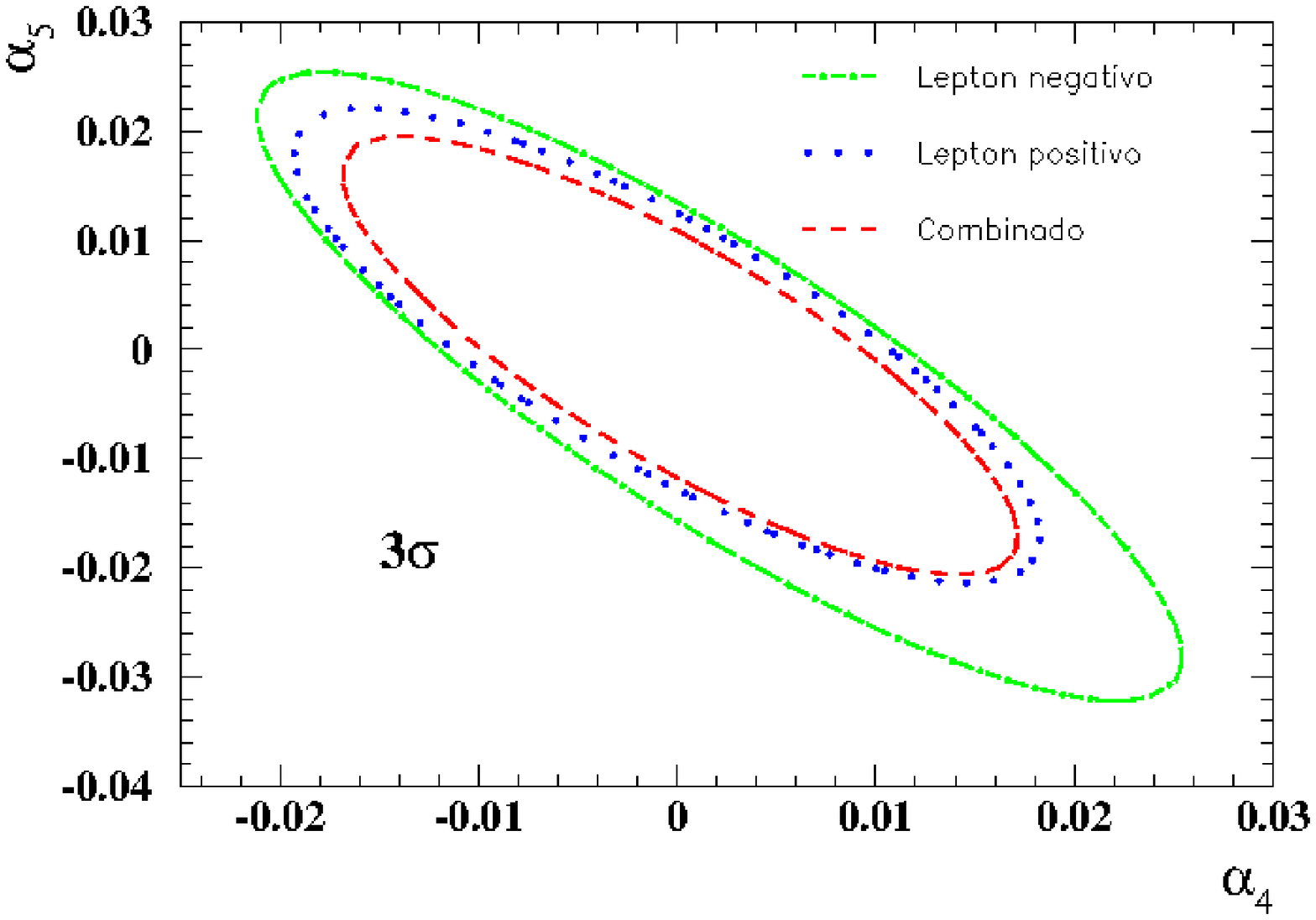}
\caption{Regiões de exclusão devido aos canais com lépton positivo,
  lépton negativo e a combinação entre ambos. Resultado obtido com a
  escolha C1~\eqref{c1}, com $\varepsilon=1$, para o LHC operando com energia de 14 TeV e
  luminosidade integradada de 100 fb${}^{-1}$.}
\label{contour2}
\end{figure}

Na Fig.~\ref{contour2}, mostramos a curva de exclusão para 3$\sigma$ com a escolha C1~\eqref{c1}, com $\varepsilon =1$, levando-se em conta as análises dos
canais $pp\to l^+\;+\not\!\!E_{T}\;+\;\mbox{4 jatos}$ e $pp\to
l^-\;+\not\!\!E_{T}\;+\;\mbox{4 jatos}$ separadamente e
combinados. Como era de se esperar, os vínculos oriundos do canal com
lépton positivo é mais restritivo no LHC visto que processos com a
produção $W^+V$ possuem seções de choque maiores do que aqueles com
$W^-V$, já que se trata de um colisor próton-próton. As curvas com
outras escolhas das escalas de renormalização e fatorização são
qualitativamente similares, não havendo portanto necessidade de mostrá-las. 

\begin{figure}[!ht]
\centering
\subfloat[$\alpha_{4}$ para C1XIR1 ]{\includegraphics[scale=0.5]{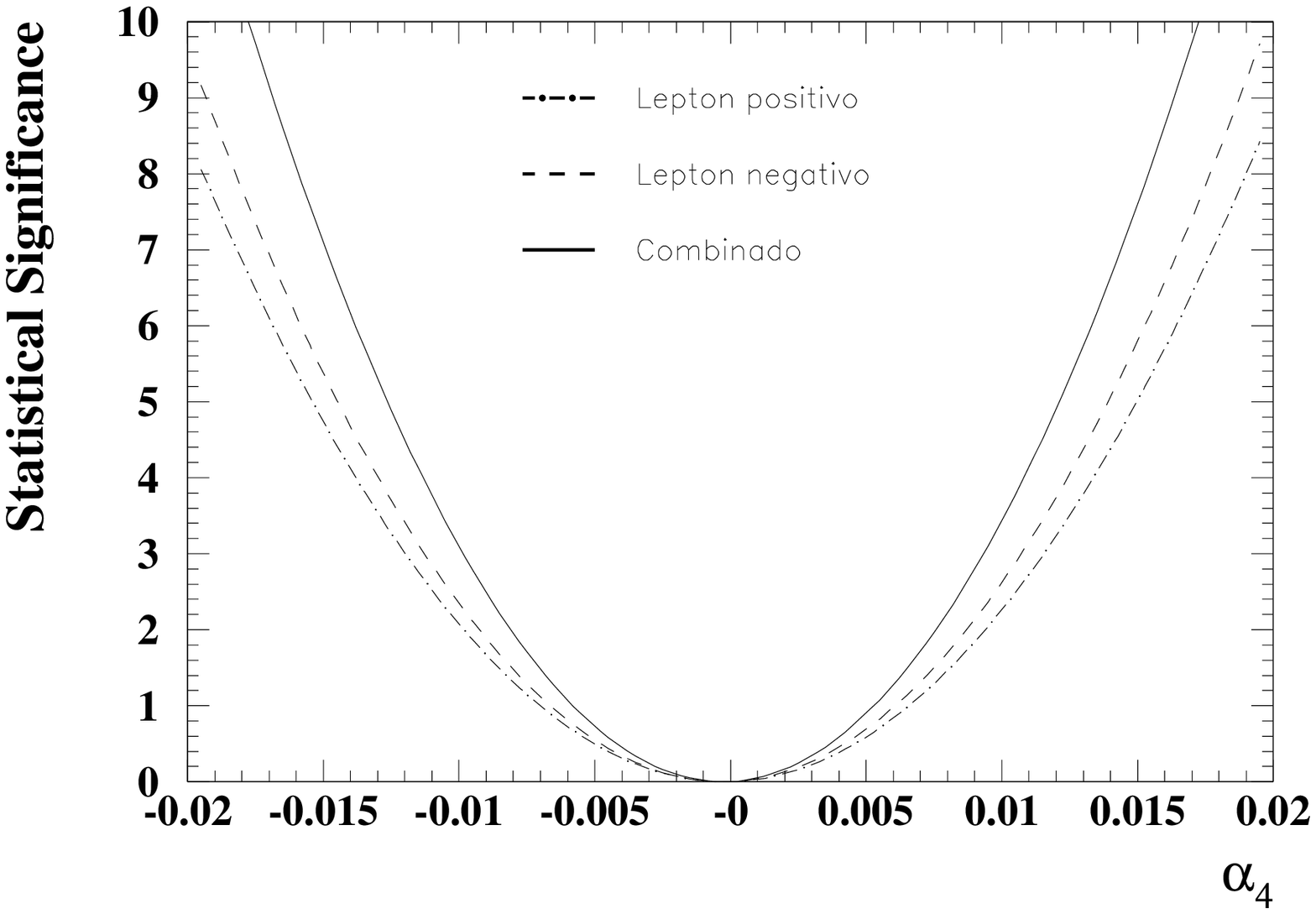}}
\hspace{0.05cm}
\subfloat[$\alpha_{5}$ para C1XIR1 ]{\includegraphics[scale=0.5]{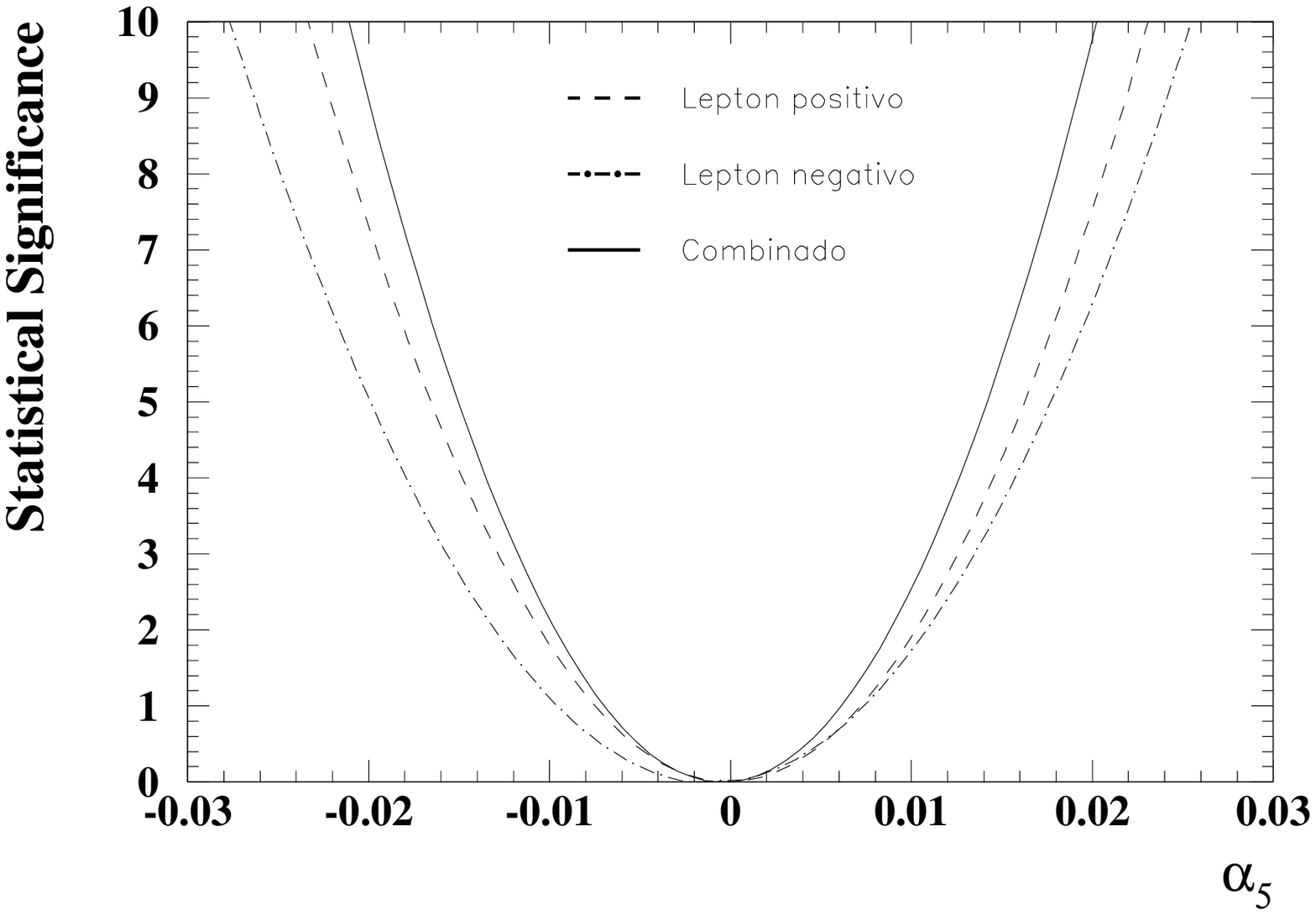}}
\caption{Significância estatística em função dos acoplamentos anômalos
  puramente quárticos (a) $\alpha_{4}$ e (b) $\alpha_{5}$, na situação
  em que somente um operador anômalo contribui por vez. Resultados
  para a escolha $C1$~\eqref{c1}, com $\varepsilon=1$, no LHC com 14 TeV de energia e
  luminosidade integrada de $L=100\;\mbox{fb}^{-1}$.}
\label{vinc_individual}
\end{figure}

Podemos também estabelecer limites individuais aos acoplamentos
anômalos, supondo a contribuição dos operadores ${\cal L}_{4}$ e
${\cal L}_{5}$ em separado para a combinação dos processos $pp\to l^+\;+\not\!\!E_{T}\;+\;\mbox{4 jatos}$ e $pp\to l^-\;+\not\!\!E_{T}\;+\;\mbox{4 jatos}$. As
Figs.~\ref{vinc_individual}a e \ref{vinc_individual}b mostram o
comportamento da significância estatística em função dos valores dos
acoplamentos anômalos $\alpha_4$ e $\alpha_5$, respectivamente. Para
efeito de ilustração mostramos as curvas para a escolha C1~\eqref{c1}, com
$\varepsilon=1$. Não exibimos as curvas com as outras escolhas das escalas de
fatorização e renormalização por serem qualitativamente
similares.

Na Tabela~\ref{tab_anom} exibimos os vínculos em 3$\sigma$ aos acoplamentos
$\alpha_4$ e $\alpha_5$, quando somente um operador contribui por vez.

\begin{table}[h!tb]
{\footnotesize
\begin{center}
\begin{tabular}{||c|c|c|c||}
\hline
& & & \\
& C1XIR025 & C1XIR1 & C1XIR4      \\
& & &  \\
\hline
& & &  \\
$\alpha_{5}=0$& $  -0.009 \leq \alpha_{4} \leq 0.009$ &  $ -0.01\leq \alpha_{4} \leq 0.009$   &  $ -0.009\leq \alpha_{4} \leq 0.009 $  \\
& & & \\
$\alpha_{4}=0$& $ 0.01\leq \alpha_{5} \leq 0.013 $ &   $ -0.012\leq \alpha_{5} \leq 0.011$  &  $ -0.011\leq \alpha_{5} \leq 0.013$   \\
& & & \\
\hline
\hline
& & & \\
& C2XIR025 & C2XIR1 & C2XIR4 \\
& & & \\
\hline
& & & \\
$\alpha_{5}=0$&  $ -0.016\leq \alpha_{4} \leq 0.014 $   &  $ -0.013\leq \alpha_{4} \leq 0.014$   &  $ -0.013 \leq \alpha_{4} \leq 0.013$ \\
& & & \\
$\alpha_{4}=0$& $ -0.018 \leq \alpha_{5} \leq 0.017 $    &  $ -0.014 \leq \alpha_{5} \leq 0.02 $ & $ -0.016\leq \alpha_{5} \leq 0.015$  \\
& & & \\
\hline
\end{tabular}
\caption{Vínculos em 3$\sigma$ para os acoplamentos anômalos puramente
  quárticos $\alpha_{4}$ e $\alpha_{5}$, na situação em que somente um
  operador anômalo contribui por vez, para as escolhas de escalas de
  renormalização e fatorização C1~\eqref{c1} e C2~\eqref{c2} com variações $\varepsilon=0.5, 1, 4$. Resultados
  para o LHC operando com energia de 14 TeV e luminosidade integrada
  de 100 fb$^{-1}$.}
\label{tab_anom}
\end{center}}
\end{table}

De acordo com os valores da Tabela~\ref{tab_anom}, concluimos que os nossos
vínculos aos acoplamentos anômalos são compatíveis com aqueles
obtidos pela análise do canal leptônico da Tabela~\ref{concha}.

Tendo em mãos os vínculos dos acoplamentos anômalos $\alpha_4$ e
$\alpha_5$ na representação não-linear, podemos traduzí-los aos
acoplamentos $f_{0}$ e $f_{1}$ da representação linear. Como
incluímos uma partícula escalar leve na descrição do modelo sigma-não
linear, a tradução dos vínculos é relativamente direta. Além disto,
considerando que a maior contribuição anômala provém do vértice
quártico $WWWW$, a relação entre os acoplamentos se simplifica ainda
mais. É fácil ver, a partir das regras de Feynman para o
vértice $WWWW$ listadas no Apêndice \ref{ap:regrasfeynman} para as
duas representações, as relações
\[
\dfrac{f_{0,1}}{\Lambda^{4}}=\dfrac{8}{v^{4}}\alpha_{4,5}.
\]

Adotando o valor $v=246$ GeV como o \emph{vev} do bóson de Higgs
padrão, os vínculos aos acoplamentos $\alpha_4$ e $\alpha_5$ 
para a escolha C1XIR1 se traduzem em
\[
-22 \leq \dfrac{f_{0}}{\Lambda^{4}}\;(\mbox{TeV}^{-4}) \leq 20, \quad 
\mbox{e} \quad -26 \leq \dfrac{f_{1}}{\Lambda^{4}}\;(\mbox{TeV}^{-4}) \leq 24.
\]


Vale lembrar que nossos vínculos aos acoplamentos $f_{0}$ e
$f_{1}$ são apenas aproximados. Para se obter uma análise mais
precisa desses acoplamentos seria necessário adaptar a
Eq.~\eqref{cross_section} para os operadores na representação linear,
visto que as estruturas de \textit{gauge} dos acoplamentos $WWWW$ e
$WWZZ$ nessa representação diferem daquelas dadas pelos operadores não-lineares.

Mesmo adotando esse procedimento, a correspondência dos
acoplamentos $\alpha_4$ e $\alpha_5$ com $f_{0}$ e $f_{1}$ não é
exata, pois processos $pp\to l^{\pm}\nu_{l}jjjj$ envolvem vértices
anômalos $WWWW$ e $WWZZ$ que não podem ser analisados separadamente, já que experimentalmente não há como distinguir $W$ do $Z$ no contexto de decaimento hadrônico. 

Embora nossos vínculos aos acoplamentos anômalos sejam comparáveis aos obtidos a partir de canais leptônicos, é importante mencionar
que utilizamos uma abordagem bem diferente daquela empregada na
Ref.~\cite{eboli2}. Nessa referência, os autores estabeleceram uma região de
``controle'', completamente dominada pelos \textit{backgrounds},
com o propósito de avaliar nessa região as incertezas teóricas nas escolhas das escalas de normalização e fatorização. Esse erros foram
extrapolados para a região do sinal adicionando-os às incertezas
estatísticas. Nas nossas análises, mostramos explicitamente o efeito das diferentes escolhas de escalas e obtivemos um vínculo para cada uma delas.

Quando os experimentos do LHC alcançarem 100 fb$^{-1}$ de dados coletados, as incertezas teóricas dos
\textit{backgrounds} deverão ser minimizadas posto que na região de
controle as previsões teóricas estarão condicionadas aos resultados
experimentais. Com isto, incertezas nas escolhas das escalas de fatorização e
renormalização, assim como correções de \textit{next-leading order} em
QCD, estarão sob controle na região do sinal.

Para finalizar, vale uma breve explicação sobre a motivação de nossas
análises. Como já mencionado na Seção~\eqref{quiral}, os acoplamentos
quárticos entre os bósons de \emph{gauge} ainda não foram medidos
experimentalmente e portanto aqui não possuímos dados para a obtenção
dos vínculos experimentais aos acoplamentos anômalos $\alpha_4$ e
$\alpha_5$, se consideramos a realização não-linear. Sendo assim,
nessa tese apenas supomos que podemos observar (ou não) algo diferente
do previsto pelo MP, mensurando essa suposição em termos da
significância estatística~\eqref{significancia2}, em um cenário futuro
onde o LHC opera com energia $\sqrt{s}=14$ TeV e já atingiu sua
luminosidade máxima proposta de 100fb$^{-1}$.

\chapter{Correntes Neutras com Troca de Sabor no Modelo $331$ com Neutrinos de Mão-Direita}\label{chap:331}

Processos que induzem a troca de sabor de um férmion (quark ou lépton)
através do acoplamento com um bóson neutro, seja ele escalar ou
vetorial, são conhecidos por processos \textit{flavor-changing neutral
  currents} (FCNC).

Processos FCNC, como a mistura partícula-antipartícula e decaimentos com violação CP, são importantes para os testes do MP e são classificados como transições $\Delta F=1$, onde o número quântico de sabor $F=S,C,B$ de um méson muda em uma unidade, ou como transições $\Delta F=2$, que modificam o número quântico de sabor $F$ em duas unidades, características das misturas méson-antiméson.

\begin{figure}[!ht]
\centering
\includegraphics[scale=0.65]{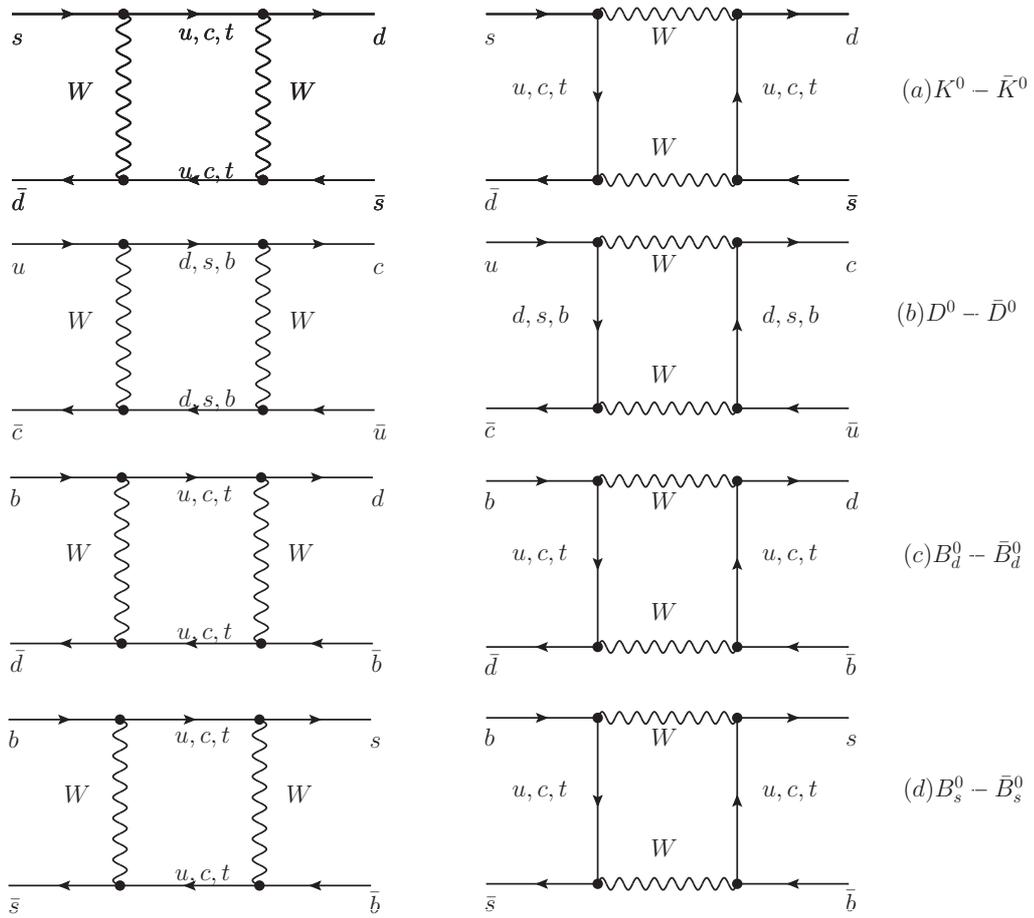}
\caption{Diagramas do tipo ``caixa'' descrevem as interações com transições $|\Delta F|=2$, onde $F=S,C,B$ são os números quânticos de sabor apropriados: $S$ para sistemas $K^{0}-\bar{K}^{0}$, $C$ para os sistemas $D^{0}-\bar{D}^{0}$ e $B$ para os sistemas $B^{0}-\bar{B^0}$.}
\label{mistura}
\end{figure}

Sabe-se que os mésons neutros $K^{0}$, $D^{0}$, $B^{0}_{d}$ e
$B^{0}_{s}$ são os únicos hádrons que se misturam com suas
antipartículas. No MP tais processos com transições $\Delta F=2$ são
suprimidos em nível de árvore através do mecanismo de
Glashow–Iliopoulos–Maiani (GIM)~\cite{charm}.  Historicamente, o quark
``charm'' foi a primeira partícula prevista teoricamente, como
consequência do mecanismo de GIM, com o intuito de eliminar
acoplamentos significativos de FCNC em nível de árvore na
teoria~\cite{charm}. Na Fig.~\ref{mistura} esboçamos os diagramas de
Feynman que contribuem para processos com transições $|\Delta F|=2$,
onde $F=S,C,B$ são os números quânticos de sabor apropriados para cada
sistema.

O experimento LHCb no CERN observou recentemente dois casos de violação CP direta~\cite{lhcb}: nos decaimentos de mésons $D^{0}$ uma diferença na assimetria CP entre os decaimentos $D^{0}\to K^+ K^-$ $D^{0}\to \pi^+ \pi^-$, em relação à hipótese de conservação CP, na ordem de $3.5\sigma$; e nos decaimentos de mésons $B^{0}_{s}\to K^{\pm}\pi^{\mp}$, em relação à hipótese de conservação CP, com significância de $3.3\sigma$.

Como forma de medição de violação CP indireta, os processos envolvendo sistemas de mésons neutros $K^{0}-\bar{K}^{0}$, $D^{0}-\bar{D}^{0}$ e $B^{0}_d-\bar{B^0_d}$ despertam grande interesse. O cálculo teórico de observáveis como a diferença de massa $\Delta M_{F}$ onde $F=S,C,B$ são os números quânticos de sabor apropriados para cada sistema, não é trivial. De forma bastante simplificada considera-se a evolução temporal de um sistema $F^{0}\bar{F}^{0}$, considerando a possibilidade de violação CP,
\begin{equation}
|\phi(t)\rangle = a(t)|F^{0}\rangle + \bar{a}(t)|\bar{F}^{0}\rangle,
\end{equation}
tal que
\begin{equation}
i\;\dfrac{d}{dt}|\phi(t)\rangle = {\cal H} |\phi(t)\rangle = \left(\begin{tabular}{cc}
${\cal H}_{11}$ & ${\cal H}_{12}$ \\
${\cal H}_{21}$ & ${\cal H}_{22}$
\end{tabular}
\right )|\phi(t)\rangle .
\end{equation}

A hamiltoniana ${\cal H}$ tem a forma
\begin{equation}
H=M-i\dfrac{\Gamma }{2}=\left( \begin{tabular}{cc}
$M_{11}-i\dfrac{\Gamma_{11} }{2}$ & $M_{12}-i\dfrac{\Gamma_{12} }{2}$ \\
$M_{21}-i\dfrac{\Gamma_{21} }{2}$ &  $M_{22}-i\dfrac{\Gamma_{22} }{2}$,
\end{tabular}  \right)
\end{equation}
onde $M$ e $\Gamma$ são as matrizes hermitianas de massa e decaimento, respectivamente.

Em geral~\cite{renton} temos que $\Delta M_{F} = -2M_{12}$ com
\begin{equation}
M_{12}=M_{21}^{*}= \langle F^{0}|{\cal H}|\bar{F}^{0} \rangle + \sum_{n}\;\dfrac{\langle F^{0}|{\cal H}|n \rangle \langle n|{\cal H}|\bar{F}^{0} \rangle }{m_{F^{0}}-E_{n}},
\label{m12}
\end{equation}
onde a soma se estende aos $n$ possíveis estados intermediários com energia $E_{n}$ e ${\cal H}=-{\cal L}$.

Note que no caso do MP, o termo $\langle F^{0}|{\cal H}|\bar{F}^{0} \rangle =0$ tendo em vista que transições $|\Delta F|=2$ em nível de árvore não são permitidas.

De fato, a lagrangiana do MP que descreve os acoplamentos das correntes carregadas (CC) com os quarks tem a forma~\cite{renton}
\begin{equation}
{\cal L}_{quarks}^{CC}=\dfrac{g}{\sqrt{2}}\;\sum_{j,k=1,2,3}\;\left[ V_{jk}\bar{u}_{jL}\gamma^{\mu}d_{kL}W^{+}_{\mu}+ V^{*}_{jk}\bar{d}_{kL}\gamma^{\mu}u_{jL}W^{-}_{\mu}\right ],
\label{lw}
\end{equation}
onde usamos a notação $(u_{1},u_{2},u_{3})=(u,c,t)$ e
$(d_{1},d_{2},d_{3})=(d,s,b)$ com
$V_{jk}=(V_{L}^{j})^{\dagger}(V_{L}^{k})$, $j\neq k$, sendo os termos
a matriz unitária $3\times 3$ de Cabibbo-Kobayashi-Maskawa (CKM),
\begin{align}
V_{CKM}&=\left( \begin{tabular}{ccc}
$V_{ud}$ & $V_{us}$  & $V_{ub}$ \\
 $V_{cd}$   & $V_{cs}$   & $V_{cb}$ \\
$V_{td}$  & $V_{ts}$  & $V_{tb}$
\end{tabular}
\right )
 \nonumber \\
&= \left( \begin{tabular}{ccc}
$c_{12}c_{13}$ & $s_{12}c_{13}$  & $s_{13}e^{-i\delta_{13}}$ \\
 $-s_{12}c_{23}-c_{12}s_{23}s_{13}e^{i\delta_{13}}$ & $c_{12}c_{23}-s_{12}s_{23}s_{13}e^{i\delta_{13}}$ & $s_{23}c_{13}$ \\
$s_{12}s_{23}-c_{12}c_{23}s_{13}e^{i\delta_{13}}$  & $ -c_{12}s_{23}-s_{12}c_{23}s_{13}e^{i\delta_{13}} $  & $c_{23}c_{13}$
\end{tabular}
\right ) 
\end{align}
onde $c_{ij}=\cos \theta_{ij}$ e $s_{ij}=\sin \theta_{ij}$ com o
ângulo $0\leq \theta_{ij} \leq \pi/2$ e a fase $0\leq \delta_{13} <
2\pi$, cujo valor diferente de zero é responsável pela violação CP nas
interações fracas. Nota-se explicitamente pela forma da
lagrangiana~\eqref{lw} que no MP a corrente carregada mistura famílias
através da matriz $V_{CKM}$ em nível de árvore.

Por outro lado, a lagrangiana padrão referente aos acoplamentos dos
quarks com correntes neutras (NC) tem a forma geral
\begin{equation}
{\cal L}_{quarks}^{NC}\approx \bar{u}_{L}\gamma^{\mu}(V_{L}^{u})^{\dagger}(V_{L}^{u})u_{L}Z_{\mu}+ \bar{d}_{L}\gamma^{\mu}(V_{L}^{d})^{\dagger}(V_{L}^{d})d_{L}Z_{\mu},
\label{lw_neutra}
\end{equation}
onde usamos a notação $u_{L}\equiv (u,c,t)$ e $d_{L}\equiv
(d,s,b)$. Nesse caso, o produto $(V_{L}^{u,d})^{\dagger}(V_{L}^{u,d})$
é a matriz identidade $\mathbf{I}_{3\times 3}$ e consequentemente no
MP as correntes neutras não misturam famílias, ou em outras palavras,
não trocam sabor.

Deste modo, no caso do MP, usa-se a lagrangiana~\eqref{lw} no segundo
termo da Eq.~\eqref{m12} para o cálculo das diferenças de massa dos sistemas de mésons neutros considerando o elemento de matriz específico para
cada caso, de acordo com os diagramas da Fig.~\ref{mistura}.

Detalhes de teoria e fenomenologia da mistura de mésons no MP estão
disponíveis em várias referências~\cite{modelo,renton}, fugindo do escopo de nosso trabalho entrarmos em detalhes desses cálculos.

Por outro lado, como extensões do MP, as diversas classes do modelo
com simetria $SU(3)_{C}\otimes SU(3)_{L}\otimes U(1)_{X}$~\cite{331}
mostram-se bastante simples e elegantes nas propostas de resolução de
alguns de seus problemas, tais como a replicação de
famílias~\cite{tese_alex,meupaper1}, candidatos à matéria
escura~\cite{331DM}, dentre outros~\cite{others, paty_fari}. 

Embora consigam preencher algumas lacunas do MP, os modelos $331$ prevêem fenômenos altamente suprimidos em nível de árvore, como citamos os processos FCNC em sistemas de mésons neutros. Desta forma, torna-se
mandatório a obtenção de vínculos aos parâmetros desses modelos considerando suas particularidades.

\section{Modelo 331 com neutrinos de mão-direita, $331_{RHN}$}
\label{sec:331}

Alguns modelos de física além do MP prevêem a existência de
novos bósons massivos que poderiam contribuir para as oscilações dos
mésons.

No modelo $331_{RHN}$, processos FCNC surgem
naturalmente em nível de árvore no setor de quarks, pois nesses modelos
um novo bóson de \textit{gauge} neutro $Z'$ acopla-se de
forma distinta com a terceira família, gerando interações
não-universais.

Além do $Z'$, o setor escalar do modelo
$331$ que apresenta neutrinos de mão-direita\footnote{Construção e
  detalhes do modelo podem ser obtidos
  em~\cite{331rhn,tese_alex,tese_diego}.} ($331_{RHN}$) introduz duas
novas fontes de violação CP: os escalares $S_2$ e $I_{3}^{0}$, como
veremos adiante.

De forma bastante resumida esboçaremos a seguir algumas características da construção do modelo $331_{RHN}$. 

Em geral, nos modelos $331$ o operador de carga
elétrica $Q$ define como os campos estão dispostos nas representações
de suas classes sendo dado por
\begin{equation}
Q=T^{3}-bT^{8}+X,
\end{equation}
onde as matrizes $T^{a}$ ($a=1,\cdots,8$) são as matrizes de
Gell-Mann, geradoras da álgebra de $SU(3)$,
$X$ é o número quântico relativo ao
grupo abeliano $U(1)_{X}$ e a constante $b$ define a classe à qual
pertence o modelo $331$~\cite{tese_alex}.

Se $b=\sqrt{3}$, temos o modelo $331$ original. Para o modelo $331_{RHN}$,
temos $b=1/\sqrt{3}$ e toda a matéria possui carga $X$ não nula.

Nos modelos $331$, os dubletos do MP fazem parte de tripletos. Em
particular, no caso $331_{RHN}$, os tripletos com $X=-1/3$ acomodam
duas componente neutras possibilitando colocar o neutrino e sua
antipartícula junto de um lépton carregado em uma estrutura de
tripleto. 

Dessa forma, o setor leptônico será composto por tripletos
de mão-esquerda (\emph{left-handed})
\begin{equation}
\Phi_{\alpha L}=\left(\nu_{\alpha}, e_{\alpha}, \nu_{\alpha}^{c} \right)_{L}^{T} \sim (\mathbf{1},\mathbf{3},-1/3),
\label{tripleto_lepton}
\end{equation}
sendo $e_{\alpha}$ ($\alpha=1,2,3$) as três gerações do lépton carregado do MP. 
O último termo mostra as propriedades de transformação em relação ao grupo 
 $SU(3)_{C}\otimes SU(3)_{L}\otimes U(1)_{X}$.

Como a conjugação de carga $c$ troca a quiralidade do férmion, então o
tripleto~\eqref{tripleto_lepton} possui as componentes direita e
esquerda do neutrino. Nesse modelo, o neutrino de mão direita se transforma
não-trivialmente sob o grupo não abeliano~\cite{tese_alex}.

As componentes de mão-direita dos léptons carregados formam os
singletos de léptons,
\begin{equation}
e_{\alpha R}\sim (\mathbf{1},\mathbf{1},-1).
\end{equation}
 
No setor de quarks as duas primeiras famílias se estruturam em
antitripletos, enquanto a terceira é representada por um tripleto, tais
que
\begin{equation}
Q_{iL}=(d_{i},u_{i},d^{\prime}_{i})^{T}_{L}\sim (\mathbf{3},\mathbf{3}^{*},0),\;\;\;Q_{3L}=(u_{3},d_{3},u^{\prime}_{3})^{T}_{L}\sim (\mathbf{3},\mathbf{3},1/3),
\end{equation}
com $i=1,2$. 

As componentes de mão-direita (\emph{right-handed}) no setor de quarks transformam-se como
singletos, tais que
\begin{align}
u_{\alpha R} &\sim (\mathbf{3},\mathbf{1},2/3),\;\;\;d_{\alpha R}\sim (\mathbf{3},\mathbf{1},-1/3), \nonumber \\
u^{\prime}_{3 R}&\sim (\mathbf{3},\mathbf{1},2/3),\;\;\;d^{\prime}_{3 R}\sim (\mathbf{3},\mathbf{1},-1/3),
\end{align}
onde $\alpha=1,2,3$.

A não-universalidade entre as três famílias no setor de quarks se deve
à necessidade do cancelamento de anomalias~\cite{tese_alex}. Como
consequência disso, aparecem fontes de FCNC em nível de árvore nesse
tipo de modelo.

Em relação ao setor escalar, o modelo $331_{RHN}$ apresenta três tripletos
escalares, necessários para quebrar espontaneamente a simetria de
\textit{gauge} $SU(3)_{L} \otimes
U(1)_{X}\rightarrow  SU(2)_{L} \otimes U(1)_{Y}
\rightarrow  U(1)_{EM}$, sendo suas respectivas
transformações~\cite{tese_alex} dadas por
\begin{align}
\label{conteudoescalar}
\chi & =  (\chi^0, \chi^-, \chi^{\prime 0})^T \approx (\mathbf{1},\mathbf{3},-1/3),\nonumber \\
\rho & =  (\rho^+, \rho^0, \rho^{\prime +})^T \approx (\mathbf{1},\mathbf{3},2/3),\nonumber \\
\eta & =  (\eta^0, \eta^-, \eta^{\prime 0})^T \approx (\mathbf{1},\mathbf{3},-1/3).
\end{align}

Ainda no setor escalar, de forma a simplificar o modelo, considerou-se um conjunto de simetrias
discretas~\cite{331DM} que tornam o potencial escalar renormalizável
e invariante de \textit{gauge}, com uma coleção de termos não-hermitianos
e hermitianos, tal que
\begin{align} 
V(\eta,\rho,\chi)=  & \mu_\chi^2 \chi^2 +\mu_\eta^2\eta^2
+\mu_\rho^2\rho^2+\lambda_1\chi^4 +\lambda_2\eta^4
\nonumber \\
& +\lambda_3\rho^4+ \lambda_4(\chi^{\dagger}\chi)(\eta^{\dagger}\eta)
+\lambda_5(\chi^{\dagger}\chi)(\rho^{\dagger}\rho) 
\nonumber \\
& +\lambda_6
(\eta^{\dagger}\eta)(\rho^{\dagger}\rho) + \lambda_7(\chi^{\dagger}\eta)(\eta^{\dagger}\chi)
\nonumber \\
& +\lambda_8(\chi^{\dagger}\rho)(\rho^{\dagger}\chi)+\lambda_9
(\eta^{\dagger}\rho)(\rho^{\dagger}\eta)
\nonumber \\
&-\dfrac{f}{\sqrt{2}}\epsilon^{ijk}\eta_i \rho_j \chi_k +\mbox{H.c}.
\label{potencial331}
\end{align}

Vale salientar que o potencial~\eqref{potencial331} não é o mais geral
possível, pois nele não consideramos termos de mistura entres os
escalares $\chi$ e $\eta$. Maiores detalhes podem ser obtidos na Ref.~\cite{tese_alex}.

Para quebrar espontaneamente a simetria, somente um escalar neutro de
cada tripleto~\footnote{As componentes neutras $\eta^{\prime 0}$ e
  $\chi^{0}$ carregam número leptônico -2 e 2, respectivamente. Caso
  esses escalares adquiram \textit{vev} diferente de zero, haveria
  quebra espontânea do número leptônico concomitante à geração de
  massa de Majorana para os neutrinos e ao surgimento de um bóson de
  Goldstone, o chamado Majoron~\cite{tese_alex}. Nesse trabalho
  consideramos, como no MP, que o número leptônico é conservado.}
desenvolve um \textit{vev} não-nulo tal que, escolhendo e expandindo
$\eta^{0}$, $\rho^{0}$ e $\chi^{\prime 0}$ em torno de seus
\textit{vev}'s, temos
\begin{align} 
\eta^{0} &\to \dfrac{1}{\sqrt{2}}(v_{\eta} + R_{\eta} + iI_{\eta}), \nonumber \\
\rho^{0} &\to \dfrac{1}{\sqrt{2}}(v_{\rho} + R_{\rho } + iI_{\rho }), \nonumber \\
\chi^{\prime 0} &\to  \dfrac{1}{\sqrt{2}}(v_{\chi^{\prime}} + R_{\chi^{\prime}} + iI_{\chi^{\prime}}).
\label{expansao}
\end{align}

Substituindo a expansão dos campos da Eq.~\eqref{expansao} na expressão do
potencial escalar~\eqref{potencial331}, encontram-se vínculos para este potencial que
possibilitam obter os termos quadráticos que originam as matrizes de
massa dos pseudo-escalares (escalares com CP ímpar) e dos escalares
(escalares com CP par). Maiores detalhes ver em~\cite{tese_diego}.

Sendo assim, a matriz de massa $M^{2}_{I,3\times 3}$ escrita na base $(I_{\chi^{\prime}},I_{\eta},I_{\rho})$ adquire a forma
\begin{equation}
M^{2}_{I,3\times 3} = \left( \begin{tabular}{ccc}
$-\dfrac{f}{4}\dfrac{v_{\eta}v_{\rho}}{v_{\chi^{\prime}}}$ & $-\dfrac{f}{4}v_{\rho}$ & $-\dfrac{f}{4}v_{\eta}$ \\
$-\dfrac{f}{4}v_{\rho}$ &  $-\dfrac{f}{4}\dfrac{v_{\chi^{\prime}}v_{\rho}}{v_{\eta}}$ & $-\dfrac{f}{4}v_{\chi^{\prime}}$ \\
$-\dfrac{f}{4}v_{\eta}$  & $-\dfrac{f}{4}v_{\chi^{\prime}}$  & $-\dfrac{f}{4}\dfrac{v_{\eta}v_{\chi^{\prime}}}{v_{\rho}}$
\end{tabular}
\right ).
\label{mass1}
\end{equation}

Da mesma maneira, a matriz de massa $M^{2}_{R,3\times 3}$ escrita na
base $(R_{\chi^{\prime}},R_{\eta},R_{\rho})$ adquire a forma
\begin{equation}
M^{2}_{R,3\times 3} = \left( \begin{tabular}{ccc}
$\lambda_{1}v^{2}_{\chi^{\prime}}-\dfrac{f}{4}\dfrac{v_{\eta}v_{\rho}}{v_{\chi^{\prime}}}$ & $\dfrac{\lambda_{4}}{2}v_{\chi^{\prime}}v_{\eta}-\dfrac{f}{4}v_{\rho}$ & $\dfrac{\lambda_{5}}{2}v_{\chi^{\prime}}v_{\eta}\dfrac{f}{4}v_{\eta}$ \\
$\dfrac{\lambda_{4}}{2}v_{\chi^{\prime}}v_{\eta}+\dfrac{f}{4}v_{\rho}$ &  $\lambda_{2}v^{2}_{\eta}-\dfrac{f}{4}\dfrac{v_{\chi^{\prime}}v_{\rho}}{v_{\eta}}$ & $ \dfrac{\lambda_{6}}{2}v_{\eta}v_{\rho}+\dfrac{f}{4}v_{\chi^{\prime}}$ \\
$\dfrac{\lambda_{5}}{2}v_{\chi^{\prime}}v_{\rho}+\dfrac{f}{4}v_{\eta}$  & $\dfrac{\lambda_{6}}{2}v_{\eta}v_{\rho}+ \dfrac{f}{4}v_{\chi^{\prime}}$  & $\lambda_{3}v^{2}_{\rho}-\dfrac{f}{4}\dfrac{v_{\eta}v_{\chi^{\prime}}}{v_{\rho}}$
\end{tabular}
\right ).
\label{mass2}
\end{equation}

De acordo com a Ref.~\cite{tese_diego}, depois do procedimento de
diagonalização das matrizes~\eqref{mass1} e~\eqref{mass2}, e usando
$v_{\eta}= v_{\rho}\equiv v$ (com $v=v_{SM}/\sqrt{2}$) da ordem da
escala de quebra da simetria eletrofraca e $f\sim
\dfrac{v_{\chi^{\prime}}}{2}$, onde $v_{\chi^{\prime}}$ representa a
escala de quebra da simetria do $331$ (para facilitar o procedimento
de diagonalização), obtém-se os autoestados físicos $(S_{1},S_{2},H)$
na base de CP-par
\begin{equation}
\label{escalaresfisicos}
S_{1} = R_{\chi^{\prime}},\ \ S_{2} = \dfrac{1}{\sqrt{2}}(R_{\eta}-R_{\rho}),\ \ H = \dfrac{1}{\sqrt{2}}(R_{\eta}+R_{\rho}).
\end{equation}
sendo
\begin{align}
M^{2}_{S_{1}} & =  \dfrac{v^{2}}{4}+2v_{\chi^\prime}^{2}\lambda_{1},
\nonumber \\
M^{2}_{S_{2}} & =  \dfrac{1}{2}(v_{\chi^\prime}^{2}+2v^{2}(2\lambda_{2}-\lambda_{6})),\nonumber \\
M^{2}_{H} & =  v^{2}(2\lambda_{2}+\lambda_{6})
\label{massashiggs}
\end{align}
e os autoestados $(I^0_1,I^0_2,I^0_3)$ na base de CP-ímpar
\begin{equation}
\label{psescalaresfisicos}
I_{1}^{0} \sim -I_{\chi^{\prime}},\ \ I_{2}^{0} \sim \dfrac{1}{\sqrt{2}}(I_{\rho}-I_{\eta}),\ \ I_{3}^{0} \sim \dfrac{1}{\sqrt{2}}(I_{\rho}+I_{\eta}),
\end{equation}
sendo
\begin{eqnarray}
\label{escalaresfisicos2}
M^{2}_{I_{1}^{0}} = 0,\
M^{2}_{I_{2}^{0}} = 0,\
M^{2}_{I_{3}^{0}} = \dfrac{1}{2}(v_{\chi^\prime}^{2}+\dfrac{v^{2}}{2}),
\end{eqnarray} onde $v$ é o \textit{vev} dos escalares neutros $\rho^0$ e $\eta^0$, enquanto que  $v_{\chi^\prime}$ é o \textit{vev} do campo neutro $\chi^{\prime 0}$.

Na Eq.~\eqref{escalaresfisicos}, $H$ corresponde ao que seria o bóson de
Higgs padrão. Na Eq.~\eqref{psescalaresfisicos}, $I_{1}^{0}$ e
$I_{2}^{0}$ são bósons de Goldstone que são absorvidos pelos bósons
$Z$ e $Z^{\prime}$, enquanto $I_{3}^{0}$ é um pseudoescalar pesado.

O tripleto de escalares~\eqref{conteudoescalar} é o responsável por gerar as massas dos férmions, exceto neutrinos, através da lagrangiana de Yukawa
\begin{eqnarray}
{-\cal{L}}^{Yuk} & = & \lambda_{2ij}\bar{Q}_{iL}\chi^{*}D^{\prime}_{jR} + \lambda_{1}\bar{Q}_{3L}\chi U^{\prime}_{3R} + \lambda_{4ia}\bar{Q}_{iL}\eta^{*}d_{aR} \nonumber\\
                & + & \lambda_{3a}\bar{Q}_{3L}\eta u_{aR}  + \lambda_{1a}\bar{Q}_{3L}\rho d_{aR} + \lambda_{2ia}\bar{Q}_{iL}\rho^{*}u_{aR}\nonumber\\
                & + & G_{aa}\bar{f}_{aL}\rho e_{aR} + H.C.
\label{yukawafermions}
\end{eqnarray} 

Em relação ao setor de \textit{gauge}, o modelo $331_{RHN}$ adiciona
cinco bósons, a saber $V^{+}$, $V^{-}$, $U^{0}$, $U^{0\dagger}$ e
$Z^{\prime}$, ao conteúdo do MP. Maiores detalhes em~\cite{tese_alex}.

Aqui nos interessa particularmente o novo bóson neutro $Z^{\prime}$,
cuja massa, no limite $Z_1 \equiv Z$ e $Z_2\equiv Z^{\prime}$ para os
bósons de \textit{gauge} físicos $Z_{1}\;\mbox{e}\;Z_{2}$, é
aproximadamente dada por
\begin{equation}
m^2_{Z^\prime} = \dfrac{g^{2}}{4(3-4s_W^2)}\left[4c^{2}_{W}v_{\chi^\prime}^2 +\dfrac{v^{2}}{ c^{2}_{W}}+\dfrac{v^{2}(1-2s^{2}_{W})^2}{c^{2}_{W}}\right ].
\label{massvec}
\end{equation}

Enfim, apresentado o espectro do modelo $331_{RHN}$ que nos interessa aqui, e tendo em vista seu desenvolvimento teórico bem
documentado em diversos trabalhos~\cite{tese_alex,331rhn,tese_diego}, nos restringiremos agora a expor as lagrangianas diretamente relacionadas
aos nossos cálculos. 

Seguindo esse protocolo, as lagrangianas que
contribuem para os processos FCNC no setor escalar têm a forma 
\begin{align}
\mathcal{L}^{|\Delta S|=2}_{S_2,I_3} = & \biggl[ \dfrac{\lambda_{413}}{2}(V^d_L)^{\ast}_{11}(V^d_R)_{32}+ \dfrac{\lambda_{423}}{2} (V^d_L)^{\ast}_{21}(V^d_R)_{32}  
\nonumber \\
 &  -\dfrac{\lambda_{13}}{2}(V^d_L)^{\ast}_{31}(V^d_R)_{32}+ \dfrac{\lambda_{422}}{2} (V^d_L)^{\ast}_{21}(V^d_R)_{22}  
\nonumber \\
 &   -\dfrac{\lambda_{12}}{2}(V^d_L)^{\ast}_{31}(V^d_R)_{22} - \dfrac{\lambda_{11}}{2} (V^d_L)^{\ast}_{31}(V^d_R)_{12} \biggr] [\bar{d}^{\prime}_{1L} d^{\prime}_{2R}]\times (S_2,I_3),
\label{FCNC4}
\end{align}

\begin{align}
\mathcal{L}^{|\Delta C|=2}_{S_2,I_3} =& \biggl[
  \dfrac{\lambda_{31}}{2}(V^u_L)^{\ast}_{31}(V^u_R)_{12}+\dfrac{\lambda_{222}}{2}(V^u_L)^{\ast}_{21}(V^u_R)_{22}
  \nonumber \\ 
&  +\dfrac{\lambda_{32}}{2}(V^u_L)^{\ast}_{31}(V^u_R)_{22}+
  \dfrac{\lambda_{213}}{2} (V^u_L)^{\ast}_{11}(V^u_R)_{32} 
\nonumber  \\ 
& +\dfrac{\lambda_{223}}{2}
  (V^u_L)^{\ast}_{21}(V^u_R)_{32}+ \dfrac{\lambda_{33}}{2}
  (V^u_L)^{\ast}_{31}(V^u_R)_{32} \biggr][\bar{u}^{\prime}_{1L}
  u^{\prime}_{2R}]\times (S_2,I_3),
\label{FCNC5}
\end{align}
e
\begin{align}
\mathcal{L}^{|\Delta B|=2}_{S_2,I_3} =&  \biggl[ \dfrac{\lambda_{413}}{2}(V^d_L)^{\ast}_{11}(V^d_R)_{33}+ \dfrac{\lambda_{423}}{2} (V^d_L)^{\ast}_{21}(V^d_R)_{33}  
\nonumber \\
  & -\dfrac{\lambda_{13}}{2}(V^d_L)^{\ast}_{31}(V^d_R)_{33}+ \dfrac{\lambda_{422}}{2} (V^d_L)^{\ast}_{21}(V^d_R)_{23}  
\nonumber \\
  &  -\dfrac{\lambda_{12}}{2}(V^d_L)^{\ast}_{31}(V^d_R)_{23} - \dfrac{\lambda_{11}}{2} (V^d_L)^{\ast}_{31}(V^d_R)_{13} \biggr] [\bar{d}^{\prime}_{1L} d^{\prime}_{3R}]\times (S_2,I_3)
\label{FCNC6}
\end{align}
onde os $\lambda$'s são os acoplamentos que aparecem na lagrangiana de
Yukawa~\eqref{yukawafermions} e os $V_{L,R}^{u,d}$ são as matrizes $3\times 3$ que
diagonalizam as matrizes de massa para os quarks \textit{up} e
\textit{down} padrão~\footnote{ Lembramos que a matriz CKM usual é definida como
$V_{CKM}=(V_{L}^{u})^{\dagger}(V_{L}^{d})$ e seguimos a notação
$(u^{\prime}_{1},u^{\prime}_{2},u^{\prime}_{3})=(u,c,t)$ e
$(d^{\prime}_{1},d^{\prime}_{2},d^{\prime}_{3})=(d,s,b)$.}. Maiores detalhes no Apêndice C da Ref.~\cite{paty_fari}.

As expressões~\eqref{FCNC4}-\eqref{FCNC6} fornecem as relações entre os escalares e os elementos da matriz de mistura mostrando as novas fontes de FCNC no modelo $331_{RHN}$.

Por outro lado, as lagrangianas que descrevem as interações entre os quarks padrão com o novo bóson de \emph{gauge} neutro $Z^{\prime}$ têm a
forma
\begin{equation}
\mathcal{L}^{|\Delta S|=2}_{Z'}= \left(  \dfrac{ -g\ c_W }{  \sqrt{ 3-4 s_{W}^{2} } }\right)\{ (V^d_L)^{\ast}_{31}(V^d_L)_{32}\}[\bar{d}^{\prime}_{1L} \gamma_\mu d^{\prime}_{2L}]Z^{\prime},
\label{FCNC1}
\end{equation}
\begin{equation}
\mathcal{L}^{|\Delta C|=2}_{Z'}= \left(  \dfrac{ -g\ c_W }{  \sqrt{ 3-4 c_{W}^{2} } }\right)\{(V^u_L)^{\ast}_{31}(V^u_L)_{32}\}[\bar{u}^{\prime}_{1L} \gamma_\mu u^{\prime}_{2L}]Z^{\prime},
\label{FCNC2}
\end{equation}
e
\begin{equation}
\mathcal{L}^{|\Delta B|=2}_{Z'}= \left(  \dfrac{ -g\ c_W }{  \sqrt{ 3-4 s_{W}^{2} } }\right)\{(V^d_L)^{\ast}_{31}(V^d_L)_{33}\}[\bar{d}^{\prime}_{1L} \gamma_\mu d^{\prime}_{3L}]Z^{\prime},
\label{FCNC3}
\end{equation}
onde $s_W \equiv \mbox{sen}\,\theta_W$ e $c_W = \cos\theta_W$.

Utilizando as lagrangianas~\eqref{FCNC4}-\eqref{FCNC3}, que permitem
transição $\Delta F=2$, podemos obter
as contribuições dos bósons escalares $S_2$, $I_3$ e do
bóson de \emph{gauge} $Z^{\prime}$ em nível de árvore para as diferenças de massa dos
sistemas de mésons $K_{0}-\bar{K_{0}}$ (transições $|\Delta S|=2$) ,
$D^{0}-\bar{D^{0}}$ (transições $|\Delta C|=2$) e
$B^{0}_d-\bar{B^{0}_d}$ (transições $|\Delta B|=2$). 

De fato, aqui vale observar que no cálculo das diferenças de massa dos sistemas em questão, considerando a contribuição do novo bóson de \emph{gauge} $Z'$, usa-se o primeiro termo da expressão~\eqref{m12} em conjunto com as lagrangianas~\eqref{FCNC1}-\eqref{FCNC3} para obtermos explicitamente que 
\begin{eqnarray}
\ensuremath{\mathcal{L}}^{K_{0}-\bar{K_{0}}}_{Z'\ eff} & = & \frac{4 \sqrt{2} G_F c^4_W}{(3-4s^2_W)}\frac{M_{Z}^{2}}{M_{Z^{\prime}}^{2}}((V_{L}^{d})_{31}^{\ast}(V_{L}^{d})_{32})^{2}(\bar{d}_{1L}^{\prime}\gamma_{\mu}d_{2L}^{\prime})^{2},\nonumber \\
& & \nonumber \\
\ensuremath{\mathcal{L}}^{D_{0}-\bar{D_{0}}}_{Z'\ eff} & = & \frac{4 \sqrt{2} G_F c^4_W}{(3-4s^2_W)}\frac{M_{Z}^{2}}{M_{Z^{\prime}}^{2}}((V_{L}^{u})_{31}^{\ast}(V_{L}^{u})_{32})^{2}(\bar{u}_{1L}^{\prime}\gamma_{\mu}u_{2L}^{\prime})^{2},\nonumber \\
& & \nonumber \\
\ensuremath{\mathcal{L}}^{B^{0}_d-\bar{B^{0}_d}}_{Z'\ eff} & = & \frac{4 \sqrt{2} G_F c^4_W}{(3-4s^2_W)}\frac{M_{Z}^{2}}{M_{Z^{\prime}}^{2}}((V_{L}^{d})_{31}^{\ast}(V_{L}^{d})_{33})^{2}(\bar{d}_{1L}^{\prime}\gamma_{\mu}d_{3L}^{\prime})^{2},\nonumber \\
\end{eqnarray}   
de onde resultam, após algumas parametrizações de acordo com as Refs.~\cite{renton,PDG}, as expressões para as diferenças de massa no limite em que $Z_1 \equiv Z$ e $Z_2\equiv Z^{\prime}$
\begin{equation}
(\Delta m_{K})_{Z^{\prime}}=\dfrac{4 \sqrt{2} G_F c^4_W}{(3-4s^2_W)}\dfrac{M_{Z}^{2}}{M_{Z^{\prime}}^{2}}((V_{L}^{d})_{31}^{\ast}(V_{L}^{d})_{32})^{2}f_{K}^{2}B_{K}\eta_{K}m_{k},
\label{massdifKZlinha}
\end{equation}
\begin{equation}
(\Delta m_{D})_{Z^{\prime}}=\dfrac{4 \sqrt{2} G_F c^4_W}{(3-4s^2_W)}\dfrac{M_{Z}^{2}}{M_{Z^{\prime}}^{2}}((V_{L}^{u})_{31}^{\ast}(V_{L}^{u})_{32})^{2}f_{D}^{2}B_{D}\eta_{D}m_{D},
\label{massdifDZlinha}
\end{equation}
\begin{equation}
(\Delta m_{B_d})_{Z^{\prime}}=\dfrac{4 \sqrt{2} G_F c^4_W}{(3-4s^2_W)}\dfrac{M_{Z}^{2}}{M_{Z^{\prime}}^{2}}|(V_{L}^{d})_{31}^{\ast}(V_{L}^{d})_{33}|^{2}f_{B}^{2}B_{B}\eta_{B}m_{B},
\label{massdifBZlinha}
\end{equation}
onde $B$ e $f$'s são o \textit{bag parameter} e as constantes de
decaimento dos mésons, respectivamente. $\eta$'s são as correções de
QCD em \textit{leading order}.

Por outro lado, para obtermos as contribuições referentes aos escalares $S_{2}$ e $I_{0}^{3}$ usamos o resultado da Ref.~\cite{mohapatra}, em conjunto com as expressões~\eqref{FCNC4}-\eqref{FCNC6}, gerando as diferenças de massa 
\begin{equation}
(\Delta m_{K})_{S_{2},I_{0}^{3}}=\dfrac{A_1}{4M_{S_{2},I_{3}^{0}}^{2}}\dfrac{m_{K}^{3}f_{k}^{2}}{(m_{d}+m_{s})^{2}},
\label{massdifKscalars}
\end{equation}
\begin{equation}
(\Delta m_{D})_{S_{2},I_{0}^{3}}= \dfrac{A_2}{4M_{S_{2},I_{3}^{0}}^{2}}\dfrac{m_{D}^{3}f_{D}^{2}}{(m_{u}+m_{c})^{2}},
\label{massdifDscalars}
\end{equation}
\begin{equation}
(\Delta m_{B_d})_{S_{2},I_{0}^{3}}=\dfrac{A_3}{4M_{S_{2},I_{3}^{0}}^{2}}\dfrac{m_{B}^{3}f_{B}^{2}}{(m_{d}+m_{b})^{2}},
\label{massdifBscalars}
\end{equation}
onde os parâmetros $A_1$, $A_2$ e $A_3$
representam o quadrado da soma dos coeficientes
entre colchetes das lagrangianas~\eqref{FCNC4}-\eqref{FCNC6}, respectivamente.

Tendo em mãos todo esse arcabouço teórico mostraremos a seguir nossos resultados.

\section{Resultados}
\label{resultados:331}

Agora apresentamos os resultados obtidos com base na Seção~\ref{sec:331}. Usando $G_F= 1.166 \times 10^{-5}\;\mbox{GeV}^{-2}$ e os valores numéricos para massas e parâmetros dos mésons $K$, $D$ e $B_{d}$ catalogados pelo Particle Data Group~\cite{PDG},
\begin{eqnarray}
&  m_{K} = 497.6\ \mbox{MeV}, \; \sqrt{B_{K}}f_{K} = 135\ \mbox{MeV}, \; \eta_K= 0.57,&
\nonumber \\ 
&  m_{D}=1865.0\ \mbox{MeV}, \; \sqrt{B_{D}}f_{D} =187\ \mbox{MeV}, \eta_D= 0.57,&
\nonumber \\
& m_{B_d} = 5279.5\ \mbox{MeV}, \; \sqrt{B_{B_d}}f_{B_d} = 208\ \mbox{MeV}, \; \eta_{B_d} = 0.55, &
\end{eqnarray}
aplicados às Eqs.\eqref{massdifKZlinha}-\eqref{massdifBscalars}, podemos escrever as expressões para diferença de massa dos sistemas de mésons somente em função da
massa do bóson vetorial $Z'$ e dos bósons escalares $S_2$ e $I_3$ tais que
\begin{equation}
\label{finalEq1}
(\Delta m_{K})_{Z^{\prime}}= \dfrac{2.066\times 10^{-9}}{M_{Z^{\prime}}^2}\mbox{(GeV)}
\end{equation}
\begin{equation}
\label{finalEq2}
(\Delta m_{K})_{S2,I_{0}^{3}}= \dfrac{1.47725\times 10^{-10}}{M_{S_2}^2,M_{I_3}^2}\mbox{(GeV)}
\end{equation}
\begin{equation}
\label{finalEq3}
(\Delta m_{D})_{Z^{\prime}}= \dfrac{1.48657\times 10^{-8}}{M_{Z^{\prime}}^2}\mbox{(GeV)}
\end{equation}
\begin{equation}
\label{finalEq4}
(\Delta m_{D})_{S2,I_{0}^{3}}= \dfrac{2.53\times 10^{-12}}{M_{S_2}^2,M_{I_3}^2}\mbox{(GeV)}
\end{equation}
\begin{equation}
\label{finalEq5}
(\Delta m_{B_d})_{Z^{\prime}}= \dfrac{5.66828\times 10^{-6}}{M_{Z^{\prime}}^2}\mbox{(GeV)}
\end{equation}
\begin{equation}
\label{finalEq6}
(\Delta m_{B_d})_{S2,I_{0}^{3}}= \dfrac{1.8304\times 10^{-8}}{M_{S_2}^2,M_{I_3}^2}\mbox{(GeV)}.
\end{equation}

Vinculamos as massas $M_{Z^{\prime}}$, $M_{S_2}$ e $M_{I_3}$ utilizando os seguintes valores experimentais~\cite{PDG},
\begin{align*}
\Delta m_{K} &= (3.484\pm 0.006) \times 10^{-15}\ \mbox{GeV},
\\
 \Delta m_{D} &= 4.607 \times 10^{-14}\ \mbox{GeV}, 
\\
\Delta m_{B_d} &= (3.337\pm 0.033) \times 10^{-13}\ \mbox{GeV}.
\end{align*}

\begin{figure}[h!tb]
\centering
\subfloat[]{\includegraphics[scale=0.5]{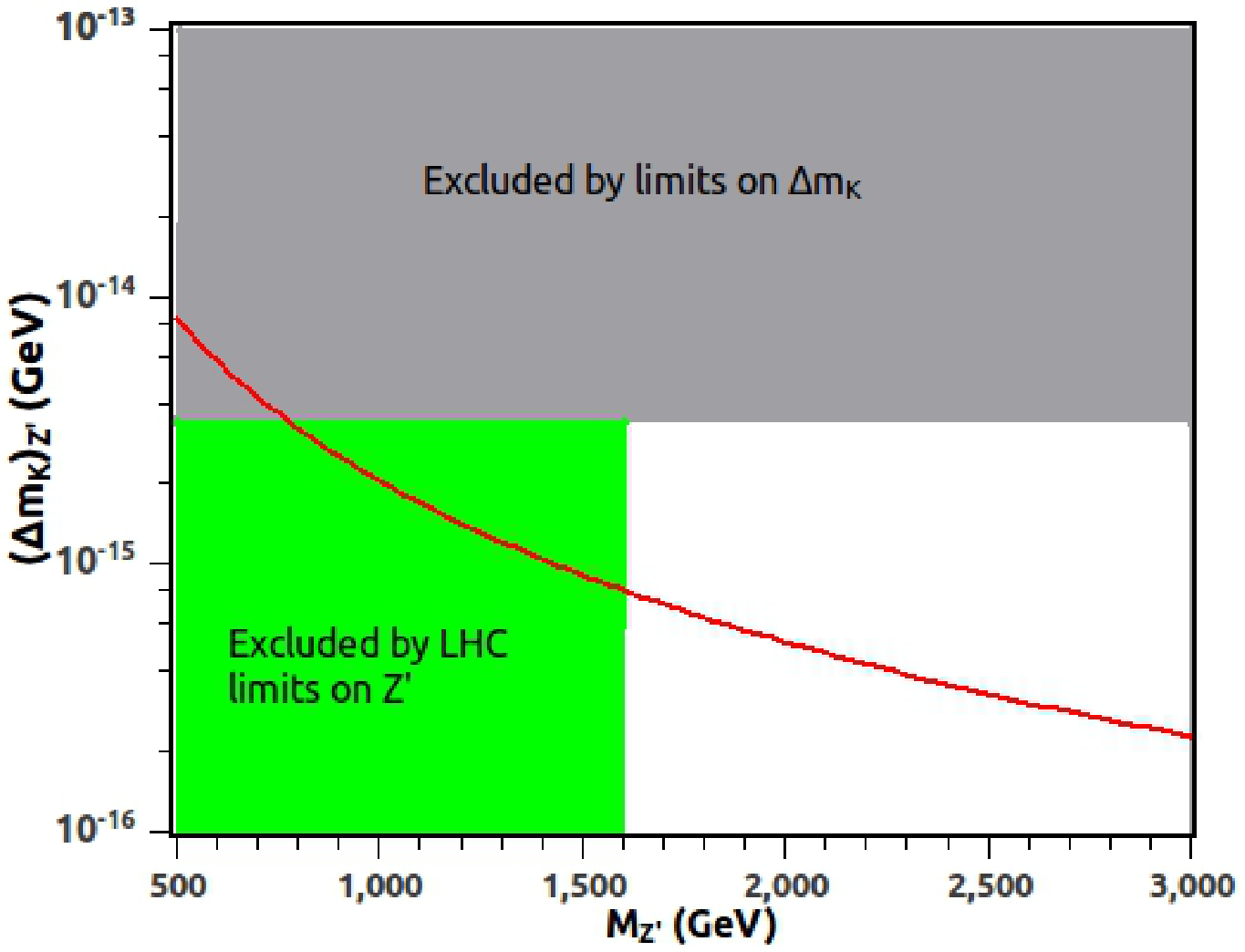}\label{fig1}}
\hspace{0.2cm}
\subfloat[]{\includegraphics[scale=0.5]{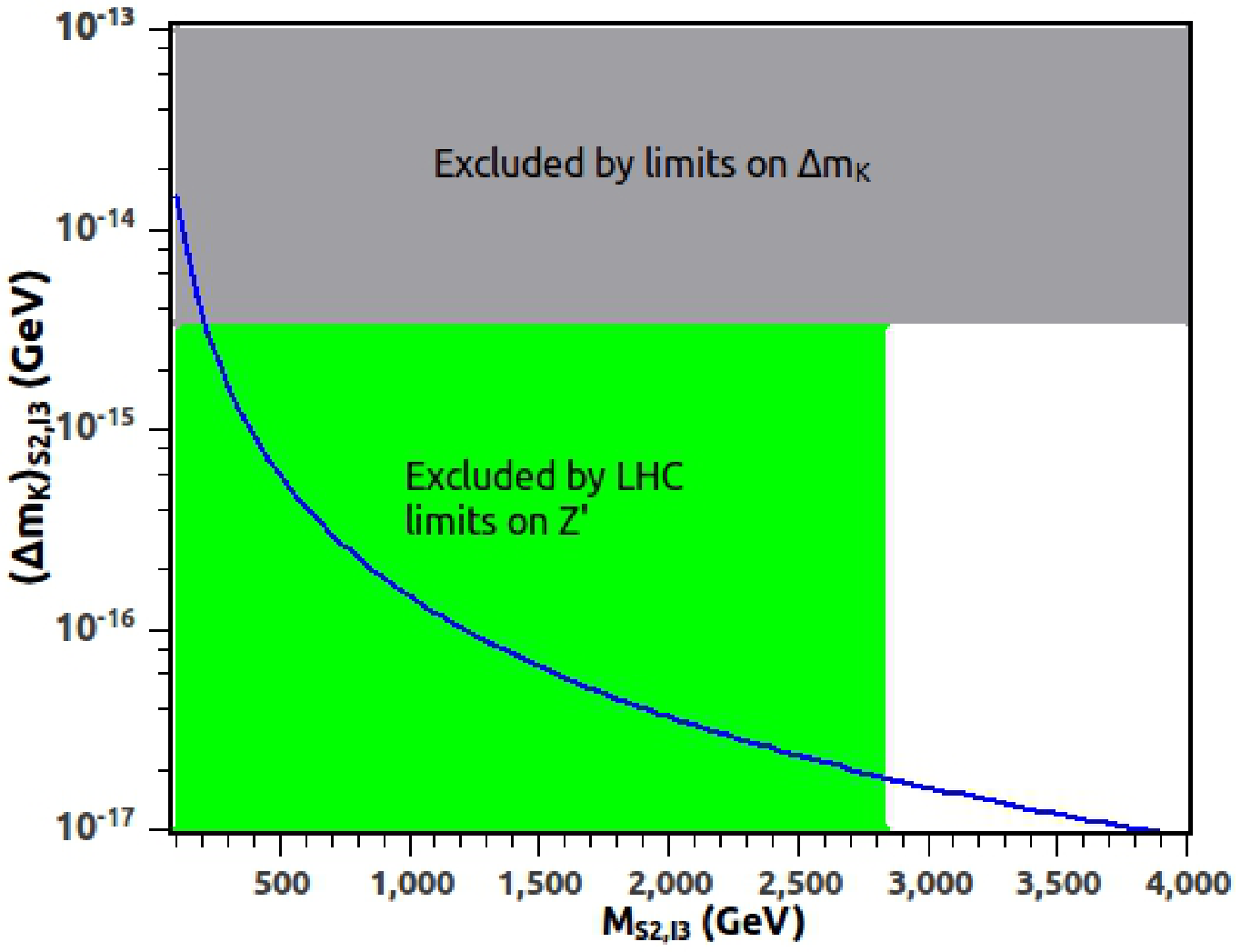}\label{fig2}}
\caption{Em (a) a linha vermelha mostra a contribuição do $Z^{\prime}$
  e em (b) a curva azul mostra a contribuição dos
  escalares $S_2$ e $I_3$ para $\Delta m_K$.  Note que as
  contribuições de $S_2$ e $I_3$ são iguais e portanto somente uma
  curva está visível. A região verde em (b) significa a exclusão
  das massas de $S_2$ e $I_3$ a partir da exclusão da massa de
  $M_{Z^\prime}$ pelo LHC, representada pela região verde em (a).}
\label{331result}
\end{figure}

\begin{figure}[h!tb]
\centering
\subfloat[]{\includegraphics[scale=0.5]{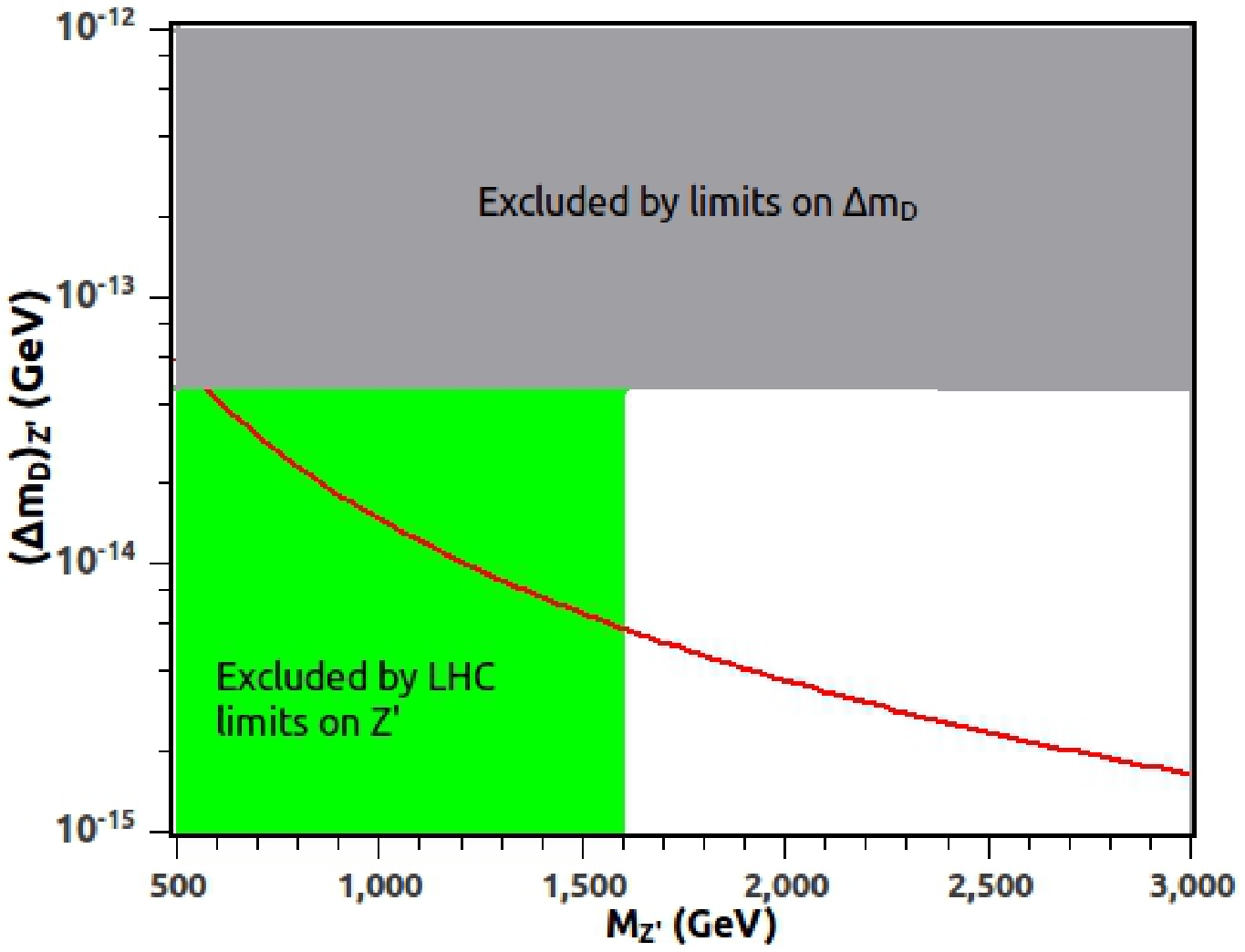}\label{fig3}}
\hspace{0.2cm}
\subfloat[]{\includegraphics[scale=0.5]{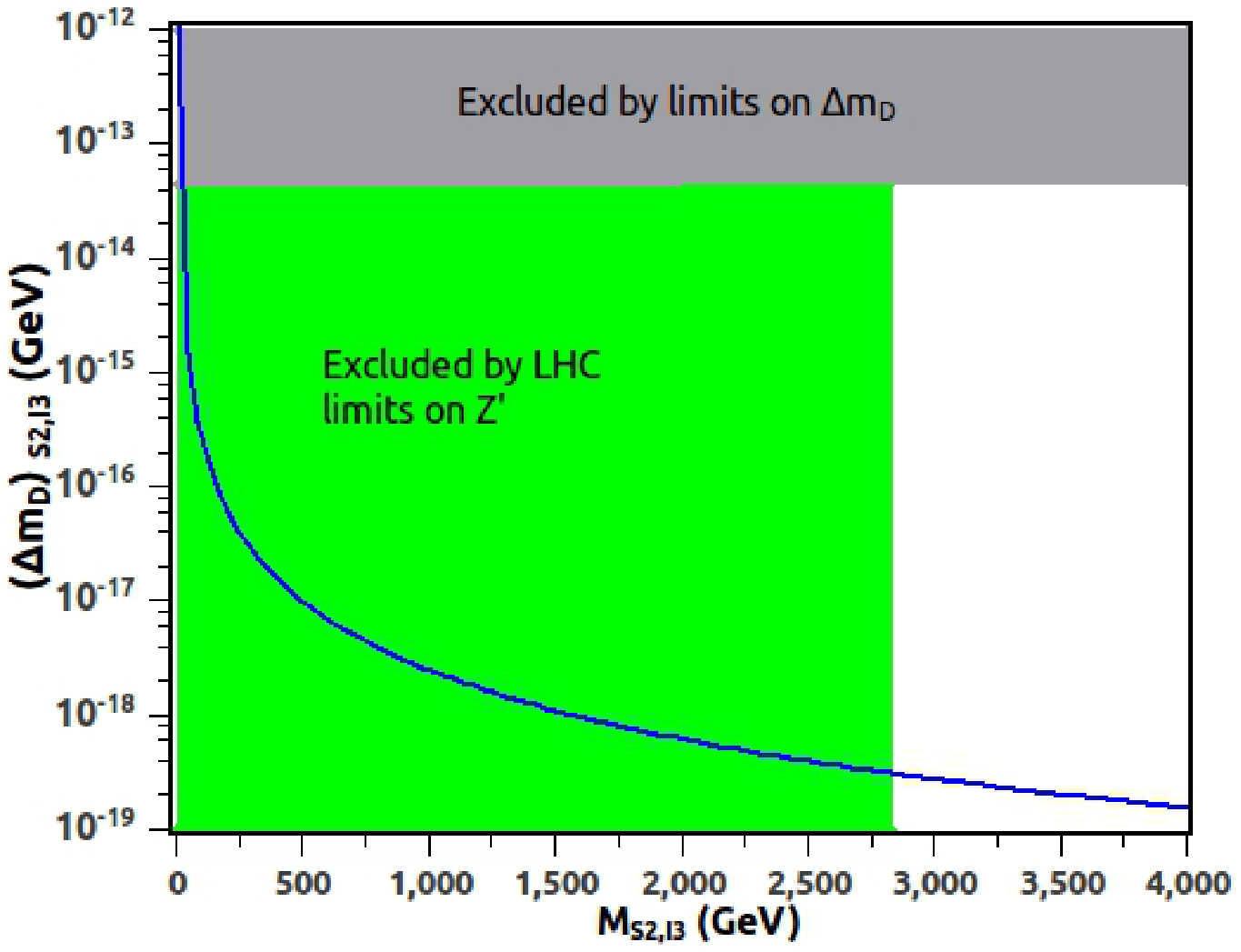}\label{fig4}}
\caption{Idem à legenda da Fig.~\ref{331result} para $\Delta m_D$.}
\end{figure}

\begin{figure}[h!tb]
\centering
\subfloat[]{\includegraphics[scale=0.5]{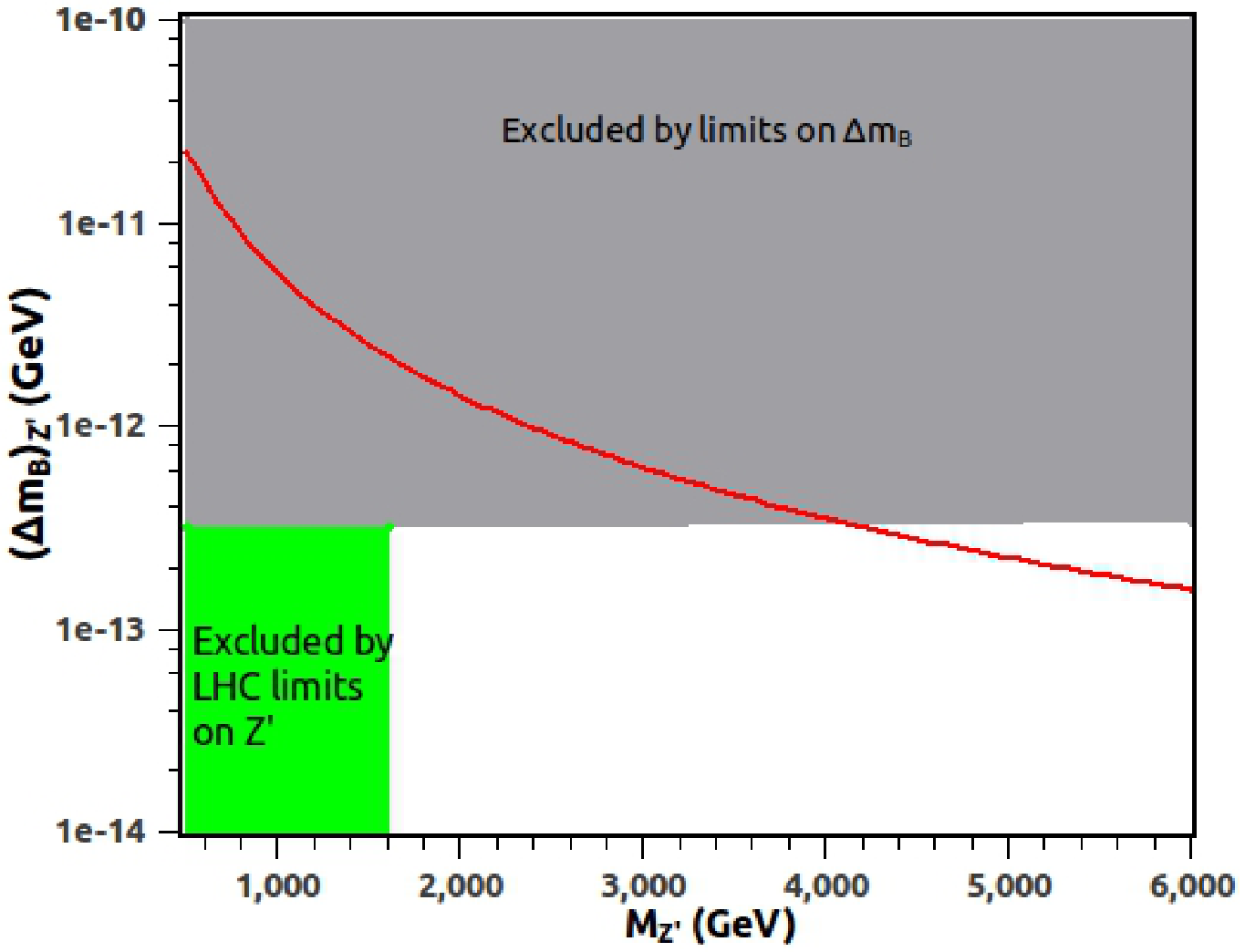}\label{fig6}}
\hspace{0.2cm}
\subfloat[]{\includegraphics[scale=0.5]{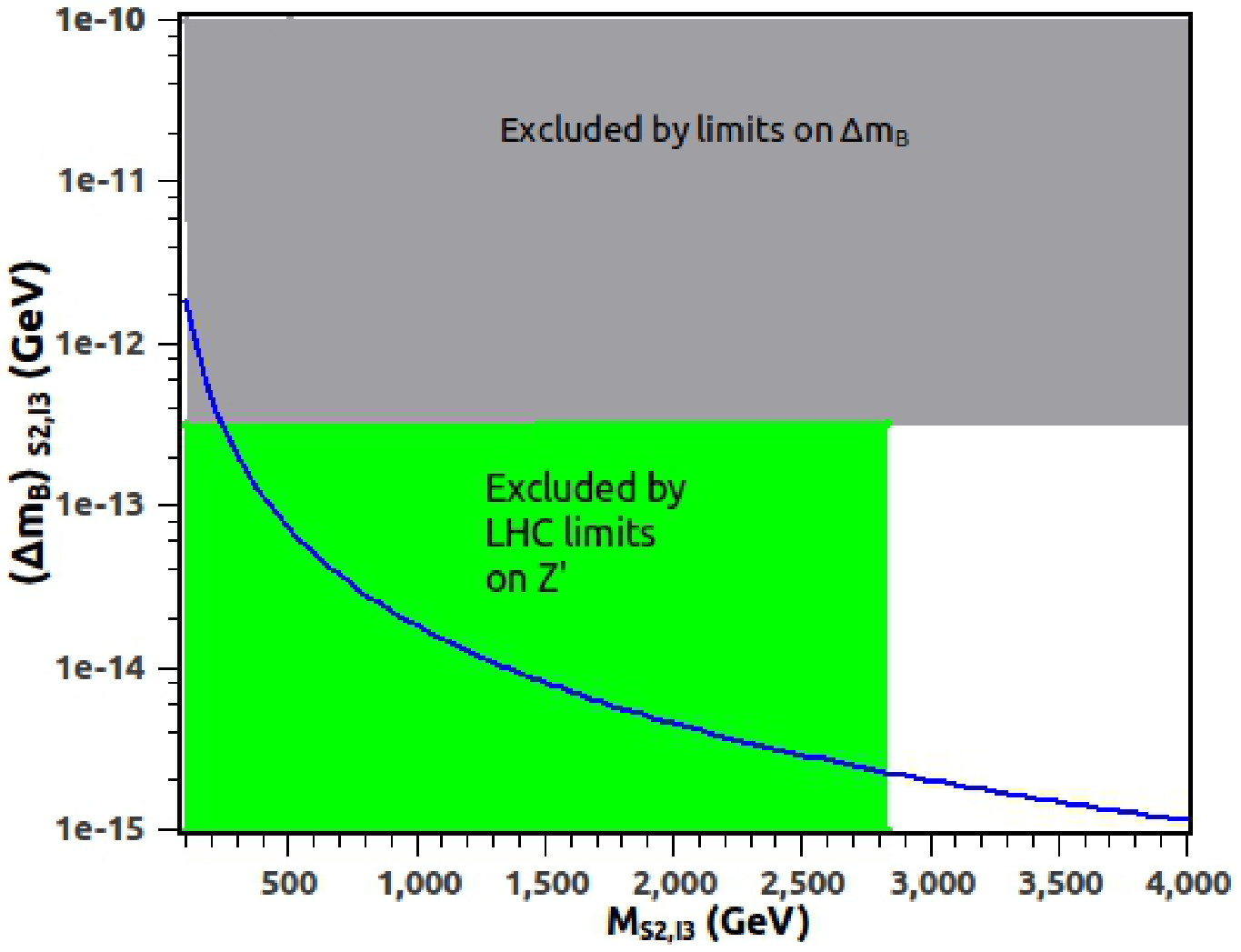}\label{fig5}}
\caption{Idem à legenda da Fig.~\ref{331result} para $\Delta m_B$.}
\end{figure}

Nas Figs.\ref{fig1}-\ref{fig5} as regiões em cinza representam regiões de exclusão definidas pelos valores experimentais ($\Delta M_{P}$), onde $P=K,D,B_{d}$, acima descritos. As regiões verdes refletem a influência do vínculo experimental $M_Z^{\prime} \geq 1.6$ TeV, de acordo os resultados obtidos recentemente pelo CMS e ATLAS~\cite{limitZlinhaLHC}. 

Lembrando que no modelo $331_{RHN}$ as massas dos novos bósons estão
vinculadas entre si pelo parâmetro $v_{\chi^{\prime}}$ através das expressões
\eqref{massashiggs}, \eqref{escalaresfisicos2} e \eqref{massvec}, onde usamos que $M_{H}=120$ GeV, $\lambda_{2}=\lambda_{6}$ e $v= v_{SM}/\sqrt{2}$, então um limite
na massa do $Z^\prime$ implica em vínculos para os escalares $S_2$ e $I_3$.

A Fig.~\ref{fig1} mostra $\Delta m_{K}$ em termos da massa do bóson
$Z^{\prime}$ e a Fig.~\ref{fig2} em termos da massa dos escalares
$S_2$ e $I_3$. Observa-se que a contribuição maior provém do
$Z^{\prime}$. Utilizando-se o valor experimental de $\Delta m_{K}$,
temos que $M_{Z^{\prime}} \gtrsim 770$ GeV e $M_{S_2,I_3}
\gtrsim 200$~GeV. Por outro lado, pelos vínculos impostos pelo modelo, o valor $M_{Z^{\prime}} \gtrsim 770$ GeV implica $v_{\chi^{\prime}} \gtrsim
1945$~GeV e consequentemente $M_{S_2},M_{I_3} \gtrsim 1376$~GeV. Ou
seja, os vínculos mais severos às massas dos escalares são
provenientes da relação existente elas e $M_{Z^\prime}$.

Medidas de precisão para diferença de massa do sistema $D^0-\bar{D^0}$ impõem que $M_{Z^{\prime}} \gtrsim 550$~GeV, conforme
mostra a Fig.~\ref{fig3}. Da Fig.~\ref{fig4}, temos
$M_{S_2},M_{I_3} \gtrsim 1$~GeV. Pela correlação existente entre as
massas desses bósons, $M_{Z^{\prime}} \gtrsim 550$~GeV nos leva ao
valor $M_{S_2},M_{I_3} \gtrsim 980$~GeV para a massa dos novos
escalares.

Finalmente, a partir do valor experimental de $\Delta m_{B_d}$, obtemos das
Figs.~\ref{fig6} e \ref{fig5} os vínculos $M_Z^{\prime} \gtrsim 4.2$ TeV
e $M_{S_2},M_{I_3} \gtrsim 230$ GeV. Este valor de massa do $Z^\prime$
implica que $v_\chi^{\prime} \gtrsim 10.6$ TeV e consequentemente
$M_{S_2},M_{I_3} \gtrsim 7.5$ TeV.

Nota-se da Fig.~\ref{fig6} que esse limite à massa do $Z^{\prime}$ é
ainda mais forte do que o imposto pelo LHC. Importante enfatizar que
acreditamos ser este o vínculo mais forte já obtido na literatura no
que diz respeito ao modelo $331_{RHN}$. Deste modo, inferimos que as
medidas de precisão das oscilações $B^0_d-\bar{B^0_d}$ descartam uma
grande região do espaço de parâmetros do modelo tornando a detecção do
$Z^{\prime}$ do modelo $331_{RHN}$ inviável na escala de energia do
LHC.

\chapter{Conclusões Gerais}

Vivemos uma época muito especial na Física das Partículas Elementares.
Com apenas dois anos de funcionamento e pouco mais da metade de sua energia de operação projetada, o LHC já
encontrou aquela que pode ser a última peça do quebra-cabeças para
consolidar de vez o MP das interações eletrofraca e forte.

Embora outros testes ainda sejam necessários, um excesso de eventos em
forma de ressonância em pares de fótons e no canal $ZZ^\ast \to 4$
léptons, revelou recentemente uma partícula (comprovadamente pelo
decaimento em pares de fóton) leve que possui algumas das características
do bóson de Higgs padrão.

Em clima científico empolgante e desafiador finalizamos essa tese. Se pela
perspectiva experimental podemos comemorar mais um sucesso do MP, do ponto de vista teórico ele ainda é visto com certo ceticismo, algo assaz salutar pois a incredulidade é essencial para o surgimento de novas idéias.

É exatamente sob este prisma que baseamos nosso trabalho. Ou
seja, acreditamos que o MP se trata apenas de uma teoria efetiva,
válida em escalas de energia da ordem de 1 TeV. Discutimos esse
conceito, e sua validade, guiando-nos por um exemplo
bem-sucedido: a teoria de Fermi.

Motivados pelo fato de que o setor de quebra espontânea de simetria do MP ainda não foi bem testado experimentalmente (apesar da
descoberta dessa partícula leve), nos baseamos em teorias
efetivas que descrevem esse setor, com ou sem a presença de um escalar com as características do Higgs padrão. 

Nessa abordagem efetiva, os vértices quárticos entre os bósons de \textit{gauge}
massivos, $WWWW$ e $WWZZ$, recebem uma contribuição extra, com estrutura de Lorentz e intensidade de acoplamentos distintos dos previstos pelo MP.

Embora exista uma rica literatura sobre o potencial do LHC em
testar esses vértices anômalos, geralmente esses trabalhos direcionam suas análises aos canais leptônicos em processos com topologia de VBF, que são os mais promissores para o estudo dos vértices quárticos. 

Aqui, considerando o cenário futuro onde o LHC opera com energia no CM de 14 TeV, fizemos um estudo detalhado
dos processos semileptônicos $pp\to l^{\pm}\nu_{l}\;\mbox{+ 4 jatos}$ onde $l^{\pm}$
são os léptons $e^{\pm}\;\mbox{e}\;\mu^{\pm}$, e $j$ são jatos
compostos por (anti)quarks leves, considerando os
\textit{backgrounds} do MP classificados como irredutíveis, QCD
em ${\cal O}(\alpha_{em}^{4}\alpha_{s}^{2})$ e ${\cal
  O}(\alpha_{em}^{2}\alpha_{s}^{4})$, e processos envolvendo quarks
\textit{top}: $pp\to t\bar{t} + 0\;\mbox{até}\;2\;\mbox{jatos}$.

Atingindo nesse cenário sua luminosidade nominal de 100 fb$^{-1}$, impomos vínculos 
aos acoplamentos anômalos $\alpha_4$ e $\alpha_5$, considerando evidência de 3$\sigma$ de significância estatística. 

Sabendo da influência grande nos processos, sobretudo nos
\textit{backgrounds} da QCD, das escolhas para as escalas de fatorização
e renormalização, investigamos os efeitos dos vínculos anômalos para
diferentes variações.
 
Embora inicialmente com cortes relativamente suaves haja predomínio
do \textit{background}, conseguimos definir um conjunto de cortes
relativamente duros e obtivemos dessa forma vínculos interessantes para os
acoplamentos anômalos, a saber $|\alpha_{4,5}| \lesssim 0.01$ a $0.02$ em 3$\sigma$, quando somente um operador anômalo atua por vez.

Até então, nossa teoria efetiva era o modelo $\sigma$ não-linear,
onde o bóson de Higgs está ausente ou, se ele existe, não
possui as características do Higgs padrão. Por outro lado, em uma realização linear, o bóson de Higgs padrão faz parte da
teoria efetiva e discutimos brevemente como os nossos vínculos podem
ser traduzidos para os acoplamentos anômalos supondo a presença do Higgs
padrão. Esses vínculos também são comparáveis aos existentes na
literatura.

Baseados nos resultados obtidos, inferimos que processos semileptônicos 
$pp\to l^{\pm}\nu_{l}\;\mbox{+ 4 jatos}$ são interessantes para sondar a estrutura dos vértices quárticos entre os bósons de
\textit{gauge}. Evidentemente, esses canais impõem um trabalho bem maior para controlar
os \textit{backgrounds} mas, acumulando 100
fb$^{-1}$ de dados, o LHC disponibilizará uma quantidade enorme de eventos na
região de controle, e portanto as incertezas devido à escolha
das escalas de fatorização e renormalização serão minimizadas.

Apesar de não ter sido foco nessa tese, vale comentar que as contribuições dos vértices tríplices não são menos importantes já que, como os acoplamentos
quárticos entre bósons de \emph{gauge} com massa ainda não foram
testados diretamente, especula-se que acoplamentos tríplices anômalos
influenciariam a medição dos acoplamentos quárticos do MP. Em outras
palavras, pode-se verificar se o sinal do anômalo tríplice e o sinal
do padrão quártico estão relacionados e deste modo medir o padrão
quártico com maior precisão. Tal situação não foi testada neste
trabalho, ficando para futura oportunidade, mas os acoplamentos tríplices padrão entram nos cálculos
como \emph{background} irredutível.

Em outro contexto, na contramão das teorias efetivas, exploramos a
fenomenologia de um modelo considerado como uma extensão do MP. Mais especificamente, analisamos o modelo com a
simetria do grupo $SU(3)_C \otimes SU(3)_L \otimes U(1)_X$, com
neutrinos de mão-direita, conhecido como modelo $331_{RHN}$. Este modelo prevê a existência de corrente neutra com a violação de sabor
(FCNC), fenômeno altamente suprimido na Natureza.

Nos modelos $331_{RHN}$, os processos FCNC ocorrem em nível de árvore,
devido a não-universalidade dos acoplamentos entre o
novo bóson de \textit{gauge} neutro ($Z^\prime$) previsto nesse modelo
e os quarks ordinários. Nesta tese, identificamos mais duas
outras fontes de FCNC, a saber as partículas escalares $S_2$ e $I_3^0$,
que possuem $CP$-par e $CP$-ímpar respectivamente.

Calculamos analiticamente as contribuições dessas novas partículas
para as diferenças de massa $\Delta M$, observável importante na medição das oscilações em sistemas de mésons neutros $K_{0}-\bar{K_{0}}$, $D^{0}-\bar{D^0}$ e $B^0_d-\bar{B^0_d}$, que permitem transições $\Delta F =2$, onde $F=S,C,B$
são os números quânticos de sabor. 

Considerando dados experimentais recentes de $\Delta M$ para os três
mésons neutros, impomos vínculos às massas do novo bóson de\textit{gauge}
neutro $Z^{\prime}$ e dos bósons escalares $S_2$ e $I_3$. As
oscilações $B^0_d-\bar{B^0_d}$ forneceram os vínculos mais restritivos
dentre os casos analisados já que, estando as massas desses
três bósons vinculadas pelo parâmetro $v_{\chi^{\prime}}$, o vínculo $M_Z^{\prime} \gtrsim 4.2$ TeV implica que a escala de
quebra de simetria do modelo $331_{RHN}$ será de $v_\chi^{\prime} \gtrsim
10.6$ TeV e ,consequentemente, a massa dos bósons escalares alcançaria $M_{S_2},M_{I_3} \gtrsim 7.5$ TeV. 

Esses resultados tornam a detecção desses bósons, tanto do $Z^\prime$ quanto dos escalares $S_2$ e $I_3^0$, inviável na escala de energia do LHC e acreditamos serem estes os vínculos mais fortes já obtidos na literatura no que diz respeito
ao modelo $331_{RHN}$.

Conforme discutimos na Introdução desta tese, se a Natureza escolheu
a simetria $SU(3)_C \otimes SU(3)_L \otimes U(1)_X$ com neutrinos de mão-direita para
representar a simetria de interação das partículas elementares, o
cenário mais provável para o LHC nos próximos anos será a opção (a), ou
seja nenhum indício de ressonâncias novas, tendo em vista que detetar esses bósons do modelo $331$ superar seu alcance tecnológico.

Enfatizamos que os acoplamentos anômalos puramente quárticos
entre os bósons de \textit{gauge} também podem estar nesse cenário,
caso a escala de massa das novas ressonâncias, que surgem para
amenizar o setor interagindo fortemente, seja suficientemente grande
[${\cal O}$(1 TeV)].

Finalmente, aguardamos com muitas expectativas o estabelecimento de
dois números ``mágicos'' no LHC: a energia no centro de massa de 14 TeV e
uma luminosidade integrada de 100 fb$^{-1}$.

\appendix

\chapter{Espaço de Fase, Seção de Choque e Integração por Monte Carlo}
\label{apendice1}
\section{Espaço de Fase e Seção de Choque}

Um processo de espalhamento do tipo~\footnote{No que se refere à cinemática e à dinâmica de um experimento podemos medir dois tipos de reações: \textit{exclusivas} (todas as partículas e seus respectivos momentos são conhecidos e portanto o estado final é bem estabelecido) e \textit{inclusivas} (somente algumas partículas e seus momentos são identificados e o estado final não é bem estabelecido. A colisão de duas partículas é um exemplo de processo \textit{exclusivo}.}
\begin{equation}
p_{a}+p_{b}\to p_{1} + p_{2} +\cdots + p_{n},
\label{reacao}
\end{equation}
onde $n$ é o número de partículas no estado final, têm como condição essencial a conservação de momento e energia de forma que, escrevendo explicitamente o quadrimomento $p_{j}=(E,\mathbf{p}_{j})$, deve-se impor o seguinte:
\begin{eqnarray}
E_{a}+E_{b}&=&\sum_{i=1}^{n}\;E_{i} \nonumber \\
& &  \\
\mathbf{p}_{a}+\mathbf{p}_{b}&=&\sum_{i=1}^{n}\;\mathbf{p}_{i} \nonumber
\label{conservacao}
\end{eqnarray}
com
\begin{equation}
E_{j}^{2}=\mathbf{p}_{j}^{2}+m_{j}^{2},
\end{equation}
onde $j=a,b,1,2,...,n$ e $m_{j}$ são as massas das partículas.

Sendo assim, considerando que o estado inicial seja fixo, os trimomentos $\mathbf{p}_{i}$ das $n$ partículas finais não podem variar arbitrariamente e formam o espaço dos momentos de dimensão $3n$.

O conjunto de condições~\eqref{conservacao} define no espaço dos momentos uma superfície de dimensão $3n-4$ denominada \textit{espaço de fase} cuja descrição verificaremos adiante. Deste modo o processo~\eqref{reacao} possui $3n-4$ variáveis independentes no estado final e o eixo do feixe (a direção do momento incidente $\mathbf{p}_{a}$) define uma direção no espaço.

A transição de probabilidade de um estado inicial $p_{a}+p_{b}$ para um estado final $p_{1} + p_{2} +\cdots + p_{n}$ é obtida através do elemento de matriz
\begin{equation}
\langle \mathbf{p}_{1},\cdots ,\mathbf{p}_{n}|{\cal M}|\mathbf{p}_{a},\mathbf{p}_{b}\rangle \equiv {\cal M}(\mathbf{p}_{i}).
\end{equation}

Para obtermos a taxa de transição, ou em outras palavras a seção de choque total $\sigma$, para um determinado processo precisamos integrar o elemento de matriz quadrado, $|{\cal M}(\mathbf{p}_{i})|^2$, no espaço de fase $3n-4$ dimensional. Restringindo a integração à um subconjunto desse espaço de fase obtém-se a seção de choque diferencial, $\dfrac{d\sigma}{dx}$. A distribução normalizada $w(x)$ correspondente à $\dfrac{d\sigma}{dx}$ é definida por $w(x)=\dfrac{1}{\sigma}\dfrac{d\sigma}{dx}$.

De forma geral, um ponto no espaço de fase pode ser descrito minimamente por um conjunto $\Phi$ de $3n-4$ variáveis que consistem em:
\begin{itemize}
\item $n-2$ massas invariantes, $M_{i}$, definidas como as massas das partículas intermediárias;
\item $2(n-1)$ ângulos $\theta_{i}$ e $\phi_{i}$ que especificam a direção das partículas.
\end{itemize}

Além disso, no caso de colisores hadrônicos, adicionamos ao conjunto $\Phi$ a fração de momento longitudinal dos partons incidentes, $x_{a}$ e $x_{b}$, nos hadrons que colidem. No nosso trabalho, onde lidamos com processos $2\to 6$ manipulamos portanto um espaço de fase de dimensão 16, ou seja com 16 variáveis independentes.

Nosso trabalho se baseia no LHC, um colisor próton-próton. Deste modo a seção de choque total $\sigma$ para um processo $pp\to X$, onde $X$ representa qualquer estado final, no LHC será a soma das seções de choque $\hat{\sigma}$ para todos os subprocessos relacionados.
\begin{equation}
\sigma = \sum_{\mbox{Sub}}\;\int\;dx_{a}\;dx_{b}\;\left[f_{a/A}(x_{a})f_{b/B}(x_{b})+(A\leftrightarrow B\;\mbox{se}\; a\neq b) \right]\;\hat{\sigma}(ab\to X)
\label{cs}
\end{equation}

No regime de altíssimas energias, quando as massas das partículas são desprezíveis se comparadas aos seus trimomentos, expressando os quadrimomentos dos partons incidentes como $a=x_{a}A$ e $b=x_{b}B$ obtém-se a relação
\begin{equation}
\hat{s}=x_{a}x_{b}s=\tau s
\end{equation}
onde $\sqrt{\hat{s}}$ é a massa invariante do sistema de partons $ab$ enquanto$\sqrt{s}$ é a massa invariante do sistema de prótons $AB$.

A variável $\tau =x_{a}x_{b}$ é bastante conveniente e sendo assim, mudando as variáveis independentes para $x_{a}$ e $\tau$, escreve-se~\eqref{cs} como
\begin{equation}
\sigma = \sum_{\mbox{Sub}}\;\int_{0}^{1}\;d\tau\;\int_{\tau}^{1}\;\dfrac{dx_{a}}{x_{a}}\;\left[f_{a/A}(x_{a})f_{b/B}(x_{b})+(A\leftrightarrow B\; \mbox{se}\; a\neq b) \right]\;\hat{\sigma}(\hat{s}).
\label{cs2}
\end{equation}

Para obter $\hat{\sigma}(\hat{s})$ na expressão~\eqref{cs2}, o elemento de matriz quadrado  $|{\cal M}(\mathbf{p}_{i})|^2$ de um dado subprocesso deve ser somado sobre os números quânticos do estado final (helicidade e/ou cor), tira-se a média sobre os mesmos números quânticos para o estado inicial (caso este se componha de partículas incidentes não polarizadas), e integra-se no espaço de fase correspondente às partículas iniciais e finais.

Traduzindo em linguagem formal matemática teremos
\begin{equation}
\sigma = \dfrac{1}{2\lambda^{1/2}(\hat{s},m_{a}^{2},m_{b}^{2})}\;I_{n}(\hat{s}),
\label{cs3}
\end{equation}
onde $\lambda (x,y,z)=x^{2}+y^{2}+z^{2}-2xy-2xz-2yz$ e
\begin{equation}
I_{n}(\hat{s})=\int\;d\Phi_{2\to n}\;\Theta (cortes)\;\bar{\sum}|{\cal M}(ab\to 1\cdots n)|^2
\end{equation}
sendo
\begin{equation}
d\Phi_{2\to n} = \prod_{i=1}^{n} \left(\dfrac{d^{3}p_{i}}{(2\pi^{3})2E_{i}} \right)\;2\pi^{4}\delta^{4}(p_{a}+p_{b}-\sum_{i}p_{i})
\label{phix}
\end{equation}
a expressão para cálculo do espaço de fase e o quadrado do elemento de matriz
\begin{equation}
\overline{\sum} |{\cal M}(ab\to 1\cdots n)|^2 = \dfrac{1}{4}\;C_{ab}\sum_{cor}\sum_{pol}|{\cal M}|^{2},
\end{equation}
onde o fator $1/4$ provém dos partons incidentes $a$ e $b$ serem partons que possuem duas polarizações (transversal e longitudinal) e $C_{ab}$ representa o fator de cor para quarks e gluons cujos valores são $C_{qq}=C_{q\bar{q}}=\dfrac{1}{9}$, $C_{qg}=\dfrac{1}{24}$,  $C_{gg}=\dfrac{1}{64}$.

A condição de conservação do quadrimomento~\eqref{conservacao} está incluída na função $\delta$ quadridimensional, que se trata de um produto de quatro funções $\delta$ correspondentes às quatro componentes de $p^{\mu}$.

Para o estado final de duas partículas~\cite{barger_collider} temos a expressão analítica
\begin{equation}
\Phi_{2\to 2} = \int\;\delta^{4}(P-p_{1}-p_{2})\dfrac{d^{3}p_{1}}{2E_{1}}\dfrac{d^{3}p_{2}}{2E_{2}}=\dfrac{\lambda^{1/2}(P^{2},p_{1}^{2},p_{2}^{2})d\Omega}{8P^{2}}
\label{phix2}
\end{equation}
onde $d\Omega$ é o elemento de ângulo sólido no referencial do CM e $P^{2},p^{2}_{i}$ são as massas invariantes.

Uma escolha apropriada de variáveis de integração para tratar o espaço de fase no cálculo das seções de choque de processos envolvendo singularidades colineares em colisores elétron-pósitron foi indicada por Hagiwara et al.~\cite{hagiwara}. Esta escolha se mostrou útil em outros trabalhos~\cite{tese_jkm} e por isso optamos por usá-la em nossa análise. Abaixo descrevemos o procedimento.

\subsection{Espaço de fase com jatos \textit{tagging}}
\label{subsec: eftagging}

Em~\cite{hagiwara} os autores tratam processos de produção de um único bóson de \textit{gauge} em colisores $e^{-}e^{+}$ na escala TeV. Fazendo uma analogia para o nosso caso temos
\begin{equation}
q1(p_{1})+q2(p_{2})\to q5(p_{5})+q8(p_{8})+X(p_{X}),
\end{equation}
onde $q1, q2$ são os partons incidentes, $q5,q8$ são os partons \textit{tagging} e $X$ é um produto final, sem necessariamente possuir uma massa invariante fixa, que em nosso caso será o processo de fusão dos bósons vetoriais $V^{*}V^{*}\to VV\to l^{\pm}\nu_{l}jj$, com $V=W^{\pm},Z$ descrito por
\begin{equation}
V(p_{6})+V(p_{7})\to l^{\pm}(p_{6a})\nu_{l}(p_{6b})+q7a(p_{7a})+q7b(p_{7b}).
\end{equation}
\begin{figure}
\centering 
\includegraphics[scale=0.75]{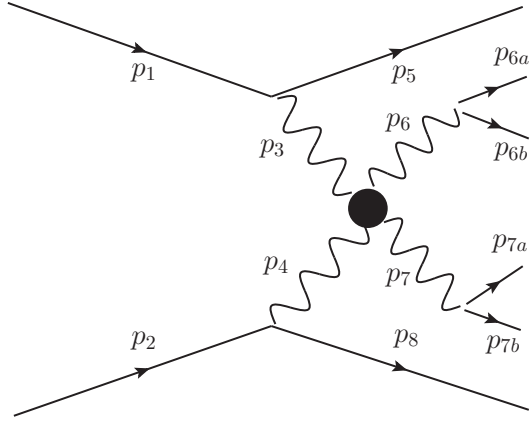} 
\caption{Esquema de Fusão de Bósons Vetoriais (VBF) usado em nossos cálculos.} 
\label{fig:vbf} 
\end{figure} 

De forma explícita na Fig.~\eqref{fig:vbf} tratamos numericamente o processo
\begin{equation}
q1(p_{1})+q2(p_{2})\to l^{\pm}(p_{6a})\nu_{l}(p_{6b})+q7a(p_{7a})+q7b(p_{7b})+ q5(p_{5})+q8(p_{8})
\end{equation}
sendo o espaço de fase invariante escrito de acordo com Hagiwara~\cite{hagiwara} como
\begin{equation}
d\Phi = (2\pi)^{4}\delta^{4}(p_{1}+p_{2}-p_{5}-p_{8}-\sum_{i\in X}p_{i})\dfrac{d^{3}p_{5}}{(2\pi)^{3}2p_{5}^{0}}\;\dfrac{d^{3}p_{8}}{(2\pi)^{3}2p_{8}^{0}}\;\prod_{i\in X}\dfrac{d^{3}p_{i}}{(2\pi)^{3}2p_{i}^{0}}
\end{equation}
podendo ser reestruturado como
\begin{equation}
d\Phi = \dfrac{d^{3}p_{5}}{(2\pi)^{3}2p_{5}^{0}}\;\dfrac{d^{3}p_{8}}{(2\pi)^{3}2p_{8}^{0}}\;\delta((p_{1}+p_{2}-p_{5}-p_{8})^{2}-M_{67}^{2})\;dM_{67}^{2}\;d\Phi_{X}
\label{phi4}
\end{equation}
onde $M_{67}^{2}$ é a massa invariante do sistema de bósons de \textit{gauge}, oriundos do processo de VBF, com
\begin{equation}
d\Phi_{X} = (2\pi)^{4}\delta^{4}(p_{1}+p_{2}-p_{5}-p_{8}-\sum_{i\in X}p_{i})\;\prod_{i\in X}\dfrac{d^{3}p_{i}}{(2\pi)^{3}2p_{i}^{0}}
\label{phix3}
\end{equation}
que é análogo à expressão \eqref{phix} e pode ser calculada com a ajuda do resultado \eqref{phix2}.

Como nossa análise envolve ressonâncias no canal $s$, devido aos decaimentos dos bósons de \textit{gauge}, precisamos tomar cuidado com o fator de Breit-Wigner quando integramos na massa invariante $M_{VV}$ do sistema composto por esses bósons. Para evitar singularidades no integrando podemos absorver o denominador do propagador do bóson de \textit{gauge} de massa $M$ e largura $\Gamma$ fazendo a seguinte mudança de variáveis:
\begin{equation}
M_{67}^{2}-M^{2}=M\Gamma \;\tan{\vartheta} \Longrightarrow dM_{67}^{2}=M\Gamma\;\sec^{2}{\theta}d\vartheta = M\Gamma (1+\tan^{2}{\vartheta}) d\vartheta
\end{equation}
de onde resulta que
\begin{equation}
\int\; \dfrac{dM_{67}^{2}}{ (M_{67}^{2}-M^{2})^{2}+(M\Gamma )^{2} }= \int\;\dfrac{d\vartheta}{M\Gamma}
\label{breit}
\end{equation}
e assim integramos na nova variável $\vartheta$ que adaptamos ao limite de integração $(0,1)$ do VEGAS~\cite{vegas}.

Para os cálculos numéricos os quadrimomentos dos partons incidentes e dos \textit{taggings} no laboratório foram parametrizados da seguinte forma:
\begin{eqnarray}
p_{1}^{\mu}&=&\dfrac{\sqrt{\hat{s}}}{2}(1,0,0,\beta) \nonumber \\
& & \nonumber \\
p_{2}^{\mu}&=&\dfrac{\sqrt{\hat{s}}}{2}(1,0,0,-\beta) \nonumber \\
& & \nonumber \\
p_{5}^{\nu}&=&\dfrac{\sqrt{\hat{s}}}{2}x_{5}(1,\beta_{5}\sin{\theta_{5}}\cos{\phi_{5}},\beta_{5}\sin{\theta_{5}}\sin{\phi_{5}},\beta_{5}\cos{\theta_{5}}) \nonumber \\
& & \nonumber \\
p_{8}^{\nu}&=&\dfrac{\sqrt{\hat{s}}}{2}x_{8}(1,\beta_{8}\sin{\theta_{8}}\cos{\phi_{8}},\beta_{8}\sin{\theta_{8}}\sin{\phi_{8}},\beta_{8}\cos{\theta_{8}}) \nonumber \\
& &
\label{parameter}
\end{eqnarray}
onde
\begin{equation}
\beta = \left(1-4\dfrac{m_{j}^{2}}{\hat{s}}\right)^{\dfrac{1}{2}},
\end{equation}
onde $m_{j}$ é a massa do parton incidente e
\begin{equation}
\beta_{i}=\left(1-4\dfrac{m_{i}^{2}}{\hat{s}x_{i}^{2}}\right)^{\dfrac{1}{2}}
\end{equation}
onde $m_{i}$ é a massa do parton \textit{tagging}.

Os elementos de volume dos partons \textit{tagging} no espaço de fase~\eqref{phi4} foram parametrizados no referencial do laboratório~\eqref{parameter}, usando o limite $\dfrac{m_{i}^{0}}{\hat{s}}\to 0$, como
\begin{eqnarray}
\dfrac{d^{3}p}{(2\pi)^{3}2E}&=&\dfrac{\mathbf{p}^{2}dp\;\sin{\theta}d\theta\; d\phi}{(2\pi)^{3}2E} \nonumber \\
& & \nonumber \\
&=& \dfrac{(E^{2}-m_{i}^{2})dp}{(2\pi)^{3}2E}\;d\cos{\theta} d\phi \nonumber \\
& & \nonumber \\
&\stackrel{m_{i}\to 0}{\rightarrow}& \dfrac{Edp}{2(2\pi)^{3}}\;d\cos{\theta} d\phi \nonumber \\
& & \nonumber \\
&=&\dfrac{1}{2}\dfrac{\sqrt{\hat{s}}}{2}\;x\;\dfrac{\sqrt{\hat{s}}}{2(2\pi)^{3}}\;dx\;d\cos{\theta} d\phi \nonumber \\
& & \nonumber \\
&=&\dfrac{\hat{s}}{(4\pi)^{3}}\;x\;dx\;d\cos{\theta} d\phi
\end{eqnarray}
de onde resulta
\begin{equation}
d\Phi = \dfrac{\hat{s}^{2}}{(4\pi)^{6}}\delta((p_{1}+p_{2}-p_{5}-p_{8})^{2}-M_{VV}^{2})x_{5}x_{8}dx_{5}dx_{8}d\cos{\theta_{5}}d\cos{\theta_{8}}d\phi_{5}d\phi_{8}d\hat{s}d\Phi_{X}.
\label{phi5}
\end{equation}

Resolvendo a função $\delta$ para $x_{8}$ em~\eqref{phi5} obtém-se
\begin{equation}
d\Phi = \dfrac{\hat{s}}{(4\pi)^{6}}\dfrac{x_{5}(1-x_{5}-M_{67}^{2}/\hat{s})}{[1-x_{5}(1-\cos{\theta_{58}})/2]^{2}}dx_{5}d\cos{\theta_{5}}d\cos{\theta_{8}}d\phi_{5}d\phi_{8}dM_{67}d\Phi_{X}.
\label{phi6}
\end{equation}

O ângulo $\theta_{58}$ é o ângulo entre os partons \textit{tagging} no referencial do laboratório~\eqref{parameter}.

As singularidades colineares em $\cos{\theta_{i}}=\pm 1$ podem ser contornadas trabalhando com a variável pseudorapidez $\eta$ devidamente regulada por um fator que depende do processo. Vejamos como foi feito em~\cite{hagiwara}. Em termos do momento, a pseudorapidez $\eta$ de uma partícula é definida como
\begin{equation}
\eta = \dfrac{1}{2}\ln{\dfrac{|\mathbf{p}|+p_{L}}{|\mathbf{p}|-p_{L}}}
\label{eta}
\end{equation}
onde $p_{L}$ é o momento longitudinal (momento ao longo da direção do feixe, usualmente definido como o eixo $z$) da partícula. Por outro lado $\cos{\theta}= p_{z}/\sqrt{p_{x}^{2} + p_{y}^{2} + p_{z}^{2}}$, expressão que nos permite escrever \eqref{eta} em função de $\cos{\theta}$
\begin{equation}
\eta = \dfrac{1}{2}\ln{\left(\dfrac{1+\cos{\theta}}{1-\cos{\theta}}\right)}
\end{equation}

Hagiwara et al.~\cite{hagiwara} introduzem um regulador $\Delta$ nessa expressão, definindo outra variável denominada $y$, tal que~\footnote{Ver expressão B.13 em \cite{hagiwara}} para cada parton \textit{tagging} $i=5,8$ temos
\begin{equation}
y_{i} = \dfrac{1}{2}\ln{\left(\dfrac{1+\cos{\theta_{i}}+2\Delta_{i}^{+}}{1-\cos{\theta}+2\Delta_{i}^{-}}\right)}.
\end{equation}

No nosso caso, $\Delta_{i}^{\pm}={\cal O}(m_{V}^{2}/\hat{s})$ onde $m_{V}$ é a massa do bóson de \textit{gauge} irradiado pelo parton \textit{tagging} $i$ no momento do espalhamento.

Finalmente, com todas essas modificações visando aprimorar os cálculos numéricos, o espaço de fase que usamos tem a forma
\begin{eqnarray}
d\Phi &=& \dfrac{\hat{s}}{(4\pi)^{6}}\dfrac{x_{5}(1-x_{5}-M_{67}^{2}/\hat{s})}{[1-x_{5}(1-\cos{\theta_{58}})/2]^{2}} \nonumber \\
& & \nonumber \\
& &\times \dfrac{1+\Delta_{5}^{+}+\Delta_{5}^{-}}{\cosh^{2}y_{5}}\dfrac{1+\Delta_{8}^{+}+\Delta_{8}^{-}}{\cosh^{2}y_{8}}dx_{5}dy_{5}dy_{8}d\phi_{5} d\phi_{8} dM_{67}^{2}d\Phi_{X}.
\label{phifinal}
\end{eqnarray}
com
\begin{eqnarray}
\cos{\theta_{i}}&=&(1+\Delta_{i}^{+}+\Delta_{i}^{-})\tanh(y_{i})-\Delta_{i}^{+}+\Delta_{i}^{-},\;\;i=5,8, \\
\phi &=& \phi_{5}-\phi_{8}, \\
\cos{\theta_{58}}&=&\cos{\theta_{5}}\cos{\theta_{8}}-\sin{\theta_{5}}\sin{\theta_{8}}\cos{\phi}, \\
x_{8}&=&\dfrac{1-x_{5}-M_{67}^{2}/\hat{s}}{1-x_{5}(1-\cos{\theta_{58}})/2}.
\end{eqnarray}

A região de integração e as escolhas adequadas para as variáveis $\Delta_{i}$ e para a função $f(x_{5})$, que parametrizam a intergral de $x_{5}$, no nosso caso foram as seguintes\footnote{Ver Apêndice B em~\cite{hagiwara}}:
\begin{eqnarray}
\left(m_{6a}+m_{6b}+m_{7a}+m_{7b}\right)^{2}&< M_{67}^{2}<& (\sqrt{\hat{s}}-m_{5}-m_{8})^{2}, \\
& & \nonumber\\
\dfrac{2m_{5}}{\sqrt{\hat{s}}}&<x_{5}<&1+\dfrac{m_{5}^{2}}{\hat{s}}-\dfrac{(\sqrt{M_{67}^{2}}+m_{8}^{2})^{2}}{\hat{s}}, \\
& & \nonumber\\
&x_{8}>&\dfrac{2m_{8}}{\sqrt{s}} \\
& & \nonumber \\
-\dfrac{1}{2}\ln{\dfrac{1+\Delta_{i}^{-}}{\Delta_{i}^{+}}}&<y_{i}<&\dfrac{1}{2}\ln{\dfrac{1+\Delta_{i}^{+}}{\Delta_{i}^{-}}}\;\;\mbox{para}\;\;i=5,8, \\
& & \nonumber \\
0&<\phi_{5},\phi<&2\pi \\
& & \nonumber \\
dx_{5}&=&\dfrac{1}{|f'(x_{5})|}df(x_{5}) \\
& & \nonumber \\
\Delta_{5}^{-}&=&\Delta_{8}^{+}=\dfrac{m_{V}^{2}}{\hat{s}} \\
& & \nonumber \\
\Delta_{5}^{+}&=&\Delta_{8}^{-}=1 \\
& & \nonumber \\
f(x_{5})&=&\ln(1-x_{5})
\end{eqnarray}

Em suma, o espaço de fase apresentado por Hagiwara~\cite{hagiwara}, e adaptado à nossa análise com fusão de bósons vetoriais, se mostrou bastante eficiente quando aplicado à intergração numérica pelo método de Monte Carlo no processo leptônico e o aplicamos convenientemente em nossos cálculos~\cite{tese_jkm}.

\subsection{Espaço de fase tradicional}
\label{subsec:eftradicional}

A Eq.~\eqref{phix} nos mostra que o espaço de fase $\Phi_{2\to n}$  para um processo de espalhamento com $n$ partículas no estado final é dado por
\begin{equation}
\Phi_{2\to n} = \int\;\delta^{4}(p_{a}+p_{b}-\sum_{i}p_{i})\;\prod_{i=1}^{n} \left(\dfrac{d^{3}p_{i}}{(2\pi^{3})2E_{i}} \right).
\end{equation}

\begin{figure}
\begin{center} 
\includegraphics[scale=0.75]{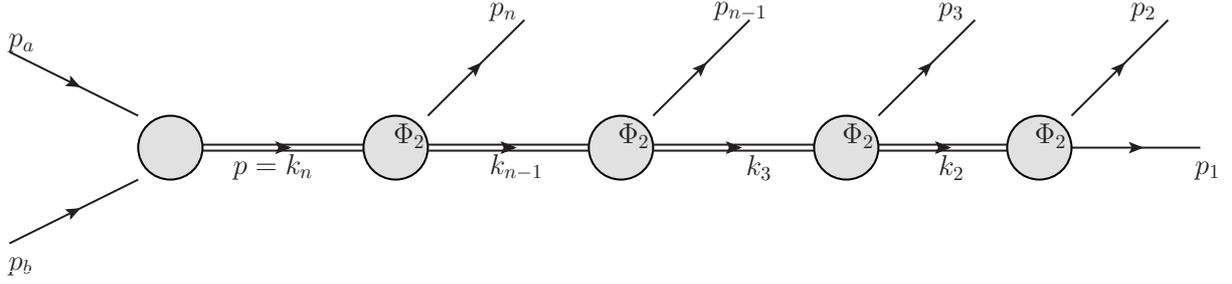} 
\end{center}
\caption{Processo $p_{a}+p_{b}\to p_{1} + p_{2} +\cdots + p_{n}$ como uma sequência de produção e decaimento em duas partículas} 
\label{recursao} 
\end{figure}

De forma geral, tal processo pode ser visualizado através de uma topologia que envolve produção de particulas instáveis e seus decaimentos em duas partículas. Sendo assim, um estado final com $n$ partículas sempre poderá ser subdividido em processos mais simples através de uma relação de recursão~\cite[Eq. VI.2.3]{kajantie}
\begin{eqnarray}
\Phi_{2\to n}(p) &=& \int\;\dfrac{d^{3}p_{n}}{(2\pi^{3})2E_{n}}\;\prod_{i=1}^{n-1}\;\dfrac{d^{3}p_{i}}{(2\pi^{3})2E_{i}}\;\delta^{4}\left\{(p-p_{n})-\sum_{i}^{n-1}\;p_{i}\right\} \nonumber \\
&=&   \int\;\dfrac{d^{3}p_{n}}{(2\pi^{3})2E_{n}}\;\Phi_{n-1}(p-p_{n}),
\label{recursao2}
\end{eqnarray}
baseada no esquema da Fig.~\eqref{recursao}.

Após algumas manipulações algébricas~\cite[Cap. VI Seção 2]{kajantie}, e usando a Eq.~\eqref{phix2} para $n=2$, a Eq.~\eqref{recursao2} retorna
\begin{eqnarray}
\Phi_{2\to n}(M_{n}^{2}) &=& \int_{\mu_{n}^{2}-1}^{(M_{n}-m_{n})^{2}}\;dM_{n-1}^{2}\;\Phi_{2\to 2}(k_{n};k_{n-1}^{2},p_{n}^{2})\Phi_{2\to n-1}(M_{n-1}^{2}) \nonumber \\
&=&  \int_{\mu_{n}^{2}-1}^{(M_{n}-m_{n})^{2}}\;dM_{n-1}^{2}\;\int d\Omega_{n-1}\dfrac{\lambda^{\dfrac{1}{2}}(M_{n}^{2},M_{n-1}^{2},m_{n}^{2})}{8M_{n}^{2}}\Phi_{2\to n-1}(M_{n-1}^{2}),
\label{recursao3}
\end{eqnarray}
onde foram definidos os quadrivetores
\begin{equation}
k_{i}=p_{1}+p_{2}+\cdots + p_{i}\;\;\;\mbox{sendo}\;\;\;k_{n}=p,
\end{equation}
e a massa invariante do sistema formado pelas $1,\cdots,n-1$ partículas será dada por
\begin{eqnarray}
M_{n-1}^{2}&=&(p-p_{n})^{2} \nonumber \\
&=& (p_{1}+p_{2}+\cdots +p_{n-1})^{2} \nonumber \\
&\sim & k^{2}_{n-1},
\end{eqnarray}
com limites, definindo $\mu_{i}=m_{1}+m_{2}+\cdots + m_{i}$, tais que\footnote{$\Phi_{2\to n-1}$ é nulo para valores abaixo de $M_{n-1}=\mu_{n-1}$ e $\Phi_{2\to 2}$ só é definida se $M_{n}$ for tal que $M_{n} \geq M_{n-1} + m_{n}$ }
\begin{equation}
\mu_{n-1}\leq M_{n-1} \leq M_{n}-m_{n}.
\end{equation}

Em suma a Eq.~\eqref{recursao3} fornece o espaço de fase tradicional para $n$ partículas, $\Phi_{2\to n}$, como um produto entre o espaço de fase de duas partículas $\Phi_{2\to 2}$ (que descreve o decaimento $p\to p_{n}+k_{n-1}$) e $\Phi_{2\to n-1}$ (que descreve o decaimento $k_{n-1}\to p_{1}+p_{2}+\cdots +p_{n-1}$) integrado sobre todos os valores possíveis da massa invariante $M_{n-1}$.

\section{Integração pelo Método de Monte Carlo}
\label{sec:montecarlo}

Como relatado na seção anterior, o processo~\eqref{reacao} possui um conjunto $\Phi$ de $3n-4$ variáveis independentes no estado final e, no caso de colisores hadrônicos, adicionamos a esse conjunto a fração de momento longitudinal dos hadrons incidentes, $x_{a}$ e $x_{b}$, carregada por cada parton na colisão.

Deste modo nosso trabalho, lidando com processos $2\to 6$, manipula um espaço de fase de dimensão 16, ou seja com 16 variáveis independentes.

Métodos de integração tradicionais baseados em regras de quadratura numérica são eficientes no cálculo de integrais unidimensionais. No entanto, a eficiência desses métodos não é satisfatória quando os aplicamos às integrais multidimensionais. De fato, suponha uma integral de uma função $d-$dimensional $f(x_{1},\cdots,x_{d})$ avaliada no hipercubo $[0,1]^{d}$. Esta integral pode ser abordada como uma interação de $d$ integrais unidimensionais calculadas por exemplo pela regra dos trapézios\footnote{A regra dos trapézios é o exemplo mais simples dentre as fórmulas fechadas de Newton-Cotes para integração numérica baseadas na substituição da função $f(x)$ por um polinômio que a aproxime razoavelmente no intervalo $[a,b]$. De acordo com essa regra, para estimar uma integral no intervalo finito $[a,b]$ dividimos esse intervalo em $n$ sub-intervalos de comprimento $\Delta x$ e aplicamos a regra dos trapézios para cada sub-intervalo. Um detalhe importante é que o erro neste cálculo será proporcional à $1/n^{2}$ e a função $f(x)$ deverá ser calculada $n$ vezes dentro do intervalo $[a,b]$.}. Esse cálculo supõe então avaliar a função aproximadamente $N=n^{d}$ vezes, fator esse que influencia proporcionalmente o tempo estimado assim como o erro no cálculo (que aumenta de acordo com a dimensão da integral na ordem de $(1/N^{2/d})$). Esses são pontos importantes onde a integração pelos métodos de Monte Carlo atua de modo bastante satisfatório, já que desvincula a estimativa de erro da dimensão da integral, como veremos adiante.

De modo simples uma simulação de Monte Carlo pode ser ilustrada no cálculo da integral de uma função definida em um intervalo finito. Enquanto outros métodos de integração numérica utilizam aproximações polinomiais, a integração por Monte Carlo escolhe aleatoriamente pontos no intervalo em que se deseja obter a integral. Suponha a função $g(x)$ definida e contínua no intervalo $[\alpha,\beta]$ e queremos avaliar a integral unidimensional
\begin{equation}
I=\int_{\alpha}^{\beta}\;g(x)dx
\label{integral}
\end{equation}
por Monte Carlo. Para isso podemos manipular a integral adotando uma função de importância que facilite a abordagem probabilística~\footnote{Mais adiante veremos que esse procedimento é a base da técnica de amostragem por importância.}. Por exemplo podemos usar uma distribuição contínua de probabilidade (ou função densidade de probabilidade -- fdp) no intervalo $[\alpha,\beta]$ tal que:
\begin{equation}
f(x)=\left\{ \begin{array}{ll}
              \dfrac{1}{\beta - \alpha} & \mbox{com $\alpha \leq x \leq \beta$, $\alpha,\beta \in R$, $\alpha < \beta$} \\
              0 & \mbox{outros valores}.
              \end{array}
      \right.
\label{fdp}
\end{equation}

Lembramos que se $X$ for uma variável aleatória contínua que possui todos os seus valores no intervalo $[\alpha,\beta]$ e a função densidade de probabilidade de $X$ for dada por~\eqref{fdp} então $X$ é uniformemente distribuida sobre o intervalo $[\alpha, \beta]$ e o \textit{valor esperado} (ou \textit{valor médio}) de $X$ nesse intervalo será dado por
\begin{equation}
E(X)=\int_{\alpha}^{\beta}\;xf(x)dx.
\label{ve}
\end{equation}

Voltando à avaliação da integral de $g(x)$, como a \textit{fdp}~\eqref{fdp} é constante em $[\alpha,\beta]$ podemos reescrever~\eqref{integral} da seguinte forma:
\begin{eqnarray}
I &=& \dfrac{1}{f(x)}\int_{\alpha}^{\beta}\;g(x)f(x)dx \nonumber \\
& & \nonumber \\
 &=& (\beta-\alpha)\int_{\alpha}^{\beta}\;g(x)\dfrac{1}{\beta-\alpha}dx
\label{integral2}
\end{eqnarray}

Usando~\eqref{ve} em~\eqref{integral2} obtemos
\begin{equation}
I=(\beta-\alpha)E(g(x)) \Longrightarrow I\approx \hat{I}=(\beta-\alpha)\dfrac{1}{n}\sum_{i=1}^{n}g(x_{i})
\label{integral3}
\end{equation}
ou seja podemos estimar~\footnote{Importante ressaltar que a aproximação feita em~\eqref{integral3} está relacionada à Lei dos Grandes Números cuja idéia básica é que a frequência relativa $f_{A}$ associada a $n$ repetições de um experimento será próxima da probabilidade $P(A)$ se $f_{A}$ for baseada em um grande número de repetições. Em outras palavras, considere uma variável aleatória $X$ e sejam $(x_{1},\cdots,x_{i},x_{i+1}.\cdots,x_{n})$ os valores obtidos quando o experimento que origina $X$ for realizado $n$ vezes independentemente. A média aritmética $\bar{x}$ desses $n$ números se aproxima de $E(X)$ se $n$ for suficientemente grande. Em geral, um alto número de amostras fornece uma aproximação precisa da do valor médio e da variância, mas eleva o custo computacional.} a integral $I$ coletando $n$ amostras $(x_{1},\cdots,x_{i},x_{i+1}.\cdots,x_{n})$ da variável aleatória $X$ (ou $f(x)$), calculando $g(x_{i})$ para cada $i$ ($1\leq i \leq n$), tirando a média desses valores e multiplicando pelo comprimento do intervalo $(\beta-\alpha)$. Dada uma noção do que é feito no caso unidimensional passemos ao que nos interessa.

Seja a função $d-$dimensional $g(x_{1},\cdots,x_{d})$, quadrado-integrável~\footnote{Se a função $g(x)$ for quadrado-integrável então o erro do método de Monte Carlo pode ser quantificado pela sua variância. Caso contrário a estimativa de erro não é confiável.}, a ser avaliada no hipercubo unitário $[0,1]^{d}$. Extendendo o raciocínio anterior, podemos portanto estimar o valor da integral $d-$dimensional
\begin{equation}
I=\int_{0}^{1}\;d^{d}x\;g(x_{1},\cdots,x_{d})
\end{equation}
pela média da soma de $N$ valores aleatórios do integrando (note que nosso intervalo é unitário), ou seja
\begin{equation}
I\approx \hat{I}=\dfrac{1}{N}\sum_{i=1}^{N}g(x_{1}(i),\cdots,x_{d}(i))
\end{equation}
pois pela Lei dos Grandes Números quando $N\to \infty$ teremos
\begin{equation}
\lim_{N\to \infty}\;\hat{I}\;=\;I.
\end{equation}

Note que agora trabalhamos com um conjunto $X=(X_{1},\cdots,X_{d})$ com $d$ variáveis aleatórias independentes $X_{j}$, $j=1,\cdot,d$. Em particular, no nosso trabalho $d=3n-4=14$ mais duas variáveis referentes às frações de momento longitudinal dos hadrons incidentes totalizando $d=16$ variáveis aleatórias independentes.

Para estimar o erro na coleta das $N$ amostras de cada variável aleatória $X_{j}$ introduzimos a variância~\footnote{$\sigma^{2}(g)=\int\;dx(g(x)-I)^{2}$} de uma função $g(x)$ tal que para cada variável aleatória $X_{j}$ temos a variância
\begin{equation}
\int\;dx_{j}(1)\cdots\;\int\;dx_{j}(N)\;(\dfrac{1}{N}\sum_{i=1}^{N}g(x_{j}(i)-I)^{2} = \dfrac{\sigma^{2}(g_{j})}{N}\equiv \sigma_{N}^{2}.
\end{equation}

Isso implica que a variância do erro é inversamente proporcional ao tamanho da amostra $N$ e sendo assim podemos construir intervalos de confiança para a estimativa $\hat{I}$ tal que
\begin{equation}
\hat{I}-a\sqrt{\sigma_{N}^{2}}\leq I \leq \hat{I}+b\sqrt{\sigma_{N}^{2}}.
\end{equation}

Na prática o valor da variância não pode ser obtido de forma exata e por isso é aproximado pelo cálculo da \textit{variância amostral}~\cite{weinzierl}
\begin{equation}
\sigma_{N}^{2}\approx S^{2}=\dfrac{1}{N-1}\sum_{i=1}^{N}(g(x_{j}(i))- \hat{I})^{2}=\dfrac{1}{N}\sum_{i=1}^{N}(g(x_{j}(i))^{2}- \hat{I}^{2}
\end{equation}

Como o erro nas estimativas varia com $1/\sqrt{N}$ então para que ele seja o menos possível é preciso aumentar o número de amostras. Entretanto, quanto maior a amostra, maior será o custo computacional, lentidão na convergência e consequentemente a inviabilização dos cálculos.

Para obter um Monte Carlo eficiente sem aumentar o número de chamadas (amostras) existem algumas \textit{técnicas de redução de variância}. Dentre elas a \textit{amostragem por importância}~\footnote{Em inglês \textit{importance sampling}} utiliza uma função $d-$dimensional $f(x_{1},\cdots,x_{d})$, denominada \textit{função de importância}, para modificar adequadamente as variáveis de integração:
\begin{equation}
\int_{0}^{1}\;d^{d}x\;g(x_{1},\cdots,x_{d})=\int_{0}^{1}\;\dfrac{g(x_{1},\cdots,x_{d})}{f(x_{1},\cdots,x_{d})}\;f(x_{1},\cdots,x_{d})\;d^{d}x = \int_{0}^{1}\;\dfrac{g(x_{1},\cdots,x_{d})}{f(x_{1},\cdots,x_{d})}\;d^{d}F(x_{1},\cdots,x_{d})
\end{equation}
com o fator Jacobiano
\begin{equation}
f(x_{1},\cdots,x_{d}) = \dfrac{\partial^{d}F(x_{1},\cdots,x_{d})}{\partial x_{1}\cdots \partial x_{d}}.
\end{equation}

Se  $f(x_{1},\cdots,x_{d}) \geq 0$ e $\int f(x_{1},\cdots,x_{d})d^{d}x=1 $ então a função  $f(x_{1},\cdots,x_{d})$ pode ser interpretada como uma função densidade de probabilidade. Deste modo pode-se estimar a integral $I$ sorteando $N$ valores aleatórios distribuídos de acordo com a densidade de probabilidade $F(x_{1},\cdots,x_{d})$ tal que
\begin{equation}
I\approx \hat{I}=\dfrac{1}{N}\;\sum_{i=1}^{N}\dfrac{g(x_{1}(i),\cdots,x_{d}(i))}{f(x_{1}(i),\cdots,x_{d}(i))}
\end{equation}
onde $x_{j}(i)$ é a $i-$ésima amostra da variável aleatória $x_{j}$ distribuída no intervalo $[0,1]$.

O erro estatístico será dado por
\begin{equation}
\sigma \left( \dfrac{g}{f} \right)
\end{equation}
sendo a variância estimada por
\begin{equation}
S^{2}(\dfrac{g}{f})=\dfrac{1}{N}\sum_{i=1}^{N}\left(\dfrac{g(x_{j}(i))}{f(x_{j}(i))} \right)^{2}- \hat{I}^{2}.
\end{equation}

Como agora nosso integrando tem a forma $g/f$ então é preciso estar atento ao seu comportamento dentro da região de integração. Essa atenção necessária é automatizada pelo algoritmo VEGAS~\cite{vegas} que, usando as idéias básicas da amostragem por importância e da amostragem estratificada~\footnote{Amostragem Estratificada é outra técnica de redução de variância que consiste em dividir a região de integração em subregiões, integrar por Monte Carlo em cada subregião e somar os resultados parciais no final.}, grava as regiões onde o integrando possui valores maiores e concentra a integração nessas regiões.

Usando amostragem estratificada o VEGAS inicia subdividindo o espaço de integração em um \textit{grid} retangular e integra em cada subespaço desse \textit{grid}. Esses resultados ajustarão o \textit{grid} para a próxima interação, sendo esse novo \textit{grid} aquele onde o integrando possui maiores valores em magnitude. Então, usando amostragem por importância, o VEGAS tenta aproximar a função densidade de probabilidade $f(x)$
\begin{equation}
f(x_{1},\cdots,x_{d})=\dfrac{|g(x_{1},\cdots,x_{d})|}{\int d^{d}x  |g(x_{1},\cdots,x_{d})|}
\end{equation}
por uma função degrau (analogamente ao feito na escolha de \eqref{fdp}).

No VEGAS existem duas fases para o calculo: uma fase exploratória, quando o código ajusta o \textit{grid}, e a fase de cálculo propriamente dita. Após algumas interações (o número de interaçõe adequado depende de vários fatores que serão expostos mais adiante) na primeira fase, o \textit{grid} é ajustado e na segunda fase a integral será calculada com a precisão fornecida pelo \textit{grid} escolhido.

Em ambas as fases, cada interação $k$, com suas $N_{k}$ avaliações, gera uma estimativa $\hat{I_{k}}$ para a integral $I$ além de uma estimativa para variância $S_{k}^{2}$:
\begin{eqnarray}
\hat{I}_{k}&=&\dfrac{1}{N_{k}}\;\sum_{i=1}^{N_{k}}\dfrac{g(x_{1}(i),\cdots,x_{d}(i))}{f(x_{1}(i),\cdots,x_{d}(i))} \\
& & \nonumber \\
S^{2}_{k}(\dfrac{g}{f})&=&\dfrac{1}{N_{k}}\sum_{i=1}^{N_{k}}\left(\dfrac{g(x_{j}(i))}{f(x_{j}(i))} \right)^{2}- \hat{I}_{k}^{2}.
\end{eqnarray}

Finalmente os resultados de cada interação $k$, obtidos na fase de cálculos, são combinados e ponderados pelo número de chamadas $N_{k}$ tal que
\begin{equation}
\hat{I}=\dfrac{\sum_{k=1}^{m}\dfrac{N_{k}E_{k}}{S^{2}_{k}}}{\sum_{k=1}^{m}\dfrac{N_{k}}{S^{2}_{k}}}.
\end{equation}

Nesse ponto lembramos da necessidade de ``suavizar'' o integrando para aumentar a eficiência do Monte Carlo. No nosso caso para obter uma função mais comportada, já que nossa análise envolve ressonâncias no canal $s$ devido aos decaimentos dos bósons de \textit{gauge}, absorvemos o denominador do propagador do bóson de \textit{gauge} de massa $M$ e largura $\Gamma$ fazendo a mudança de variáveis \eqref{breit}.

\chapter{Regras de Feynmann para Vértices Quárticos}
\label{ap:regrasfeynman}

\section{Representação linear}
\begin{figure}[!ht]
\centering 
\label{VVVV_lin} 
\includegraphics[scale=1]{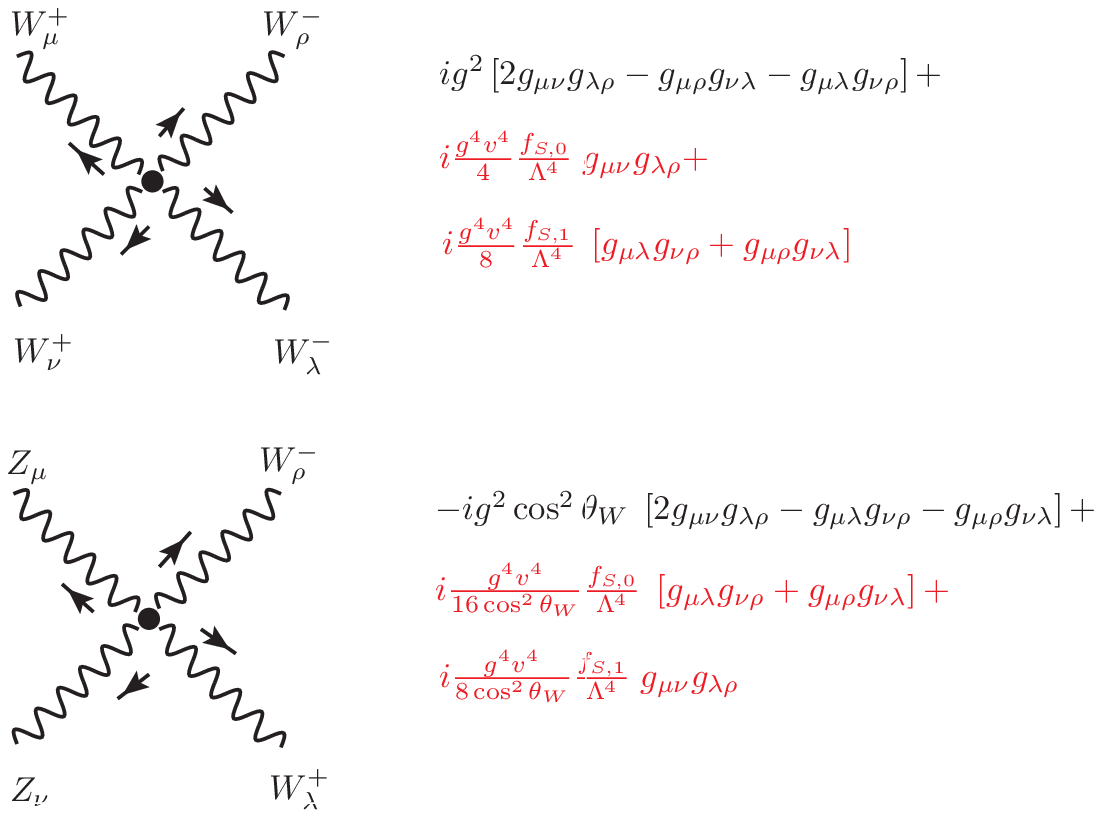} 
\caption{Vértices Quárticos: Padrão em preto e Anômalos em vermelho. Vale salientar que a construção das lagrangianas ${\cal L}_{S,0}$ e ${\cal L}_{S,1}$ não contempla contribuição de vértices tríplices nem de vértices quárticos envolvendo fóton.} 
\end{figure}
\newpage
\section{Representação não linear}
\begin{figure}[!ht]
\centering 
\label{VVVV} 
\includegraphics[scale=0.95]{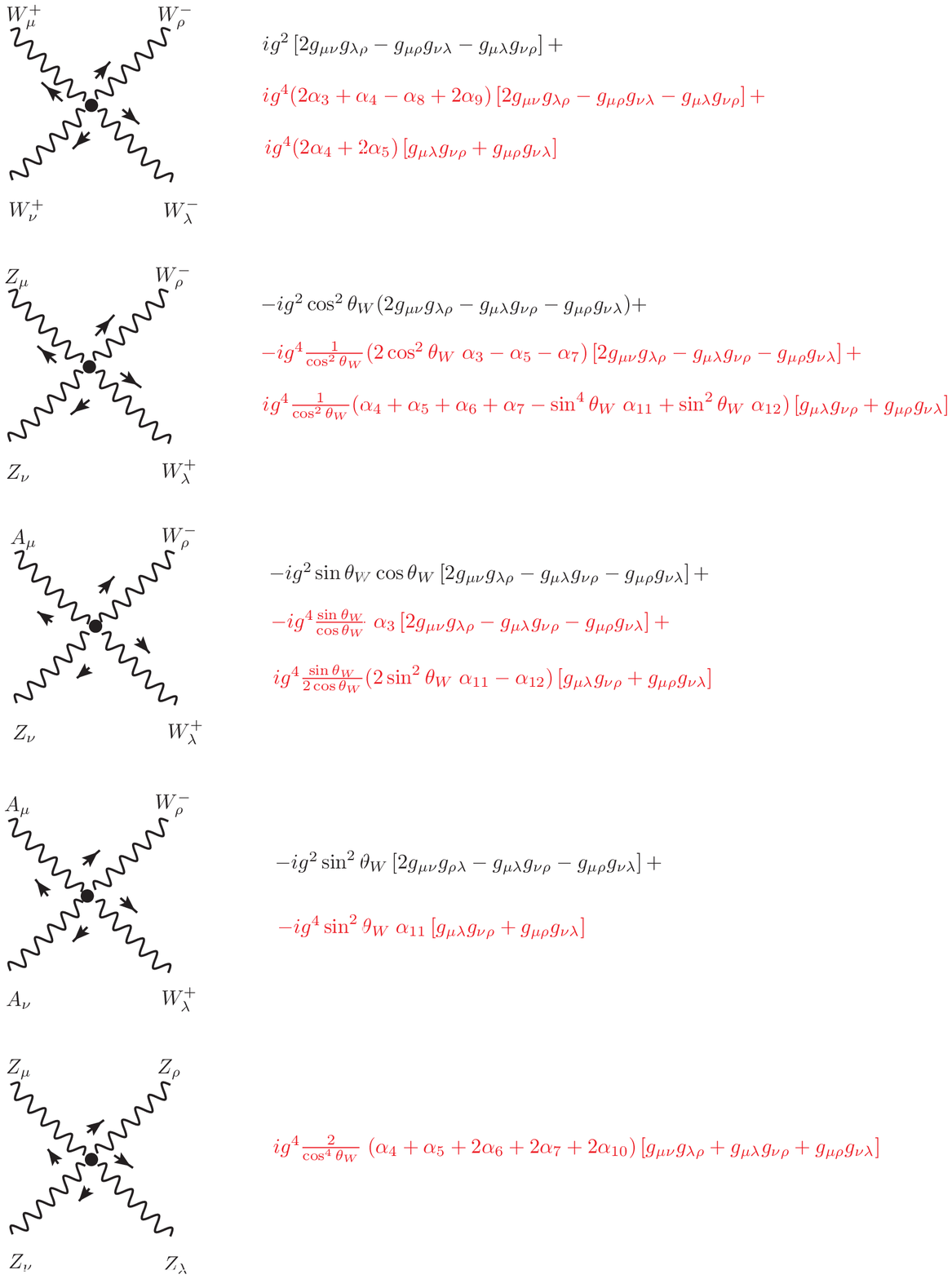} 
\caption{Vértices Quárticos: Padrão em preto e Anômalos em vermelho} 
\end{figure}

\chapter{Subprocessos para $pp\to l^{\pm}\nu_{l}+\;\mbox{4 jatos}$}
\label{subs}

\begin{table}[h!tb]
\begin{center}
\caption{Subprocessos de $pp\to l^{+}\nu_{l}jjjj$ em ${\cal O}(\alpha_{EW}^{6})$}
\label{wp_qed6}
\begin{tabular}{||l||l||l||l||l||}
\hline
$dd\to l^{+}\nu_{l}\bar{c}dds$ & $d\bar{d}\to l^{+}\nu_{l}\bar{u}ds\bar{s}$ & $\bar{d}u\to l^{+}\nu_{l}c\bar{c}d\bar{d}$ & $\bar{c}u\to l^{+}\nu_{l}\bar{u}dd\bar{s}$ & $su\to l^{+}\nu_{l}u\bar{c}ss$ \\
$dd\to l^{+}\nu_{l}\bar{u}ddd$ &$d\bar{u}\to l^{+}\nu_{l}\bar{u}\bar{c}ds$&$cd\to l^{+}\nu_{l}c\bar{c}ds$&$ds\to l^{+}\nu_{l}\bar{c}dss$&$\bar{s}u\to l^{+}\nu_{l}dd\bar{d}\bar{s}$ \\
$du\to l^{+}\nu_{l}c\bar{c}dd$&$d\bar{u}\to l^{+}\nu_{l}\bar{u}\bar{u}dd$&$cd\to l^{+}\nu_{l}c\bar{c}ds$&$ds\to l^{+}\nu_{l}\bar{u}dds$&$\bar{s}u\to l^{+}\nu_{l}c\bar{c}d\bar{s}$\\
$du\to l^{+}\nu_{l}ddd\bar{d}$&$\bar{d}u\to l^{+}\nu_{l}dd\bar{d}\bar{d}$&$cd\to l^{+}\nu_{l}dd\bar{d}s$&$d\bar{s}\to l^{+}\nu_{l}c\bar{c}\bar{c}d$&$\bar{s}u\to l^{+}\nu_{l}uc\bar{c}\bar{c}$\\
$du\to l^{+}\nu_{l}dds\bar{s}$&$\bar{d}u\to l^{+}\nu_{l}d\bar{d}s\bar{s}$&$cd\to l^{+}\nu_{l}dss\bar{s}$&$d\bar{s}\to l^{+}\nu_{l}c\bar{c}\bar{c}d$&$\bar{s}u\to l^{+}\nu_{l}u\bar{c}d\bar{d}$\\
$du\to l^{+}\nu_{l}u\bar{c}ds$&$\bar{d}u\to l^{+}\nu_{l}u\bar{c}\bar{d}s$&$cd\to l^{+}\nu_{l}u\bar{c}ss$&$d\bar{s}\to l^{+}\nu_{l}\bar{c}dd\bar{d}$&$\bar{s}u\to l^{+}\nu_{l}u\bar{c}s\bar{s}$\\
$du\to l^{+}\nu_{l}u\bar{u}dd$&$\bar{d}u\to l^{+}\nu_{l}uu\bar{u}\bar{u}$&$cd\to l^{+}\nu_{l}u\bar{u}ds$&$d\bar{s}\to l^{+}\nu_{l}\bar{c}ds\bar{s}$&$\bar{s}u\to l^{+}\nu_{l}uu\bar{u}\bar{c}$\\
$uu\to l^{+}\nu_{l}cdd\bar{s}$&$\bar{d}u\to l^{+}\nu_{l}u\bar{u}d\bar{d}$&$cd\to l^{+}\nu_{l}\bar{u}cdd$&$d\bar{s}\to l^{+}\nu_{l}u\bar{c}\bar{c}s$&$\bar{s}u\to l^{+}\nu_{l}u\bar{u}d\bar{s}$ \\
$uu\to l^{+}\nu_{l}uc\bar{c}d$&$\bar{d}u\to l^{+}\nu_{l}u\bar{u}s\bar{s}$&$cu\to l^{+}\nu_{l}cds\bar{s}$&$d\bar{s}\to l^{+}\nu_{l}u\bar{u}\bar{c}d$&$uc\to l^{+}\nu_{l}cc\bar{c}d$			\\
$uu\to l^{+}\nu_{l}udd\bar{d}$&$\bar{d}u\to l^{+}\nu_{l}\bar{u}cd\bar{s}$&$cu\to l^{+}\nu_{l}uc\bar{c}s$&$d\bar{s}\to l^{+}\nu_{l}\bar{u}dd\bar{s}$&$uc\to l^{+}\nu_{l}cdd\bar{d}$\\
$uu\to l^{+}\nu_{l}uds\bar{s}$&$u\bar{u}\to l^{+}\nu_{l}\bar{c}d\bar{d}s$&$cu\to l^{+}\nu_{l}ud\bar{d}s$&$su\to l^{+}\nu_{l}c\bar{c}ds$&$uc\to l^{+}\nu_{l}uss\bar{s}$\\
$uu\to l^{+}\nu_{l}uu\bar{c}s$&$u\bar{u}\to l^{+}\nu_{l}u\bar{u}\bar{c}s$&$cu\to l^{+}\nu_{l}u\bar{u}cd$&$su\to l^{+}\nu_{l}dd\bar{d}s$&$uc\to l^{+}\nu_{l}uu\bar{u}s$ \\
$uu\to l^{+}\nu_{l}uu\bar{u}d$&$u\bar{u}\to l^{+}\nu_{l}u\bar{u}\bar{u}d$&$\bar{c}d\to l^{+}\nu_{l}\bar{c}\bar{c}ds$&$su\to l^{+}\nu_{l}dss\bar{s}$&$u\bar{c}\to l^{+}\nu_{l}c\bar{c}\bar{c}d$\\
$d\bar{d}\to l^{+}\nu_{l}\bar{c}d\bar{d}s$&$u\bar{u}\to l^{+}\nu_{l}\bar{u}c\bar{c}d$&$\bar{c}d\to l^{+}\nu_{l}\bar{u}\bar{c}dd$&$su\to l^{+}\nu_{l}u\bar{u}ds$& $\bar{c}u\to l^{+}\nu_{l}u\bar{u}\bar{c}d$ \\
$d\bar{d}\to l^{+}\nu_{l}u\bar{u}\bar{u}d$ & $u\bar{u}\to l^{+}\nu_{l}\bar{u}dd\bar{d}$ & $\bar{c}u\to l^{+}\nu_{l}\bar{c}ds\bar{s}$ & $su\to l^{+}\nu_{l}\bar{u}cdd$ & $u\bar{c}\to l^{+}\nu_{l}\bar{c}dd\bar{d}$\\
$d\bar{d}\to l^{+}\nu_{l}\bar{u}c\bar{c}d$ & $u\bar{u}\to l^{+}\nu_{l}\bar{u}ds\bar{s}$ & $\bar{c}u\to l^{+}\nu_{l}u\bar{c}\bar{c}s$ & $\bar{s}u\to l^{+}\nu_{l}ds\bar{s}\bar{s}$ & $d\bar{d}\to l^{+}\nu_{l}\bar{u}dd\bar{d}$ \\
\hline
\end{tabular}
\end{center}
\end{table}

\begin{table}[h!tb]
\begin{center}
\caption{Subprocessos da reação $pp\to l^{+}\nu_{l}jjj$ em ${\cal O}(\alpha_{EW}^{4}, \alpha_{S}^{2})$}
\begin{tabular}{||l||l||l||l||l||}
\hline
$dd\to l^{+}\nu_{l}\bar{c}dds$&$\bar{d}u\to l^{+}\nu_{l}u\bar{u}d\bar{d}$&$\bar{c}u\to l^{+}\nu_{l}u\bar{c}\bar{c}s$&$\bar{s}u\to l^{+}\nu_{l}u\bar{u}d\bar{s}$&$\bar{d}g\to l^{+}\nu_{l}u\bar{u}\bar{u}g$ \\
$dd\to l^{+}\nu_{l}\bar{u}ddd$&$\bar{d}u\to l^{+}\nu_{l}u\bar{u}s\bar{s}$&$\bar{c}u\to l^{+}\nu_{l}u\bar{u}\bar{c}d$&$uc\to l^{+}\nu_{l}cc\bar{c}d$&$\bar{d}g\to l^{+}\nu_{l}\bar{u}c\bar{c}g$ \\
$du\to l^{+}\nu_{l}c\bar{c}dd$&$\bar{d}u\to l^{+}\nu_{l}\bar{u}cd\bar{s}$&$\bar{c}u\to l^{+}\nu_{l}\bar{u}dd\bar{s}$&$uc\to l^{+}\nu_{l}cdd\bar{d}$&$\bar{d}g\to l^{+}\nu_{l}\bar{u}d\bar{d}g$\\
$du\to l^{+}\nu_{l}ddd\bar{d}$&$u\bar{u}\to l^{+}\nu_{l}\bar{c}d\bar{d}s$&$ds\to l^{+}\nu_{l}\bar{c}dss$&$uc\to l^{+}\nu_{l}uss\bar{s}$&$\bar{d}g\to l^{+}\nu_{l}\bar{u}s\bar{s}g$\\
$du\to l^{+}\nu_{l}dds\bar{s}$&$u\bar{u}\to l^{+}\nu_{l}u\bar{u}\bar{c}s$&$ds\to l^{+}\nu_{l}\bar{u}dds$&$uc\to l^{+}\nu_{l}uu\bar{u}s$&$gg\to l^{+}\nu_{l}\bar{c}d\bar{d}s$\\
$du\to l^{+}\nu_{l}u\bar{c}ds$&$u\bar{u}\to l^{+}\nu_{l}u\bar{u}\bar{u}d$&$d\bar{s}\to l^{+}\nu_{l}c\bar{c}\bar{c}d$&$u\bar{c}\to l^{+}\nu_{l}c\bar{c}\bar{c}d$&$du\to l^{+}\nu_{l}u\bar{u}dd$ \\
$u\bar{u}\to l^{+}\nu_{l}\bar{u}c\bar{c}d$&$d\bar{s}\to l^{+}\nu_{l}\bar{c}dd\bar{d}$&$u\bar{c}\to l^{+}\nu_{l}\bar{c}dd\bar{d}$&$uu\to l^{+}\nu_{l}cdd\bar{s}$&$u\bar{u}\to l^{+}\nu_{l}\bar{u}dd\bar{d}$\\
$d\bar{s}\to l^{+}\nu_{l}\bar{c}ds\bar{s}$&$ud\to l^{+}\nu_{l}ddgg$&$gg\to l^{+}\nu_{l}u\bar{u}\bar{c}s$&$uu\to l^{+}\nu_{l}uc\bar{c}d$&$u\bar{u}\to l^{+}\nu_{l}\bar{u}ds\bar{s}$ \\
$d\bar{s}\to l^{+}\nu_{l}u\bar{c}\bar{c}s$&$uu\to l^{+}\nu_{l}udgg$&$gg\to l^{+}\nu_{l}u\bar{u}\bar{u}d$&$uu\to l^{+}\nu_{l}udd\bar{d}$&$\bar{d}u\to l^{+}\nu_{l}u\bar{u}c\bar{c}$\\
$d\bar{s}\to l^{+}\nu_{l}u\bar{u}\bar{c}d$&$d\bar{d}\to l^{+}\nu_{l}\bar{u}dgg$&$gg\to l^{+}\nu_{l}\bar{u}c\bar{c}d$&$uu\to l^{+}\nu_{l}uds\bar{s}$&$\bar{d}u\to l^{+}\nu_{l}c\bar{c}d\bar{d}$\\
$d\bar{s}\to l^{+}\nu_{l}\bar{u}dd\bar{s}$&$\bar{d}u\to l^{+}\nu_{l}c\bar{c}gg$&$gg\to l^{+}\nu_{l}\bar{u}dd\bar{d}$&$uu\to l^{+}\nu_{l}uu\bar{c}s$&$cd\to l^{+}\nu_{l}c\bar{c}ds$\\
$su\to l^{+}\nu_{l}c\bar{c}ds$&$\bar{d}u\to l^{+}\nu_{l}d\bar{d}gg$&$gs\to l^{+}\nu_{l}\bar{u}dsg$&$uu\to l^{+}\nu_{l}uu\bar{u}d$&$cd\to l^{+}\nu_{l}dd\bar{d}s$\\
$su\to l^{+}\nu_{l}dd\bar{d}s$&$\bar{d}u\to l^{+}\nu_{l}s\bar{s}gg$&$g\bar{s}\to l^{+}\nu_{l}u\bar{u}\bar{c}g$&$d\bar{d}\to l^{+}\nu_{l}\bar{c}d\bar{d}s$&$cd\to l^{+}\nu_{l}dss\bar{s}$\\
$\bar{d}u\to l^{+}\nu_{l}u\bar{u}gg$&$g\bar{s}\to l^{+}\nu_{l}\bar{u}d\bar{s}g$&$d\bar{d}\to l^{+}\nu_{l}u\bar{u}\bar{u}d$&$cd\to l^{+}\nu_{l}u\bar{c}ss$&$su\to l^{+}\nu_{l}u\bar{c}ss$\\
$u\bar{u}\to l^{+}\nu_{l}\bar{c}sgg$&$gu\to l^{+}\nu_{l}c\bar{c}dg$&$d\bar{d}\to l^{+}\nu_{l}\bar{u}c\bar{c}d$&$cd\to l^{+}\nu_{l}u\bar{u}ds$&$su\to l^{+}\nu_{l}u\bar{u}ds$\\
$u\bar{u}\to l^{+}\nu_{l}\bar{u}dgg$&$gu\to l^{+}\nu_{l}dd\bar{d}g$&$d\bar{d}\to l^{+}\nu_{l}\bar{u}dd\bar{d}$&$cd\to l^{+}\nu_{l}\bar{u}cdd$&$su\to l^{+}\nu_{l}\bar{u}cdd$\\
$cd\to l^{+}\nu_{l}dsgg$&$gu\to l^{+}\nu_{l}ds\bar{s}g$&$d\bar{d}\to l^{+}\nu_{l}\bar{u}ds\bar{s}$&$cu\to l^{+}\nu_{l}cds\bar{s}$&$\bar{s}u\to l^{+}\nu_{l}c\bar{c}d\bar{s}$\\
$cg\to l^{+}\nu_{l}d\bar{d}sg$&$gu\to l^{+}\nu_{l}u\bar{c}sg$&$d\bar{u}\to l^{+}\nu_{l}\bar{u}\bar{c}ds$&$cu\to l^{+}\nu_{l}uc\bar{c}s$&$\bar{s}u\to l^{+}\nu_{l}dd\bar{d}\bar{s}$\\
$cg\to l^{+}\nu_{l}\bar{u}cdg$&$gu\to l^{+}\nu_{l}u\bar{u}dg$&$d\bar{u}\to l^{+}\nu_{l}\bar{u}\bar{u}dd$&$cu\to l^{+}\nu_{l}ud\bar{d}s$&$\bar{s}u\to l^{+}\nu_{l}ds\bar{s}\bar{s}$\\
$\bar{c}g\to l^{+}\nu_{l}\bar{u}\bar{c}dg$&$g\bar{u}\to l^{+}\nu_{l}\bar{u}\bar{c}sg$&$\bar{d}u\to l^{+}\nu_{l}dd\bar{d}\bar{d}$&$cu\to l^{+}\nu_{l}u\bar{u}cd$&$\bar{s}u\to l^{+}\nu_{l}uc\bar{c}\bar{c}$\\
$dg\to l^{+}\nu_{l}\bar{c}dsg$&$g\bar{u}\to l^{+}\nu_{l}\bar{u}\bar{u}dg$&$\bar{d}u\to l^{+}\nu_{l}d\bar{d}s\bar{s}$&$\bar{c}d\to l^{+}\nu_{l}\bar{c}\bar{c}ds$&$\bar{s}u\to l^{+}\nu_{l}u\bar{c}d\bar{d}$\\
$dg\to l^{+}\nu_{l}\bar{u}ddg $&$su\to l^{+}\nu_{l}dsgg$&$\bar{d}u\to l^{+}\nu_{l}u\bar{c}\bar{d}s$&$\bar{c}d\to l^{+}\nu_{l}\bar{u}\bar{c}dd$&$\bar{s}u\to l^{+}\nu_{l}u\bar{c}s\bar{s}$\\
$d\bar{s}\to l^{+}\nu_{l}\bar{c}dgg$&$\bar{s}u\to l^{+}\nu_{l}d\bar{s}gg$&$\bar{d}u\to l^{+}\nu_{l}uu\bar{u}\bar{u}$&$\bar{c}u\to l^{+}\nu_{l}\bar{c}ds\bar{s}$&$\bar{s}u\to l^{+}\nu_{l}uu\bar{u}\bar{c}$\\
$\bar{d}g\to l^{+}\nu_{l}\bar{c}\bar{d}sg$&$\bar{s}u\to l^{+}\nu_{l}u\bar{c}gg$&$uc\to l^{+}\nu_{l}cdgg$&$uc\to l^{+}\nu_{l}usgg$ & $su\to l^{+}\nu_{l}dss\bar{s}$ \\
\hline
\end{tabular}
\end{center}
\label{wp_qcd2}
\end{table}

\begin{table}[h!tb]
\begin{center}
\caption{Subprocessos da reação $pp\to l^{+}\nu_{l}jjj$ em ${\cal O}(\alpha_{EW}^{2}, \alpha_{S}^{4})$}
\begin{tabular}{||l||l||l||l||l||}
\hline
$dd\to l^{+}\nu_{\to l}\bar{u}ddd$&$du\to l^{+}\nu_{l}c\bar{c}dd$&$du\to l^{+}\nu_{l}ddd\bar{d}$&$du\to l^{+}\nu_{l}dds\bar{s}$&$du\to l^{+}\nu_{l}u\bar{c}ds$\\
$du\to l^{+}\nu_{l}u\bar{u}dd$&$uu\to l^{+}\nu_{l}uc\bar{c}d$&$uu\to l^{+}\nu_{l}udd\bar{d}$&$uu\to l^{+}\nu_{l}uds\bar{s}$&$uu\to l^{+}\nu_{l}uu\bar{c}s$\\
$uu\to l^{+}\nu_{l}uu\bar{u}d$&$ud\to l^{+}\nu_{l}ddgg$&$uu\to l^{+}\nu_{l}udgg$&$d\bar{d}\to l^{+}\nu_{l}u\bar{u}\bar{u}d$&$d\bar{d}\to l^{+}\nu_{l}\bar{u}c\bar{c}d$\\
$d\bar{d}\to l^{+}\nu_{l}\bar{u}dd\bar{d}$&$d\bar{d}\to l^{+}\nu_{l}\bar{u}ds\bar{s}$&$d\bar{u}\to l^{+}\nu_{l}\bar{u}\bar{c}ds$&$d\bar{u}\to l^{+}\nu_{l}\bar{u}\bar{u}dd$&$\bar{d}u\to l^{+}\nu_{l}dd\bar{d}\bar{d}$\\
$\bar{d}u\to l^{+}\nu_{l}d\bar{d}s\bar{s}$&$\bar{d}u\to l^{+}\nu_{l}u\bar{c}\bar{d}s$&$\bar{d}u\to l^{+}\nu_{l}uu\bar{u}\bar{u}$&$\bar{d}u\to l^{+}\nu_{l}u\bar{u}d\bar{d}$&$\bar{d}u\to l^{+}\nu_{l}u\bar{u}s\bar{s}$\\
$u\bar{u}\to l^{+}\nu_{l}\bar{c}d\bar{d}s$&$u\bar{u}\to l^{+}\nu_{l}u\bar{u}\bar{c}s$&$u\bar{u}\to l^{+}\nu_{l}u\bar{u}\bar{u}d$&$u\bar{u}\to l^{+}\nu_{l}\bar{u}c\bar{c}d$&$u\bar{u}\to l^{+}\nu_{l}\bar{u}dd\bar{d}$\\
$u\bar{u}\to l^{+}\nu_{l}\bar{u}ds\bar{s}$&$\bar{d}u\to l^{+}\nu_{l}u\bar{u}c\bar{c}$&$\bar{d}u\to l^{+}\nu_{l}c\bar{c}d\bar{d}$&$\bar{d}d\to l^{+}\nu_{l}\bar{u}dgg$&$\bar{d}u\to l^{+}\nu_{l}c\bar{c}gg$\\
$\bar{d}u\to l^{+}\nu_{l}d\bar{d}gg$&$\bar{d}u\to l^{+}\nu_{l}gggg$&$\bar{d}u\to l^{+}\nu_{l}u\bar{u}gg$&$u\bar{u}\to l^{+}\nu_{l}\bar{c}sgg$&$u\bar{u}\to l^{+}\nu_{l}\bar{u}dgg$\\
$cd\to l^{+}\nu_{l}u\bar{c}ss$&$cd\to l^{+}\nu_{l}u\bar{u}ds$&$cd\to l^{+}\nu_{l}\bar{u}cdd$&$cu\to l^{+}\nu_{l}cds\bar{s}$&$cu\to l^{+}\nu_{l}uc\bar{c}s$\\
$cu\to l^{+}\nu_{l}ud\bar{d}s$&$cu\to l^{+}\nu_{l}u\bar{u}cd$&$\bar{c}d\to l^{+}\nu_{l}\bar{u}\bar{c}dd$&$\bar{c}u\to l^{+}\nu_{l}\bar{c}ds\bar{s}$&$\bar{c}u\to l^{+}\nu_{l}u\bar{c}\bar{c}s$\\
$\bar{c}u\to l^{+}\nu_{l}u\bar{u}\bar{c}d$&$ds\to l^{+}\nu_{l}\bar{u}dds$&$d\bar{s}\to l^{+}\nu_{l}u\bar{u}\bar{c}d$&$d\bar{s}\to l^{+}\nu_{l}\bar{u}dd\bar{s}$&$su\to l^{+}\nu_{l}c\bar{c}ds$\\
$su\to l^{+}\nu_{l}dd\bar{d}s$&$su\to l^{+}\nu_{l}dss\bar{s}$&$su\to l^{+}\nu_{l}u\bar{c}ss$&$su\to l^{+}\nu_{l}u\bar{u}ds$&$\bar{s}u\to l^{+}\nu_{l}c\bar{c}d\bar{s}$\\
$\bar{s}u\to l^{+}\nu_{l}dd\bar{d}\bar{s}$&$\bar{s}u\to l^{+}\nu_{l}ds\bar{s}\bar{s}$&$\bar{s}u\to l^{+}\nu_{l}uc\bar{c}\bar{c}$&$\bar{s}u\to l^{+}\nu_{l}u\bar{c}d\bar{d}$&$\bar{s}u\to l^{+}\nu_{l}u\bar{c}s\bar{s}$\\
$\bar{s}u\to l^{+}\nu_{l}uu\bar{u}\bar{c}$&$\bar{s}u\to l^{+}\nu_{l}u\bar{u}d\bar{s}$&$uc\to l^{+}\nu_{l}cc\bar{c}d$&$uc\to l^{+}\nu_{l}cdd\bar{d}$&$uc\to l^{+}\nu_{l}uss\bar{s}$\\
$uc\to l^{+}\nu_{l}uu\bar{u}s$&$u\bar{c}\to l^{+}\nu_{l}c\bar{c}\bar{c}d$&$u\bar{c}\to l^{+}\nu_{l}\bar{c}dd\bar{d}$&$cg\to l^{+}\nu_{l}\bar{u}cdg$&$\bar{c}g\to l^{+}\nu_{l}\bar{u}\bar{c}dg$\\
$dg\to l^{+}\nu_{l}\bar{u}ddg$&$\bar{d}g\to l^{+}\nu_{l}u\bar{u}\bar{u}g$&$\bar{d}g\to l^{+}\nu_{l}\bar{u}c\bar{c}g$&$\bar{d}g\to l^{+}\nu_{l}\bar{u}d\bar{d}g$&$g\bar{d}\to l^{+}\nu_{l}\bar{u}ggg$\\
$gg\to l^{+}\nu_{l}u\bar{u}\bar{c}s$&$gg\to l^{+}\nu_{l}u\bar{u}\bar{u}d$&$gg\to l^{+}\nu_{l}\bar{u}c\bar{c}d$&$gg\to l^{+}\nu_{l}\bar{u}dd\bar{d}$&$gg\to l^{+}\nu_{l}\bar{u}dgg$\\
$gg\to l^{+}\nu_{l}\bar{u}ds\bar{s}$&$gs\to l^{+}\nu_{l}\bar{u}dsg$&$g\bar{s}\to l^{+}\nu_{l}u\bar{u}\bar{c}g$&$g\bar{s}\to l^{+}\nu_{l}\bar{u}d\bar{s}g$&$gu\to l^{+}\nu_{l}c\bar{c}dg$\\
$gu\to l^{+}\nu_{l}dd\bar{d}g$&$gu\to l^{+}\nu_{l}dggg$&$gu\to l^{+}\nu_{l}ds\bar{s}g$&$gu\to l^{+}\nu_{l}u\bar{c}sg$&$gu\to l^{+}\nu_{l}u\bar{u}dg$\\
$g\bar{u}\to l^{+}\nu_{l}\bar{u}\bar{c}sg$&$g\bar{u}\to l^{+}\nu_{l}\bar{u}\bar{u}dg$&$su\to l^{+}\nu_{l}dsgg$&$\bar{s}u\to l^{+}\nu_{l}d\bar{s}gg$&$\bar{s}u\to l^{+}\nu_{l}u\bar{c}gg$\\
$uc\to l^{+}\nu_{l}cdgg$&$uc\to l^{+}\nu_{l}usgg$&$u\bar{c}\to l^{+}\nu_{l}\bar{c}dgg$& &\\
\hline
\end{tabular}
\end{center}
\label{wp_qcd4}
\end{table}

\begin{table}[h!tb]
\begin{center}
\caption{Subprocessos de $pp\to l^{-}\bar{\nu}_{l}jjjj$ em ${\cal O}(\alpha_{EW}^{6})$}
\label{wm_qed6}
\begin{tabular}{||l||l||l||l||l||}
\hline
$d\bar{c}\to l^{-}\bar{\nu}_{l}\bar{u}du\bar{s}$&$ds\to l^{-}\bar{\nu}_{l}d\bar{d}cd$&$su\to l^{-}\bar{\nu}_{l}u\bar{u}uc$&$d\bar{d}\to l^{-}\bar{\nu}_{l}u\bar{d}u\bar{u}$&$dd\to l^{-}\bar{\nu}_{l}\bar{u}duu$\\
$ds\to l^{-}\bar{\nu}_{l}d\bar{d}us$&$u\bar{c}\to l^{-}\bar{\nu}_{l}d\bar{d}u\bar{s}$&$ds\to l^{-}\bar{\nu}_{l}u\bar{d}ds$&$ds\to l^{-}\bar{\nu}_{l}\bar{u}duc$&$ds\to l^{-}\bar{\nu}_{l}u\bar{u}cd$\\
$u\bar{c}\to l^{-}\bar{\nu}_{l}u\bar{u}u\bar{s}$&$d\bar{s}\to l^{-}\bar{\nu}_{l}u\bar{d}d\bar{s}$&$d\bar{u}\to l^{-}\bar{\nu}_{l}\bar{u}dc\bar{s}$&$ds\to l^{-}\bar{\nu}_{l}u\bar{u}us$&$u\bar{u}\to l^{-}\bar{\nu}_{l}d\bar{d}c\bar{s}$\\
$d\bar{s}\to l^{-}\bar{\nu}_{l}u\bar{d}u\bar{c}$&$d\bar{u}\to l^{-}\bar{\nu}_{l}\bar{u}du\bar{d}$&$d\bar{s}\to l^{-}\bar{\nu}_{l}d\bar{d}u\bar{s}$&$u\bar{u}\to l^{-}\bar{\nu}_{l}d\bar{d}u\bar{d}$&$du\to l^{-}\bar{\nu}_{l}u\bar{d}ud$\\
$cd\to l^{-}\bar{\nu}_{l}d\bar{d}uc$&$d\bar{s}\to l^{-}\bar{\nu}_{l}u\bar{u}u\bar{s}$&$u\bar{u}\to l^{-}\bar{\nu}_{l}u\bar{u}c\bar{s}$&$d\bar{u}\to l^{-}\bar{\nu}_{l}u\bar{d}\bar{c}s$&$cd\to l^{-}\bar{\nu}_{l}u\bar{u}uc$\\
$du\to l^{-}\bar{\nu}_{l}d\bar{d}uu$&$u\bar{u}\to l^{-}\bar{\nu}_{l}u\bar{u}u\bar{d}$&$d\bar{u}\to l^{-}\bar{\nu}_{l}u\bar{d}\bar{u}d$&$\bar{c}d\to l^{-}\bar{\nu}_{l}d\bar{d}d\bar{s}$&$du\to l^{-}\bar{\nu}_{l}u\bar{u}uu$\\
$cu\to l^{-}\bar{\nu}_{l}u\bar{d}uc$&$\bar{d}u\to l^{-}\bar{\nu}_{l}u\bar{d}c\bar{s}$&$\bar{c}d\to l^{-}\bar{\nu}_{l}d\bar{d}u\bar{c}$&$d\bar{u}\to l^{-}\bar{\nu}_{l}d\bar{d}c\bar{c}$&$\bar{c}u\to l^{-}\bar{\nu}_{l}u\bar{d}d\bar{s}$\\
$\bar{d}u\to l^{-}\bar{\nu}_{l}u\bar{d}u\bar{d}$&$\bar{c}d\to l^{-}\bar{\nu}_{l}u\bar{u}d\bar{s}$&$d\bar{u}\to l^{-}\bar{\nu}_{l}d\bar{d}d\bar{d}$&$\bar{c}u\to l^{-}\bar{\nu}_{l}u\bar{d}u\bar{c}$&$us\to l^{-}\bar{\nu}_{l}u\bar{d}us$\\
$\bar{c}d\to l^{-}\bar{\nu}_{l}u\bar{u}u\bar{c}$&$d\bar{u}\to l^{-}\bar{\nu}_{l}d\bar{d}s\bar{s}$&$dc\to l^{-}\bar{\nu}_{l}u\bar{d}cd$&$u\bar{s}\to l^{-}\bar{\nu}_{l}u\bar{d}u\bar{s}$&$dd\to l^{-}\bar{\nu}_{l}d\bar{d}ud$\\
$d\bar{u}\to l^{-}\bar{\nu}_{l}d\bar{d}u\bar{u}$&$dc\to l^{-}\bar{\nu}_{l}u\bar{d}us$&$uu\to l^{-}\bar{\nu}_{l}u\bar{d}uu$&$dd\to l^{-}\bar{\nu}_{l}u\bar{u}ud$&$d\bar{u}\to l^{-}\bar{\nu}_{l}u\bar{u}c\bar{c}$\\
$d\bar{c}\to l^{-}\bar{\nu}_{l}u\bar{d}\bar{c}d$&$u\bar{u}\to l^{-}\bar{\nu}_{l}u\bar{d}c\bar{c}$&$d\bar{d}\to l^{-}\bar{\nu}_{l}d\bar{d}c\bar{s}$&$d\bar{u}\to l^{-}\bar{\nu}_{l}u\bar{u}d\bar{d}$&$dd\to l^{-}\bar{\nu}_{l}u\bar{d}dd$\\
$u\bar{u}\to l^{-}\bar{\nu}_{l}u\bar{d}d\bar{d}$&$d\bar{d}\to l^{-}\bar{\nu}_{l}d\bar{d}u\bar{d}$&$d\bar{u}\to l^{-}\bar{\nu}_{l}u\bar{u}s\bar{s}$&$d\bar{d}\to l^{-}\bar{\nu}_{l}u\bar{d}c\bar{c}$&$u\bar{u}\to l^{-}\bar{\nu}_{l}u\bar{d}s\bar{s}$\\
$d\bar{d}\to l^{-}\bar{\nu}_{l}u\bar{u}c\bar{s}$&$d\bar{u}\to l^{-}\bar{\nu}_{l}u\bar{u}u\bar{u}$&$d\bar{d}\to l^{-}\bar{\nu}_{l}u\bar{d}d\bar{d}$&$u\bar{u}\to l^{-}\bar{\nu}_{l}u\bar{d}u\bar{u}$&$d\bar{d}\to l^{-}\bar{\nu}_{l}u\bar{u}u\bar{d}$\\
$su\to l^{-}\bar{\nu}_{l}d\bar{d}uc$&$d\bar{d}\to l^{-}\bar{\nu}_{l}u\bar{d}s\bar{s}$& & &\\
\hline
\end{tabular}
\end{center}
\end{table}

\begin{table}[h!tb]
\begin{center}
\caption{Subprocessos de $pp\to l^{-}\bar{\nu}_{l}jjjj$ em ${\cal O}(\alpha_{EW}^{4}\alpha_{S}^{2})$}
\begin{tabular}{||l||l||l||l||l||}
\hline
$d\bar{c}\to l^{-}\bar{\nu}_{l}\bar{u}du\bar{s}$&$dd\to l^{-}\bar{\nu}_{l}\bar{u}duu$&$ds\to l^{-}\bar{\nu}_{l}\bar{u}duc$&$d\bar{u}\to l^{-}\bar{\nu}_{l}\bar{u}dc\bar{s}$&$d\bar{u}\to l^{-}\bar{\nu}_{l}\bar{u}du\bar{d}$\\
$cd\to l^{-}\bar{\nu}_{l}d\bar{d}uc$&$cd\to l^{-}\bar{\nu}_{l}u\bar{u}uc$&$\bar{c}d\to l^{-}\bar{\nu}_{l}d\bar{d}d\bar{s}$&$\bar{c}d\to l^{-}\bar{\nu}_{l}d\bar{d}u\bar{c}$&$\bar{c}d\to l^{-}\bar{\nu}_{l}u\bar{u}d\bar{s}$\\
$dd\to l^{-}\bar{\nu}_{l}d\bar{d}ud$&$dd\to l^{-}\bar{\nu}_{l}u\bar{u}ud$&$d\bar{d}\to l^{-}\bar{\nu}_{l}d\bar{d}c\bar{s}$&$d\bar{d}\to l^{-}\bar{\nu}_{l}d\bar{d}u\bar{d}$&$d\bar{d}\to l^{-}\bar{\nu}_{l}u\bar{u}c\bar{s}$\\
$d\bar{d}\to l^{-}\bar{\nu}_{l}u\bar{u}u\bar{d}$&$ds\to l^{-}\bar{\nu}_{l}d\bar{d}cd$&$ds\to l^{-}\bar{\nu}_{l}d\bar{d}us$&$ds\to l^{-}\bar{\nu}_{l}u\bar{u}cd$&$ds\to l^{-}\bar{\nu}_{l}u\bar{u}us$\\
$d\bar{s}\to l^{-}\bar{\nu}_{l}d\bar{d}u\bar{s}$&$d\bar{s}\to l^{-}\bar{\nu}_{l}u\bar{u}u\bar{s}$&$du\to l^{-}\bar{\nu}_{l}d\bar{d}uu$&$du\to l^{-}\bar{\nu}_{l}u\bar{u}uu$&$d\bar{u}\to l^{-}\bar{\nu}_{l}d\bar{d}c\bar{c}$\\
$d\bar{u}\to l^{-}\bar{\nu}_{l}d\bar{d}d\bar{d}$&$d\bar{u}\to l^{-}\bar{\nu}_{l}d\bar{d}s\bar{s}$&$d\bar{u}\to l^{-}\bar{\nu}_{l}d\bar{d}u\bar{u}$&$d\bar{u}\to l^{-}\bar{\nu}_{l}u\bar{u}c\bar{c}$&$d\bar{u}\to l^{-}\bar{\nu}_{l}u\bar{u}d\bar{d}$\\
$d\bar{u}\to l^{-}\bar{\nu}_{l}u\bar{u}s\bar{s}$&$d\bar{u}\to l^{-}\bar{\nu}_{l}u\bar{u}u\bar{u}$&$su\to l^{-}\bar{\nu}_{l}d\bar{d}uc$&$su\to l^{-}\bar{\nu}_{l}u\bar{u}uc$&$u\bar{c}\to l^{-}\bar{\nu}_{l}d\bar{d}u\bar{s}$\\
$u\bar{c}\to l^{-}\bar{\nu}_{l}u\bar{u}u\bar{s}$&$u\bar{u}\to l^{-}\bar{\nu}_{l}d\bar{d}c\bar{s}$&$u\bar{u}\to l^{-}\bar{\nu}_{l}d\bar{d}u\bar{d}$&$u\bar{u}\to l^{-}\bar{\nu}_{l}u\bar{u}c\bar{s}$&$u\bar{u}\to l^{-}\bar{\nu}_{l}u\bar{u}u\bar{d}$\\
$cu\to l^{-}\bar{\nu}_{l}u\bar{d}uc$&$\bar{c}u\to l^{-}\bar{\nu}_{l}u\bar{d}d\bar{s}$&$\bar{c}u\to l^{-}\bar{\nu}_{l}u\bar{d}u\bar{c}$&$dc\to l^{-}\bar{\nu}_{l}u\bar{d}cd$&$dc\to l^{-}\bar{\nu}_{l}u\bar{d}us$\\
$d\bar{c}\to l^{-}\bar{\nu}_{l}u\bar{d}\bar{c}d$&$dd\to l^{-}\bar{\nu}_{l}u\bar{d}dd$&$d\bar{d}\to l^{-}\bar{\nu}_{l}u\bar{d}c\bar{c}$&$d\bar{d}\to l^{-}\bar{\nu}_{l}u\bar{d}d\bar{d}$&$d\bar{d}\to l^{-}\bar{\nu}_{l}u\bar{d}s\bar{s}$\\
$d\bar{d}\to l^{-}\bar{\nu}_{l}u\bar{d}u\bar{u}$&$ds\to l^{-}\bar{\nu}_{l}u\bar{d}ds$&$d\bar{s}\to l^{-}\bar{\nu}_{l}u\bar{d}d\bar{s}$&$d\bar{s}\to l^{-}\bar{\nu}_{l}u\bar{d}u\bar{c}$&$du\to l^{-}\bar{\nu}_{l}u\bar{d}ud$\\
$d\bar{u}\to l^{-}\bar{\nu}_{l}u\bar{d}\bar{c}s$&$d\bar{u}\to l^{-}\bar{\nu}_{l}u\bar{d}\bar{u}d$&$\bar{d}u\to l^{-}\bar{\nu}_{l}u\bar{d}c\bar{s}$&$\bar{d}u\to l^{-}\bar{\nu}_{l}u\bar{d}u\bar{d}$&$us\to l^{-}\bar{\nu}_{l}u\bar{d}us$\\
$u\bar{s}\to l^{-}\bar{\nu}_{l}u\bar{d}u\bar{s}$&$uu\to l^{-}\bar{\nu}_{l}u\bar{d}uu$&$u\bar{u}\to l^{-}\bar{\nu}_{l}u\bar{d}c\bar{c}$&$u\bar{u}\to l^{-}\bar{\nu}_{l}u\bar{d}d\bar{d}$&$u\bar{u}\to l^{-}\bar{\nu}_{l}u\bar{d}s\bar{s}$\\
$u\bar{u}\to l^{-}\bar{\nu}_{l}u\bar{d}u\bar{u}$&$cd\to l^{-}\bar{\nu}_{l}ucgg$&$\bar{c}d\to l^{-}\bar{\nu}_{l}d\bar{s}gg$&$\bar{c}d\to l^{-}\bar{\nu}_{l}u\bar{c}gg$&$\bar{c}g\to l^{-}\bar{\nu}_{l}d\bar{d}\bar{s}g$\\
$\bar{c}g\to l^{-}\bar{\nu}_{l}u\bar{c}\bar{d}g$&$u\bar{c}\to l^{-}\bar{\nu}_{l}u\bar{s}gg$&$ud\to l^{-}\bar{\nu}_{l}uugg$&$ug\to l^{-}\bar{\nu}_{l}uc\bar{s}g$&$ug\to l^{-}\bar{\nu}_{l}uu\bar{d}g$\\
$us\to l^{-}\bar{\nu}_{l}ucgg$&$u\bar{u}\to l^{-}\bar{\nu}_{l}c\bar{s}gg$&$u\bar{u}\to l^{-}\bar{\nu}_{l}u\bar{d}gg$&$\bar{u}d\to l^{-}\bar{\nu}_{l}c\bar{c}gg$&$\bar{u}d\to l^{-}\bar{\nu}_{l}d\bar{d}gg$\\
$\bar{u}d\to l^{-}\bar{\nu}_{l}s\bar{s}gg$&$\bar{u}d\to l^{-}\bar{\nu}_{l}u\bar{u}gg$&$\bar{u}g\to l^{-}\bar{\nu}_{l}c\bar{c}\bar{d}g$&$\bar{u}g\to l^{-}\bar{\nu}_{l}d\bar{d}\bar{d}g$&$\bar{u}g\to l^{-}\bar{\nu}_{l}\bar{d}s\bar{s}g$\\
$\bar{u}g\to l^{-}\bar{\nu}_{l}u\bar{u}\bar{d}g$&$\bar{u}g\to l^{-}\bar{\nu}_{l}\bar{u}c\bar{s}g$&$\bar{c}d\to l^{-}\bar{\nu}_{l}u\bar{u}u\bar{c}$&&\\
\hline
\end{tabular}
\end{center}
\label{wm_qcd2}
\end{table}

\begin{table}[h!tb]
\begin{center}
\caption{Subprocessos de $pp\to l^{-}\bar{\nu}_{l}jjjj$ em ${\cal O}(\alpha_{EW}^{2}\alpha_{S}^{4})$}
\begin{tabular}{||l||l||l||l||l||}
\hline
$d\bar{c}\to l^{-}\bar{\nu}_{l}\bar{u}du\bar{s}$&$dd\to l^{-}\bar{\nu}_{l}\bar{u}duu$&$ds\to l^{-}\bar{\nu}_{l}\bar{u}duc$&$d\bar{u}\to l^{-}\bar{\nu}_{l}\bar{u}dc\bar{s}$&$d\bar{u}\to l^{-}\bar{\nu}_{l}\bar{u}du\bar{d}$\\
$cd\to l^{-}\bar{\nu}_{l}d\bar{d}uc$&$cd\to l^{-}\bar{\nu}_{l}u\bar{u}uc$&$\bar{c}d\to l^{-}\bar{\nu}_{l}d\bar{d}d\bar{s}$&$\bar{c}d\to l^{-}\bar{\nu}_{l}d\bar{d}u\bar{c}$&$\bar{c}d\to l^{-}\bar{\nu}_{l}u\bar{u}d\bar{s}$\\
$dd\to l^{-}\bar{\nu}_{l}d\bar{d}ud$&$dd\to l^{-}\bar{\nu}_{l}u\bar{u}ud$&$d\bar{d}\to l^{-}\bar{\nu}_{l}d\bar{d}c\bar{s}$&$d\bar{d}\to l^{-}\bar{\nu}_{l}d\bar{d}u\bar{d}$&$d\bar{d}\to l^{-}\bar{\nu}_{l}u\bar{u}c\bar{s}$\\
$d\bar{d}\to l^{-}\bar{\nu}_{l}u\bar{u}u\bar{d}$&$ds\to l^{-}\bar{\nu}_{l}d\bar{d}cd$&$ds\to l^{-}\bar{\nu}_{l}d\bar{d}us$&$ds\to l^{-}\bar{\nu}_{l}u\bar{u}cd$&$ds\to l^{-}\bar{\nu}_{l}u\bar{u}us$\\
$d\bar{s}\to l^{-}\bar{\nu}_{l}d\bar{d}u\bar{s}$&$d\bar{s}\to l^{-}\bar{\nu}_{l}u\bar{u}u\bar{s}$&$du\to l^{-}\bar{\nu}_{l}d\bar{d}uu$&$du\to l^{-}\bar{\nu}_{l}u\bar{u}uu$&$d\bar{u}\to l^{-}\bar{\nu}_{l}d\bar{d}c\bar{c}$\\
$d\bar{u}\to l^{-}\bar{\nu}_{l}d\bar{d}d\bar{d}$&$d\bar{u}\to l^{-}\bar{\nu}_{l}d\bar{d}s\bar{s}$&$d\bar{u}\to l^{-}\bar{\nu}_{l}d\bar{d}u\bar{u}$&$d\bar{u}\to l^{-}\bar{\nu}_{l}u\bar{u}c\bar{c}$&$d\bar{u}\to l^{-}\bar{\nu}_{l}u\bar{u}d\bar{d}$\\
$d\bar{u}\to l^{-}\bar{\nu}_{l}u\bar{u}s\bar{s}$&$d\bar{u}\to l^{-}\bar{\nu}_{l}u\bar{u}u\bar{u}$&$su\to l^{-}\bar{\nu}_{l}d\bar{d}uc$&$su\to l^{-}\bar{\nu}_{l}u\bar{u}uc$&$u\bar{c}\to l^{-}\bar{\nu}_{l}d\bar{d}u\bar{s}$\\
$u\bar{c}\to l^{-}\bar{\nu}_{l}u\bar{u}u\bar{s}$&$u\bar{u}\to l^{-}\bar{\nu}_{l}d\bar{d}c\bar{s}$&$u\bar{u}\to l^{-}\bar{\nu}_{l}d\bar{d}u\bar{d}$&$u\bar{u}\to l^{-}\bar{\nu}_{l}u\bar{u}c\bar{s}$&$u\bar{u}\to l^{-}\bar{\nu}_{l}u\bar{u}u\bar{d}$\\
$cu\to l^{-}\bar{\nu}_{l}u\bar{d}uc$&$\bar{c}u\to l^{-}\bar{\nu}_{l}u\bar{d}d\bar{s}$&$\bar{c}u\to l^{-}\bar{\nu}_{l}u\bar{d}u\bar{c}$&$dc\to l^{-}\bar{\nu}_{l}u\bar{d}cd$&$dc\to l^{-}\bar{\nu}_{l}u\bar{d}us$\\
$d\bar{c}\to l^{-}\bar{\nu}_{l}u\bar{d}\bar{c}d$&$dd\to l^{-}\bar{\nu}_{l}u\bar{d}dd$&$d\bar{d}\to l^{-}\bar{\nu}_{l}u\bar{d}c\bar{c}$&$d\bar{d}\to l^{-}\bar{\nu}_{l}u\bar{d}d\bar{d}$&$d\bar{d}\to l^{-}\bar{\nu}_{l}u\bar{d}s\bar{s}$\\
$d\bar{d}\to l^{-}\bar{\nu}_{l}u\bar{d}u\bar{u}$&$ds\to l^{-}\bar{\nu}_{l}u\bar{d}ds$&$d\bar{s}\to l^{-}\bar{\nu}_{l}u\bar{d}d\bar{s}$&$d\bar{s}\to l^{-}\bar{\nu}_{l}u\bar{d}u\bar{c}$&$du\to l^{-}\bar{\nu}_{l}u\bar{d}ud$\\
$d\bar{u}\to l^{-}\bar{\nu}_{l}u\bar{d}\bar{c}s$&$d\bar{u}\to l^{-}\bar{\nu}_{l}u\bar{d}\bar{u}d$&$\bar{d}u\to l^{-}\bar{\nu}_{l}u\bar{d}c\bar{s}$&$\bar{d}u\to l^{-}\bar{\nu}_{l}u\bar{d}u\bar{d}$&$us\to l^{-}\bar{\nu}_{l}u\bar{d}us$\\
$u\bar{s}\to l^{-}\bar{\nu}_{l}u\bar{d}u\bar{s}$&$uu\to l^{-}\bar{\nu}_{l}u\bar{d}uu$&$u\bar{u}\to l^{-}\bar{\nu}_{l}u\bar{d}c\bar{c}$&$u\bar{u}\to l^{-}\bar{\nu}_{l}u\bar{d}d\bar{d}$&$u\bar{u}\to l^{-}\bar{\nu}_{l}u\bar{d}s\bar{s}$\\
$u\bar{u}\to l^{-}\bar{\nu}_{l}u\bar{d}u\bar{u}$&$cd\to l^{-}\bar{\nu}_{l}ucgg$&$cg\to l^{-}\bar{\nu}_{l}uc\bar{d}g$&$\bar{c}g\to l^{-}\bar{\nu}_{l}u\bar{c}\bar{d}g$&$u\bar{c}\to l^{-}\bar{\nu}_{l}u\bar{s}gg$\\
$ud\to l^{-}\bar{\nu}_{l}uugg$&$ug\to l^{-}\bar{\nu}_{l}uc\bar{s}g$&$ug\to l^{-}\bar{\nu}_{l}uu\bar{d}g$&$us\to l^{-}\bar{\nu}_{l}ucgg$&$u\bar{u}\to l^{-}\bar{\nu}_{l}c\bar{s}gg$\\
$u\bar{u}\to l^{-}\bar{\nu}_{l}u\bar{d}gg$&$\bar{u}d\to l^{-}\bar{\nu}_{l}c\bar{c}gg$&$\bar{u}d\to l^{-}\bar{\nu}_{l}d\bar{d}gg$&$\bar{u}d\to l^{-}\bar{\nu}_{l}gggg$&$\bar{u}d\to l^{-}\bar{\nu}_{l}u\bar{u}gg$\\
$\bar{u}g\to l^{-}\bar{\nu}_{l}c\bar{c}\bar{d}g$&$\bar{u}g\to l^{-}\bar{\nu}_{l}d\bar{d}\bar{d}g$&$\bar{u}g\to l^{-}\bar{\nu}_{l}\bar{d}ggg$&$\bar{u}g\to l^{-}\bar{\nu}_{l}\bar{d}s\bar{s}g$&$\bar{u}g\to l^{-}\bar{\nu}_{l}u\bar{u}\bar{d}g$\\
$\bar{u}g\to l^{-}\bar{\nu}_{l}\bar{u}c\bar{s}g$&&&&\\
\hline
\end{tabular}
\end{center}
\label{wm_qcd4}
\end{table}



\end{document}